\documentclass{article}

\usepackage{arxiv}

\usepackage[utf8]{inputenc}
\usepackage{graphicx}
\usepackage{amsmath}
\usepackage[version=4]{mhchem}
\usepackage{siunitx}
\usepackage{longtable,tabularx}
\usepackage{multirow}
\usepackage{wrapfig}
\usepackage{makeidx}
\usepackage{subfig}
\usepackage{booktabs}
\usepackage{amsfonts}

\title{Multi-fidelity hydrodynamic analysis of an autonomous surface vehicle at surveying speed in deep water subject to variable payload}

\author{
    Riccardo Pellegrini\thanks{Dr. R. Pellegrini and Mr. S. Ficini equally contributed to this paper.} \\
    CNR-INM, National Research Council-Institute of Marine Engineering\\
    Via di Vallerano 139\\
    Rome 00128, Italy \\  
    \And
    Simone Ficini$^*$ \\
    Roma Tre University, Department of Engineering \\
    Via Vito Volterra 62 \\
    Rome 00146, Italy \\
    CNR-INM, National Research Council-Institute of Marine Engineering\\
    Via di Vallerano 139\\
    Rome 00128, Italy \\
    \And
    Angelo Odetti\\
    CNR-INM, National Research Council-Institute of Marine Engineering\\
    Via di Vallerano 139\\
    Rome 00128, Italy \\
    \And
    Andrea Serani\\
    CNR-INM, National Research Council-Institute of Marine Engineering\\
    Via di Vallerano 139\\
    Rome 00128, Italy \\
    \And
    Massimo Caccia\\
    CNR-INM, National Research Council-Institute of Marine Engineering\\
    Via di Vallerano 139\\
    Rome 00128, Italy \\
    \And
    Matteo Diez\\
    CNR-INM, National Research Council-Institute of Marine Engineering\\
    Via di Vallerano 139\\
    Rome 00128, Italy \\
}

\begin{document}

\maketitle

\newpage

\begin{abstract}
Autonomous surface vehicles (ASV) allow the investigation of coastal areas, ports and harbors as well as harsh and dangerous environments such as the arctic regions. Despite receiving increasing attention, the hydrodynamic analysis of ASV performance subject to variable operational parameters is little investigated. In this context, this paper presents a multi-fidelity (MF) hydrodynamic analysis of an ASV, namely the Shallow Water Autonomous Multipurpose Platform (SWAMP), at surveying speed in calm water and subject to variable payload and location of the center of mass, accounting for the variety of equipment that the vehicle can carry. The analysis is conducted in deep water, which is the condition mostly encountered by the ASV during surveys of coastal and harbors areas. Quantities of interest are the resistance, the vehicle attitude, and the wave generated in the region between the catamaran hulls. These are assessed using a Reynolds Averaged Navier Stokes Equation (RANSE) code and a linear potential flow (PF) solver. The objective is to accurately assess the quantities of interest, along with identifying the limitation of PF analysis in the current context. Finally, a multi-fidelity Gaussian Process (MF-GP) model is obtained combining RANSE and PF solutions. The latter also include variable grid refinement and coupling between hydrodynamic loads and rigid body equations of motion. The surrogate model is iteratively refined using an active learning approach. 
Numerical results show that the MF-GP is effective in producing response surfaces of the SWAMP performance with a limited computational cost. It is highlighted how the SWAMP performance is significantly affected not only by the payload, but also by the location of the center of mass. The latter can be therefore properly calibrated to minimize the resistance and allow for longer-range operations.
\end{abstract}



\section{Introduction}

Autonomous surface vehicles (ASV) are unmanned vessels (UV) characterized by a level of autonomy in its employment, where the human operator is partially or completely replaced by the on board or remote systems \cite{schiaretti2017survey_I}. %
ASVs are receiving increasing attention because their development and operation can be performed with lower costs than larger vessels operated by humans. Furthermore, ASVs can be designed to operate in dirty, dull, harsh, and dangerous environments \cite{liu2016unmanned,odetti2020swamp}  while assuring personnel safety and security during operations.
Several ASVs models have been developed in the last years \cite{liu2016unmanned,schiaretti2017survey_II} for a significant number of scenarios. As an example, \cite{odetti2020swamp} developed the Shallow Water Autonomous Multipurpose Platform (SWAMP) to perform surveys in the remote areas of wetlands ({\it e.g.}, rivers and lakes) as well as ports and coastal zones;  \cite{bruzzone2020monitoring} developed an UV for exploration in the arctic regions;
\cite{howard2011unmanned} developed an ASV for port and harbor security and bathymetric and hydrological surveys; \cite{beck2009seawasp} developed an ASV for shallow water bathymetry to analyze the geography, ecosystem, and health of marine habitats. 
Recently, the development, design, and characterization of ASVs focused on shape optimization \cite{joung2012shape}, control and stability \cite{wirtensohn2013modelling,klinger2016control,simetti2022control}, and seakeeping \cite{brizzolara2012design}.

Despite the increasing attention on ASVs, the hydrodynamic performance analysis of the ASVs subject to variable operational parameters (such as the speed or the payload and the onboard layout) received little attention \cite{liu2016unmanned}. The hydrodynamic analysis of the ASVs performance has the potentiality to identify the best vessel's hull forms and operational parameters, highlighting undesired hydrodynamic effects and finally leading to better designs, operations, and more accurate surveys. The hydrodynamic analysis allows to identify the complex interactions between the flow field produced by the ship and the ship's dynamics \cite{sanada2021experimental}. In the case of ASVs, which manufacturing is usually simpler and more economical than larger vessels, the hydrodynamic analysis has the potentiality of highlighting hydrodynamic properties that can be detrimental for performance or surveys (\emph{e.g.}, flow separation or recirculation regions in the proximity of the instruments) indicating directions for design or operational improvements.

Although many ASVs have compact size, the hydrodynamic analysis by experimental studies would require dedicated facilities and dedicated resources that are usually not available to most designers. Computational fluid dynamics (CFD) achieved the technological maturity to provide reliable performance analysis also for extreme operating conditions \cite{serani2021hull} making it an effective tool to support and complement experimental activities. 

The use of high-fidelity computational fluid dynamics (CFD) solvers is sought to achieve accurate performance predictions, especially for complex and nonlinear fluid-dynamics. Nevertheless, the computational cost of high-fidelity physics-based solvers can easily become unaffordable for most designers. To give an example, recent research showed how an accurate Unsteady Reynolds Averaged Navier-Stokes Equations (URANSE) simulation, to evaluate statistically significant performance of ship maneuvering in irregular waves may require up to 1M CPU hours on HPC systems \cite{serani2021urans}. 

To reduce the computational cost of the performance analysis, multi-fidelity (MF) approaches can be used  \cite{beran2020comparison,brevault2020overview,peherstorfer2018survey}. A MF surrogate model exploits the possibility of using several fidelities, using different solvers and/or the same solver with different space/time discretization, as well as different coupling levels between several disciplines.

The objective of the present work is to perform an accurate multi-fidelity hydrodynamic analysis of the SWAMP performance \cite{odetti2020swamp} in calm and deep water, subject to variable payload and location of the center of mass. The motivation of considering calm and deep water condition is threefold: i) currently, the SWAMP is not intended to be operated with a significant wave height to guarantee the safety of the equipment and accuracy of measured data;  ii) although the SWAMP is developed with specific features to allow survey in shallow water, it is often if not mostly used in deep water, especially when servicing coastal and harbors areas. The hydrodynamics analysis focuses on the resistance, the vehicle attitude, and the wave generated in the region between the catamaran hulls. Furthermore, the advantages and limitations of using high- and low-fidelity CFD solvers are identified and discussed. 

The SWAMP is a fully-electric, modular, portable, lightweight, and highly controllable ASV. It is a catamaran, equipped with four azimuth Pump-Jet thrusters that are flush with the hull and specifically designed for this vehicle. The SWAMP is also characterized by small draft soft-foam, unsinkable hull structure with high modularity and a flexible hardware/software architecture. 
The SWAMP performance is assessed by multi-fidelity CFD simulations. Specifically, two solvers with a different physical model are used: a RANSE and an in-house linear potential flow (PF) solver. Furthermore, the PF solver leverages three fidelity levels based on the computational grid size and the level (tightness) of the coupling between the hydrodynamic loads and the rigid body equations of motion. 
A multi-fidelity Gaussian process (MF-GP) is used to combine the multi-fidelity CFD simulations. MF-GP is built as a low-fidelity trained surrogate model corrected with the surrogate of the errors between successive fidelities \cite{ficini2021assessing}. 
The MF-GP training sets are dynamically updated with an active learning based on the maximum uncertainty associated with the MF-GP prediction. The active learning method automatically selects the fidelity to sample, balancing between the maximum prediction uncertainty and the computational cost associated to each fidelity level \cite{liu2016sequential,serani2019adaptive}. 

The multi-fidelity hydrodynamic analysis discusses the pressure distribution on the hull, and the trim and sinkage, the wave field, and the free-surface elevation between the SWAMP hulls and their interaction. Furthermore, the vortical structures are investigated using the Q-criterion \cite{jeong1995identification}.

\section{Shallow water autonomous multipurpose platform}
%
\begin{figure}[!b] 
    \centering 
    \includegraphics[width=0.55\textwidth]{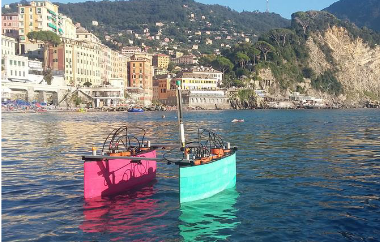}  
    \includegraphics[width=0.43\textwidth]{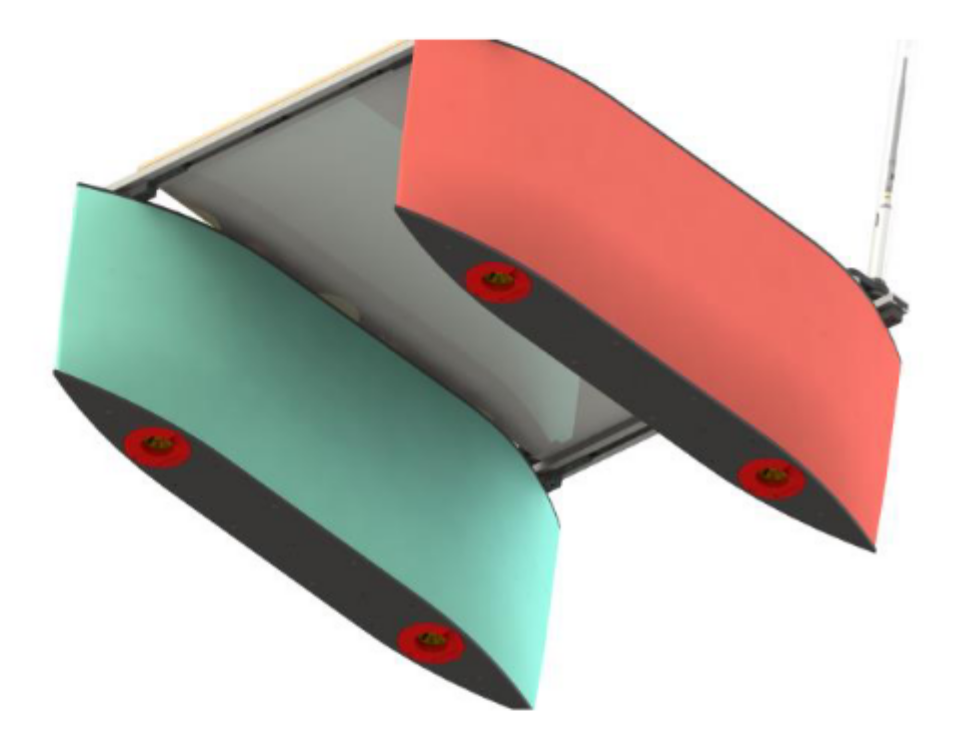}  
    \caption{The SWAMP autonomous surface vehicle \cite{odetti2020swamp}}\label{fig:SWAMP} 
\end{figure} 

The SWAMP is a full-electric modular multi-functional catamaran. The hull shape is inspired by the double-ended Wigley series and presents a flat bottom part, necessary to host the pump-jet propulsion system. The hull form and propulsion system are characterized by equally efficient sailing ahead and astern with the possibility of maneuvering in narrow spaces. The main characteristics of the SWAMP are listed in table \ref{tab:SWAMPch}.
The development of the SWAMP addresses three main concepts. First, finding new solutions for accessing and monitoring remote or difficult to access areas (such as polar regions, rivers, and alpine lakes) and harsh and dangerous environments (such as ports or polluted areas). Second, stimulate marine robotics research, {\it e.g.}, the SWAMP is made with sandwich of polyethylene (PE) foam and high-density PE plates to provide impact resistance while being easily modifiable and repairable; its thrusters are embedded within the hull for protection and to reduce their disturbance during the surveying, while effectively providing propulsion and maneuvering \cite{odetti2020swamp}. Third, developing new concepts in marine robotics, such as swarm intelligence for a fleet of ASVs and cooperation between an ASV and an aerial drone. 
Having these three concepts as basis for its development, the SWAMP can serve for various missions, such as geomorphological and bathymetric analysis, water sampling, and physical and chemical data collection. As a consequence, the payload (the instrumentation required by the survey) can vary depending on the mission and furthermore the location of the center of mass can vary depending on the payload arrangement. 

The payload and the location of the center of mass can affect the hydrodynamic resistance of the SWAMP, its attitude, and the waves that are generated during the operations. Therefore, the quantities of interest chosen to assess the performance of the SWAMP are: i) the resistance ($R_T$), which directly affects the power consumption and therefore the duration of the operations; ii) the sinkage ($T$) and trim ($\theta$), which define the attitude of the SWAMP and may affect the instrumentation settings and calibration; and iii) the root mean square (RMS) of the free-surface elevation between the hulls, which is evaluated to quantify the disturbance produced by the free-surface between the catamaran hulls that may negatively affects the measurements {\it e.g.}, sonar/video measures. 

Ideally, the SWAMP should operate with minimum resistance, with an horizontal hull (zero trim angle), and producing a minimum disturbance of the free-surface.

\begin{table}[!t]
\caption{Main characteristics of the SWAMP}
\centering
{\begin{tabular}{l*{1}{c}} 
\toprule
Parameter & Nominal value \\
\midrule
Hull length ($L_{pp}$) & 1.240 [{m}] \\ 
\midrule 
Semi-hull width & 0.245  [{m}] \\ 
\midrule 
Hull height & 1.100  [{m}] \\ 
\midrule
Nominal drought & 0.140  [{m}] \\ 
\midrule 
Nominal operating speed & 1.000  [{m/s}] \\
\midrule 
Froude number & 0.287  [-] \\ 
\bottomrule 
\end{tabular}}
\label{tab:SWAMPch} 
\end{table}


The SWAMP displacement $\nabla$ (directly connected to the payload) varies depending on the type and purpose of the survey. The nominal range of variation is set as $\nabla \sim\mathrm{unif}[35, 60]$ kg. The center of mass $x_G$ is located in the center of the hull ($x_G=0$ m) but can vary depending on the payload arrangement. The nominal range of variation is set as $x_G\sim\mathrm{unif}[-0.0372, 0.0930]$ m (corresponding to a variation $\delta x_G$ of $[-3,+7.5]\%$L of the nominal value). A negative value of $x_G$ yields a forward translation (in direction of the bow) of the center of mass, whereas a positive value yields a backward translation (in direction of the stern). 

Of interest are the performance and hydrodynamic characteristics of the SWAMP operating in calm and deep water. The calm water is considered representative of the real-world operating conditions since the SWAMP is not currently intended to operate with a significant wave height, to guarantee the safety of the equipment and accuracy of the measurements. The deep sea conditions are of interest because, although the SWAMP is developed with specific features to allow survey in shallow water (such as the flat bottom) it is often if not mostly used in deep water (such as during coastal or harbor surveys). 


%
\begin{table*}[!b] 
\caption{Percentage variations of $R_T$, $\theta$, and $T$ varying $\nabla \sim\mathrm{unif}[35, 60]$ kg for a fixed $\delta x_G$. The percentage is computed with respect to the nominal displacement $\nabla = 58$ kg. The values are reported for the RANSE and the PF solver}
\centering
{\begin{tabular}{{l}*{8}{c}} 
\toprule
              & \multicolumn{4}{c}{RANSE} & \multicolumn{4}{c}{PF} \\
$\delta x_G$ [kg] & $R_T$ & $T$ & $\theta$ & RMS & $R_T$ & $T$ & $\theta$ & RMS \\ 
\midrule 
$-3\%$L &  32.2\% &  37.3\% & 34.4\% &  18.6\% & 45.0\% &  27.6\% & 39.5\% &  30.2 \\
$0\%$L &  36.7\% &  36.0\% & 29.0\% &  19.6\% & 47.8\% &  22.1\% &   5.5\% &  32.2 \\
$3\%$L &  40.4\% &  39.6\% &  39.2\% &  24.2\% & 43.8\% &  25.7\% &  33.9\% &  32.3 \\
$7.5\%$L  & 37.6\% &  40.8\% &  37.2\% &  18.6\% & 30.3\% &  34.1\% &  36.1\% &  27.5 \\
\bottomrule 
\end{tabular}}
\label{tab:RTVariation} 
\end{table*}

\section{RANSE flow solver}

\begin{figure}[!b]
    \centering 
    \subfloat[][Boundary conditions]{\includegraphics[width=0.45\textwidth]{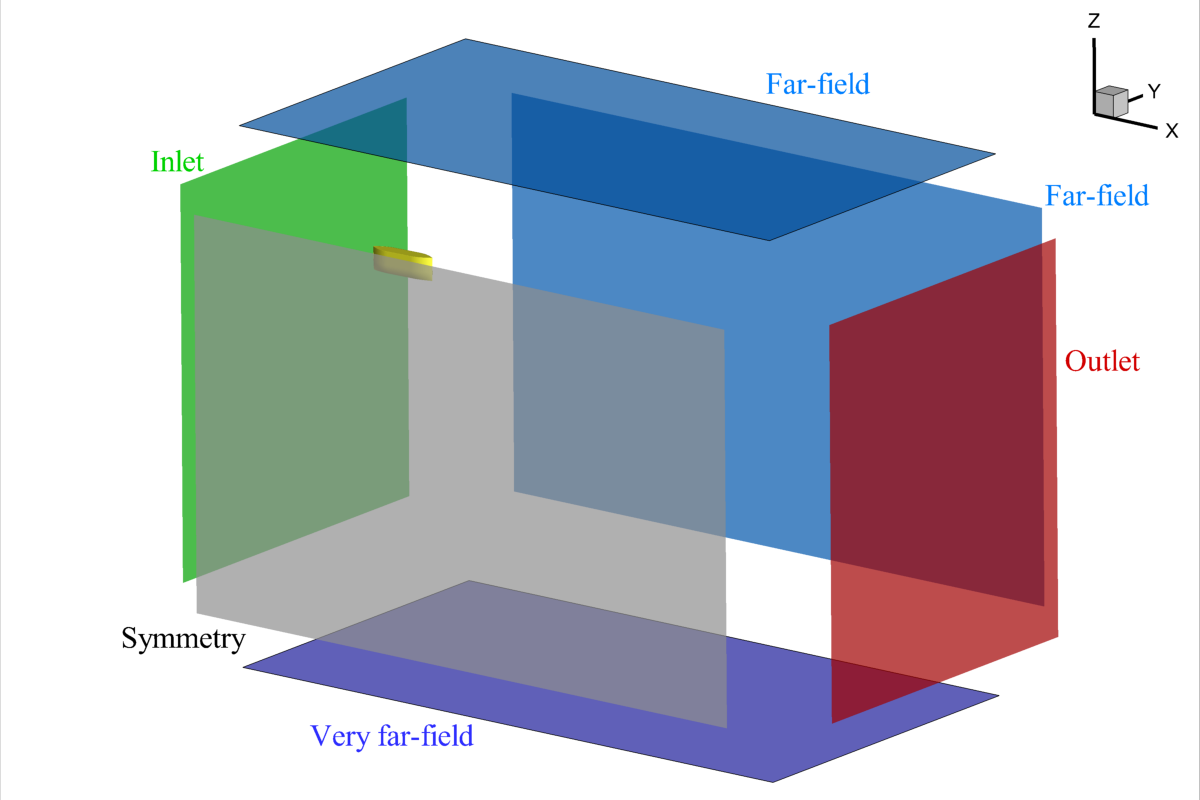} }
    \subfloat[][Body fitted multi-block grid]{\includegraphics[width=0.45\textwidth]{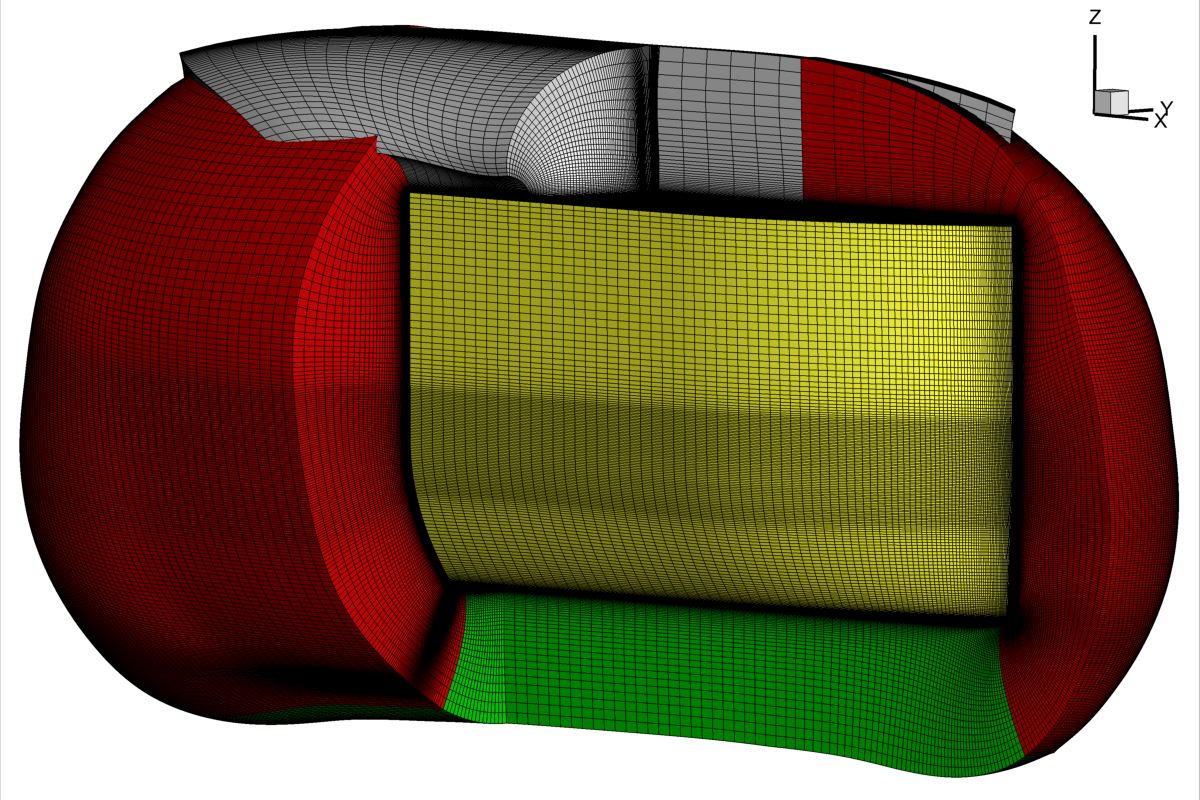} }\\
    \subfloat[][Background and refinement grids, top]{\includegraphics[width=0.45\textwidth]{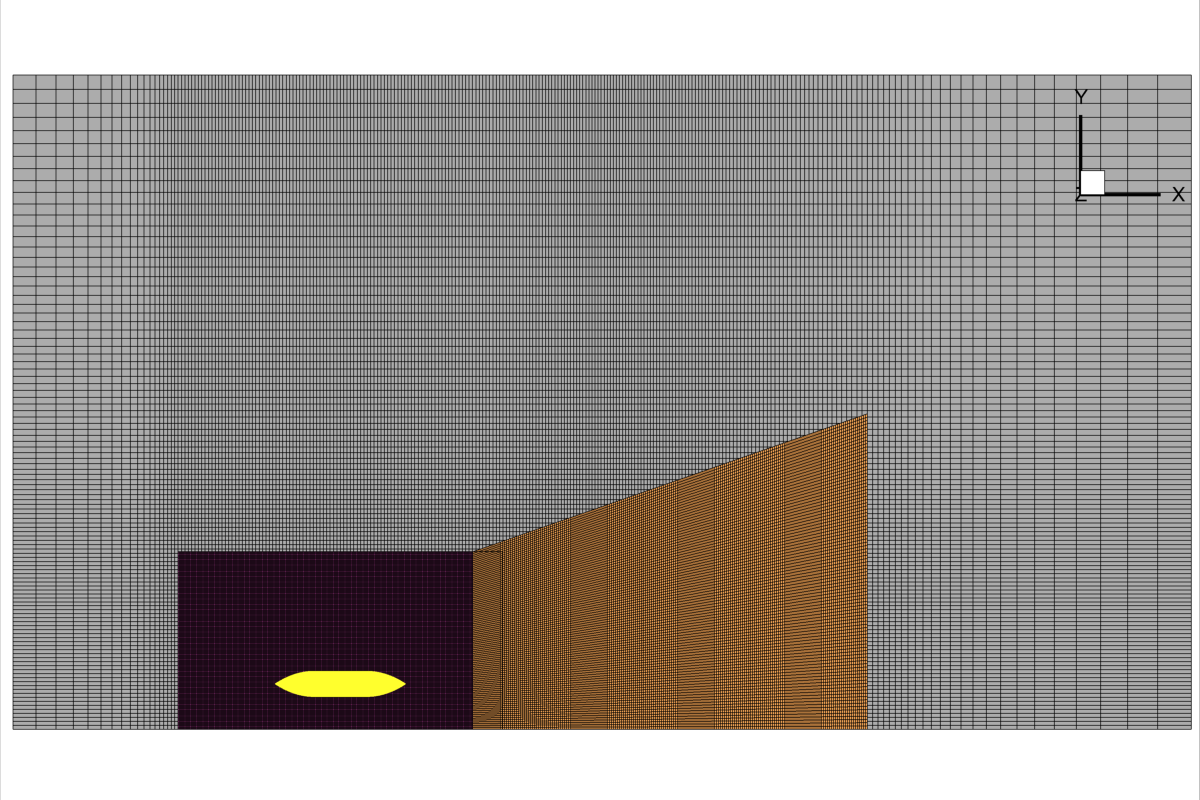} }
    \subfloat[][Background and refinement grids, side]{\includegraphics[width=0.45\textwidth]{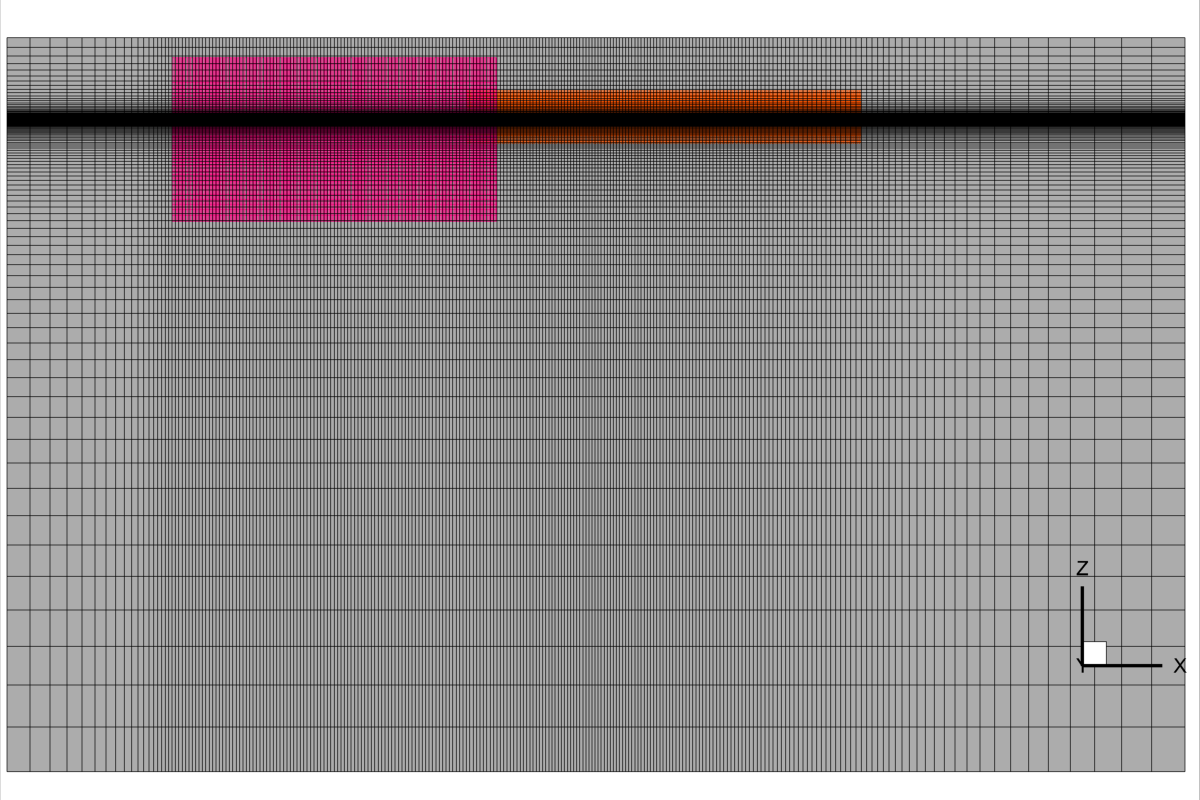} }
    \caption{Boundary conditions and grids for the RANSE solver}
    \label{fig:Grids} 
\end{figure}

\subsection{Method}

The RANSE solutions are provided by CFDShip-Iowa V4.5, an incompressible RANS/DES solver designed for ship hydrodynamics \cite{huang2008semi}. The equations are solved in an inertial coordinate system, either fixed to a ship or other frame moving at constant speed or in the earth system. The free surface is modeled with a single-phase capturing approach, which means that only the water flow is solved, enforcing kinematic and dynamic free surface boundary conditions on the interfaces. A level-set function method is used for the sharp-interface tracking of the interface between air and water (free-surface). A semi-coupled immersed boundary approach can be used for two phase flow computations in which the water flow is decoupled from the air solution, but the air flow uses the unsteady water flow as a boundary condition. The limitation of the approach is that pressurized closed air/water packets (bubbles) cannot be simulated. The code uses structured multi-block grids and has overset capabilities using the Structured, Unstructured, and Generalized overset Grid AssembleR (SUGGAR) code \cite{noack2005suggar}. The functionalities of the code include 6DOF motions, URANS/DES turbulence models, moving control surfaces, multi-objects, advanced controllers, propulsion models, incoming waves and winds, bubbly flow, and fluid-structure interaction. 

\subsection{Setup}
A second-order upwind scheme is used to discretize the convective terms of momentum equations. Pressure implicit with splitting of operators (PISO) loop for pressure/velocity coupling is used. For high-performance parallel computing, an MPI-based domain decomposition approach is used, where each decomposed block is mapped to one processor. The code SUGGAR runs as a separate process from the flow solver to compute interpolation coefficients for the overset grid, which enables CFDShip-Iowa to take care of 6DoF with a motion controller at every time step. 
For the current study, absolute inertial earth-fixed coordinates are employed. The turbulence is computed by the isotropic Menter’s blended $k-\epsilon/k-\omega$ (BKW) model with shear stress transport (SST) using no wall function.  

Simulations are performed in calm and deep water for the right demi-hull, taking advantage of symmetry about the xz-plane. The computational domain is a rectangular cuboid defined by $-2.5 \leq x/L_{pp} \leq 6.5$, $0 \leq y/L_{pp} \leq 5$, $-5.00 \leq z/L_{pp} \leq 0.6$. The ship axis is aligned with the $x$-axis, with the bow at $x/L_{pp} = 0.5$ and the stern at $x/L_{pp} = -0.5$, the free surface at rest lies at $z/L_{pp} = 0$. The boundary conditions are depicted in Fig. \ref{fig:Grids}a. The computational grid is made of 4 grids, namely the body-fitted grid with about 7.3M elements (adjacent and overset blocks show in Fig. \ref{fig:Grids}b); the background grid with about 2.9M elements (in grey, see Fig. \ref{fig:Grids}c-d); the refinement grid with about 5.2M elements (in purple, see Fig. \ref{fig:Grids}c-d); and the wake refinement with about 2.5M elements (in orange, see Fig. \ref{fig:Grids}c-d). A nondimensional time step equal to $0.005$ is used, the inflow velocity approaches the nominal value following a linear ramp with a duration of $1.0$ time units. The Reynolds number in model scale is $Re = 8.7E5$, this may suggest that the flow is not characterized by a completely developed turbulence, which is instead assumed by the turbulence model. This is not an issue for the specific application, since the environment where the SWAMP is supposed to operate are often characterized by turbulent flows (such as coastal areas or ports).

\section{Potential flow solver} 

\subsection{Method}

\begin{figure}[!b] 
    \centering 
    \subfloat[High-fidelity G2]{\includegraphics[width=0.42\textwidth]{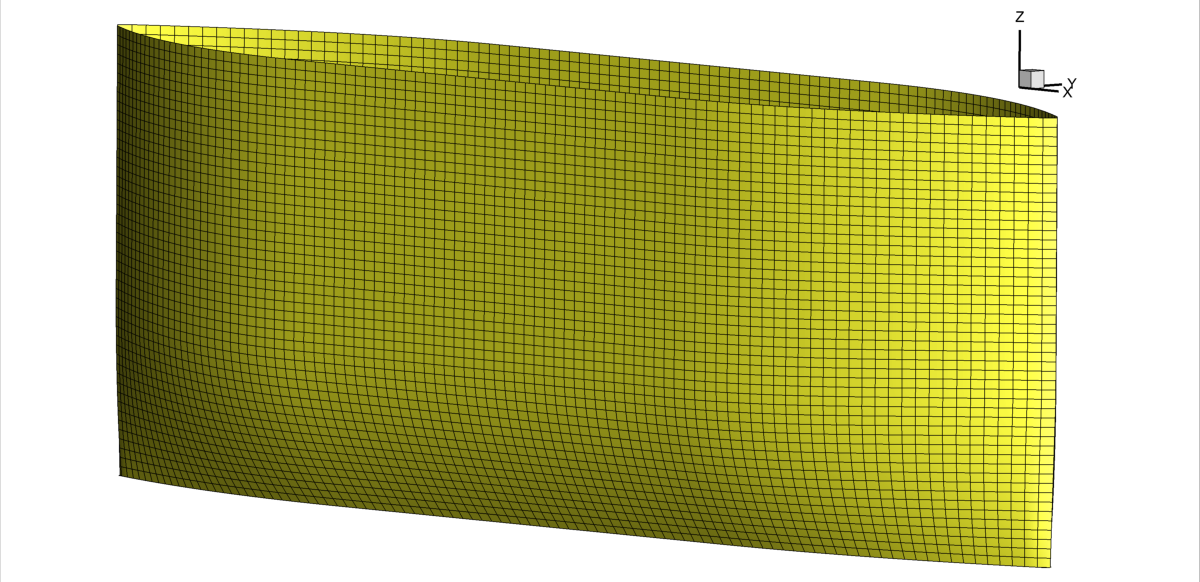} 
        \includegraphics[width=0.45\textwidth]{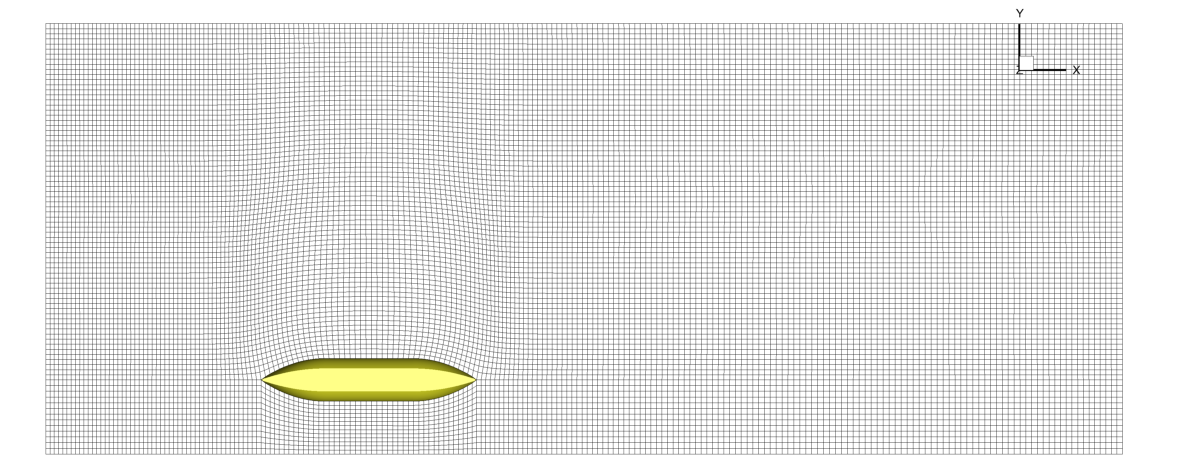}}\\
    \subfloat[Medium-fidelity G3]{\includegraphics[width=0.42\textwidth]{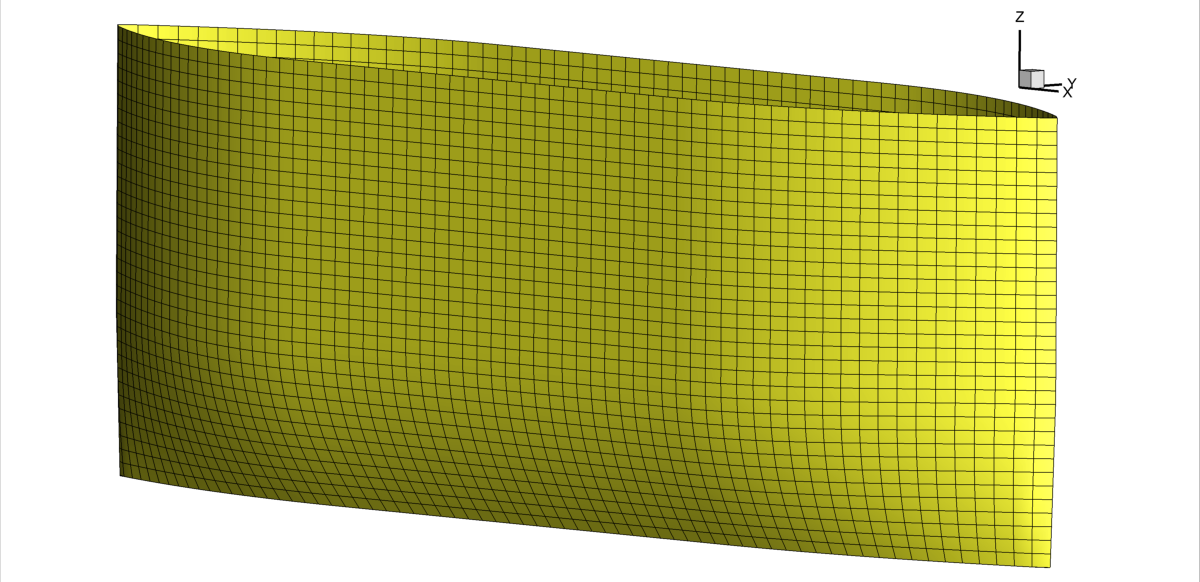}
        \includegraphics[width=0.45\textwidth]{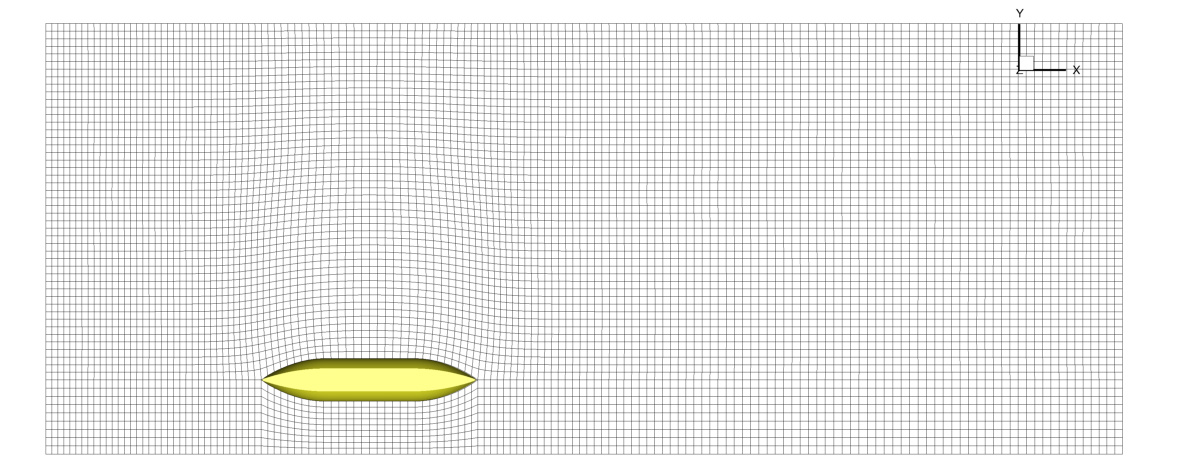}}\\
    \subfloat[Low-fidelity G4]{\includegraphics[width=0.42\textwidth]{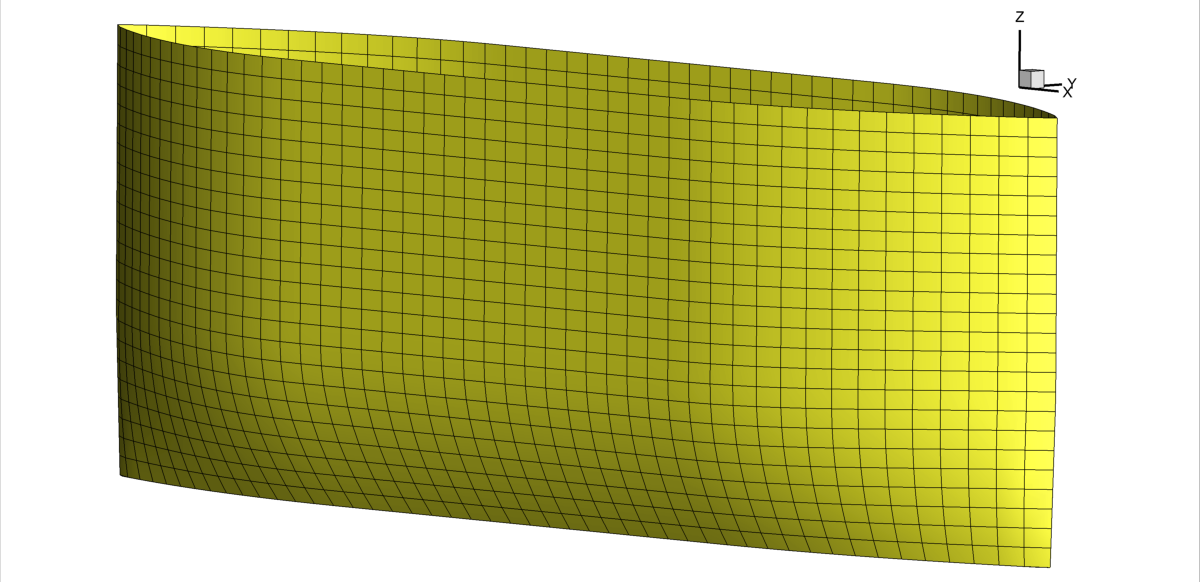} 
        \includegraphics[width=0.45\textwidth]{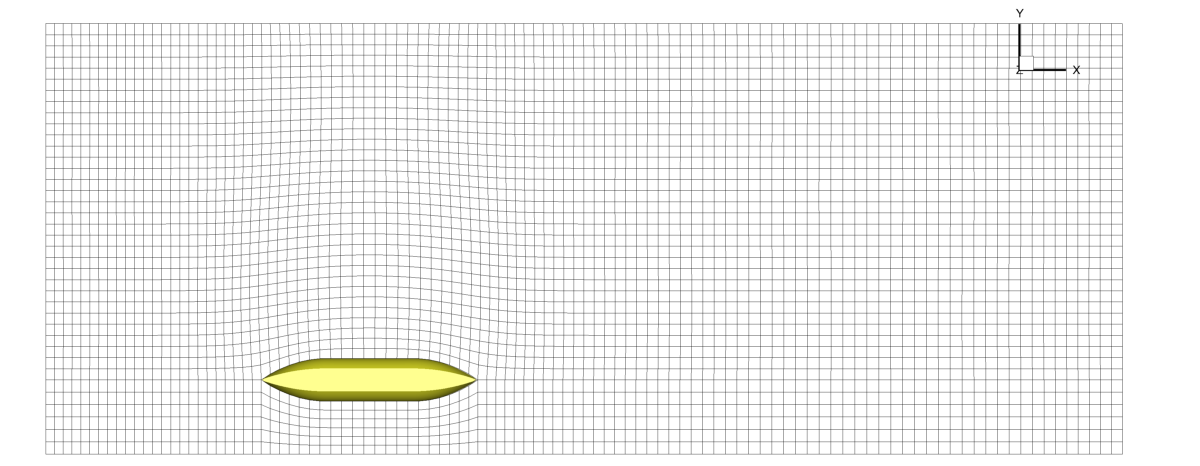}}
    \caption{Computational domain and grid of the free surface (left) and the hull (right) for the PF solver}\label{fig:PFgrids} 
\end{figure} 

The PF solutions are computed by an in-house linear potential flow code (WAve Resistance Program, WARP) \cite{Bas:94}, developed at CNR-INM. 
Wave resistance computations are based on the Dawson (double-model) linearization \cite{dawson1977practical}, whereas the frictional resistance is estimated using a flat-plate approximation, based on the local Reynolds number \cite{schlichting2000fundamentals}. Details of equations, numerical implementations and validation of the numerical solvers are given in \cite{Bas:94}.
The steady 2 DOF (sinkage $T$ and trim $\theta$) equilibrium is achieved considering iteratively the coupling between the hydrodynamic loads and the rigid body equations of motion. The equilibrium is achieved when the variation of the sinkage and the pitch angle are smaller than a user-defined threshold value $\varepsilon^{\star}$. 
%
%
%
%
%

\subsection{Setup}
The simulations are performed in calm water. The accuracy of the PF simulations for the MF analysis is varied by changing both the grid size and the value of $\varepsilon^{\star}$. Three fidelity levels are used (see Fig. \ref{fig:PFgrids}), Table \ref{tab:GridRefinements} summarizes the size of the different grids, the values of $\varepsilon^{\star}$, and the normalized computational cost $\beta_i$ associated with each fidelity. 
%
%
\begin{table*}[!t]
\caption{PF solver: grid size, convergence threshold ($\varepsilon^{\star}$), and normalized computational cost ($\beta_i$) for each fidelity $i$}
\centering
{\begin{tabular}{*{6}{c}} 
\toprule
$i$-th fidelity & Grid label & Hull panels & Free-surface panels & $\varepsilon^{\star}_{i}$ \% & $\beta_i$ \\ 
\midrule 
2 & G2 & 6.6k & 16.5k & 1.00 & 1.000 \\ 
3 & G3 & 3.2k & 8.5k & 5.00 & 0.094  \\ 
4 & G4 & 1.5k & 4.1k & 10.0 & 0.014 \\ 
\bottomrule 
\end{tabular}}
\label{tab:GridRefinements} 
\end{table*} 

The computational domain for the free surface is defined as 1.5$L$ upstream, 3.5$L$ downstream, and 2.0$L$ sideways. 

\begin{figure}[!t] 
\centering 
\mbox{} \hfill
\subfloat[G2]{\includegraphics[width=0.3\textwidth]{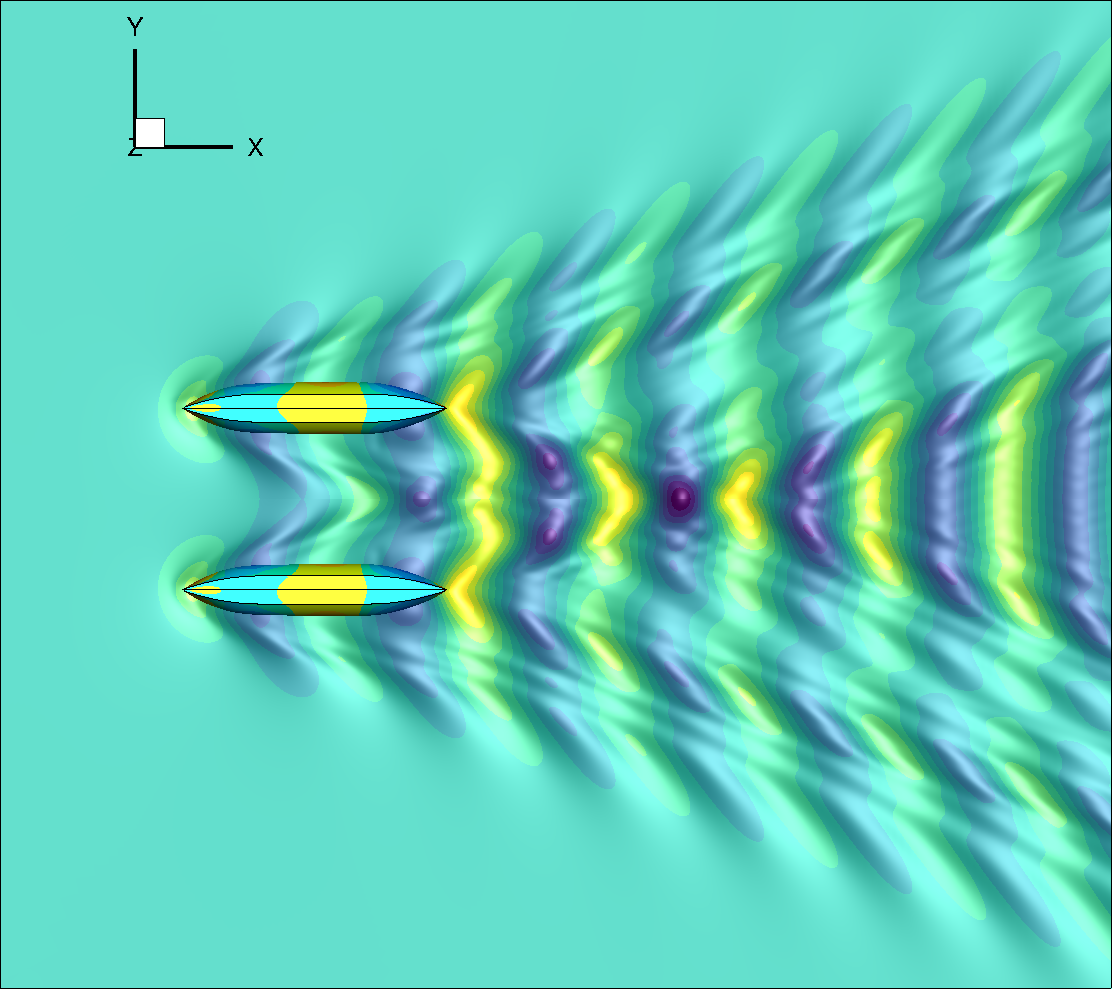}} \hfill
\subfloat[G3]{\includegraphics[width=0.3\textwidth]{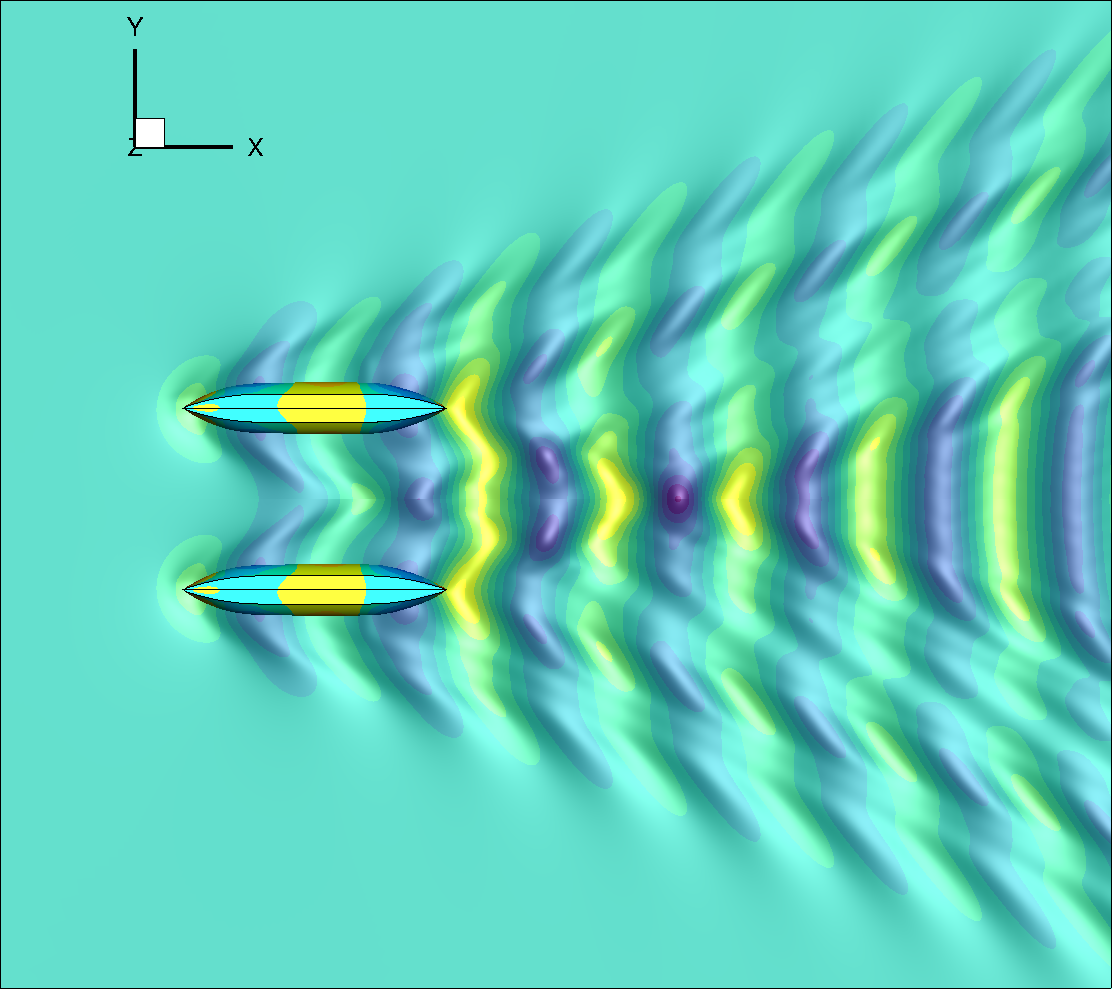}} \hfill
\subfloat[G4]{\includegraphics[width=0.3\textwidth]{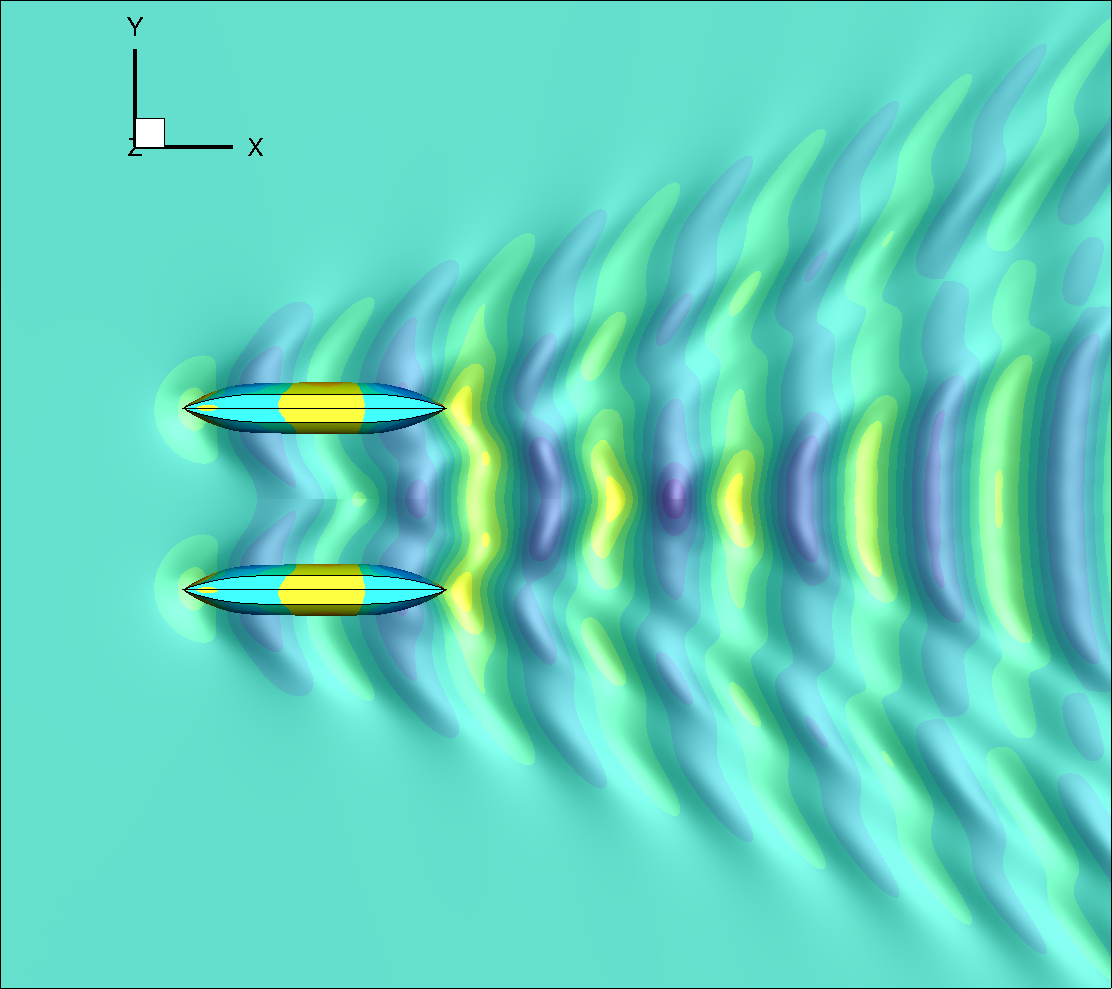}} \hfill \mbox{} 
\caption{PF solution, wave elevations and pressure fields for the different fidelities}\label{fig:WaveElevation}
\end{figure}

%
Figure \ref{fig:WaveElevation} shows the wave elevation and the pressure on the hull for the three grids, considering $\nabla= 58$ kg and $\delta x_G = 0\%$L. The wave pattern is similar for all the grids, but the local value of the wave elevation shows some differences. Differences in the pressure on the hulls can be noted in the front section.

\section{Adaptive multi-fidelity Gaussian process} 

Given a training set $\mathcal{T}=\{{\bf x^{\prime}}_i,f({\bf x^{\prime}}_i)\}_{i=1}^{J}$, where $\bf x^{\prime} \in \mathbb{R}^{\it{D}}$ is the parameters vector of dimension $D$ and $J$ the training set size, normalizing the parameters domain into a unit hypercube, the GP prediction $\tilde{f}({\bf x})$ with constant mean and its variance ${\rm Var}[\tilde{f}({\bf x})]$ can be written as \cite{williams2006gaussian} 
\begin{equation}\label{eq:GPpred} 
\begin{aligned}
\widetilde{f}(\mathbf{x}) &=  \mathbb{E}[\mathbf{f}(\mathbf{x^{\prime}})] + \mathbf{k}(\mathbf{x},\mathbf{x^{\prime}}) \mathbf{K}(\mathbf{x^{\prime}},\mathbf{x^{\prime}})^{-1}(\mathbf{f}(\mathbf{x^{\prime}})-\mathbb{E}[\mathbf{f}(\mathbf{x^{\prime}})]), \\ 
{\rm Var}[\widetilde{f}(\mathbf{x})] &=  \mathbf{K}(\mathbf{x},\mathbf{x}) - \mathbf{k}(\mathbf{x},\mathbf{x^{\prime}})^{\mathsf{T}} \mathbf{K}(\mathbf{x^{\prime}},\mathbf{x^{\prime}})^{-1}\mathbf{k}(\mathbf{x},\mathbf{x^{\prime}}),  
\end{aligned}
\end{equation}  
where $\mathbb{E}[\mathbf{f}(\mathbf{x^{\prime}})]$ is the expected value of $\{{f}(\mathbf{x^{\prime}}_i)\}_{i=1}^{J}$, $\mathbf{K}(\mathbf{x}^{\prime},\mathbf{x}^{\prime})$ is the covariance matrix with elements $K_{ij} = k(\mathbf{x}^{\prime}_i,\mathbf{x}^{\prime}_j)$, and $\mathbf{k}(\mathbf{x},\mathbf{x^{\prime}})$ is the covariance vector with elements $k_{i}=k(\mathbf{x},\mathbf{x}^{\prime}_i)$. Finally, $k({\cdot},{\cdot})$ is the covariance function defined as \cite{williams2006gaussian} 
\begin{equation} 
k(\mathbf{x},\mathbf{x^{\prime}}) = \sigma_F^{2} \exp\left(-\boldsymbol{\gamma}^{\mathsf{T}} (\mathbf{x}-\mathbf{x^{\prime}})^{\circ 2}\right) + \sigma^{2}_n \delta(\mathbf{x},\mathbf{x^{\prime}}), 
\end{equation} 
with "$\circ$" the Hadamard product, $\delta(\mathbf{x},\mathbf{x^{\prime}})$ the Kronecker delta, and $\boldsymbol{\Lambda} = \{\sigma_{n}^{2},\sigma_{F}^{2},\boldsymbol{\gamma}\}$ the set of the GP hyper-parameters \cite{williams2006gaussian}. Specifically, $\sigma_n^2$ is the variance associated to the noise in the training set, $\sigma_F^2$ is the signal variance, and $\boldsymbol{\gamma} \in \mathbb{R}^{D}$ is the vector of the inverse length scale parameters. It can be noted that the use of the parameter $\sigma_n^2$ leads to a regressive formulation of the GP, this choice is motivated by the fact that it is not possible to exclude a priori that the numerical simulations are not affected by numerical noise.
The hyper-parameters are evaluated by maximizing the log marginal likelihood $l$ \cite{williams2006gaussian} as follows
\begin{equation}\label{eq:MLE} 
\boldsymbol{\Lambda}^\star = \{\sigma_{n}^{2,\star},\sigma_{F}^{2,\star},\boldsymbol{\gamma}^{\star}\} = {\underset{\sigma_n^2,\sigma_F^2,\boldsymbol{\gamma}}{\rm argmax}}[l], 
\end{equation} 
with
\begin{equation} 
\begin{aligned}
l =& \log\left[ p(f(\mathbf{x}^{\prime})|\mathbf{x}^{\prime})\right] = \\
=& -\frac{J}{2} \log{2\pi} - \frac{1}{2} f(\mathbf{x}^{\prime})^{\mathsf{T}}{\mathbf{K}}(\mathbf{x}^{\prime},\mathbf{x}^{\prime})^{-1} f(\mathbf{x}^{\prime}) - \frac{1}{2} \log|{\bf K}(\mathbf{x}^{\prime},\mathbf{x}^{\prime})|. 
\end{aligned}
\end{equation} 
The uncertainty $U_{\widetilde{f}}$, associated to the surrogate model prediction, is here quantified as 
\begin{equation}\label{eq:GPU} 
U_{\widetilde{f}} = 4\sqrt{{\rm Var}[\widetilde{f}({\bf x})]}.
\end{equation} 
It may be noted that the present definition of $U_{\widetilde{f}}$ incorporates the variance associated with the noise in the training set. 

\subsection{Multi-fidelity integration} 

Considering a quantity of interest that can be evaluated with $N$ fidelity levels (where the first level is the highest-fidelity and the $N$-th level is the lowest-fidelity) the MF extension of the GP is built as follows \cite{wackers2020adaptive}. 
Given a training set $\mathcal{T}_i=\{{\bf x^{\prime}}_{j},f_i({\bf x^{\prime}}_{j})\}_{j=1}^{J_i}$ for $i=1,\dots,N$, the MF approximation $\hat{f}_{i}(\mathbf{x})$ of $f_{i}(\mathbf{x})$ is defined as  
\begin{equation}\label{eq:MLGeneral} 
\widehat{f}_{i}(\mathbf{x}) := \widetilde{f}_N(\mathbf{x})+\sum_{k=i}^{{N}-1}\widetilde{\varepsilon}_k(\mathbf{x}),  
\end{equation} 
where $\widetilde{{\varepsilon}}_k(\mathbf{x})$ is the inter-level error surrogate with an associate training set $\mathcal{E}_k=\{({\bf x^{\prime}}_i,f_{k}({\bf x^{\prime}}_i)-\widehat{f}_{k+1}({\bf x^{\prime}}_i))\,|\,{\bf x^{\prime}}_i \in \mathcal{T}_{k} \cap \mathcal{T}_{k+1} \}_{i=1}^{J_{k}}$.
Assuming the uncertainty associated to the prediction of the lowest-fidelity $U_{\widetilde{f}_N}$ and inter-level errors $U_{\widetilde{\varepsilon}_k}$ as uncorrelated, the MF approximation $\widehat{f}_1(\mathbf{x})$ of $f_1(\mathbf{x})$ and the associated uncertainty $U_{\widehat{f}_1}$ read
\begin{equation}\label{eq:ML-MF} 
\begin{aligned}
f_1(\mathbf{x}) \approx& \widehat{f}_1(\mathbf{x})=\widetilde{f}_N(\mathbf{x})+\sum_{k=1}^{N-1}\widetilde{\varepsilon}_k(\mathbf{x}), \\
U_{\widehat{f}_1}(\mathbf{x})=&
\sqrt{U^2_{\widetilde{f}_N}(\mathbf{x})+\sum_{k=1}^{N-1}U^2_{{\widetilde{\varepsilon}}_k}(\mathbf{x})}. 
\end{aligned}
\end{equation} 
%

\subsection{Active learning approach}\label{sec:AL}

The MF surrogate is iteratively updated adding new training points following an active learning procedure. First, the coordinates of the new training point ${\bf x}^\star$ are identified solving a single-objective minimization problem
\begin{equation}\label{eq:PSI} 
{\bf x}^{\star}={\underset{{\bf x}}{\rm argmin}}[\psi({\bf x})]. 
\end{equation}
The acquisition function $\psi({\bf x})$ is based on the uncertainty prediction \cite{serani2019adaptive}.  
%
%
%
\begin{equation}\label{eq:Uopt} 
\begin{aligned}
&\psi({\bf x})=\\
&= -\sqrt{ \max\left[U^2_{\widetilde{f}_N}(\mathbf{x})-P_{x_N},0 \right]+\sum_{k=1}^{N-1} \max\left[U^2_{{\widetilde{\varepsilon}}_k}(\mathbf{x})-P_{x_i},0\right]}. 
\end{aligned}
\end{equation} 
Where $P_{x_i}$ is a penalization value used to avoid overfitting of the training points, which would result in an ill-conditioned matrix while solving Eqs. \ref{eq:GPpred} and \ref{eq:MLE}. The penalization $P_{x_i}$ is applied only if $\mathbf{x}^\star$ lies within a minimum distance $d_{\min}$ of an already existing training point of $\mathcal{T}_i$. In such a case $P_{x_i}$ is evaluated as
\begin{equation}\label{eq:Penalization} 
P_{x_i}= \sum_{j=1}^{J_i} 
\frac{1}{\lVert \mathbf{x}^\star -\mathbf{x}_j^{\prime} \rVert + \tau}, 
\end{equation} 
where $\tau=0.01$ is a scalar used only to avoid the null value of the denominator in Eq. \ref{eq:Penalization}. The optimization problems in Eqs. \ref{eq:MLE} and \ref{eq:Uopt} are solved using the deterministic particle swarm optimization algorithm presented in \cite{serani2016-ASC}.
%

%
%
%
%
Once ${\bf x}^\star$ is identified, the training set $\mathcal{T}_{i}$ is updated with the new training point $ (\mathbf{x}^\star, f_{i}(\mathbf{x}^\star))$ with $i=k,\dots,N$, where $k$ is defined as  
\begin{equation}\label{eq:VarianceBased2} 
\begin{aligned}
&k=\mathrm{maxloc}\left[\boldsymbol{\phi}(\mathbf{x}^\star)\right], \\
%
\qquad
&\mathrm{with} 
\qquad
\boldsymbol{\phi} \equiv 
\begin{Bmatrix} 
({\rm Var}[{{\tilde{\varepsilon}}_1}]- \sigma_{n, {\tilde{\varepsilon}_1}}^{2, \star})/\beta_1 - P_{x_1}\\ 
 \vdots\\ 
({\rm Var}[{{\tilde{\varepsilon}}_{N-1}}]- \sigma_{n, {\tilde{\varepsilon}_{N-1}}}^{2, \star})/\beta_{N-1} -P_{x_{N-1}}\\ 
({\rm Var}[{{\tilde{f}}_N}]- \sigma_{n, {f_N}}^{2, \star} )/\beta_N -P_{x_N}
\end{Bmatrix} 
\end{aligned}
\end{equation} 
The subtraction of $\sigma_n^{2,\star}$ from the variance of the prediction is performed to, ideally, filter-out the noise from the training set while selecting the fidelity level to sample.
Once $\mathbf{x}^{\star}$ is added to the training set, the range of variation of the GP hyper-parameters  $\boldsymbol{\Lambda}_i$ for the $i$-th fidelity, are bounded as: 
 $\boldsymbol{\Lambda}_{i,j} = \pm \alpha \boldsymbol{\Lambda}^{\star}_{i,j-1}$,
where $j$ is the active learning iteration and $\alpha=0.1$ is a parameter used to avoid abrupt variations of the surrogate model prediction.

\section{Comparison between RANSE and PF analyses}

\begin{figure}[!h] 
\centering 
\subfloat[Total resistance $R_T$]{\includegraphics[width=0.33\textwidth]{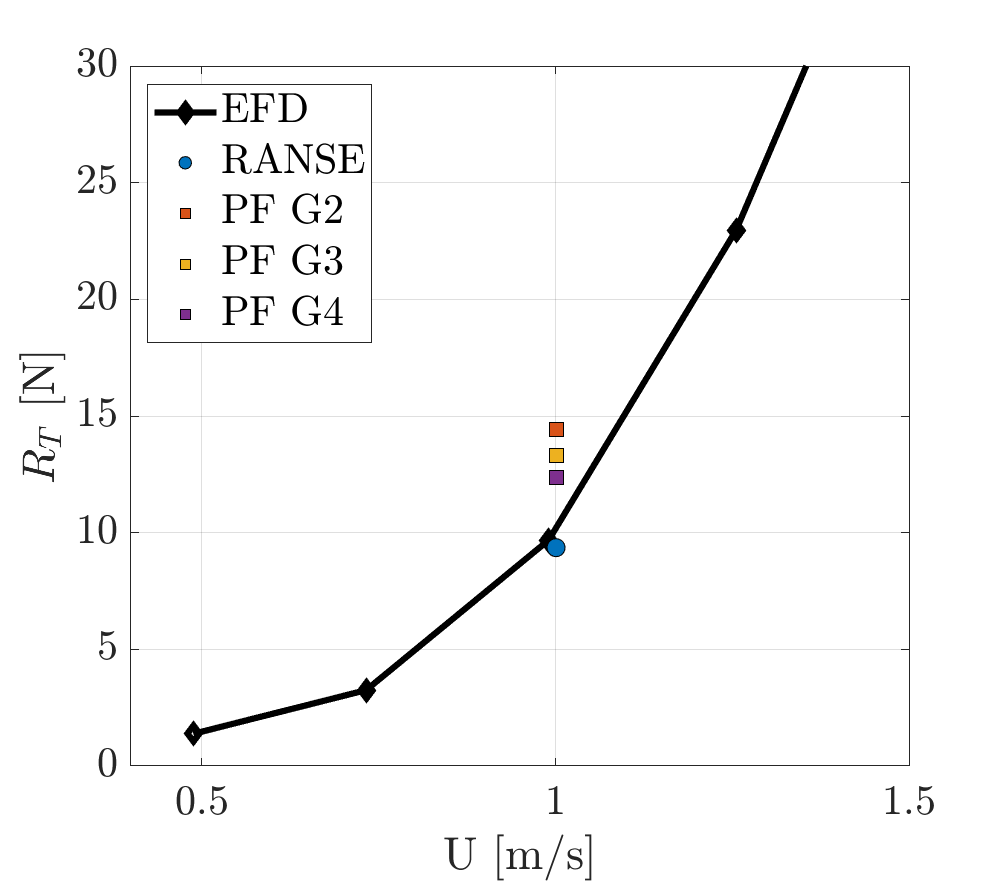}} 
\subfloat[Trim $\theta$]{\includegraphics[width=0.33\textwidth]{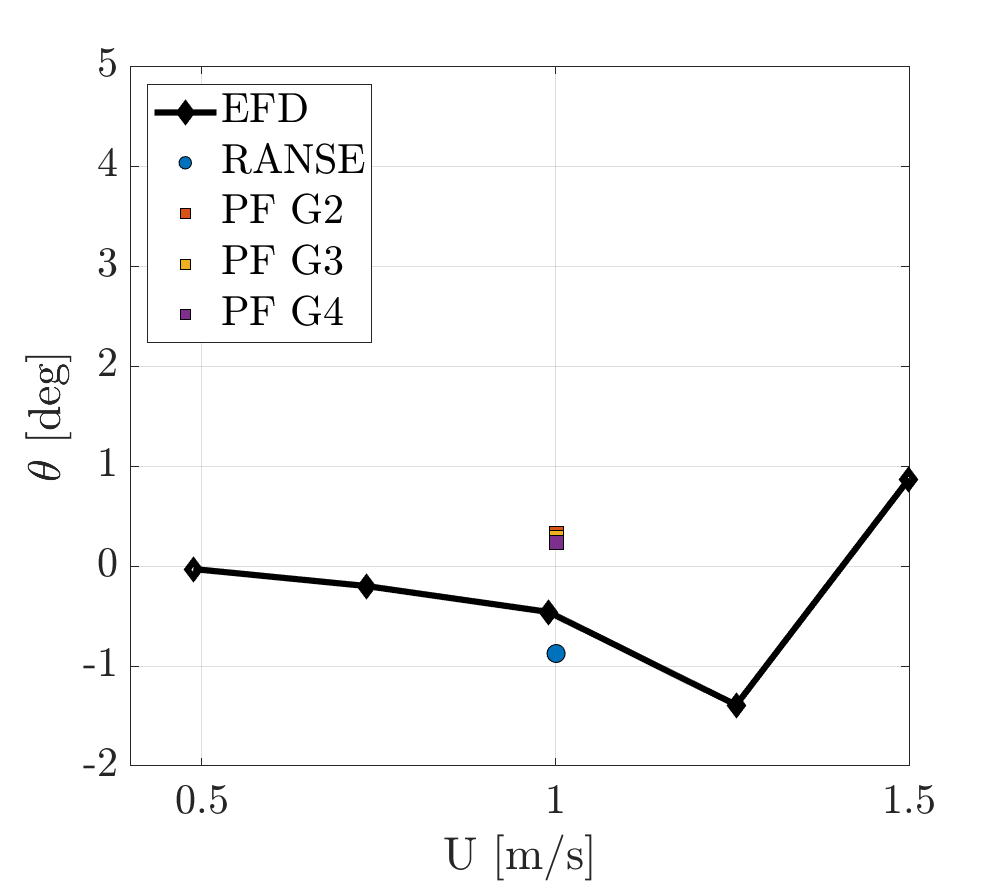}}  
\subfloat[Sinkage $T$]{\includegraphics[width=0.33\textwidth]{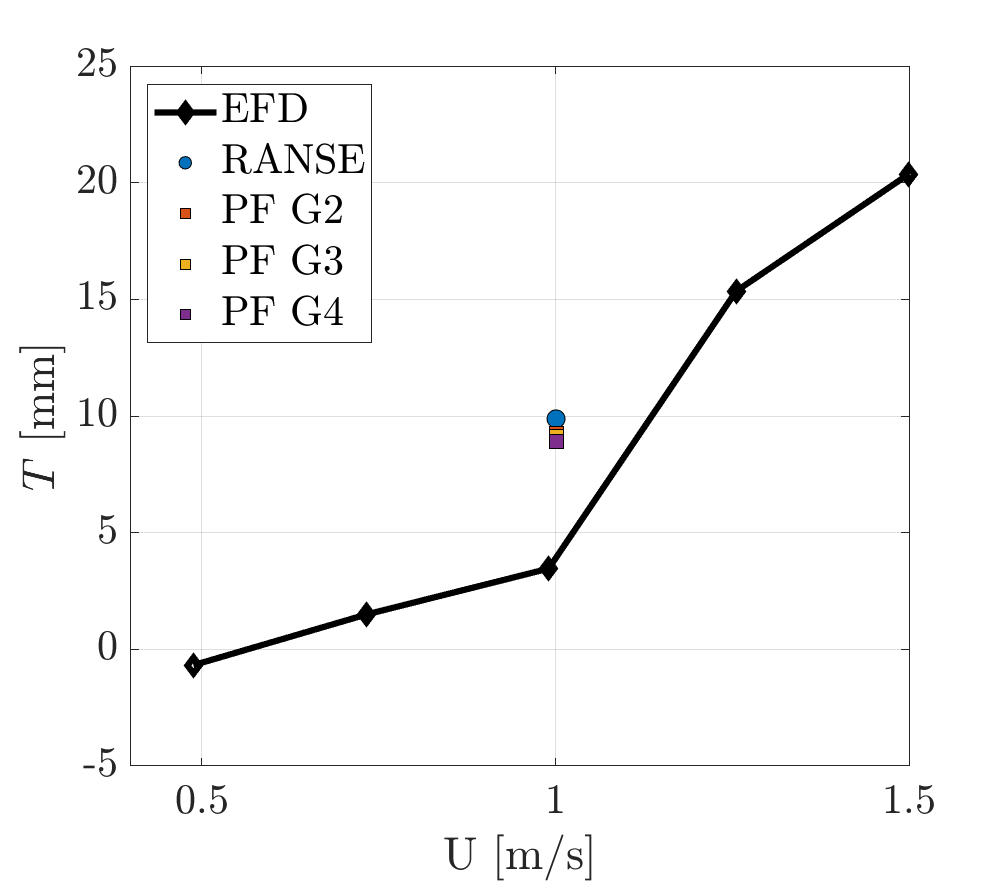}} 
\caption{Comparison between the numerical solver prediction of the SWAMP performance and experiments \cite{odetti2020swamp}}\label{fig:EFD-CFD} 
\end{figure}

Figure \ref{fig:EFD-CFD} compares the numerical solver prediction of the SWAMP performance and experiments (\cite{odetti2020swamp}). The RANSE solver achieves a good accuracy in the prediction of the resistance, whereas the PF solver overestimates the experimental value. Furthermore, the overestimation increases as the grid refinement increases. This is a strong motivation for the inclusion of the RANSE solver for this application. 
The RANSE solver overestimates the value of the trim angle $\theta$, whereas the PF solver predicts $\theta$ with an opposite sign.
Finally, the RANSE and PF solvers overestimate similarly the sinkage $T$ of the SWAMP hull.

\begin{figure}[!b]
    \centering 
    \includegraphics[width=0.45\textwidth]{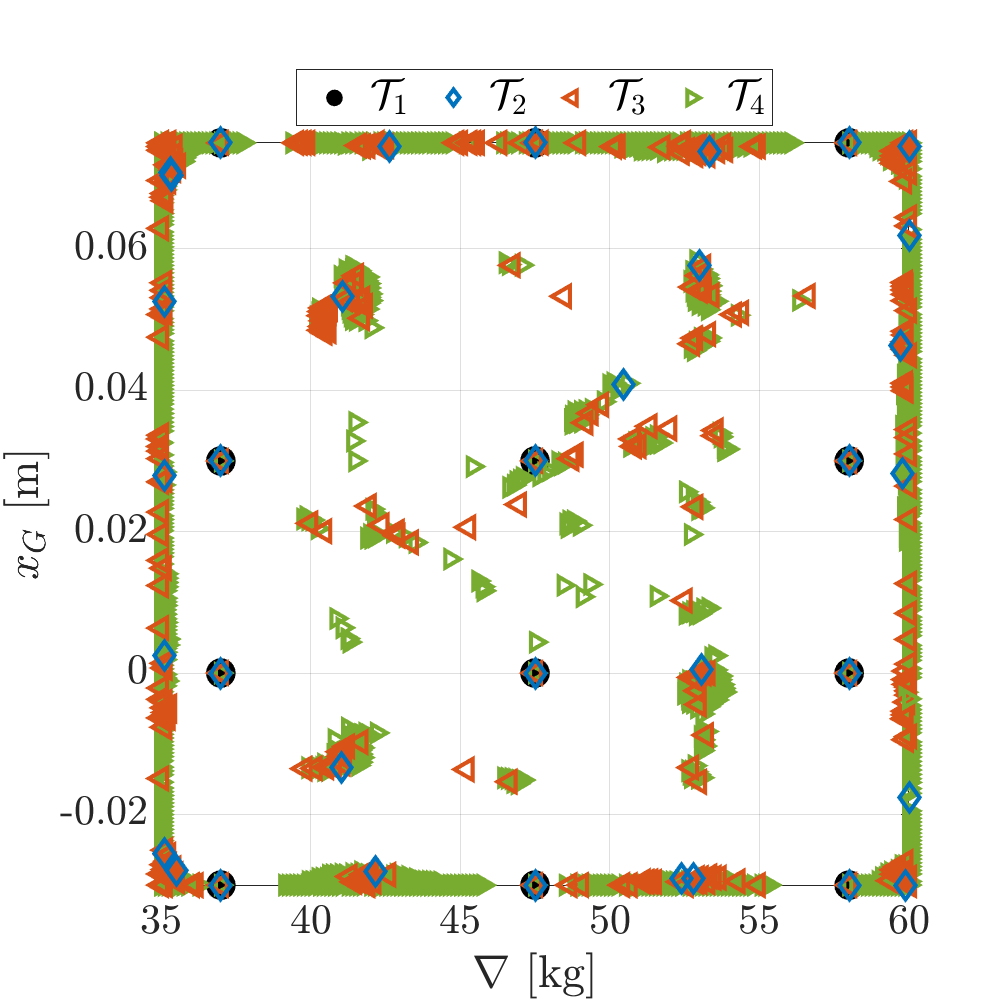}
    \caption{Final training sets distribution for the active learning procedure}
    \label{fig:FinalTS} 
\end{figure}

Twelve simulations are performed with the RANSE and PF (using the G2 grid) codes. The operating conditions are identified along directions parallel to the coordinated axes (see the black circles in Fig. \ref{fig:FinalTS}). Starting from the nominal payload (58 kg), the simulations along this direction are uniformly distributed within the payload range of variation (35-60 kg). The selected values for $\delta x_G$ are symmetrical with respect to the nominal condition ($\delta x_G = 0\%$L) plus an extreme condition at $\delta x_G = 7.5\%$L.

\begin{figure}[!t] 
\centering 
\mbox{} \hfill
\subfloat[RANSE outer, $\nabla = 37$kg]{\includegraphics[width=0.33\textwidth]{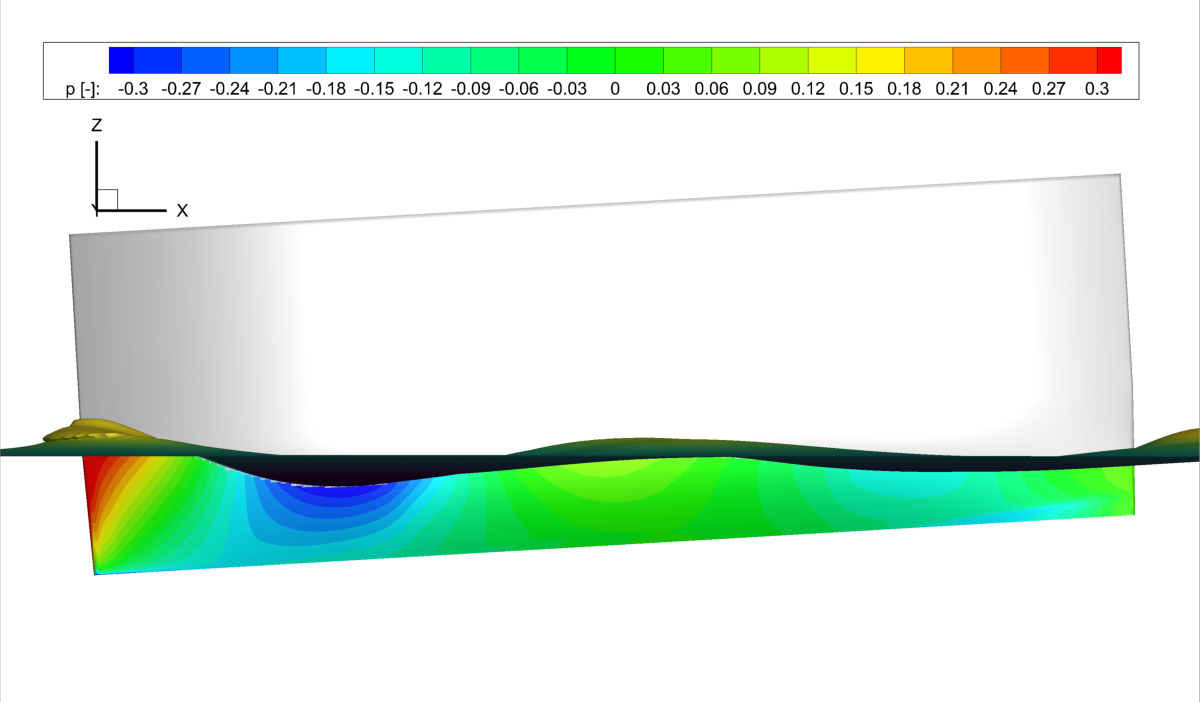}}  \hfill 
\subfloat[RANSE outer, $\nabla = 47.5$kg]{\includegraphics[width=0.33\textwidth]{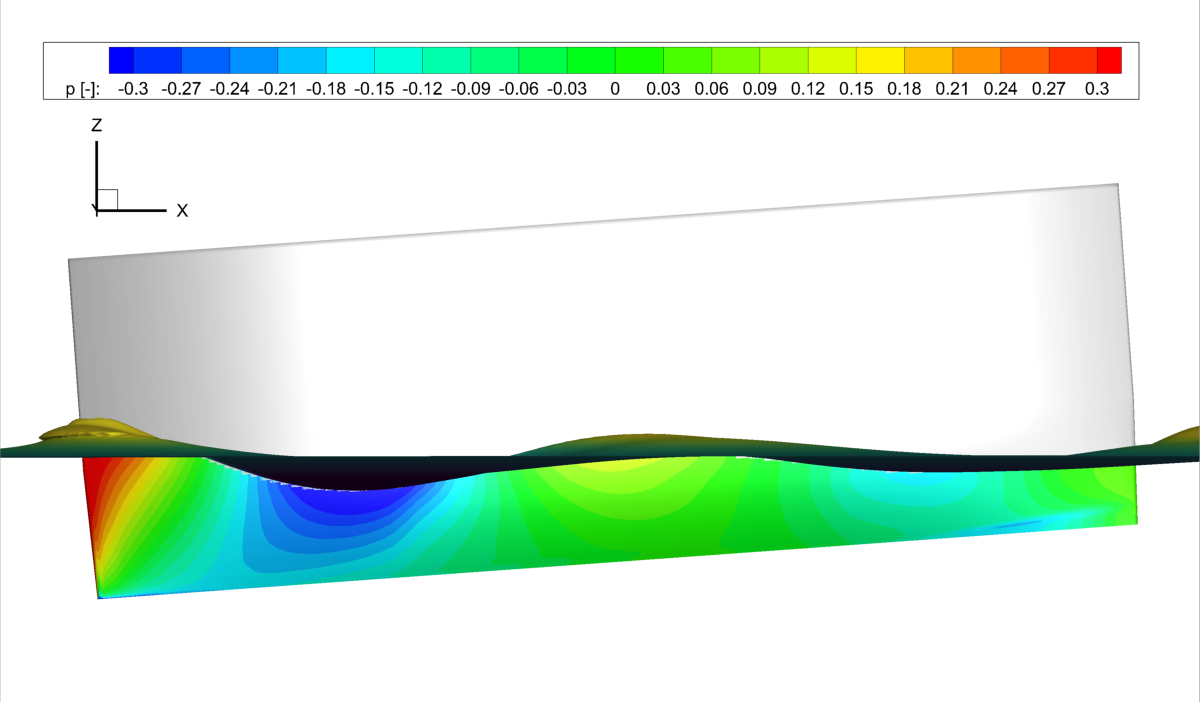}}   \hfill 
\subfloat[RANSE outer, $\nabla = 58$kg]{\includegraphics[width=0.33\textwidth]{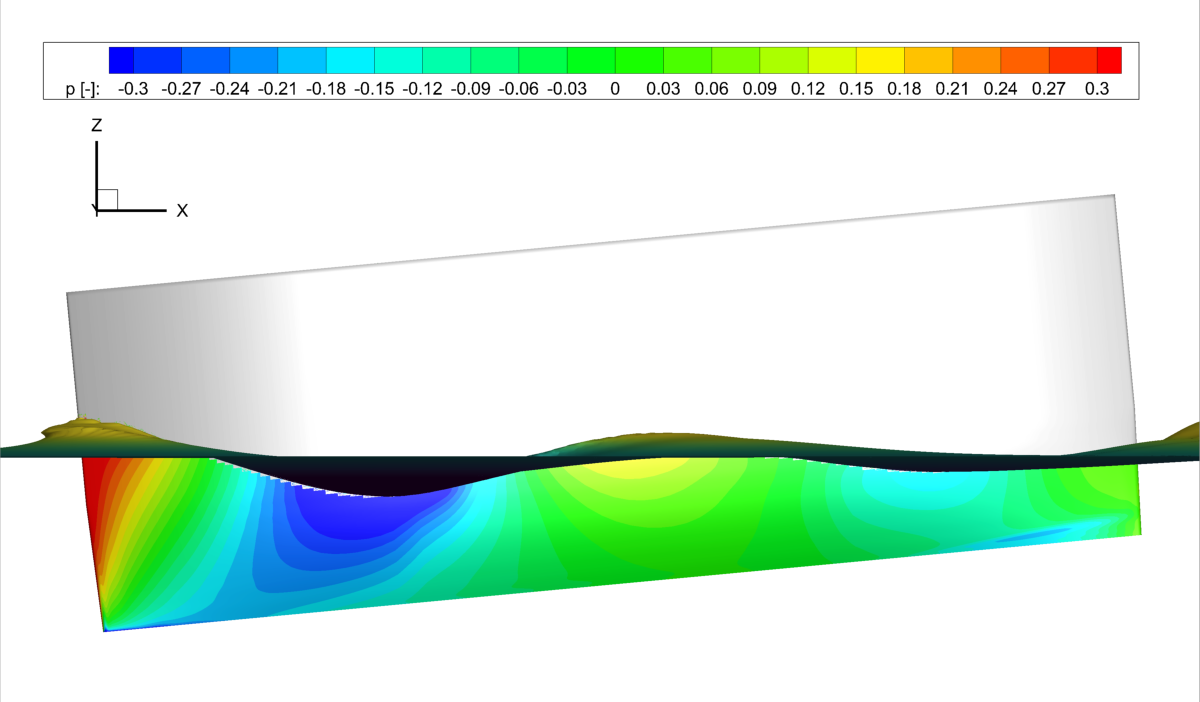}}  \hfill \mbox{}   \\
\mbox{} \hfill
\subfloat[RANSE inner, $\nabla = 37$kg]{\includegraphics[width=0.33\textwidth]{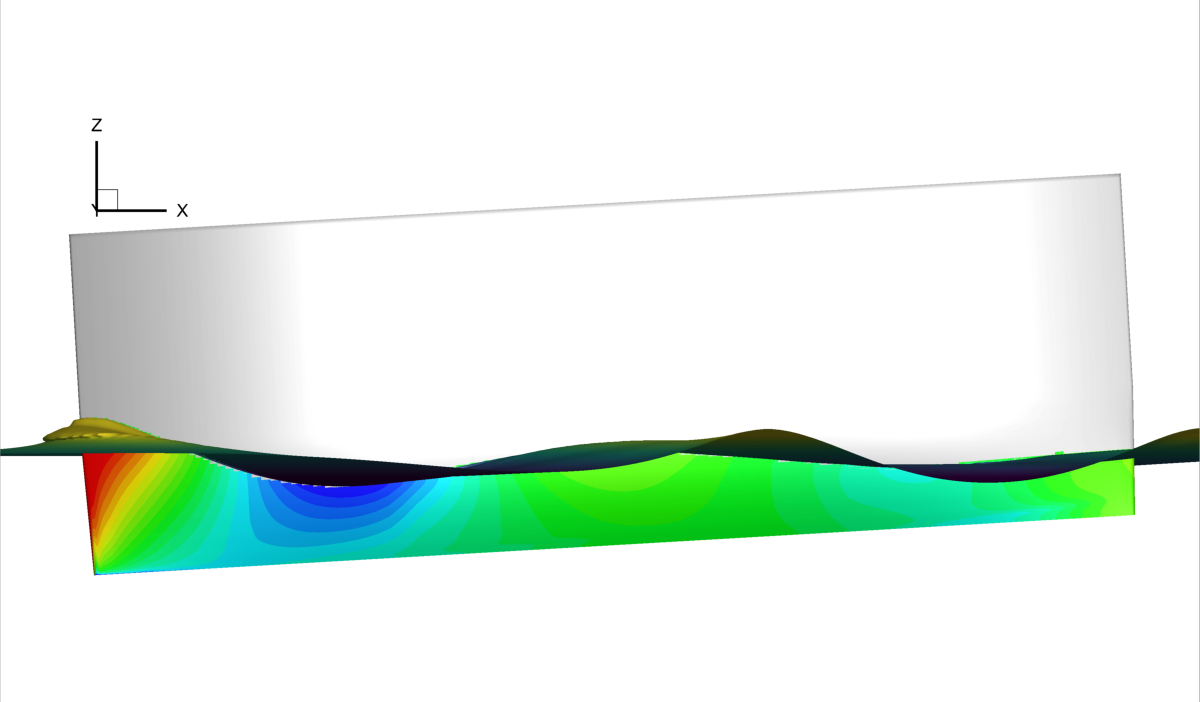}}  \hfill 
\subfloat[RANSE inner, $\nabla = 47.5$kg]{\includegraphics[width=0.33\textwidth]{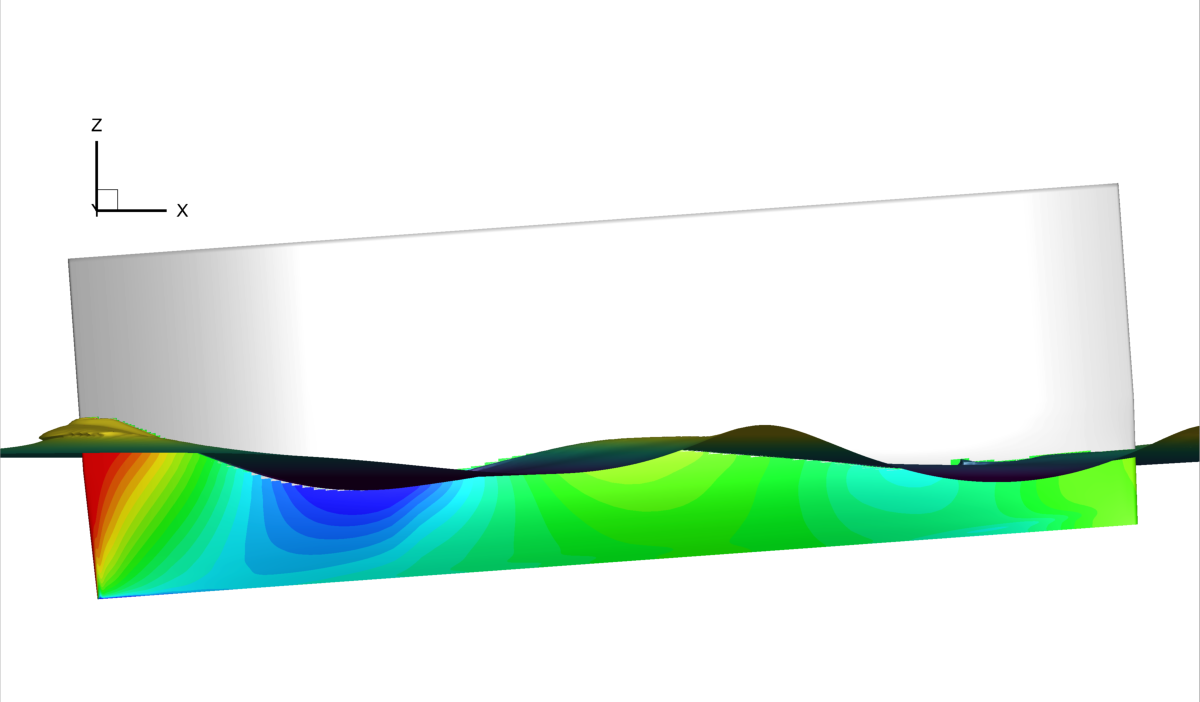}}   \hfill 
\subfloat[RANSE inner, $\nabla = 58$kg]{\includegraphics[width=0.33\textwidth]{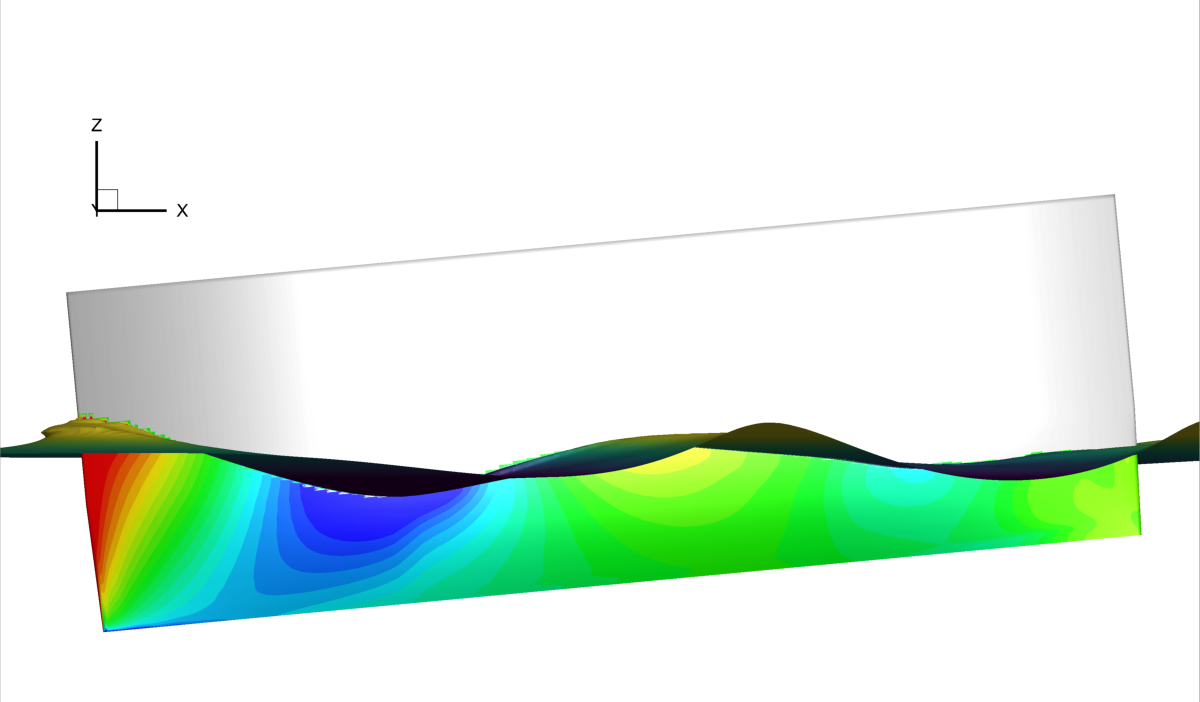}}  \hfill \mbox{}   \\
\mbox{} \hfill
\subfloat[PF outer, $\nabla = 37$kg]{\includegraphics[width=0.33\textwidth]{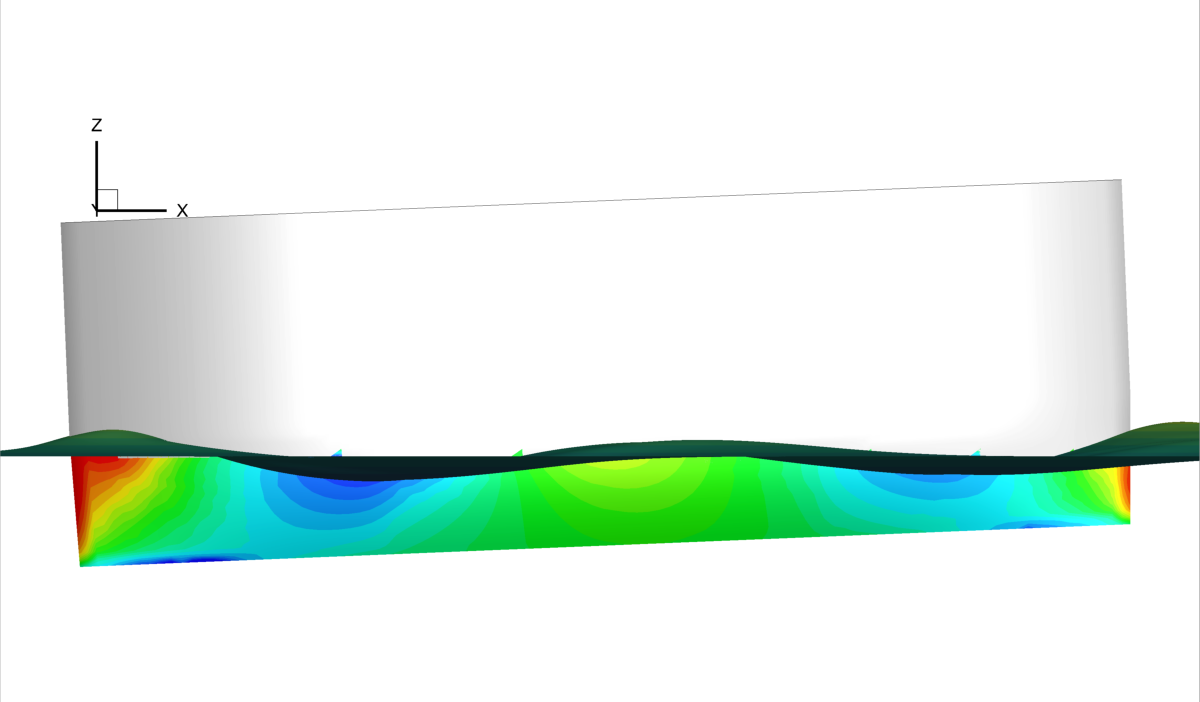}}  \hfill 
\subfloat[PF outer, $\nabla = 47.5$kg]{\includegraphics[width=0.33\textwidth]{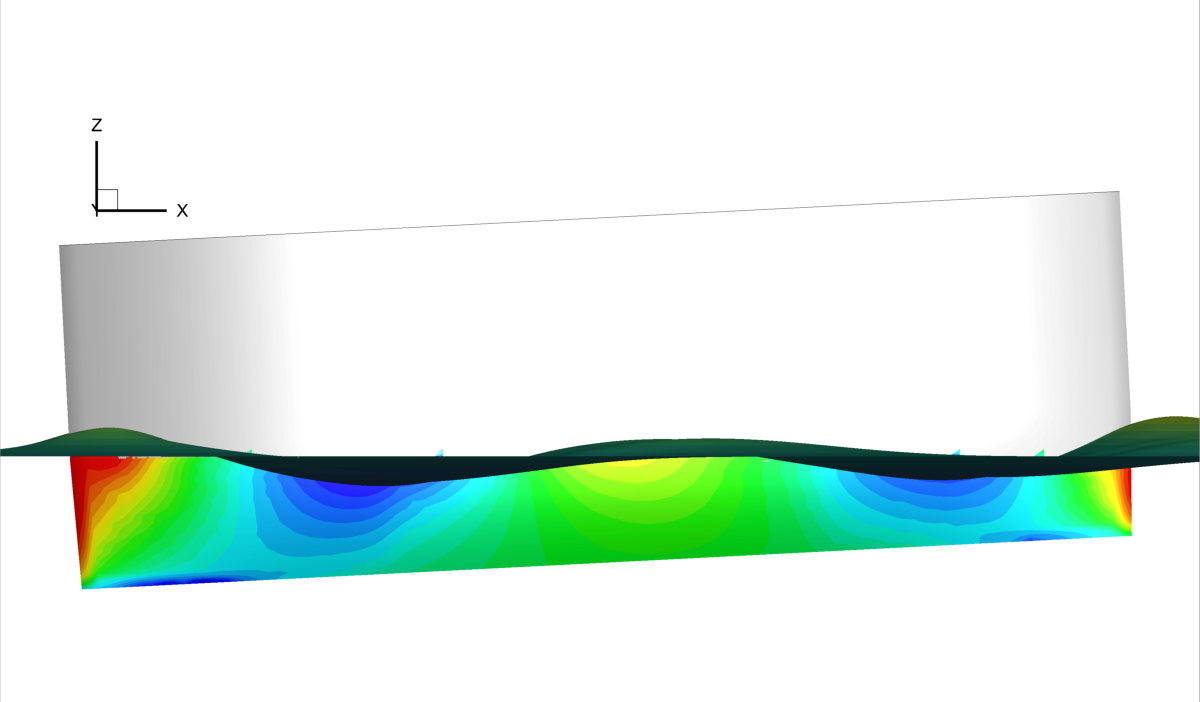}} \hfill 
\subfloat[PF outer, $\nabla = 58$kg]{\includegraphics[width=0.33\textwidth]{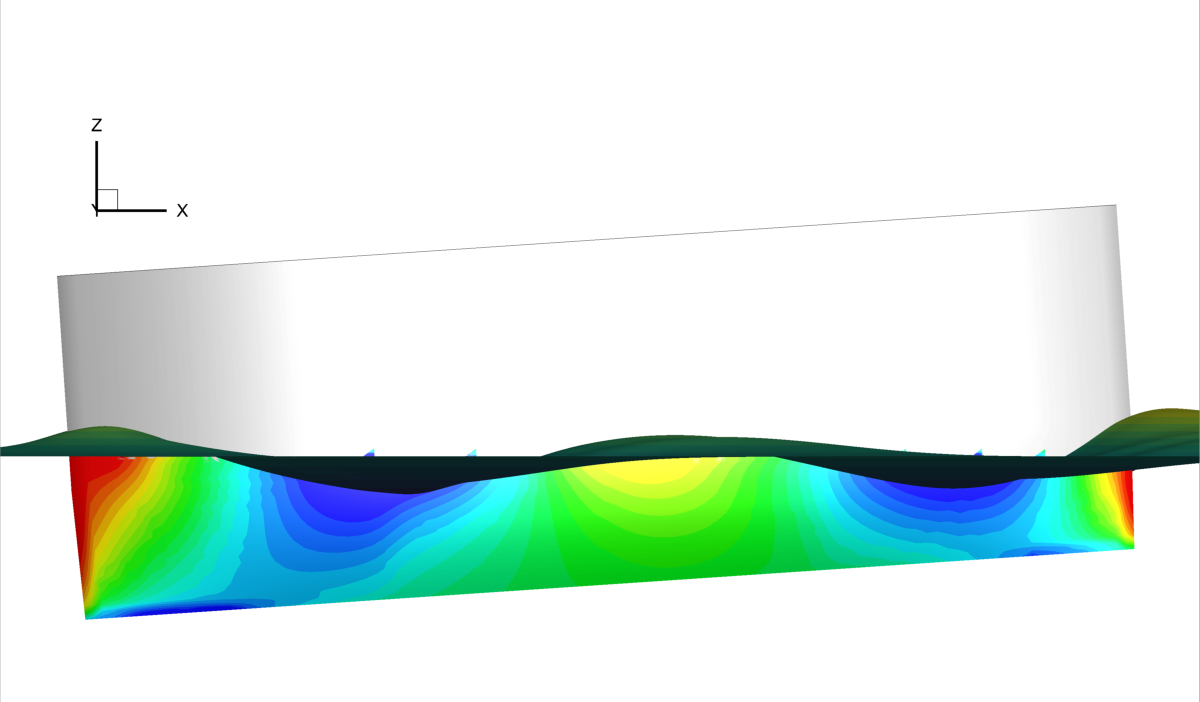}}   \hfill \mbox{}  \\
\mbox{} \hfill
\subfloat[PF inner, $\nabla = 37$kg]{\includegraphics[width=0.33\textwidth]{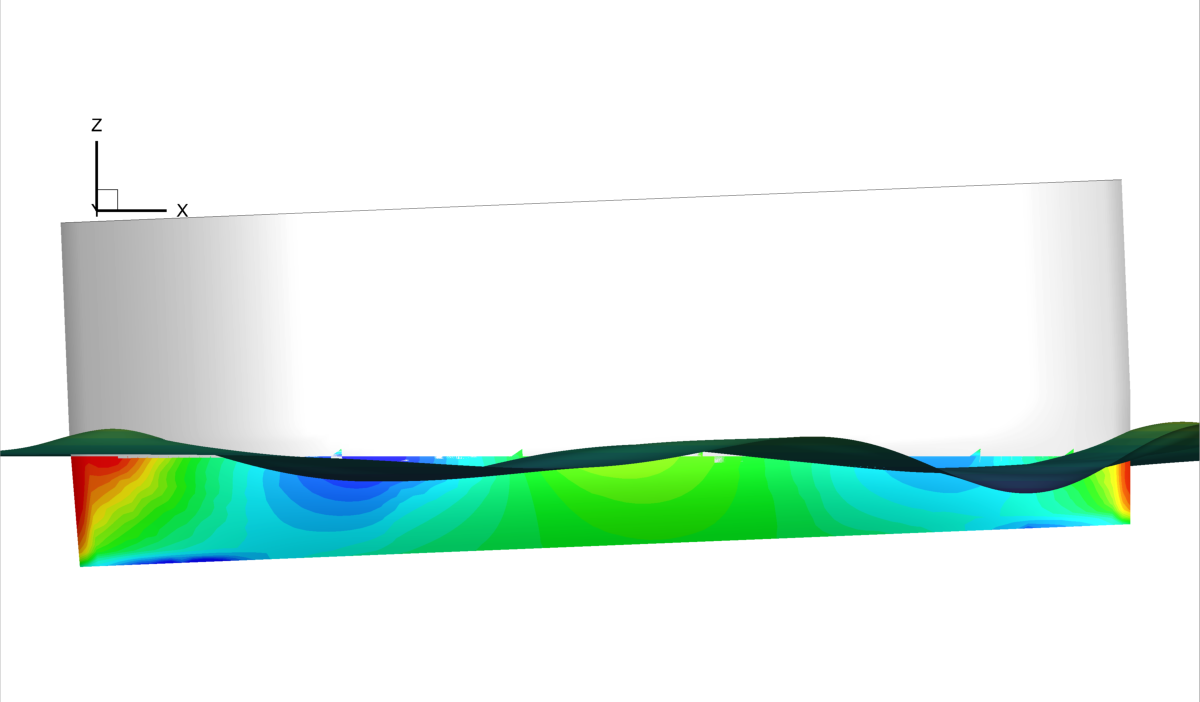}}  \hfill 
\subfloat[PF inner, $\nabla = 47.5$kg]{\includegraphics[width=0.33\textwidth]{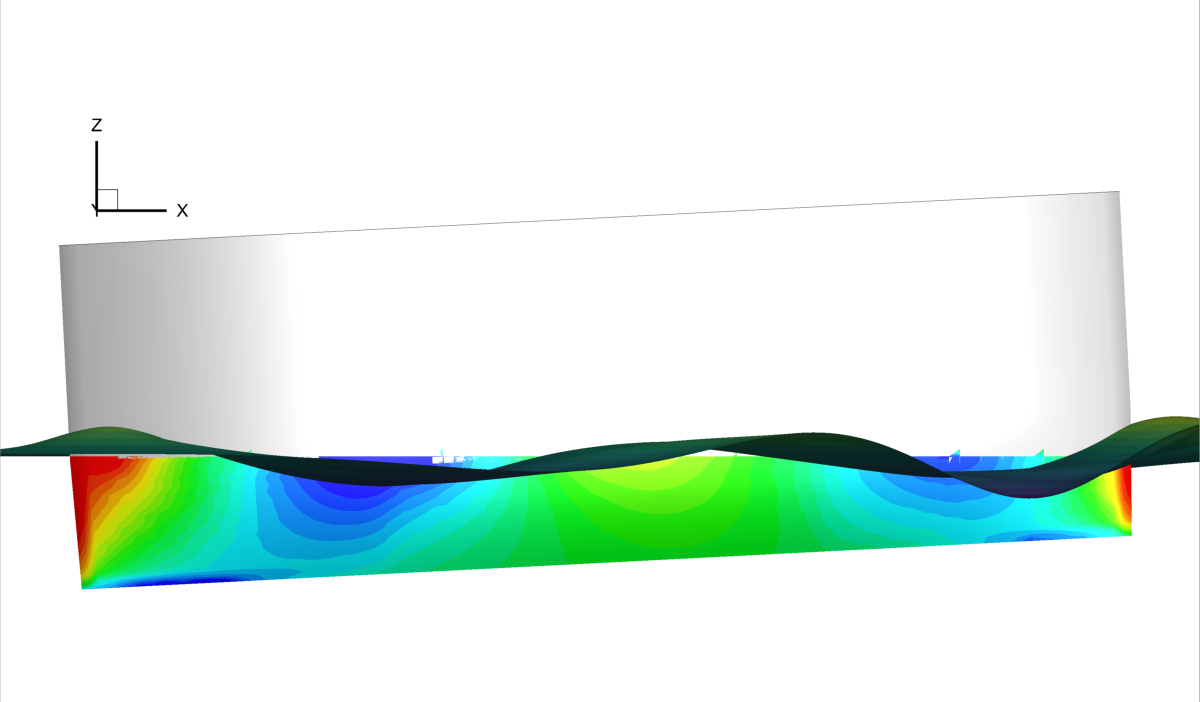}} \hfill 
\subfloat[PF inner, $\nabla = 58$kg]{\includegraphics[width=0.33\textwidth]{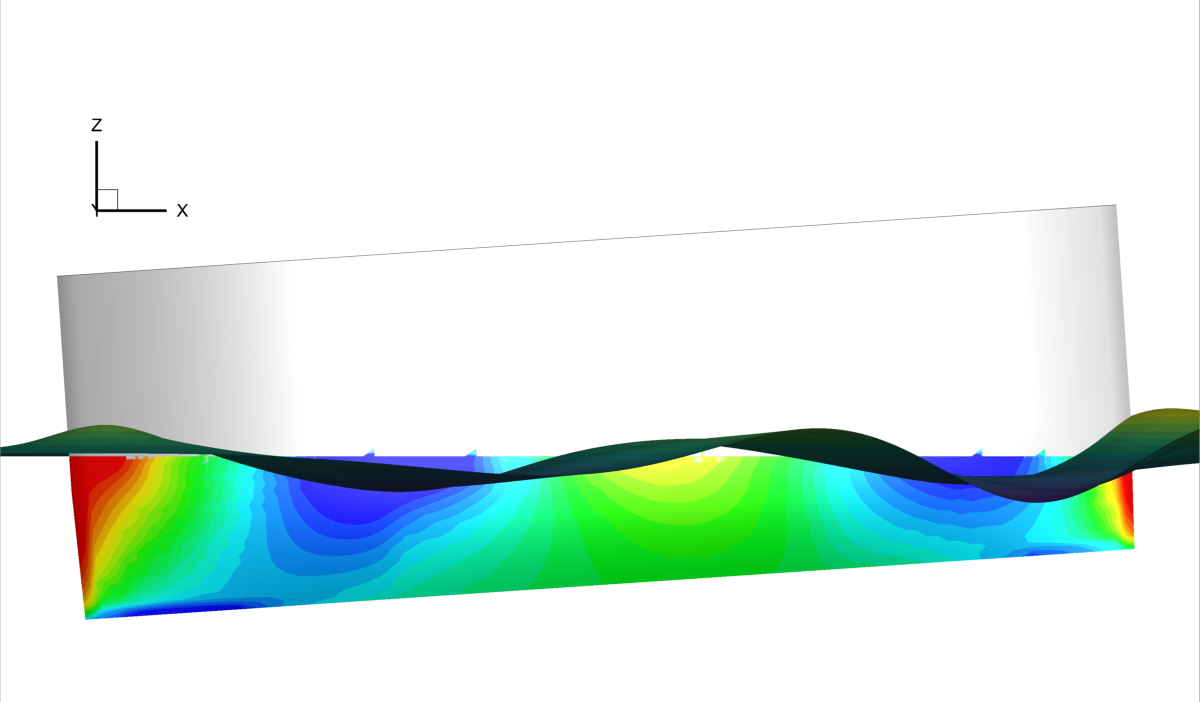}}  \hfill \mbox{}  
\caption{Pressure contour varying the SWAMP payload with $\delta x_G = -3\%$L, side views}\label{fig:PSide-3} 
\end{figure}

\begin{figure}[!t] 
\centering 
\mbox{} \hfill
\subfloat[RANSE outer, $\nabla = 37$kg]{\includegraphics[width=0.33\textwidth]{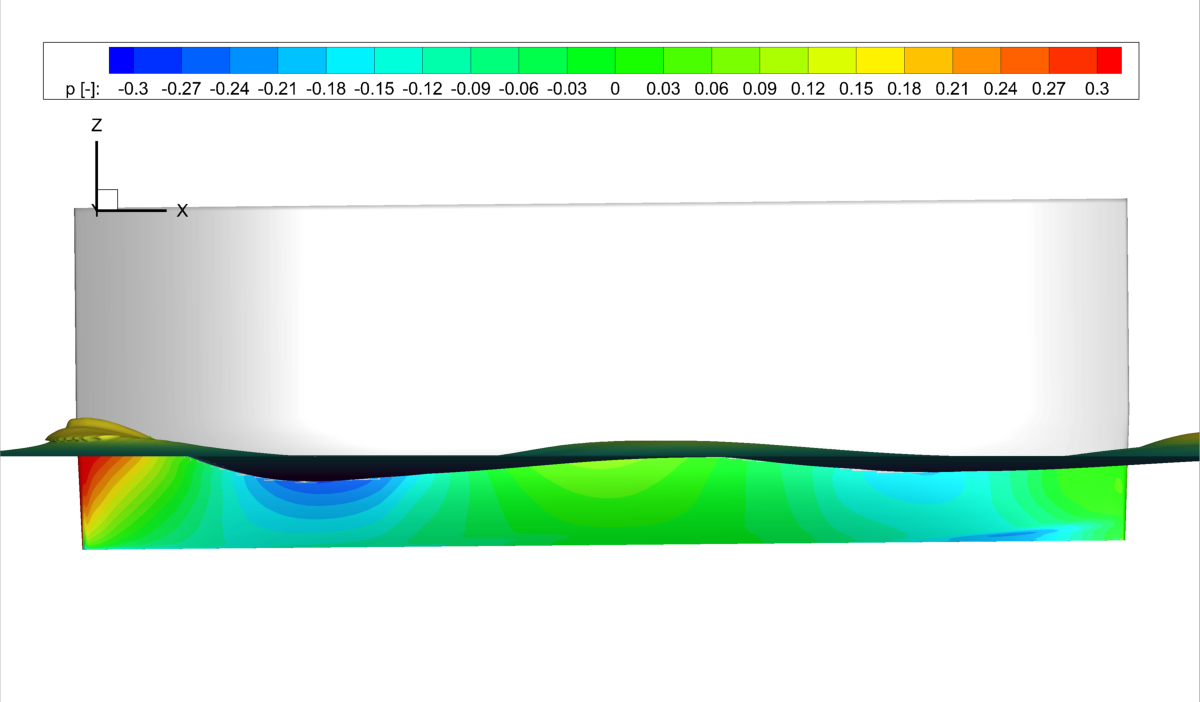}}  \hfill 
\subfloat[RANSE outer, $\nabla = 47.5$kg]{\includegraphics[width=0.33\textwidth]{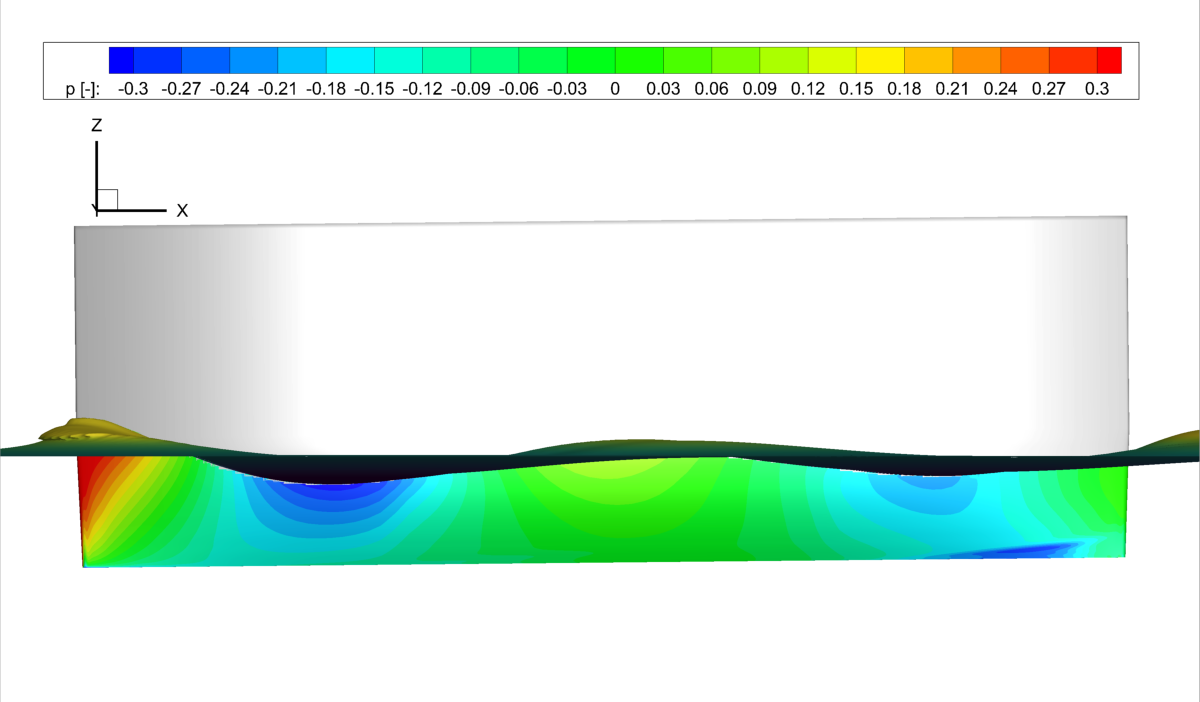}}   \hfill 
\subfloat[RANSE outer, $\nabla = 58$kg]{\includegraphics[width=0.33\textwidth]{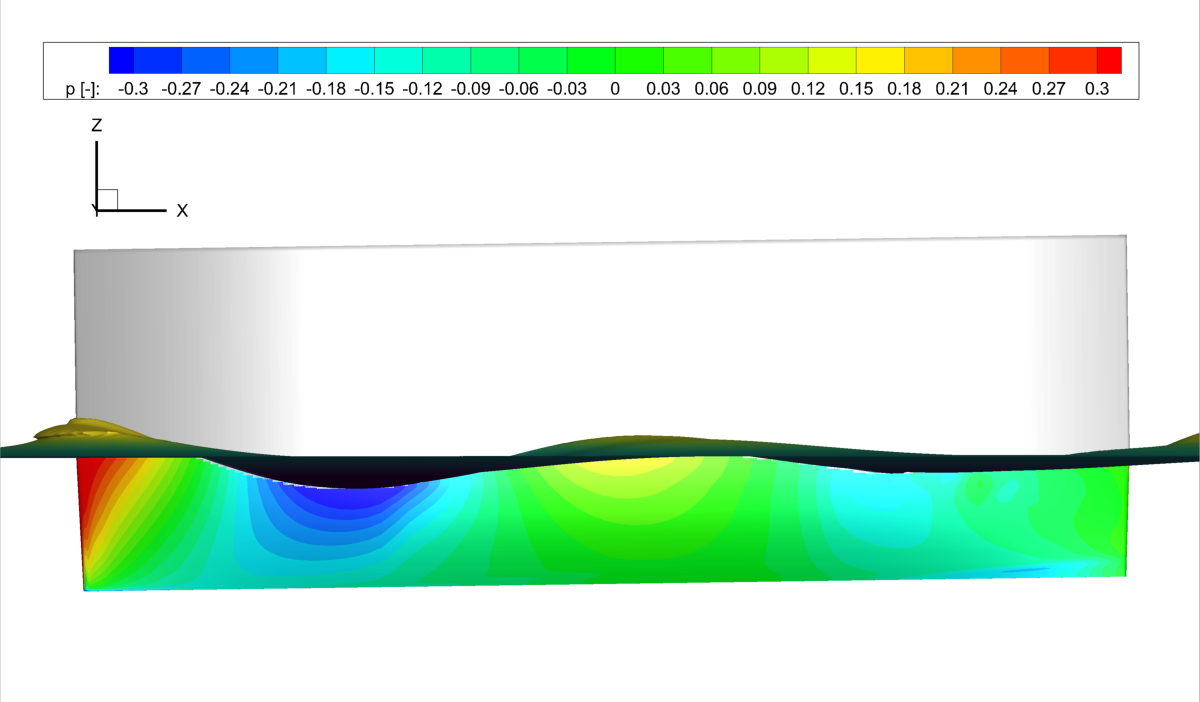}}  \hfill \mbox{}   \\
\mbox{} \hfill
\subfloat[RANSE inner, $\nabla = 37$kg]{\includegraphics[width=0.33\textwidth]{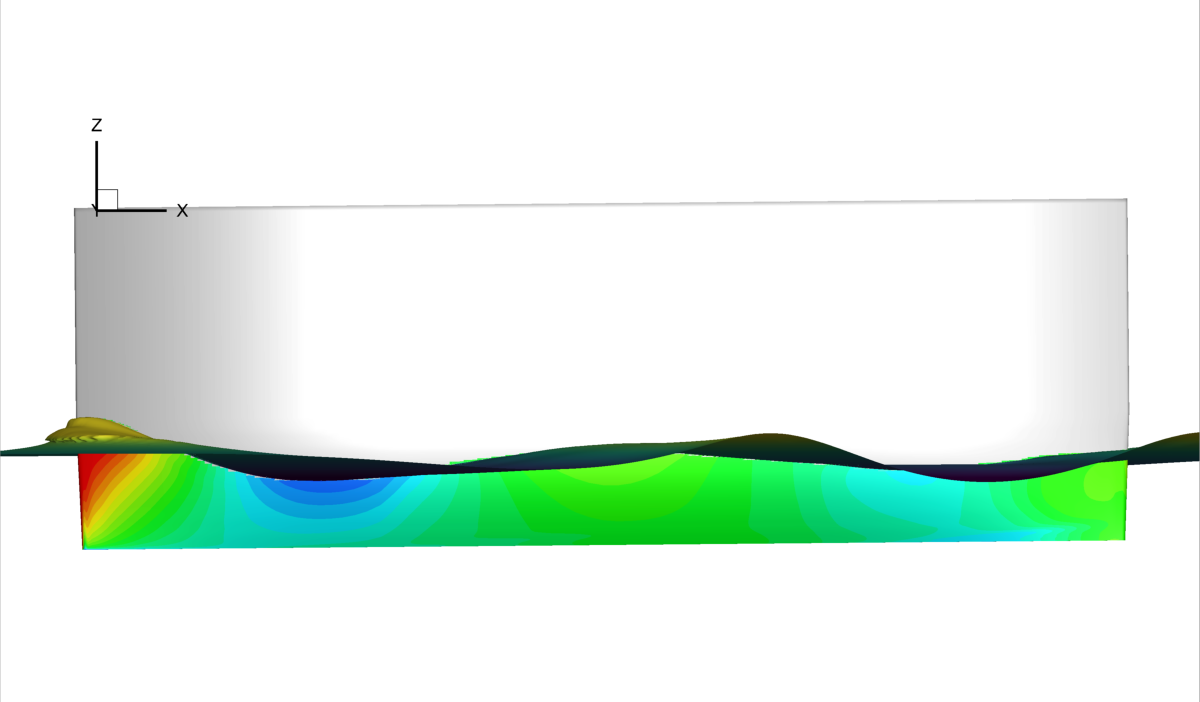}}  \hfill 
\subfloat[RANSE inner, $\nabla = 47.5$kg]{\includegraphics[width=0.33\textwidth]{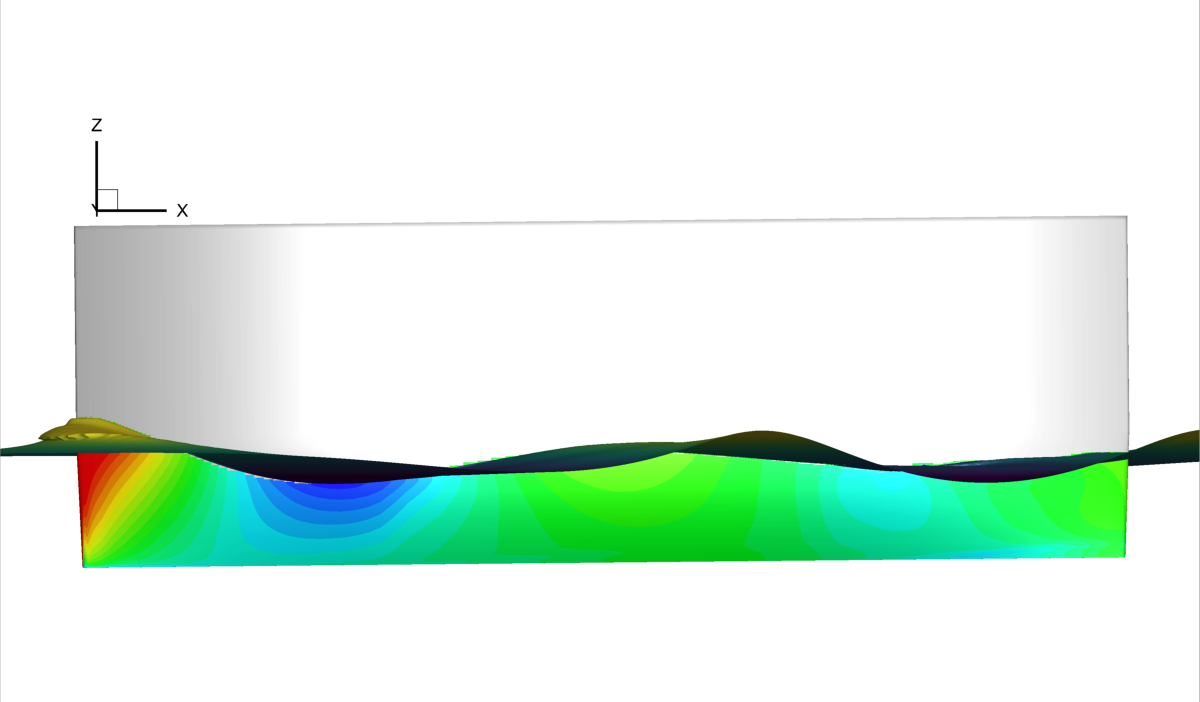}}   \hfill 
\subfloat[RANSE inner, $\nabla = 58$kg]{\includegraphics[width=0.33\textwidth]{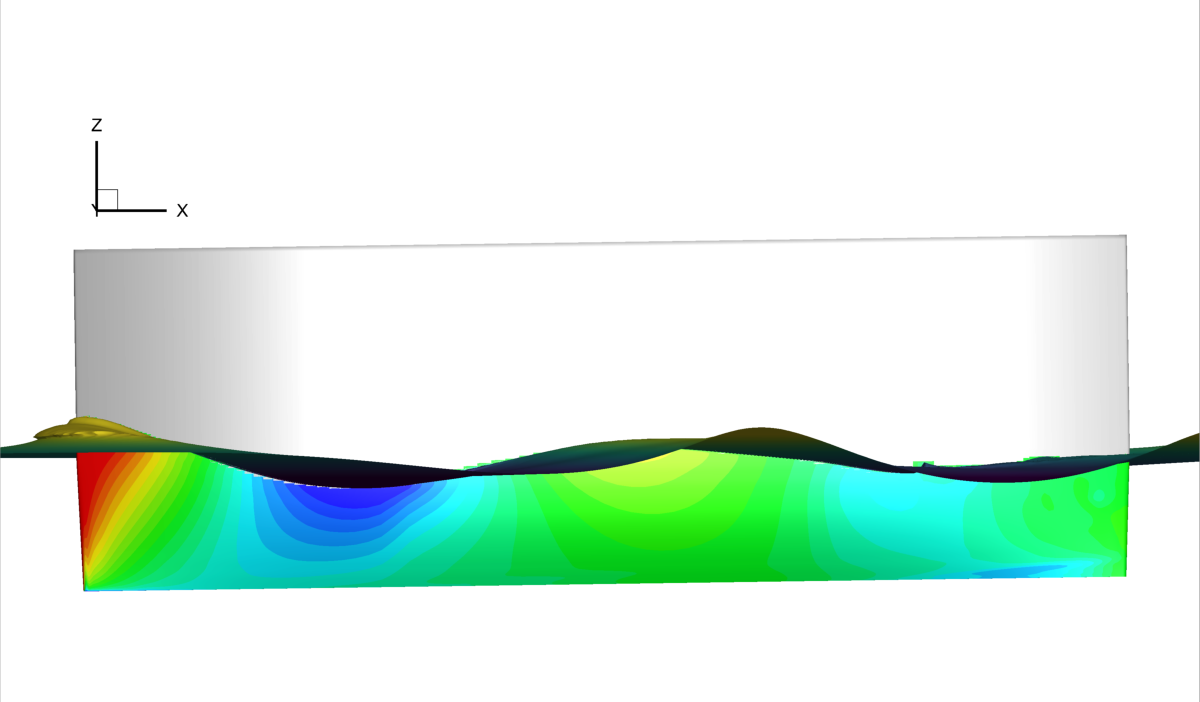}}  \hfill \mbox{}   \\
\mbox{} \hfill
\subfloat[PF outer, $\nabla = 37$kg]{\includegraphics[width=0.33\textwidth]{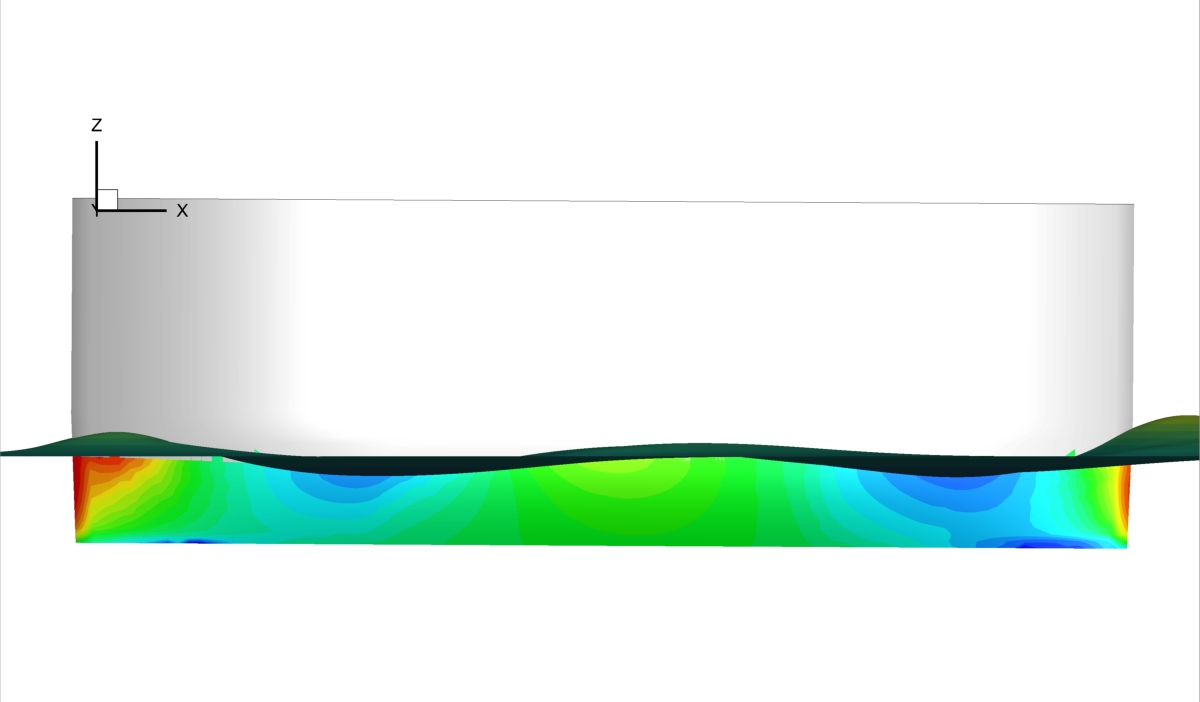}} \hfill 
\subfloat[PF outer, $\nabla = 47.5$kg]{\includegraphics[width=0.33\textwidth]{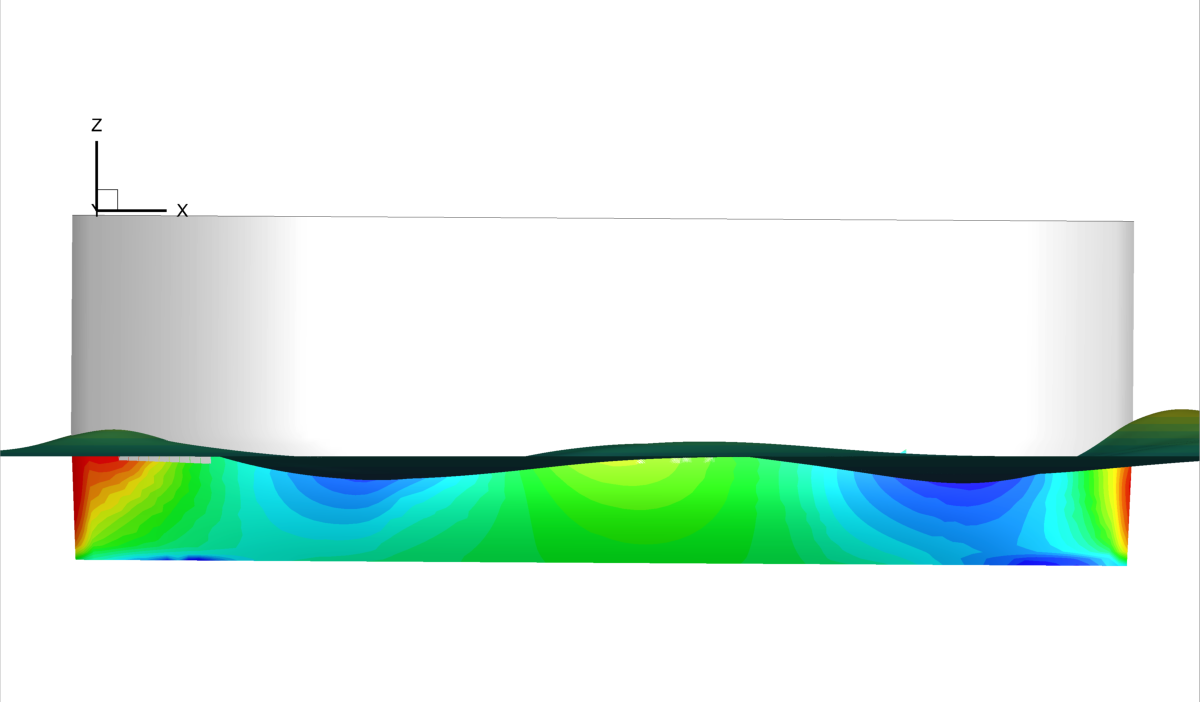}}   \hfill 
\subfloat[PF outer, $\nabla = 58$kg]{\includegraphics[width=0.33\textwidth]{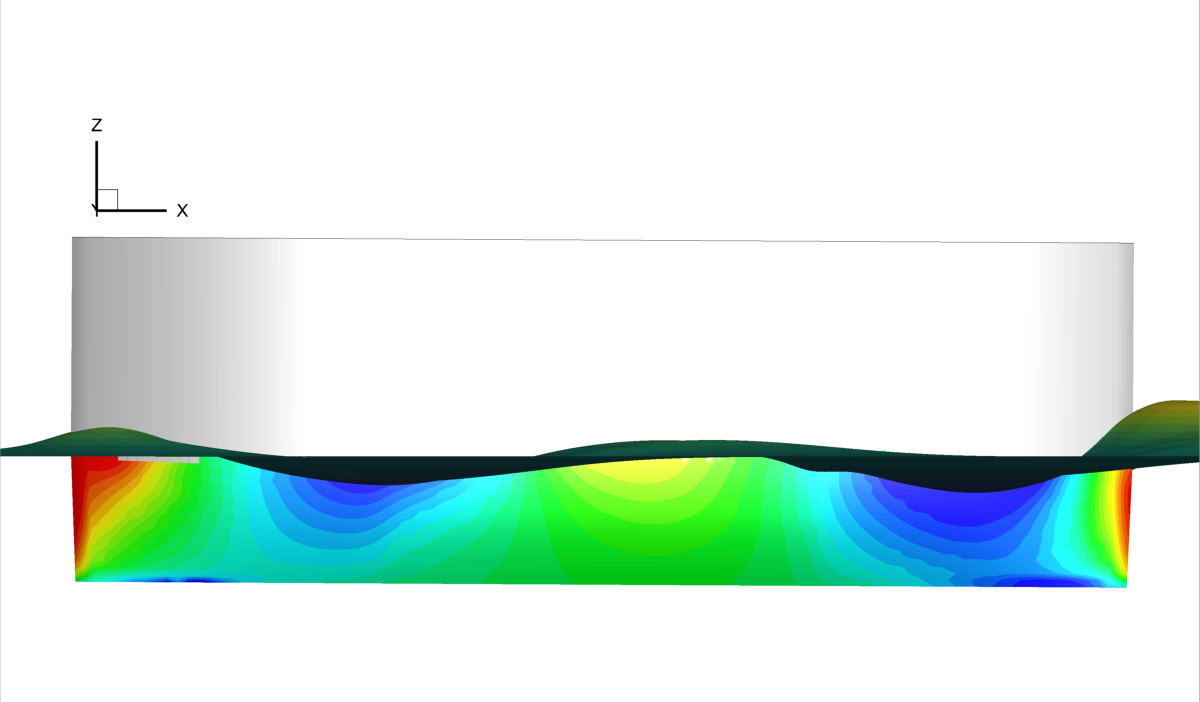}}   \hfill \mbox{}  \\
\mbox{} \hfill
\subfloat[PF inner, $\nabla = 37$kg]{\includegraphics[width=0.33\textwidth]{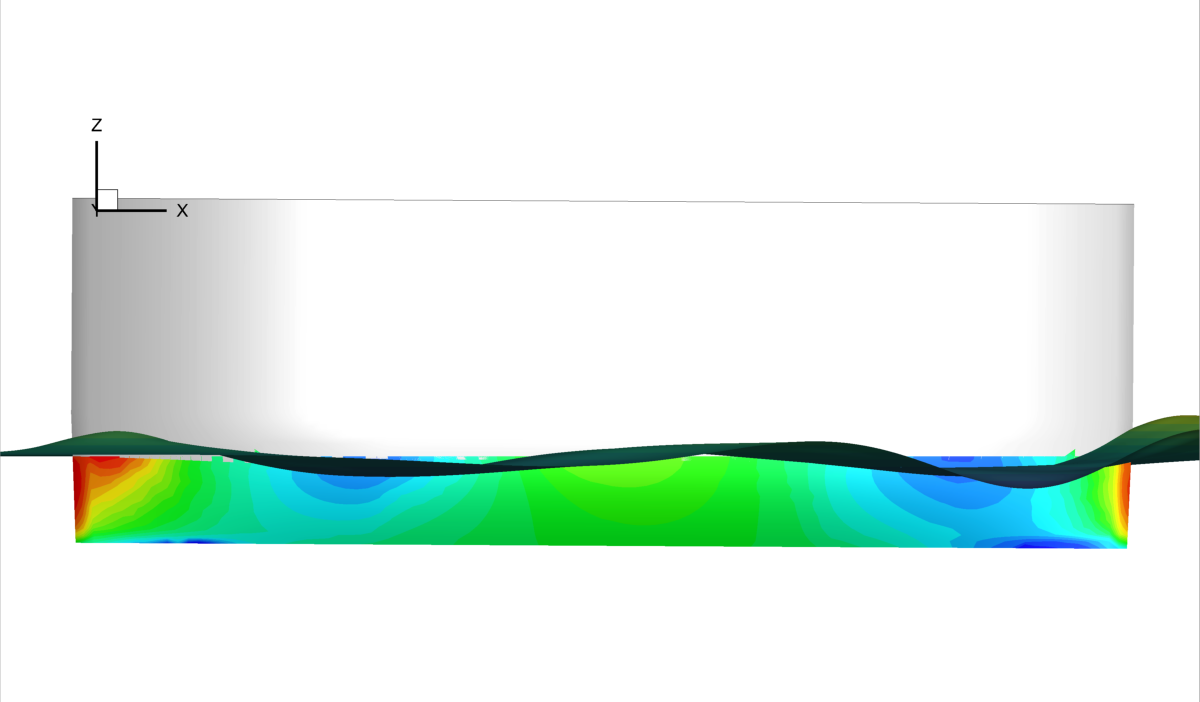}} \hfill 
\subfloat[PF inner, $\nabla = 47.5$kg]{\includegraphics[width=0.33\textwidth]{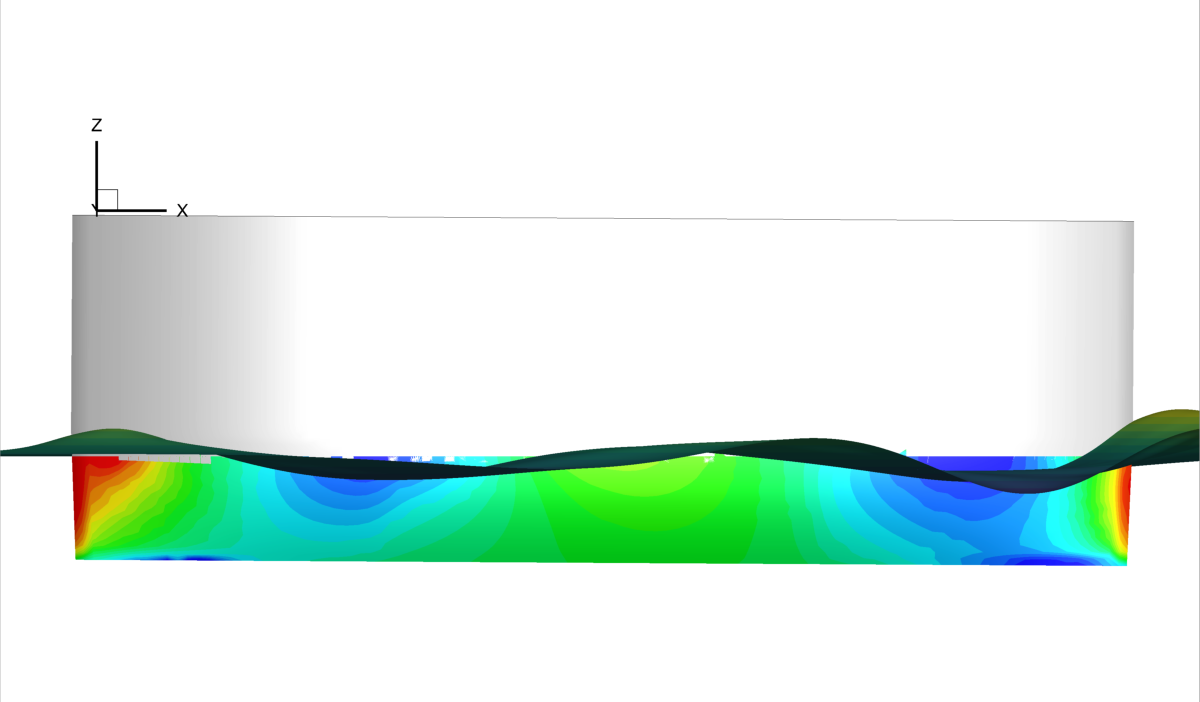}}   \hfill 
\subfloat[PF inner, $\nabla = 58$kg]{\includegraphics[width=0.33\textwidth]{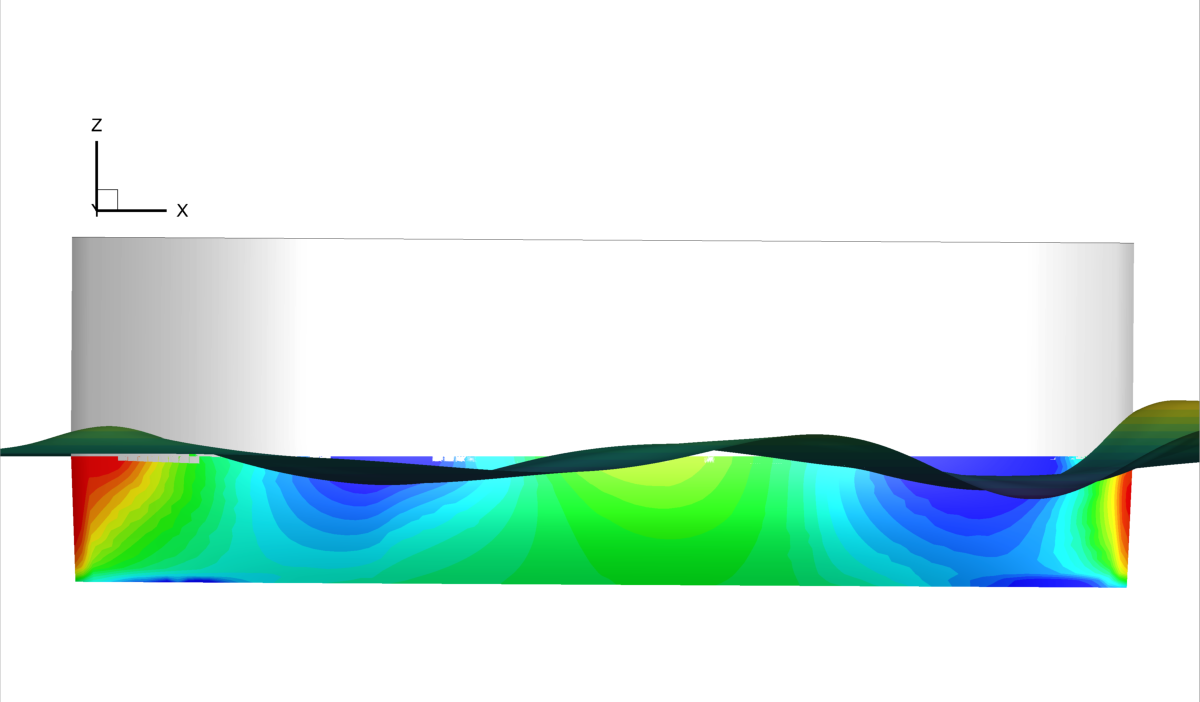}}  \hfill \mbox{}  
\caption{Pressure contour varying the SWAMP payload with $\delta x_G = 0\%$L, side views}\label{fig:PSide0} 
\end{figure}

\begin{figure}[!h] 
\centering 
\mbox{} \hfill
\subfloat[RANSE outer, $\nabla = 37$kg]{\includegraphics[width=0.33\textwidth]{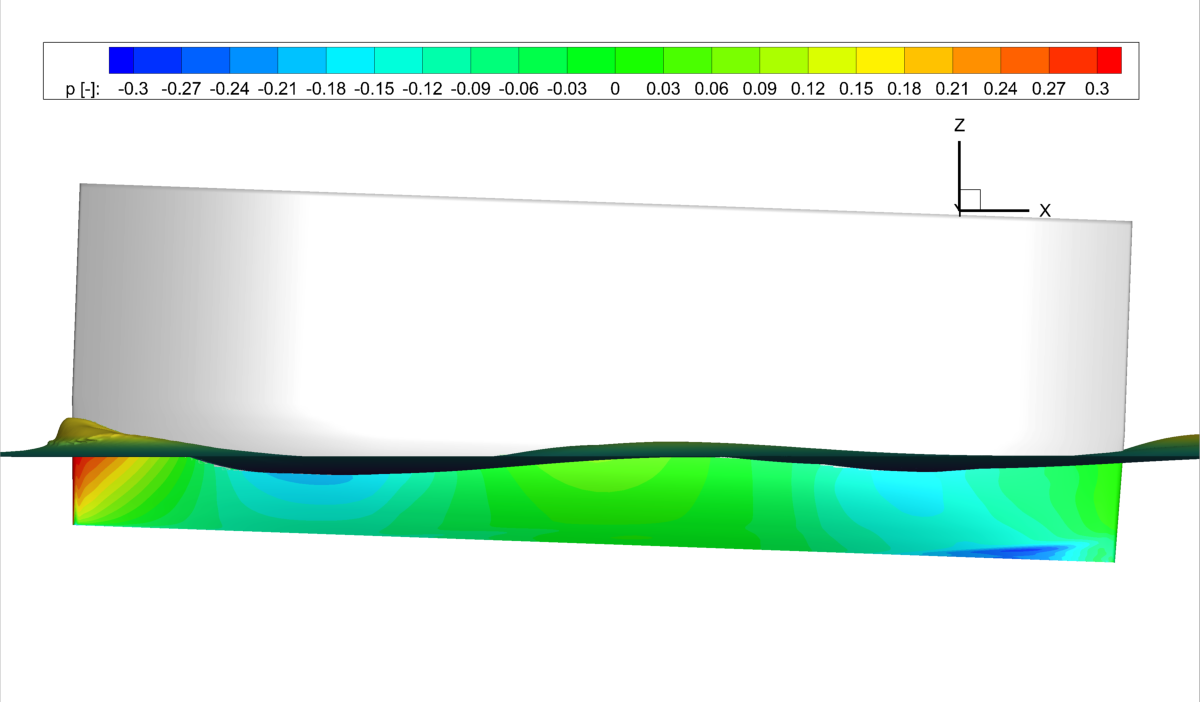}}  \hfill 
\subfloat[RANSE outer, $\nabla = 47.5$kg]{\includegraphics[width=0.33\textwidth]{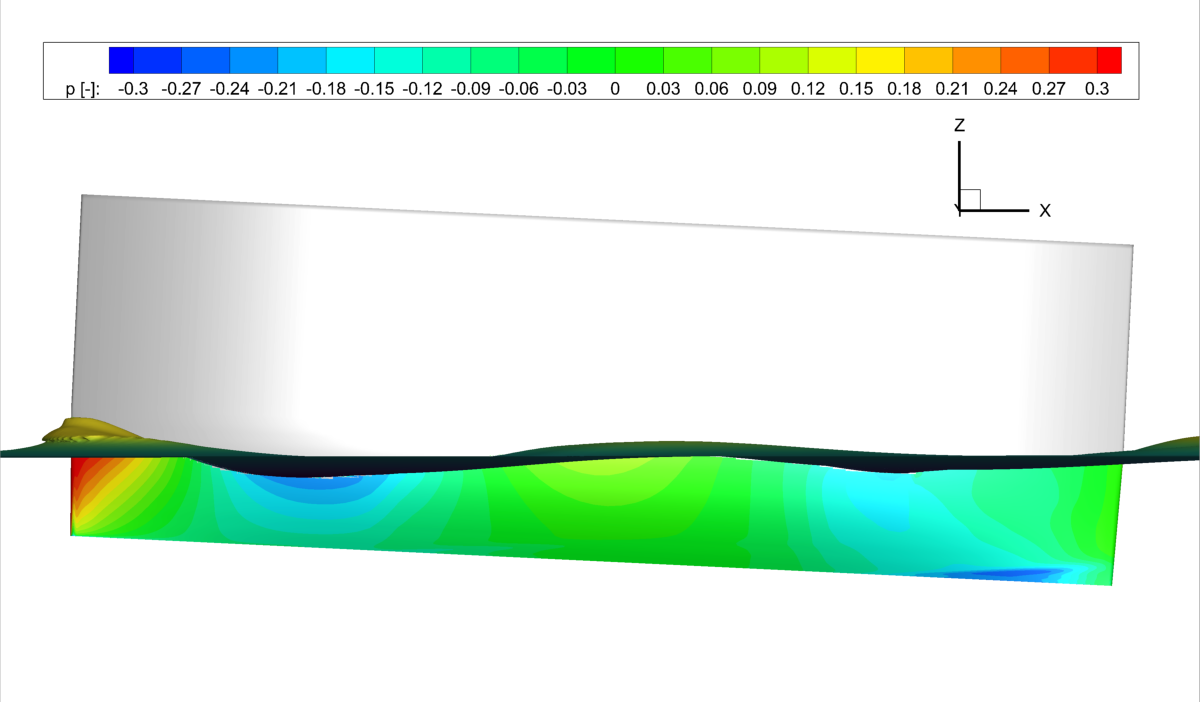}}   \hfill 
\subfloat[RANSE outer, $\nabla = 58$kg]{\includegraphics[width=0.33\textwidth]{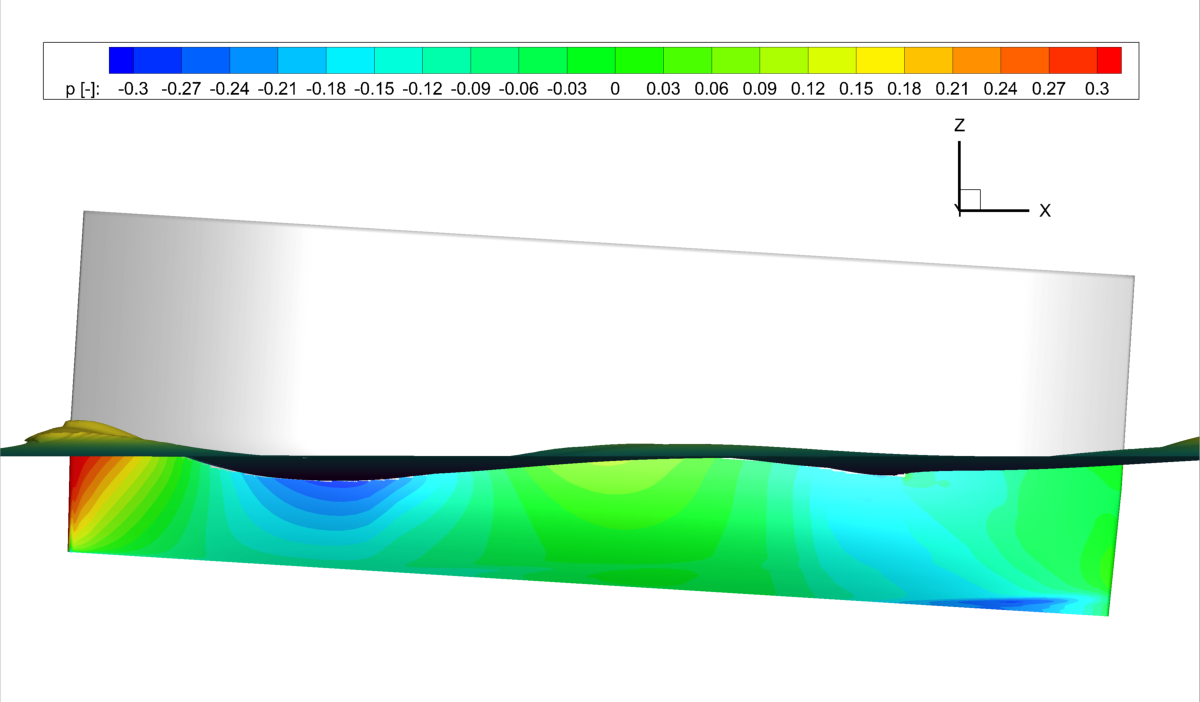}}   \hfill \mbox{}  \\
\mbox{} \hfill
\subfloat[RANSE inner, $\nabla = 37$kg]{\includegraphics[width=0.33\textwidth]{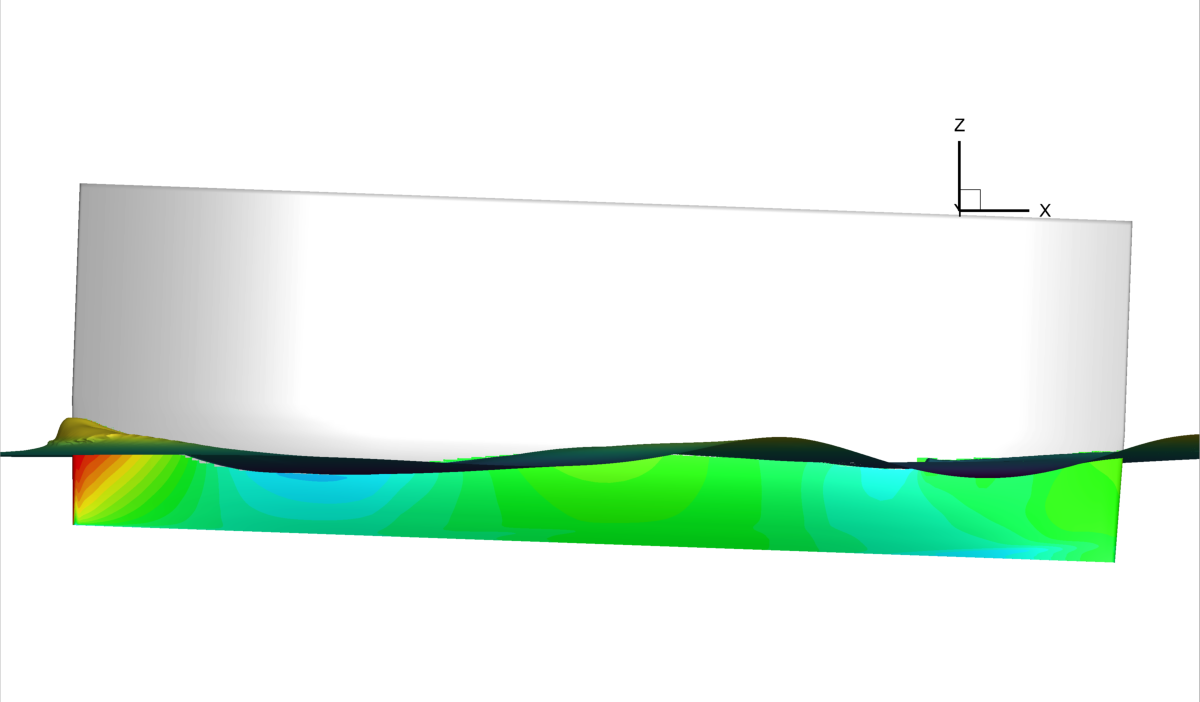}}  \hfill 
\subfloat[RANSE inner, $\nabla = 47.5$kg]{\includegraphics[width=0.33\textwidth]{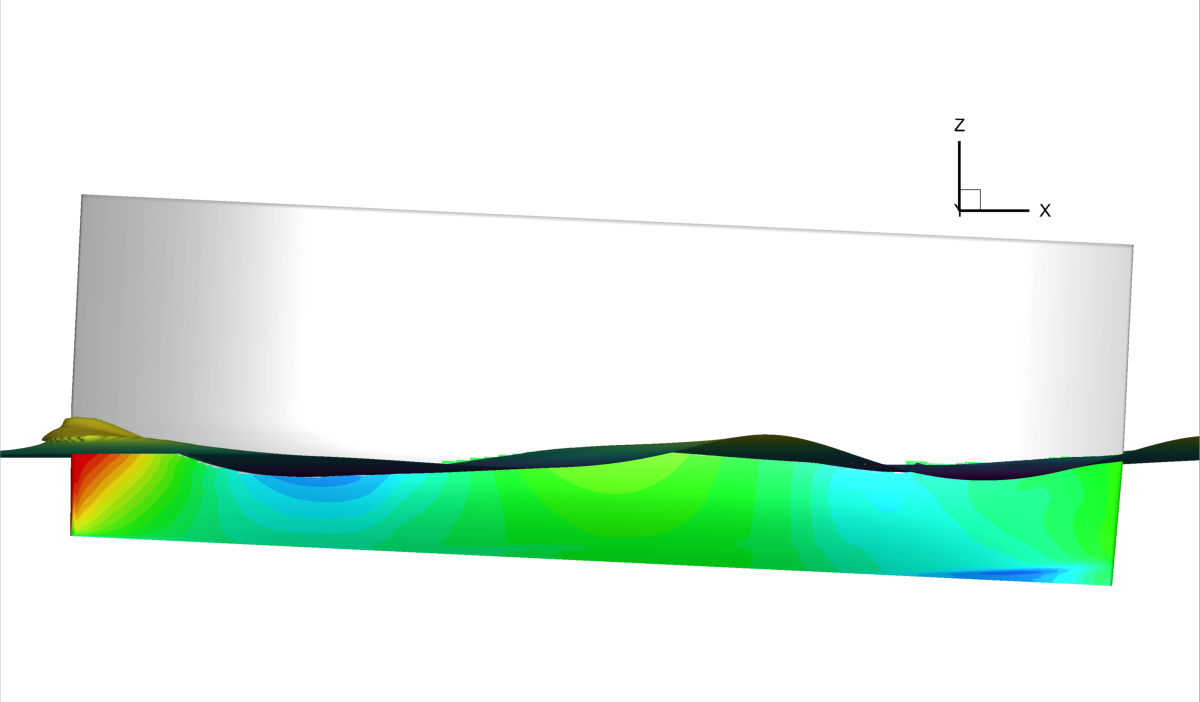}}   \hfill 
\subfloat[RANSE inner, $\nabla = 58$kg]{\includegraphics[width=0.33\textwidth]{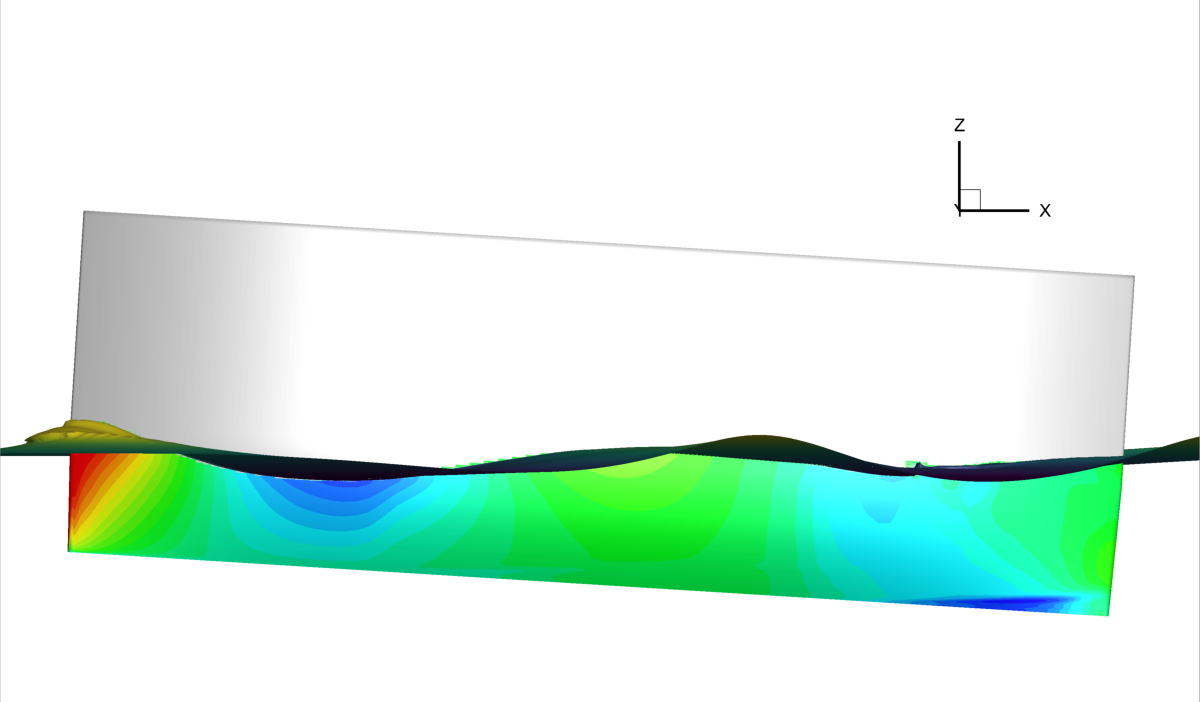}}   \hfill \mbox{}  \\
\mbox{} \hfill
\subfloat[PF outer, $\nabla = 37$kg]{\includegraphics[width=0.33\textwidth]{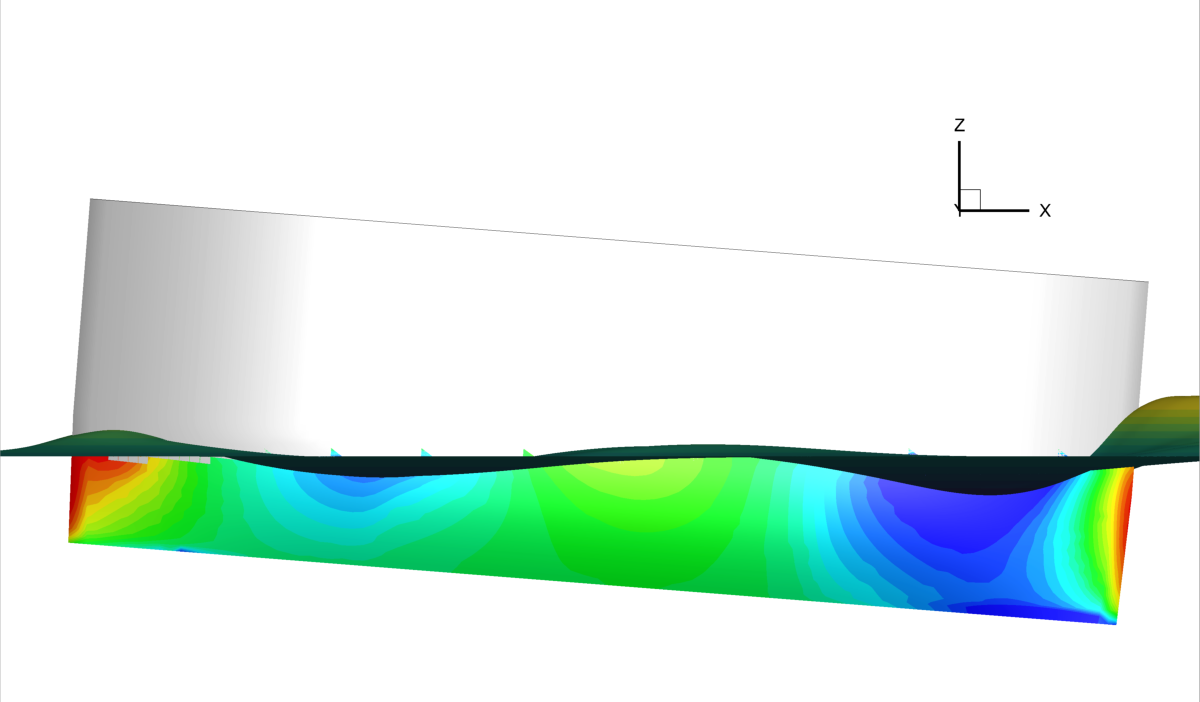}} \hfill 
\subfloat[PF outer, $\nabla = 47.5$kg]{\includegraphics[width=0.33\textwidth]{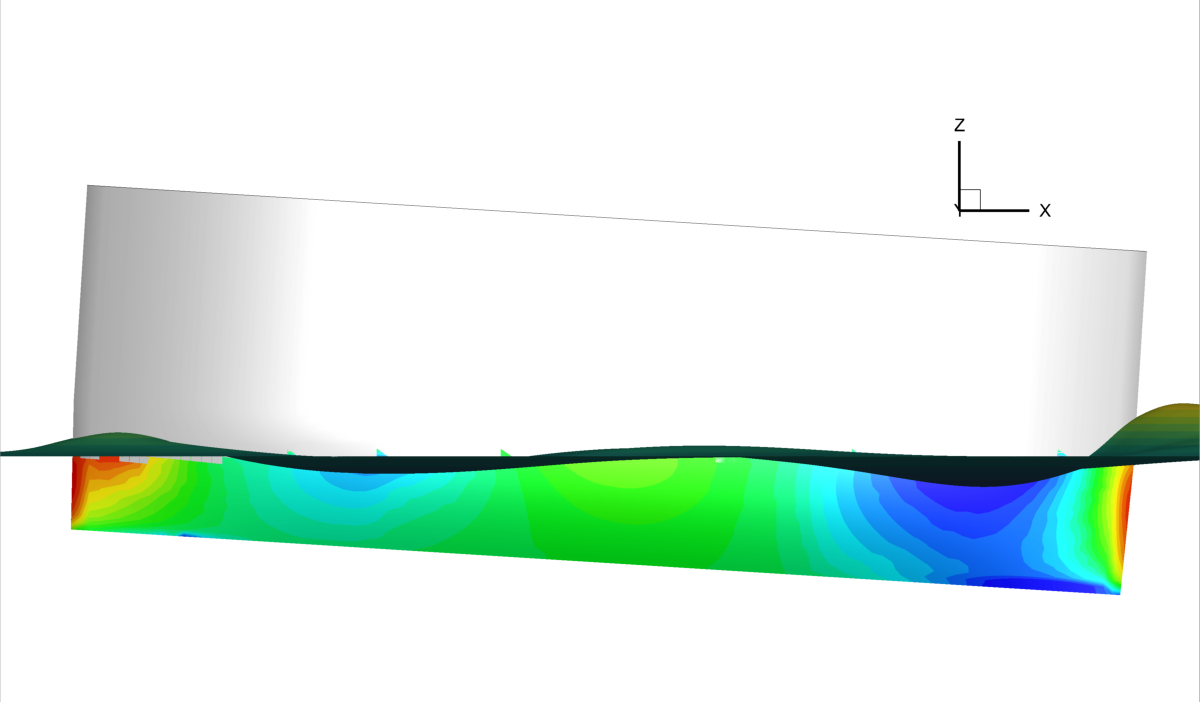}}   \hfill 
\subfloat[PF outer, $\nabla = 58$kg]{\includegraphics[width=0.33\textwidth]{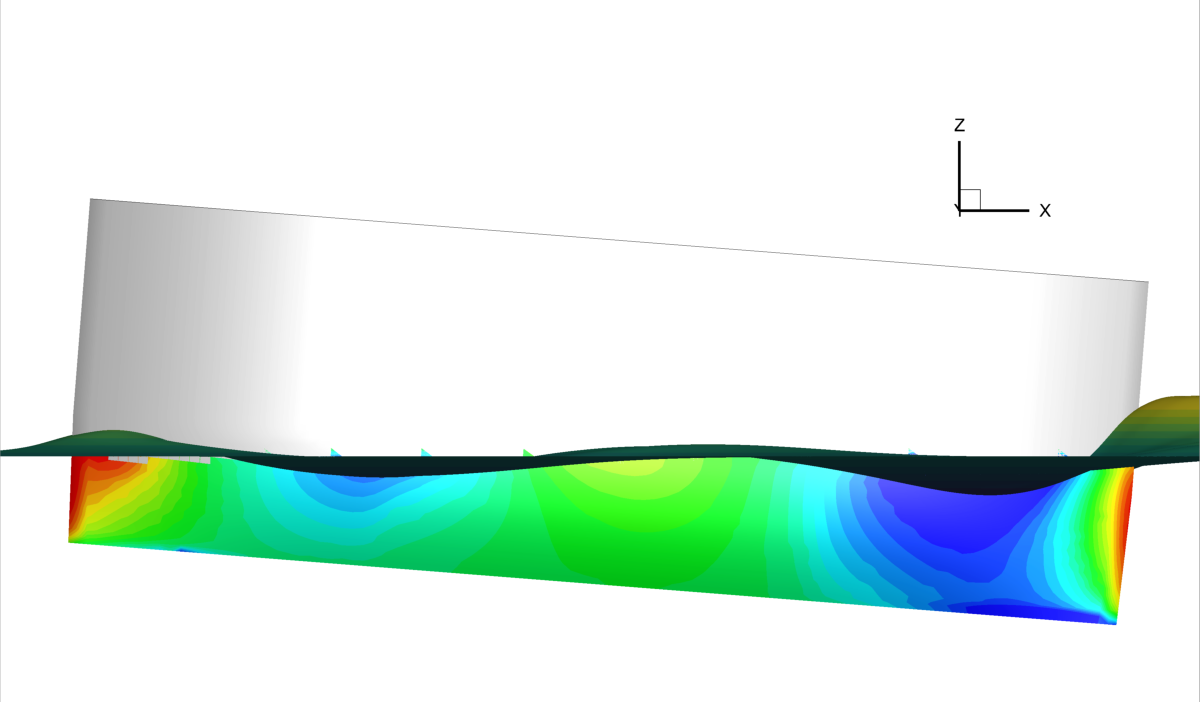}}   \hfill \mbox{}  \\
\mbox{} \hfill
\subfloat[PF inner, $\nabla = 37$kg]{\includegraphics[width=0.33\textwidth]{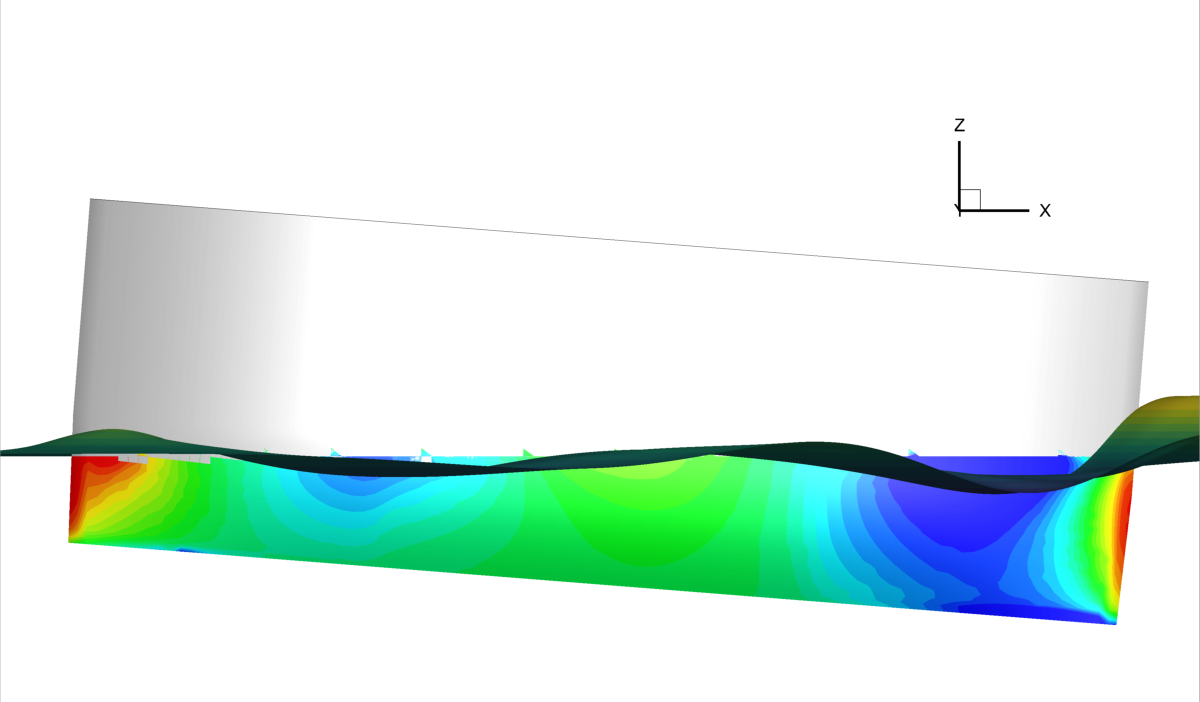}} \hfill 
\subfloat[PF inner, $\nabla = 47.5$kg]{\includegraphics[width=0.33\textwidth]{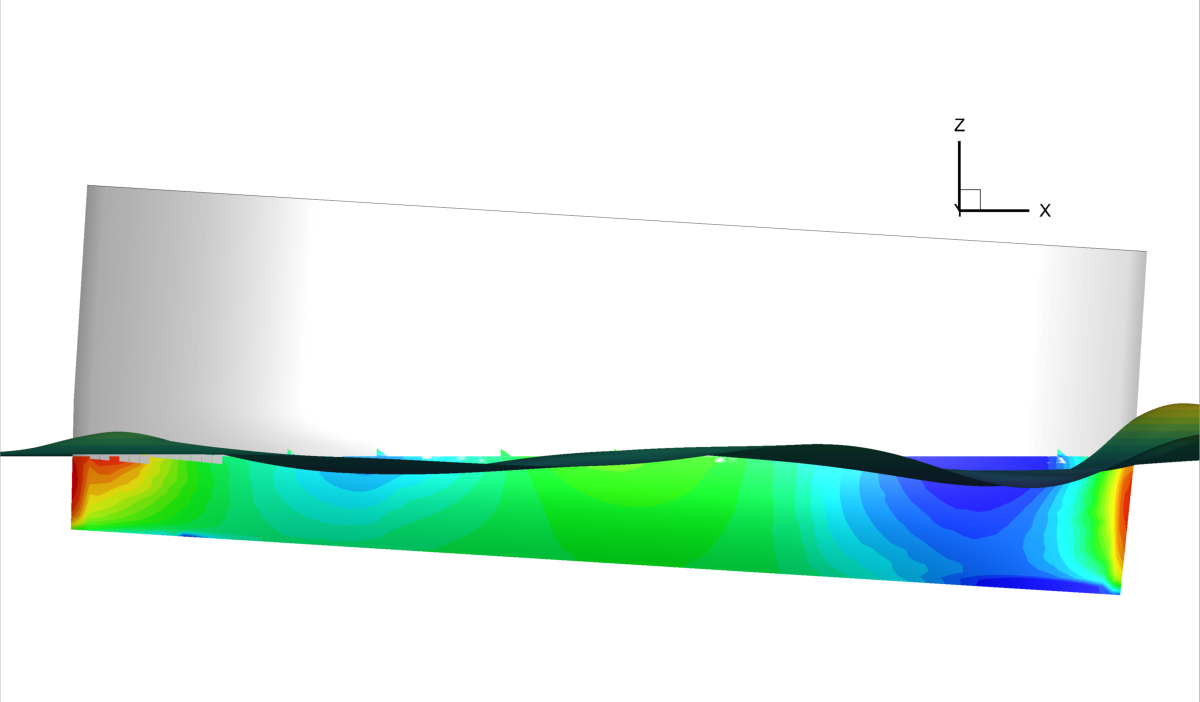}}  \hfill  
\subfloat[PF inner, $\nabla = 58$kg]{\includegraphics[width=0.33\textwidth]{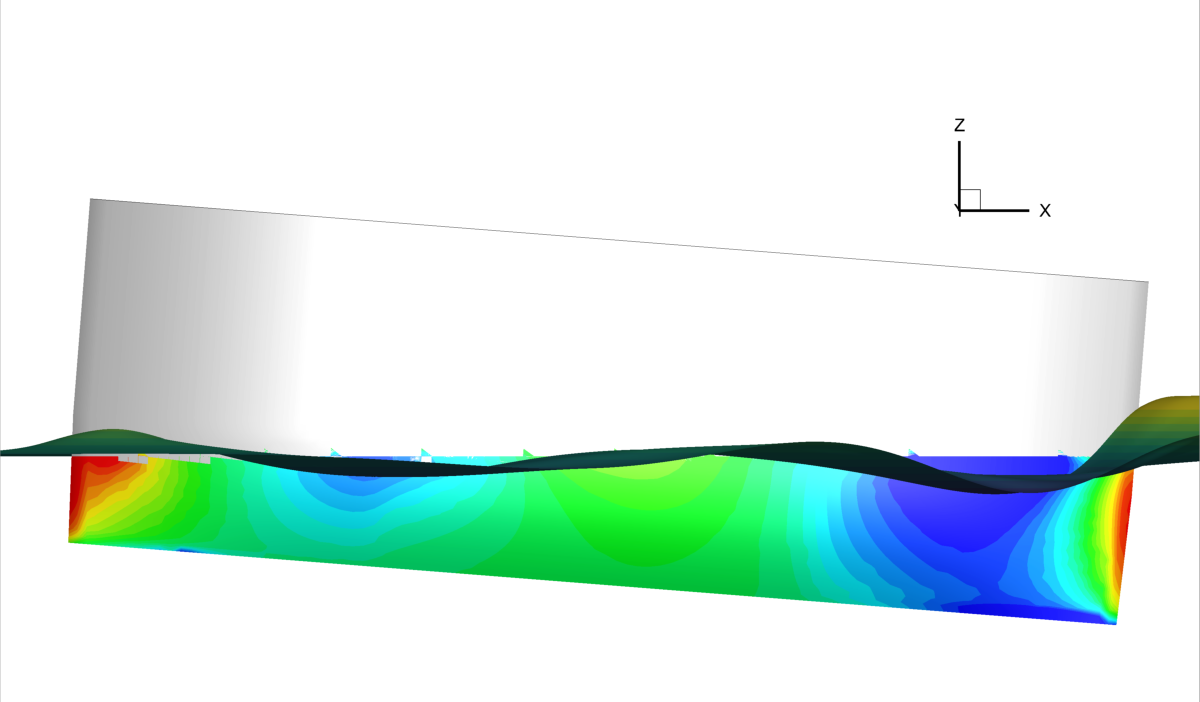}}  \hfill \mbox{}  
\caption{Pressure contour varying the SWAMP payload with $\delta x_G = 3\%$L, side views}\label{fig:PSide3} 
\end{figure}

\begin{figure}[!h] 
\centering 
\mbox{} \hfill
\subfloat[RANSE outer, $\nabla = 37$kg]{\includegraphics[width=0.33\textwidth]{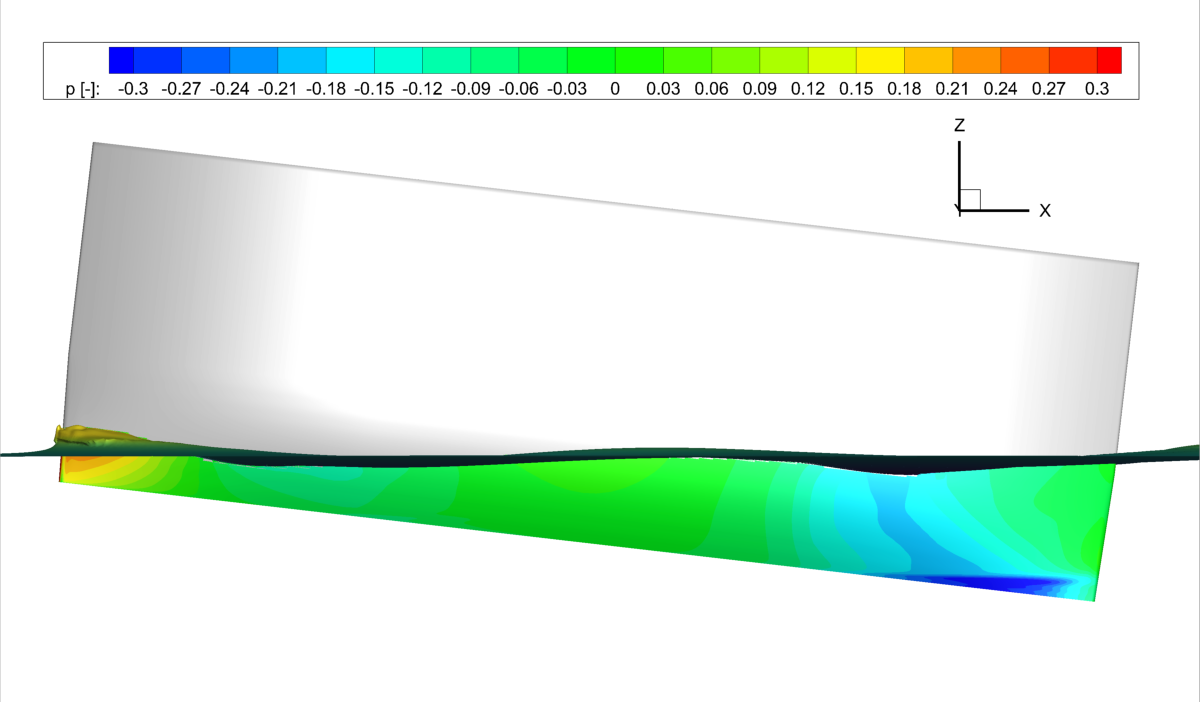}}  \hfill 
\subfloat[RANSE outer, $\nabla = 47.5$kg]{\includegraphics[width=0.33\textwidth]{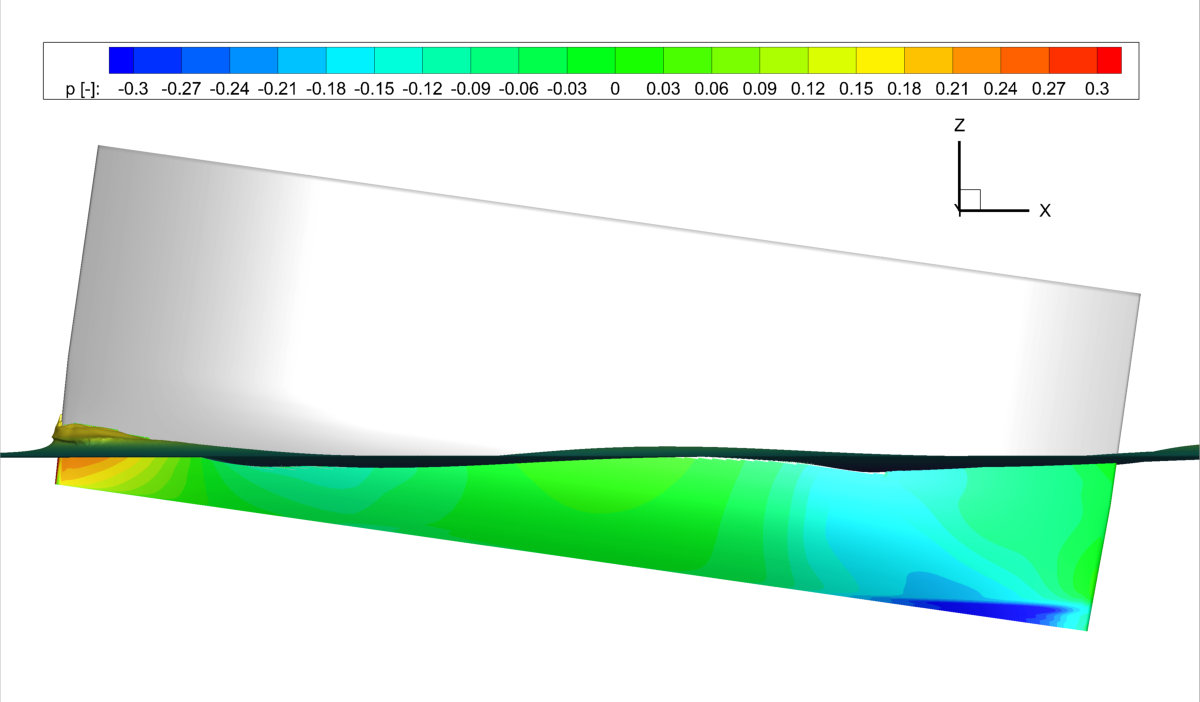}}   \hfill 
\subfloat[RANSE outer, $\nabla = 58$kg]{\includegraphics[width=0.33\textwidth]{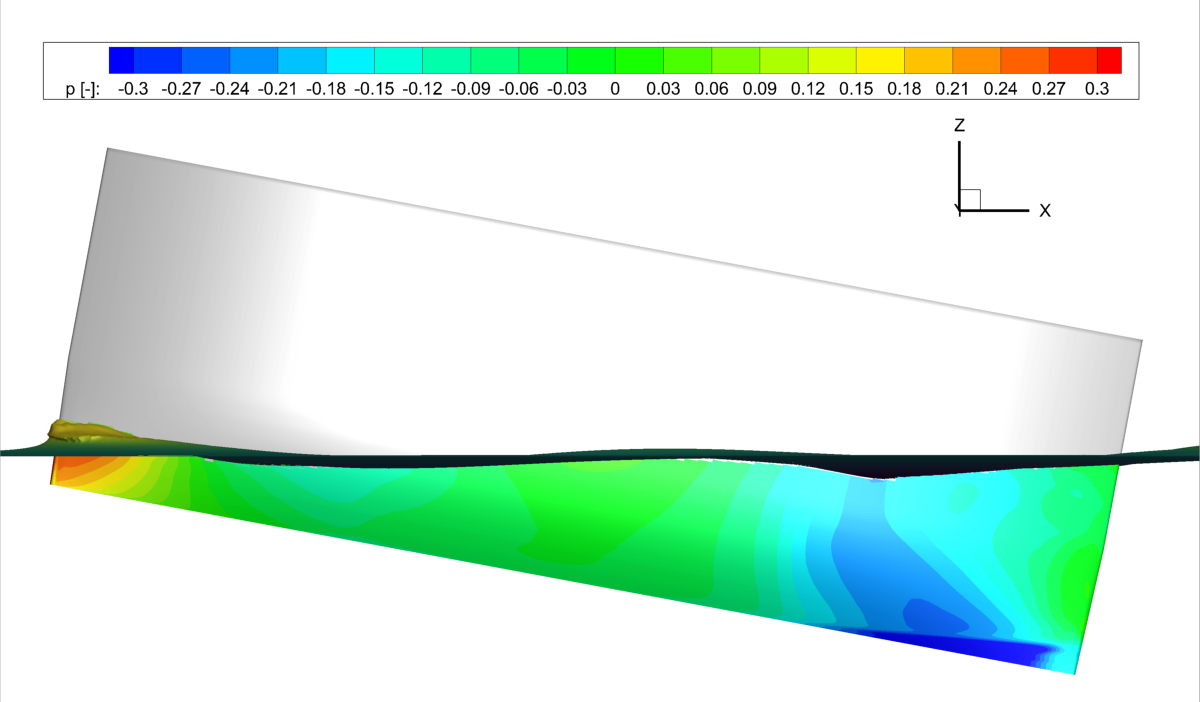}}  \hfill \mbox{}   \\
\mbox{} \hfill
\subfloat[RANSE inner, $\nabla = 37$kg]{\includegraphics[width=0.33\textwidth]{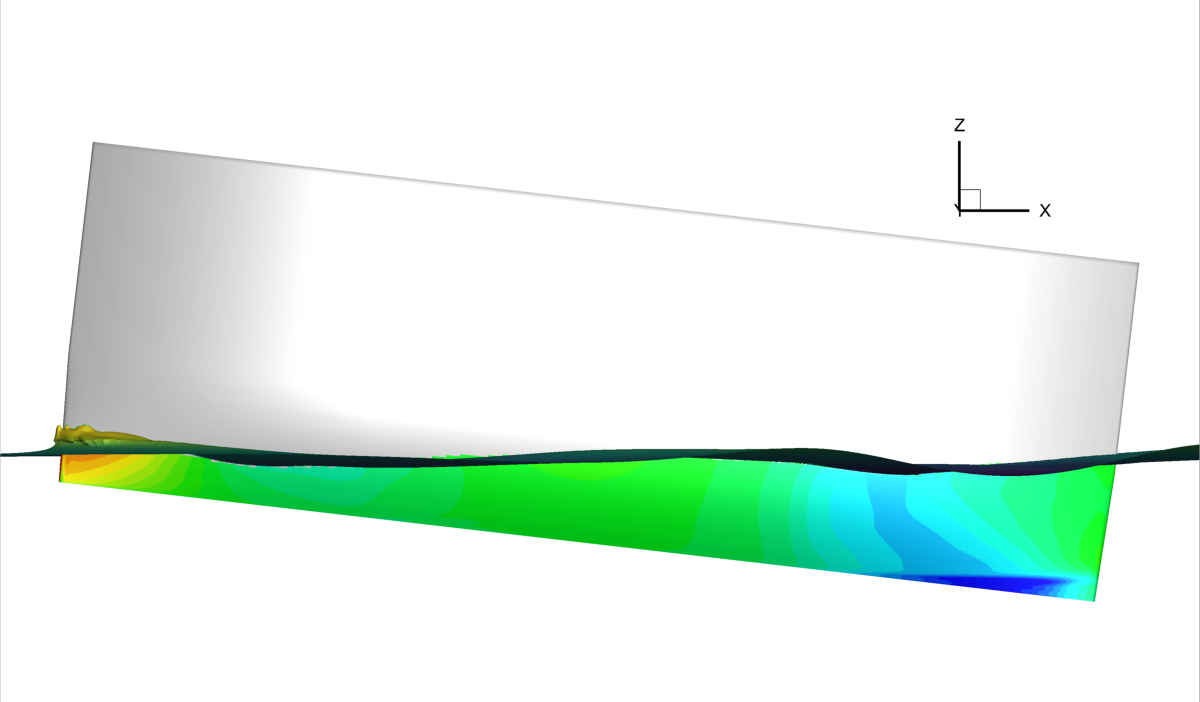}}  \hfill 
\subfloat[RANSE inner, $\nabla = 47.5$kg]{\includegraphics[width=0.33\textwidth]{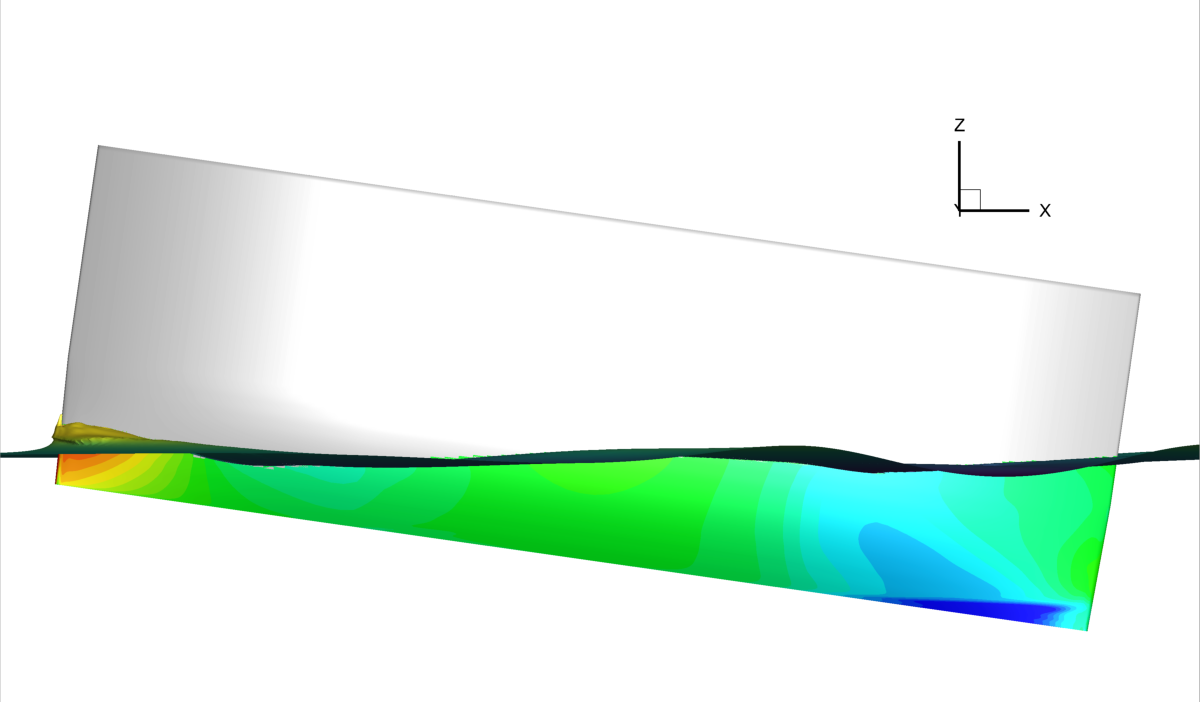}}  \hfill  
\subfloat[RANSE inner, $\nabla = 58$kg]{\includegraphics[width=0.33\textwidth]{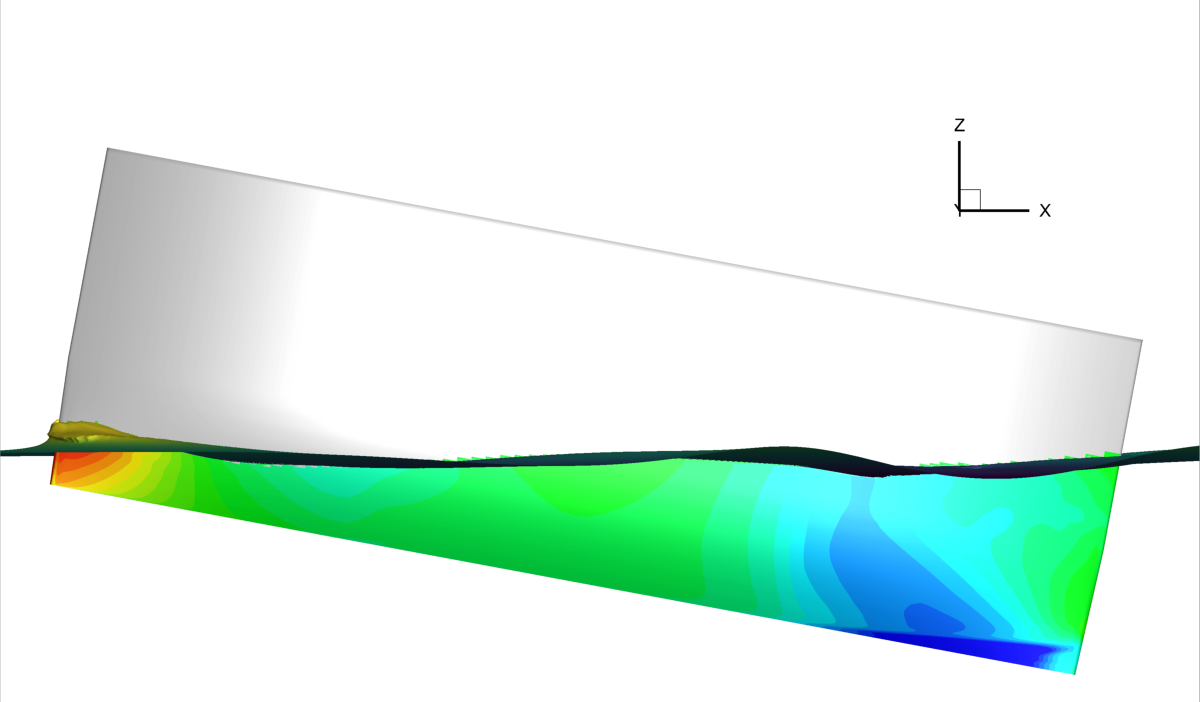}}  \hfill \mbox{}   \\
\mbox{} \hfill
\subfloat[PF outer, $\nabla = 37$kg]{\includegraphics[width=0.33\textwidth]{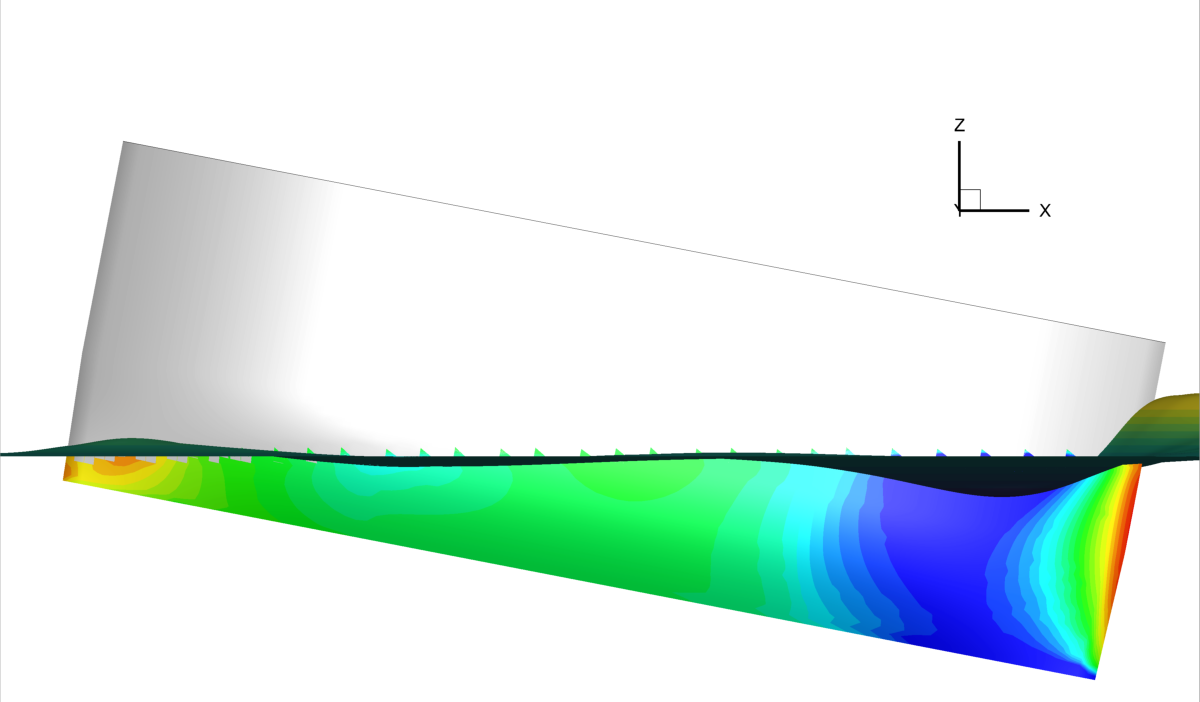}}  \hfill 
\subfloat[PF outer, $\nabla = 47.5$kg]{\includegraphics[width=0.33\textwidth]{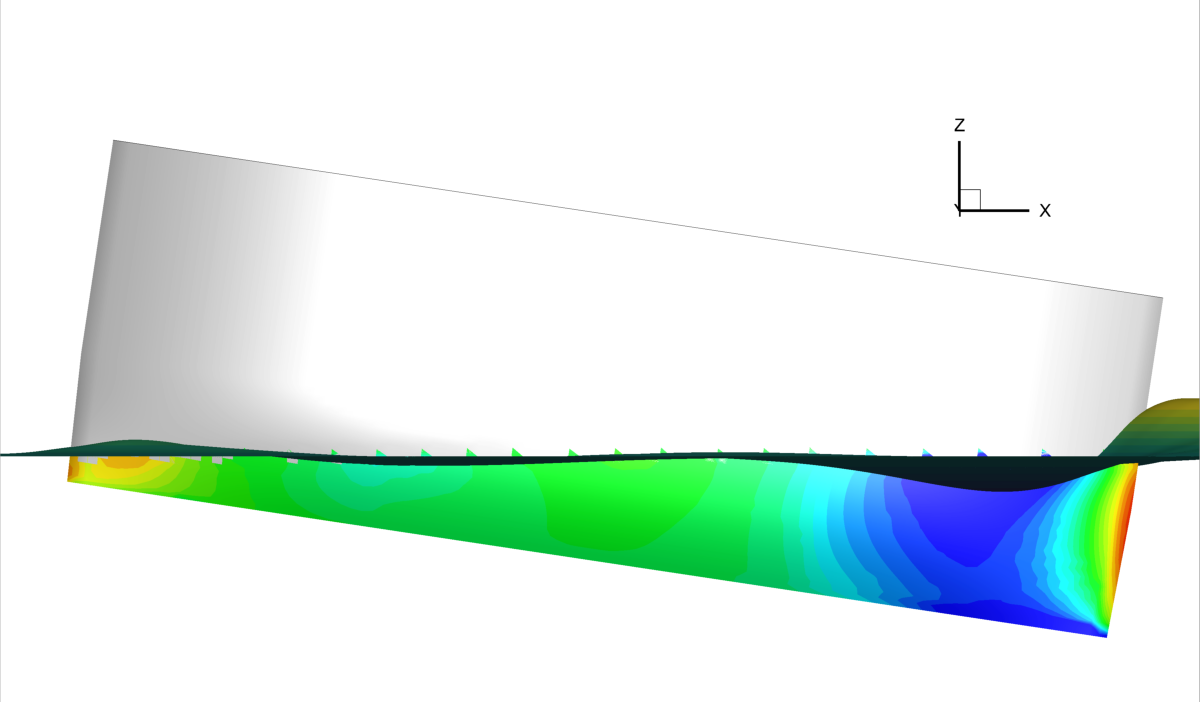}}   \hfill 
\subfloat[PF outer, $\nabla = 58$kg]{\includegraphics[width=0.33\textwidth]{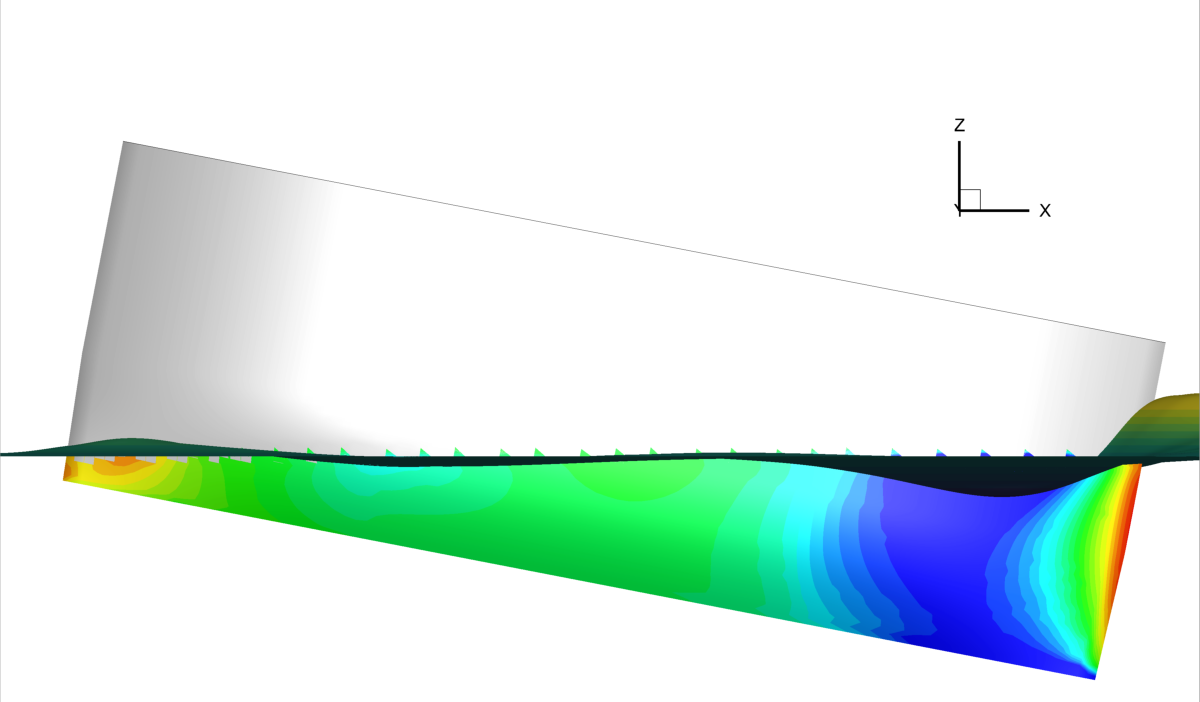}}   \hfill \mbox{}  \\
\mbox{} \hfill
\subfloat[PF inner, $\nabla = 37$kg]{\includegraphics[width=0.33\textwidth]{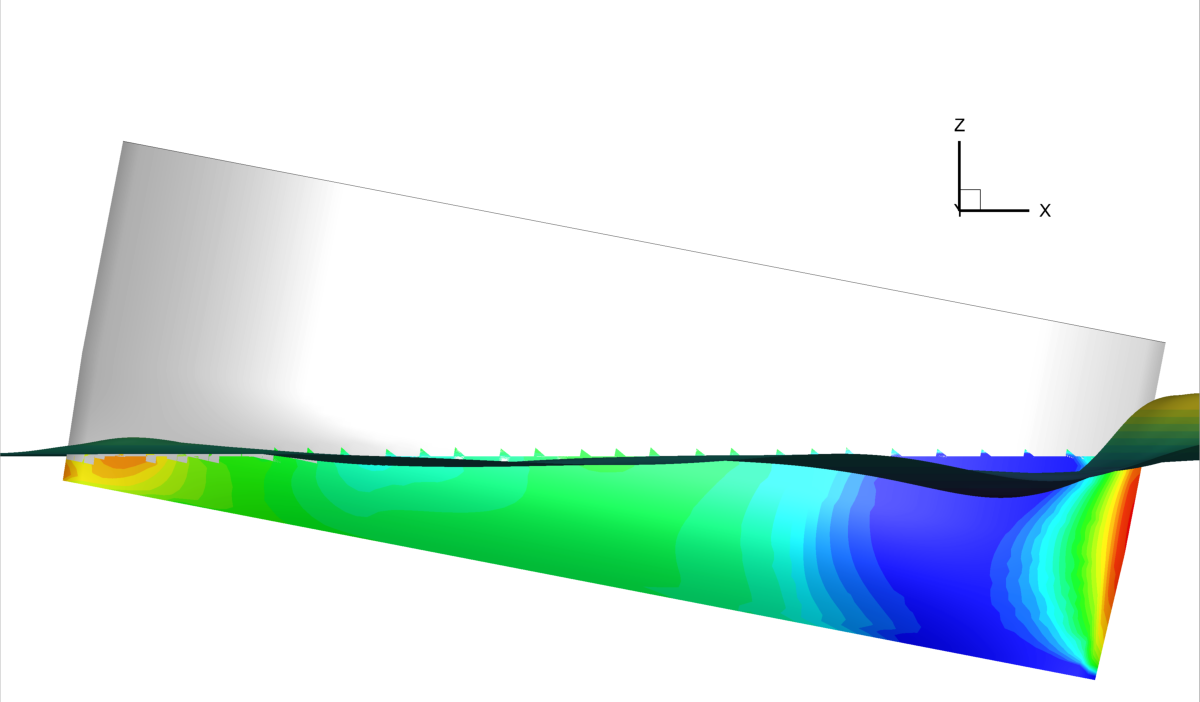}}  \hfill 
\subfloat[PF inner, $\nabla = 47.5$kg]{\includegraphics[width=0.33\textwidth]{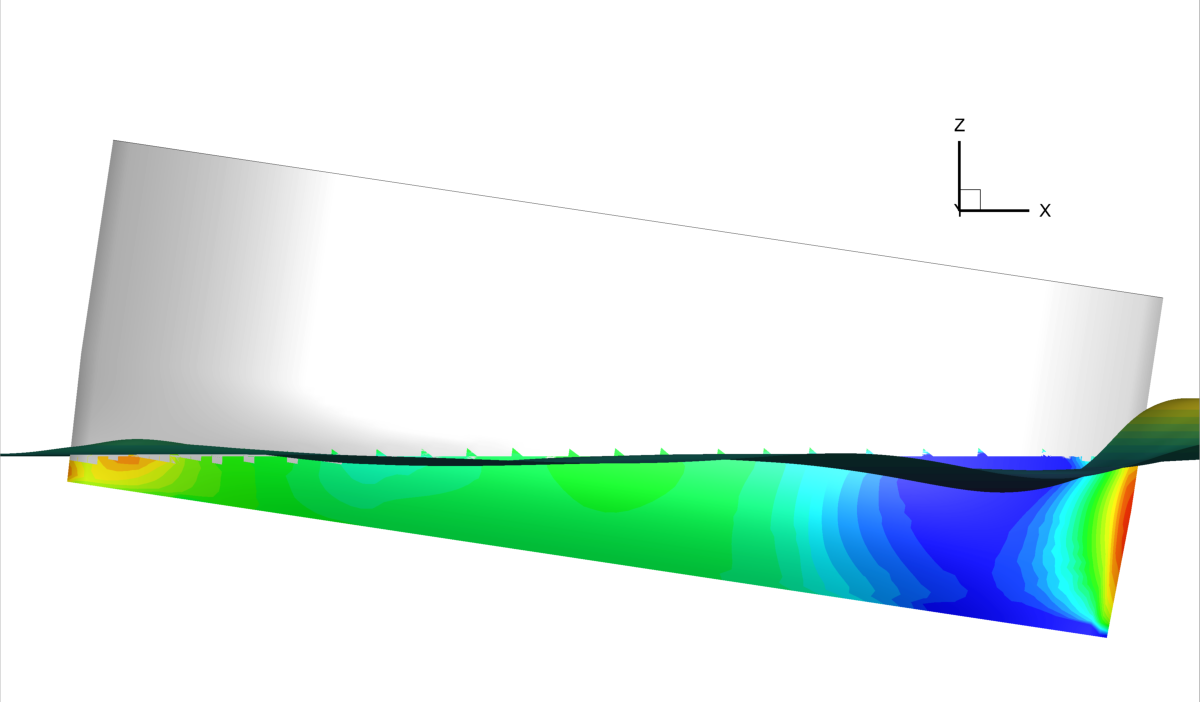}}  \hfill  
\subfloat[PF inner, $\nabla = 58$kg]{\includegraphics[width=0.33\textwidth]{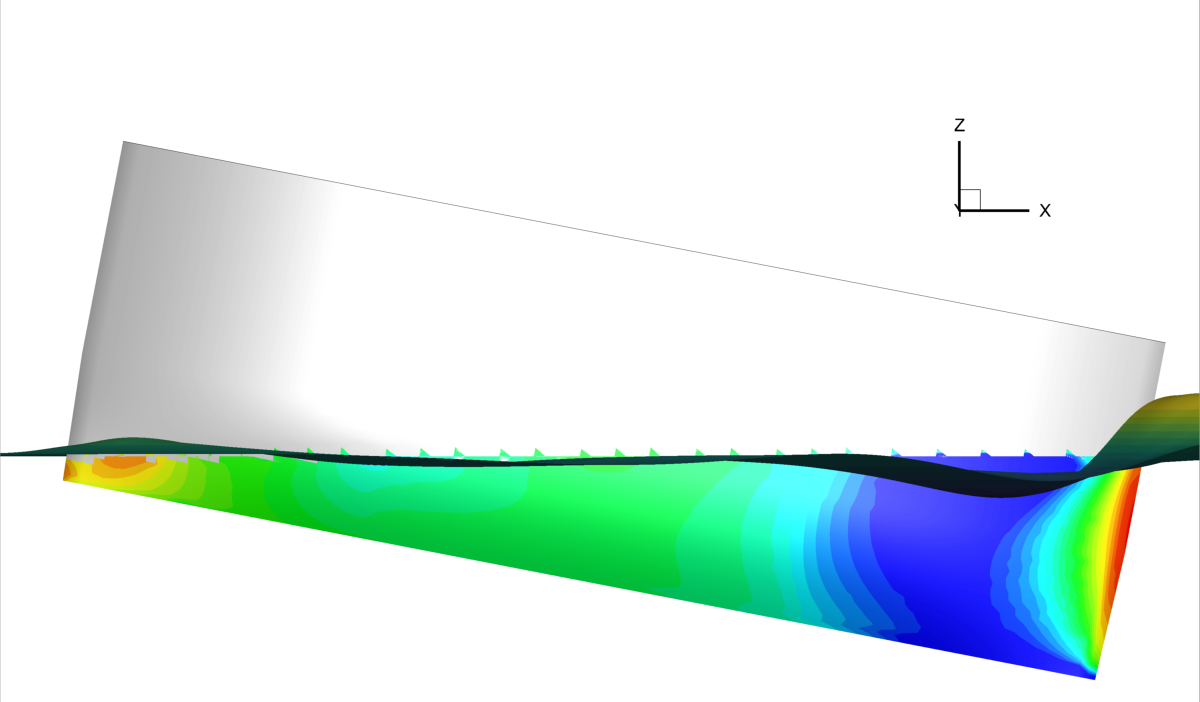}}  \hfill \mbox{}  
\caption{Pressure contour varying the SWAMP payload with $\delta x_G = 7.5\%$L, side views}\label{fig:PSide75} 
\end{figure}

Figures \ref{fig:PSide-3}-\ref{fig:PSide75} show the pressure contour on the outer and inner side of the hull, varying the SWAMP payload and keeping $\delta x_G$ constant and equal to $-3\%$L, $0\%$L, $3\%$L, and $7.5\%$L, respectively. 

Figure \ref{fig:PSide-3}a-c show that, on the outer hull, a positive peak is present at the bow, followed by a negative peak on the side of the hull, then the pressure increases and decreases again in the rear part of the hull and finally slightly increases at the stern. At the highest payload the forward negative peak expands towards the keel. Furthermore, as the payload increases a narrow stripe of low pressure forms at the keel of the hull, near the stern. For each payload the bow is lower than the stern. The inner hull shows a pressure distribution very similar to the outer, with pressure value slightly higher than the outer likely due to the interference between the hulls. The inner hull does not show the presence of the narrow low-pressure stripe as the outer.
The PF solver predicts a similar pressure distribution to the RANSE solution in the first half of the hull, but also significantly overestimate the pressure values in the center and rear part of the hull, showing a positive peak at the stern similar to the bow for both the outer and inner sides (see, Figs. \ref{fig:PSide-3}g-l). As in the RANSE solution, at the highest payload the negative peak extends towards the keel. Furthermore, both the outer and inner sides show the narrow stripe of low pressure, suggesting that the predicted interference between the hulls is smaller than in the RANSE solution. The pitch angle is similar between the RANSE and PF solutions.

Figure \ref{fig:PSide0}a-c shows that the same pressure distribution exists as per the $\delta x_G = -3\%$L case but with smaller peak values for both the inner and outer sides. At the highest payload, the negative peak no longer expands towards the keel but a narrow stripe of low pressure on the outer hull starts forming near the stern. For each payload, the bow is slightly lower than the stern.
The PF solver predicts a similar pressure distribution to the RANSE solution in the first half of the hull, but also significantly overestimate the pressure values in the center and rear part of the hull, showing a positive peak at the stern similar to the bow for both sides of the hull (see, Figs. \ref{fig:PSide0}g-l). Differently from the RANSE solution, the negative peak at rear extends towards the keel. Furthermore, the narrow stripe of low pressure is more evident than in the $\delta x_G = -3\%$L case. The pitch angle is opposite between the RANSE and PF solutions.

Figure \ref{fig:PSide3}a-c show that the same pressure distribution exists as per the previous cases but with smaller peak values for both sides of the hull. The narrow stripe of low pressure starts forming near the stern at the lowest payload value and increases as the payload increases. For each payload the bow is higher than the stern.
The PF solver predicts a similar pressure distribution to the RANSE solution in the first half of the hull, but also significantly overestimate the pressure values, especially in rear part of the hull, showing a positive peak at the stern similar to the bow for both sides of the hull (see, Figs. \ref{fig:PSide3}g-l). Furthermore, differently from the RANSE solution, the negative peak at rear extends towards the keel. The pitch value is increased with respect to the previous cases for each payload.

Figure \ref{fig:PSide75}a-c shows that the pressure is different from the previous cases, showing a positive peak at the bow and a negative peak in the rear part for both sides of the hull. The narrow stripe of low pressure is now evident for each payload value and increases as the payload increases, extending towards the free surface. For each payload, the bow is higher than the stern.
The PF solver predicts a similar pressure distribution to the RANSE solution in the first half of the hull, but also significantly overestimate the pressure values, especially in rear part of the hull, still showing a positive peak at the stern similar to the bow for both sides of the hull (see, Figs. \ref{fig:PSide75}g-l). Furthermore, differently from the RANSE solution, the negative peak at rear extends towards the keel. The pitch value is increased with respect to the previous cases for each payload.

\begin{figure}[!h] 
\centering 
\mbox{} \hfill
\subfloat[$\nabla = 37$kg,   $\delta x_G = -3\%$L]{\includegraphics[width=0.33\textwidth]{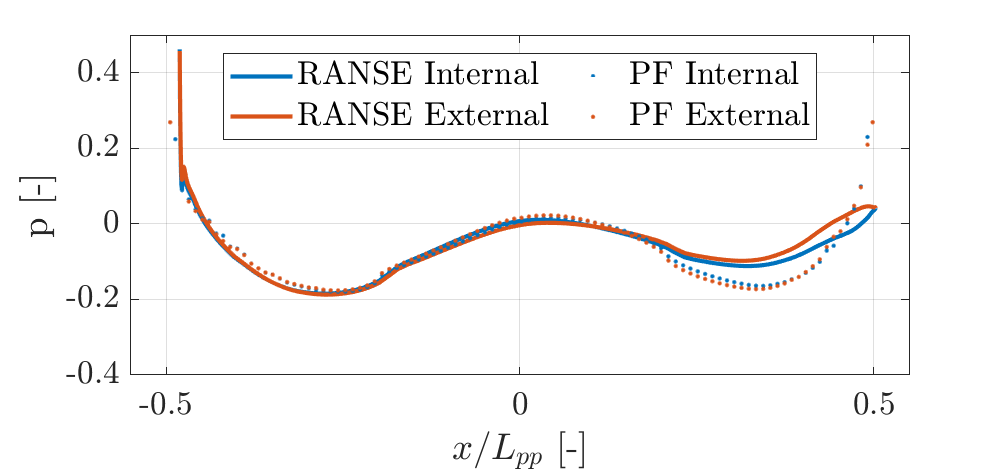}}  \hfill 
\subfloat[$\nabla = 47.5$kg, $\delta x_G = -3\%$L]{\includegraphics[width=0.33\textwidth]{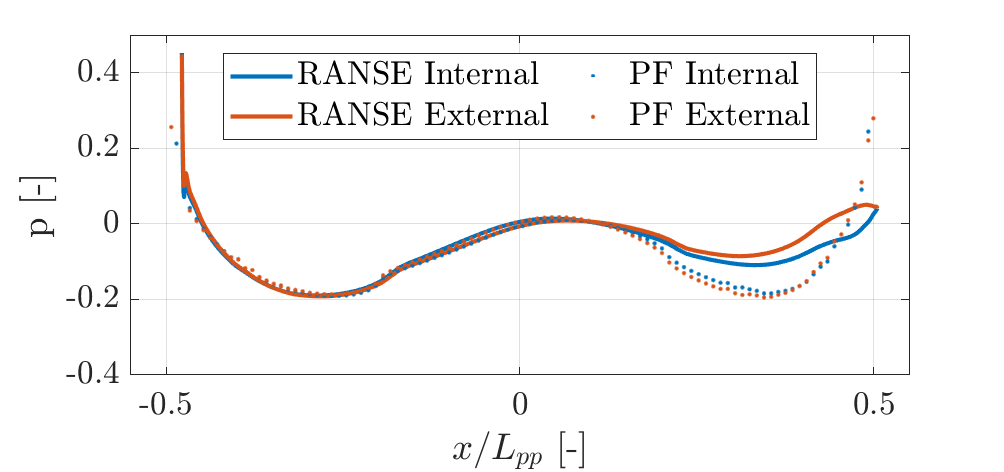}}  \hfill 
\subfloat[$\nabla = 58$kg,   $\delta x_G = -3\%$L]{\includegraphics[width=0.33\textwidth]{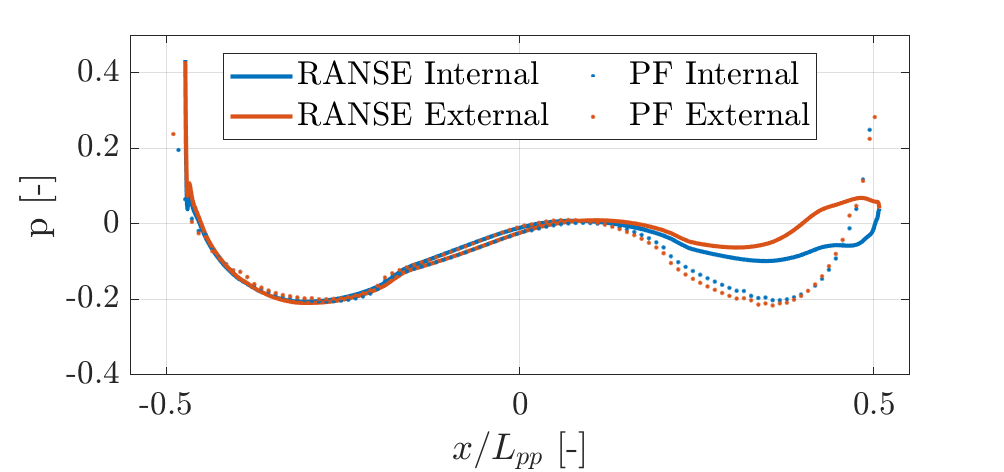}}   \hfill \mbox{}  \\
\mbox{} \hfill
\subfloat[$\nabla = 37$kg,   $\delta x_G = 0\%$L]{\includegraphics[width=0.33\textwidth]{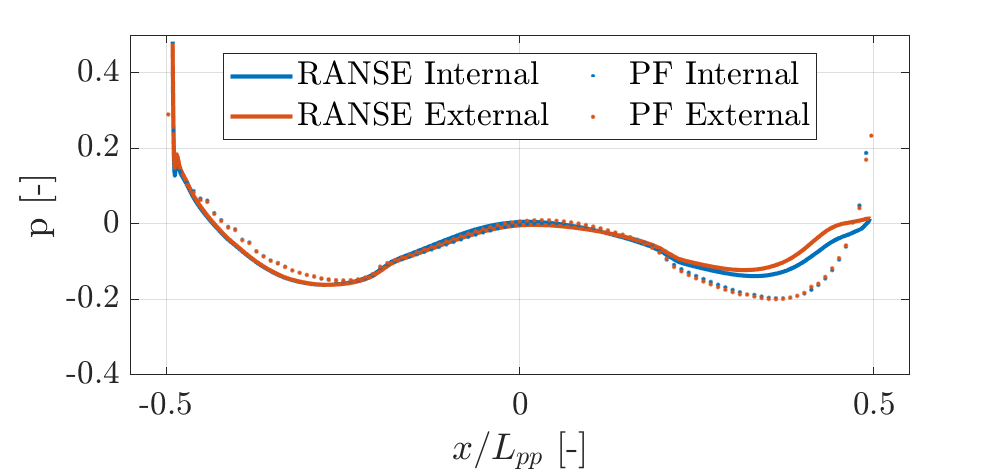}}  \hfill 
\subfloat[$\nabla = 47.5$kg, $\delta x_G = 0\%$L]{\includegraphics[width=0.33\textwidth]{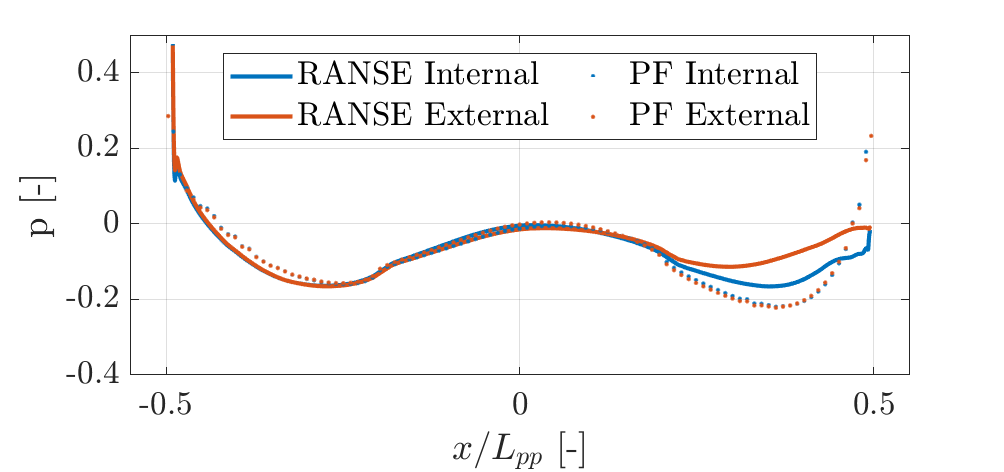}}   \hfill 
\subfloat[$\nabla = 58$kg,   $\delta x_G = 0\%$L]{\includegraphics[width=0.33\textwidth]{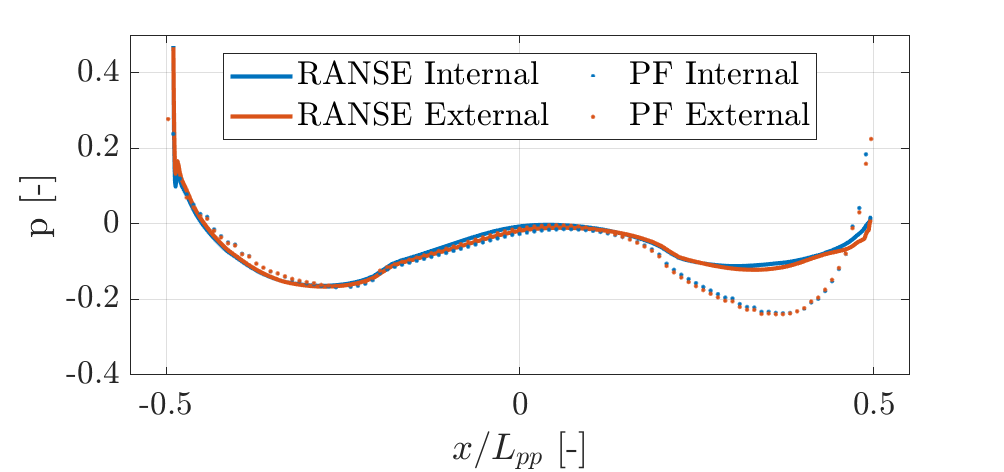}}   \hfill \mbox{}  \\
\mbox{} \hfill
\subfloat[$\nabla = 37$kg,   $\delta x_G = 3\%$L]{\includegraphics[width=0.33\textwidth]{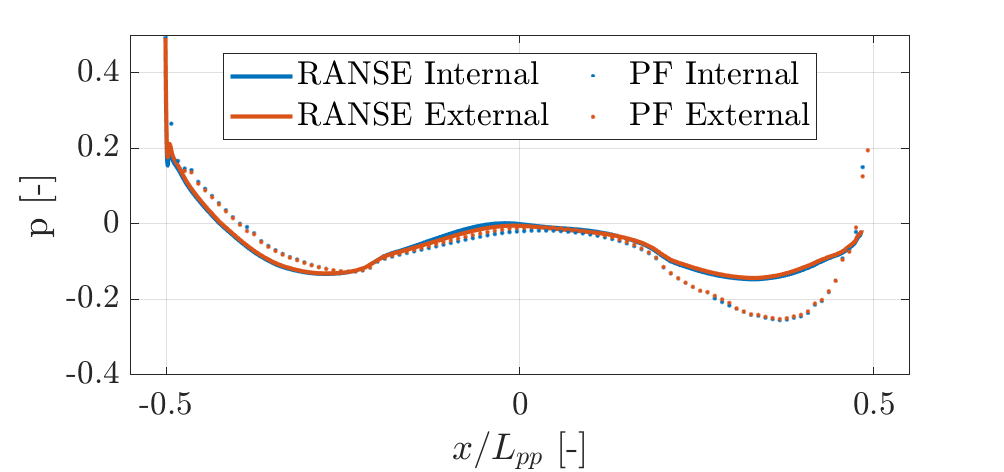}}  \hfill 
\subfloat[$\nabla = 47.5$kg, $\delta x_G = 3\%$L]{\includegraphics[width=0.33\textwidth]{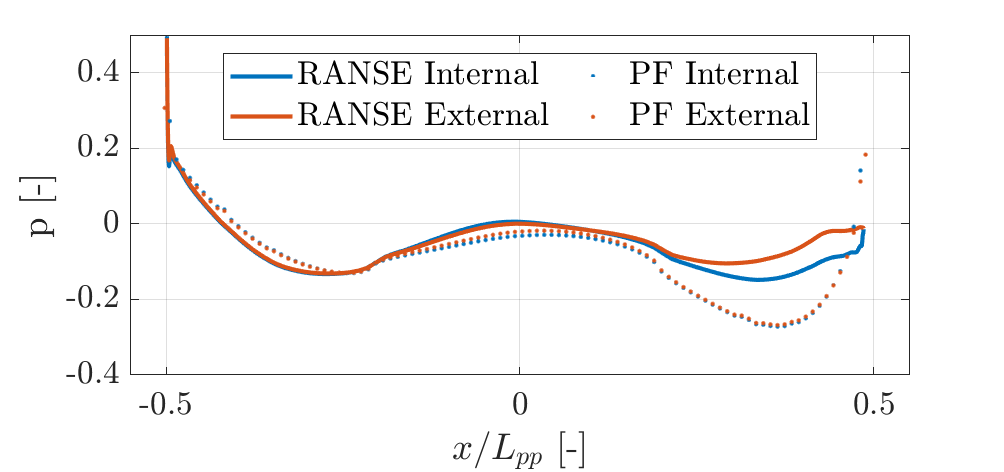}}   \hfill 
\subfloat[$\nabla = 58$kg,   $\delta x_G = 3\%$L]{\includegraphics[width=0.33\textwidth]{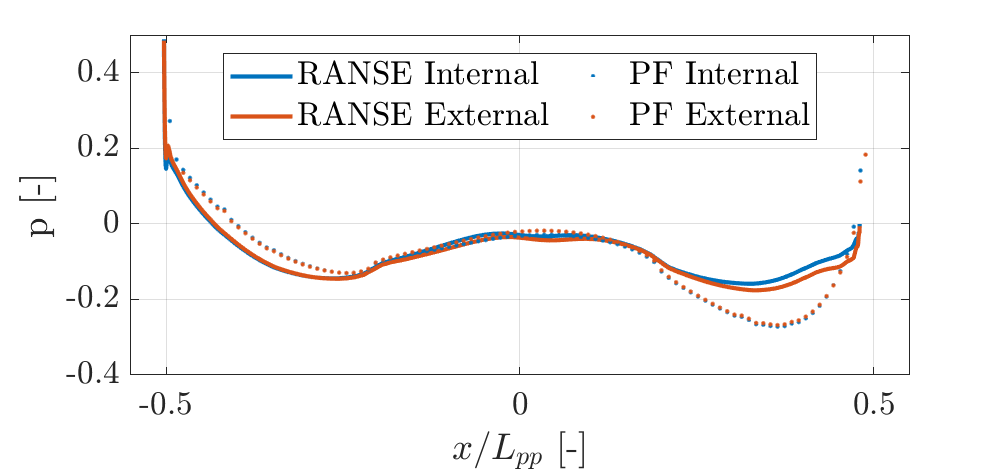}}   \hfill \mbox{}  \\
\mbox{} \hfill
\subfloat[$\nabla = 37$kg,   $\delta x_G = 7.5\%$L]{\includegraphics[width=0.33\textwidth]{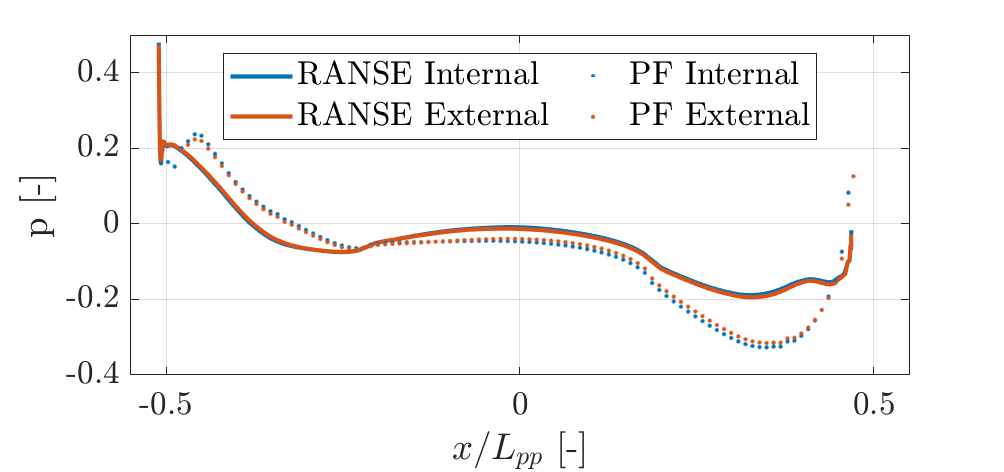}} \hfill 
\subfloat[$\nabla = 47.5$kg, $\delta x_G = 7.5\%$L]{\includegraphics[width=0.33\textwidth]{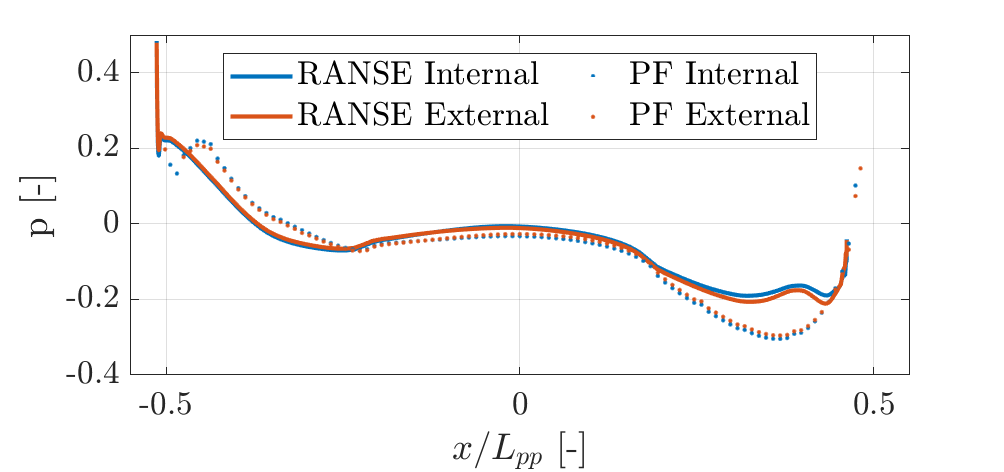}} \hfill 
\subfloat[$\nabla = 58$kg,   $\delta x_G = 7.5\%$L]{\includegraphics[width=0.33\textwidth]{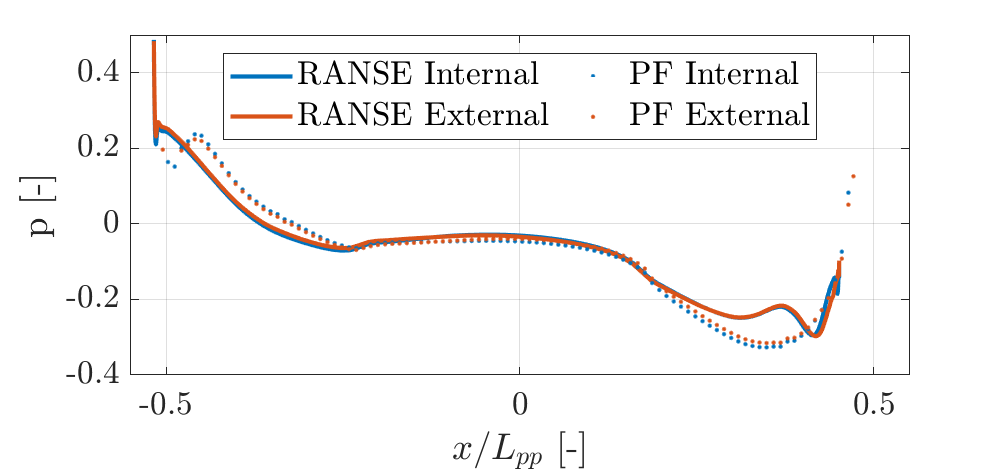}}  \hfill \mbox{}  
\caption{Dynamic pressure value along the hull for the RANSE and PF solutions. The line has a vertical distance from the keel (bottom of the hull) equal to 15\% of the draught of the SWAMP in nominal conditions ($\nabla = 58$kg,   $\delta x_G = 0\%$L)}\label{fig:PLine} 
\end{figure}

Figure \ref{fig:PLine} shows the dynamic pressure value along the hull for the RANSE and PF solutions. The line has a vertical distance from the keel (bottom of the hull) equal to 15\% of the draught of the SWAMP in nominal conditions ($\nabla = 58$kg,   $\delta x_G = 0\%$L). The pressure distributions are similar for $-3\%\mathrm{L} \leq \delta x_G \leq 3\%\mathrm{L}$, whereas for $\delta x_G = 7.5\%$L the pressure gradient is smaller in the first part of the hull and shows a greater peak in the rear. Differences between the inner and outer side of the hull are evident in the rear part of the hull, with the inner pressure being lower than the outer. This is likely due to an acceleration of the fluid between the hulls. 
RANSE and PF solvers are in reasonable agreement in the first part of the hull but predict different pressure values in the rear part. Furthermore, the PF solver shows smaller differences between the inner and outer pressure distributions.

\begin{figure}[!h] 
\centering 
\mbox{} \hfill
\subfloat[$\nabla = 37$kg,   $\delta x_G = -3\%$L]{\includegraphics[width=0.33\textwidth]{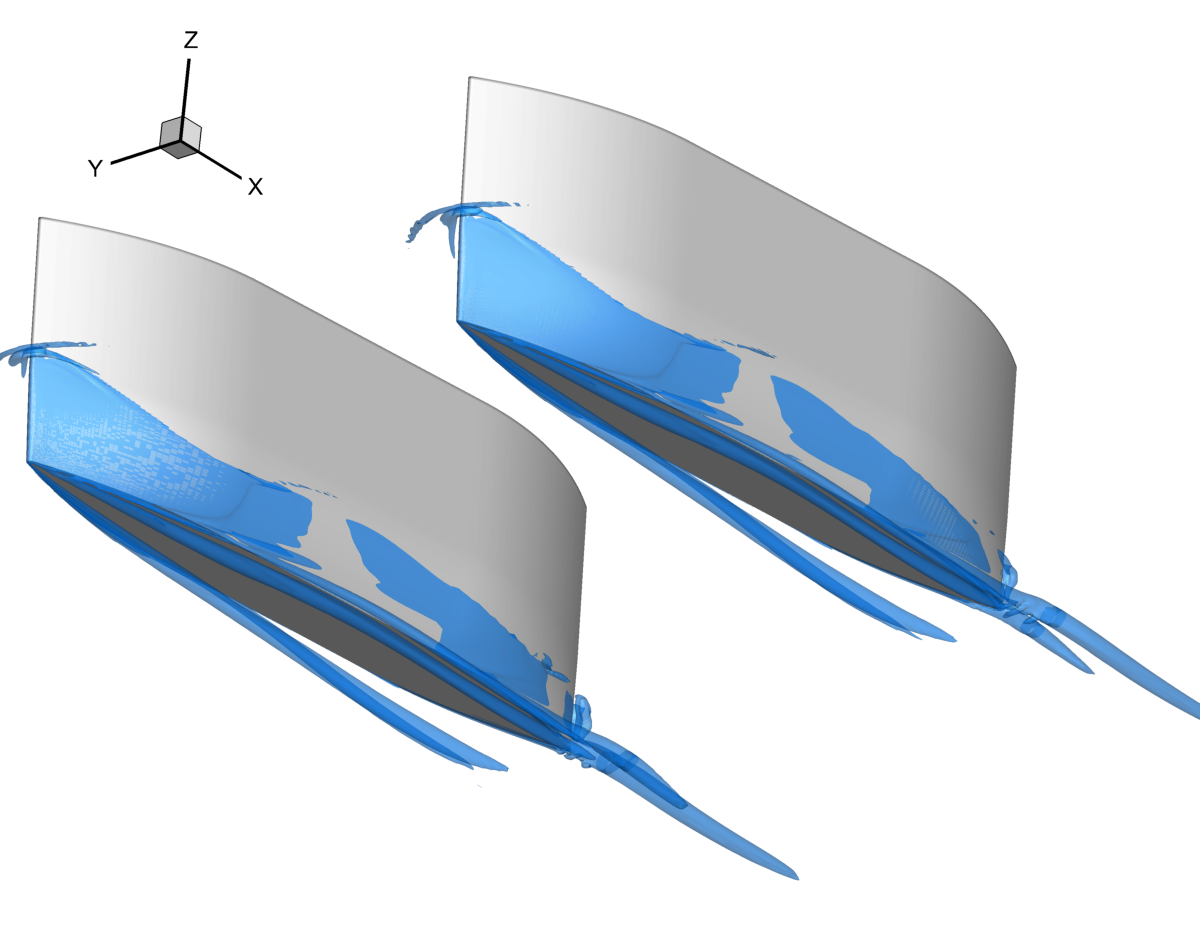}}  \hfill 
\subfloat[$\nabla = 47.5$kg, $\delta x_G = -3\%$L]{\includegraphics[width=0.33\textwidth]{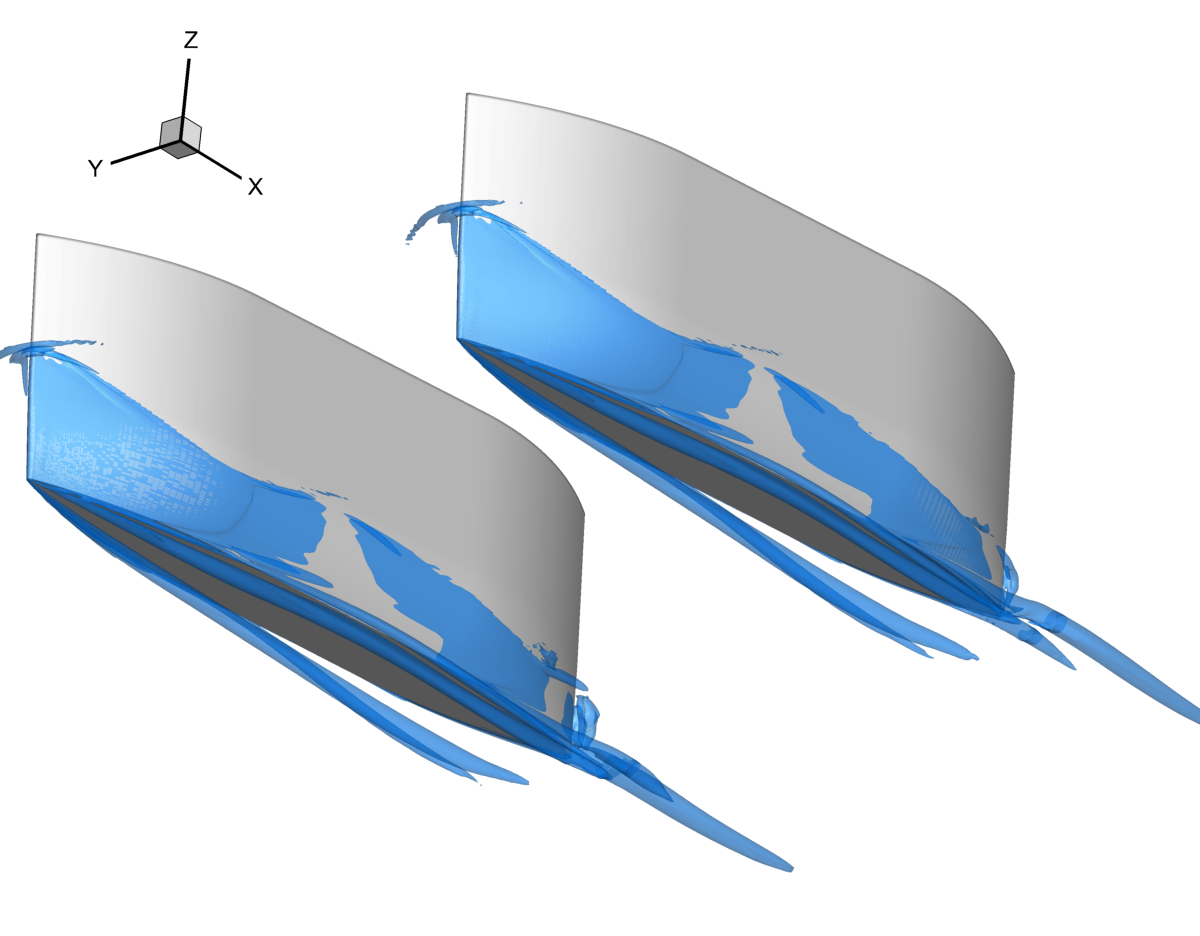}} \hfill   
\subfloat[$\nabla = 58$kg,   $\delta x_G = -3\%$L]{\includegraphics[width=0.33\textwidth]{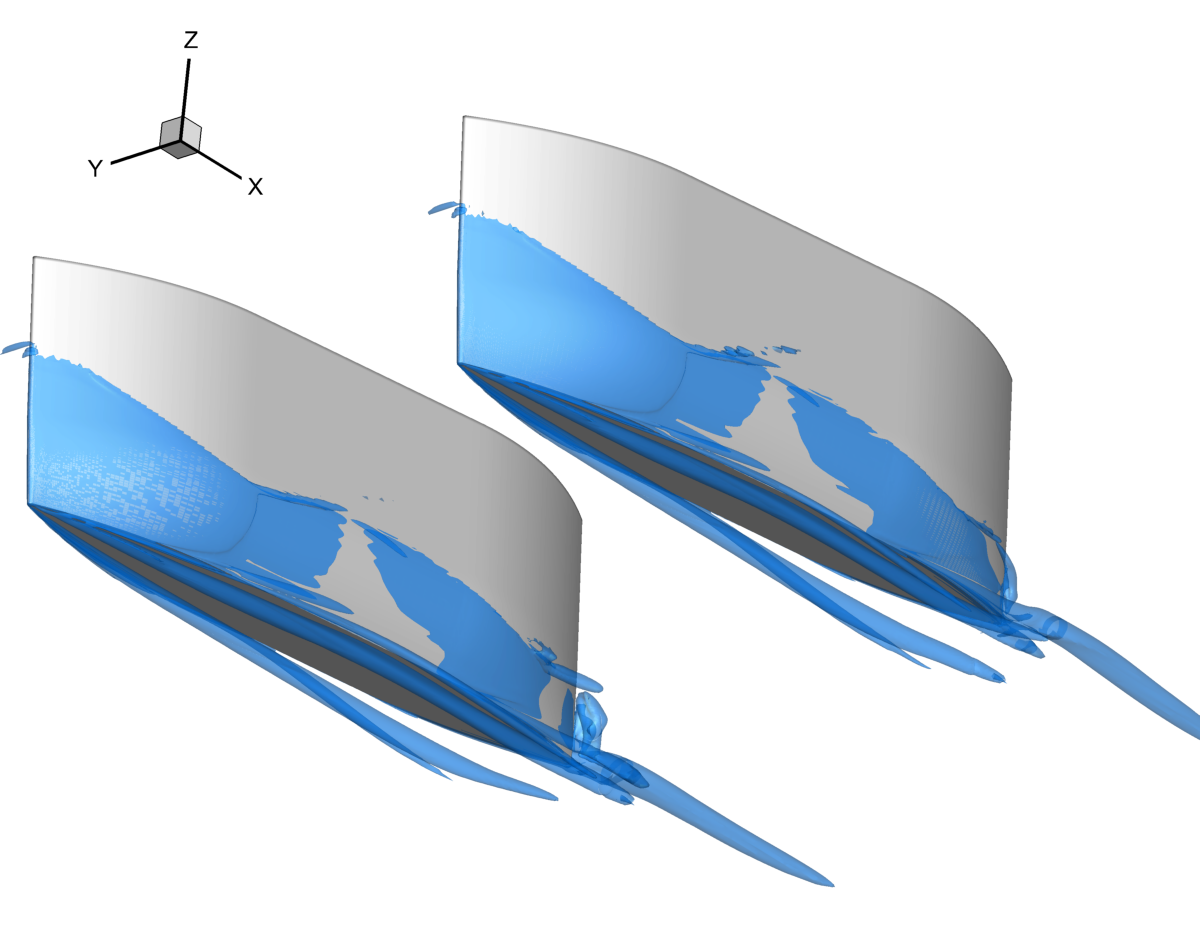}}   \hfill \mbox{}  \\\vspace{-2mm}
\mbox{} \hfill
\subfloat[$\nabla = 37$kg,   $\delta x_G = 0\%$L]{\includegraphics[width=0.33\textwidth]{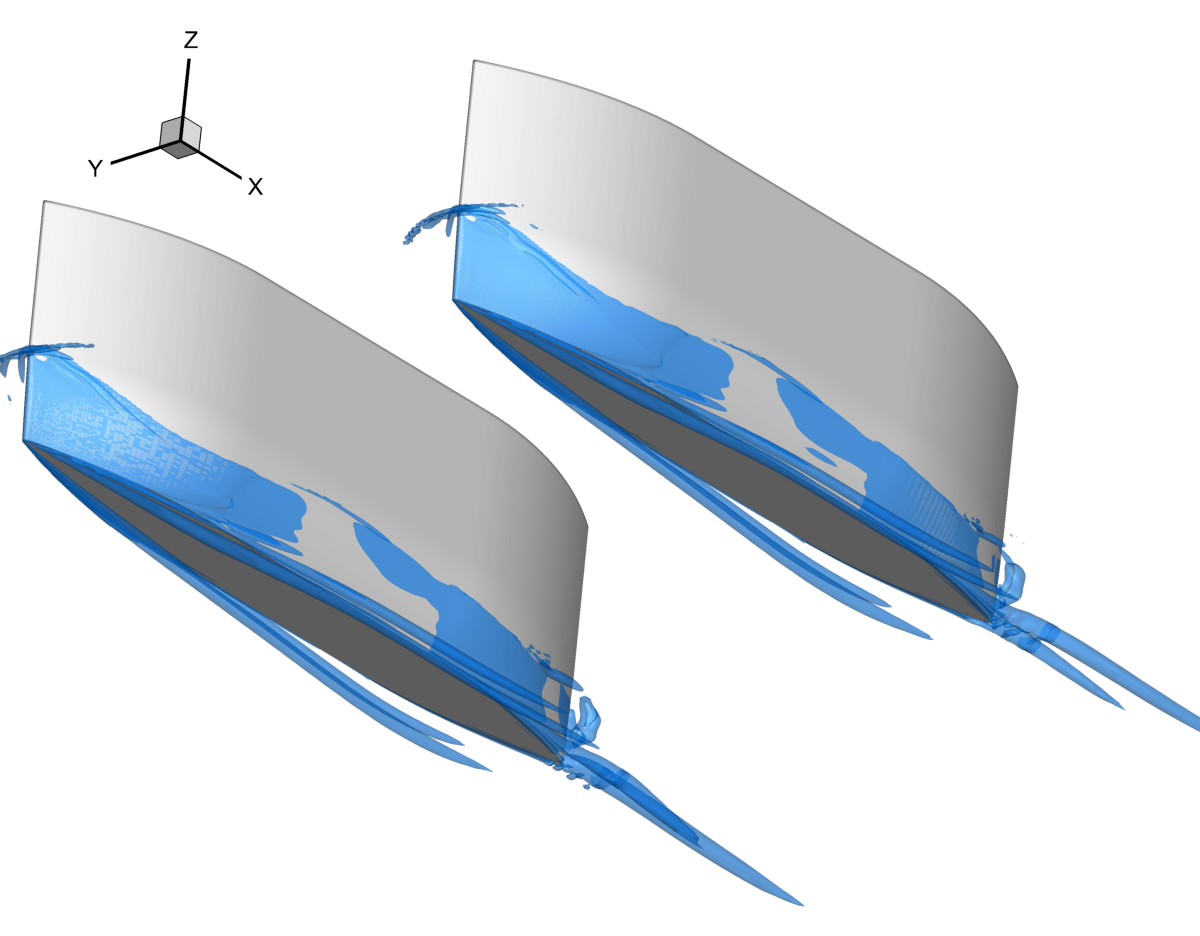}} \hfill  
\subfloat[$\nabla = 47.5$kg, $\delta x_G = 0\%$L]{\includegraphics[width=0.33\textwidth]{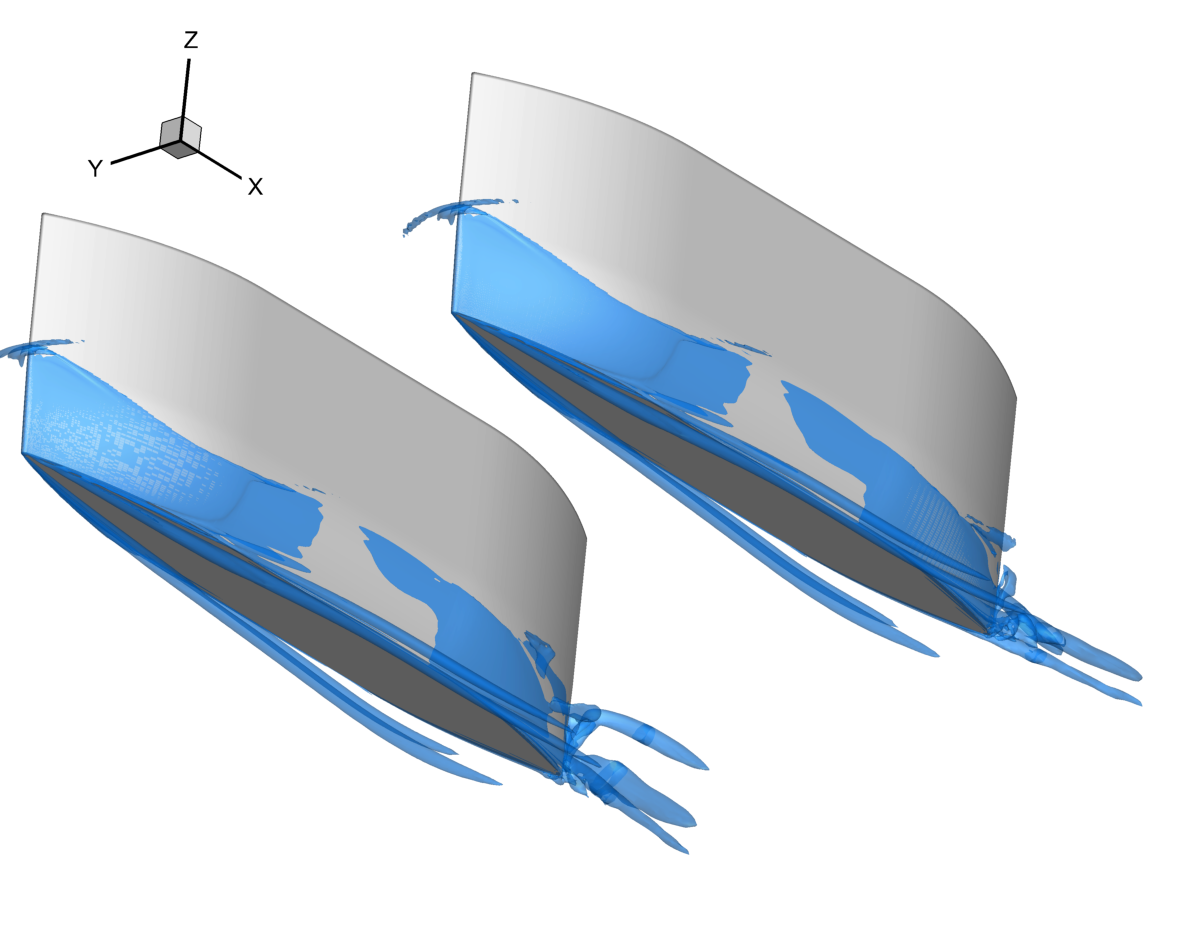}} \hfill   
\subfloat[$\nabla = 58$kg,   $\delta x_G = 0\%$L]{\includegraphics[width=0.33\textwidth]{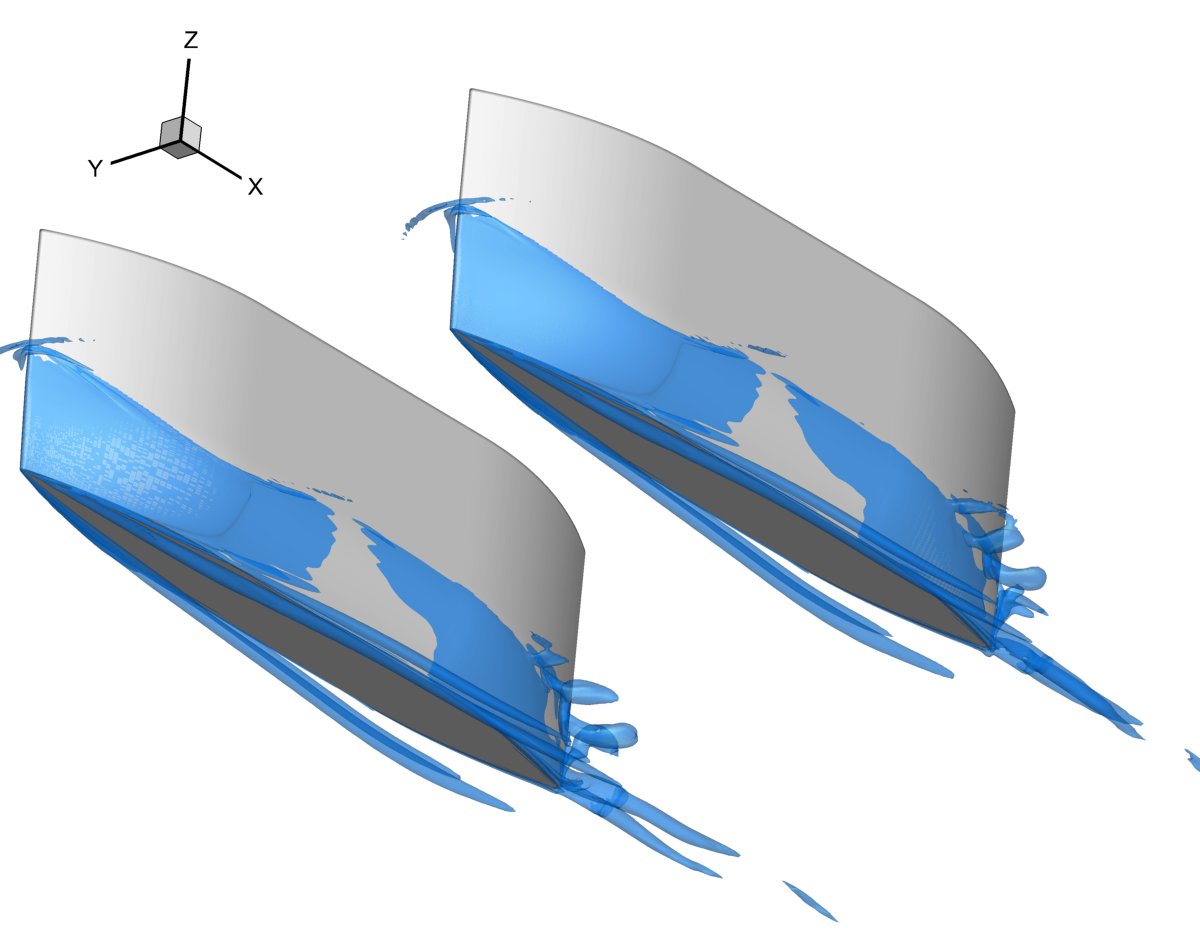}}   \hfill \mbox{}  \\\vspace{-2mm}
\mbox{} \hfill
\subfloat[$\nabla = 37$kg,   $\delta x_G = 3\%$L]{\includegraphics[width=0.33\textwidth]{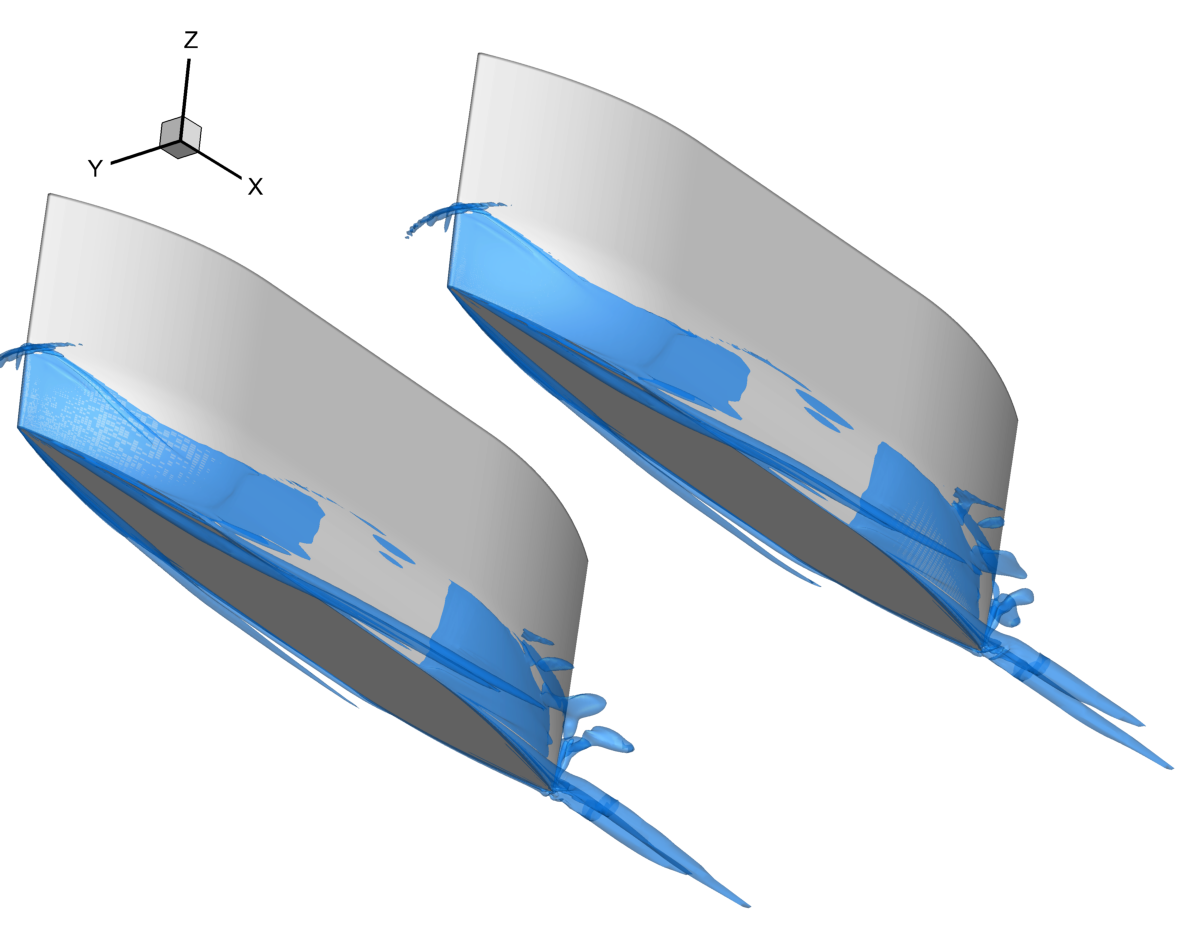}} \hfill  
\subfloat[$\nabla = 47.5$kg, $\delta x_G = 3\%$L]{\includegraphics[width=0.33\textwidth]{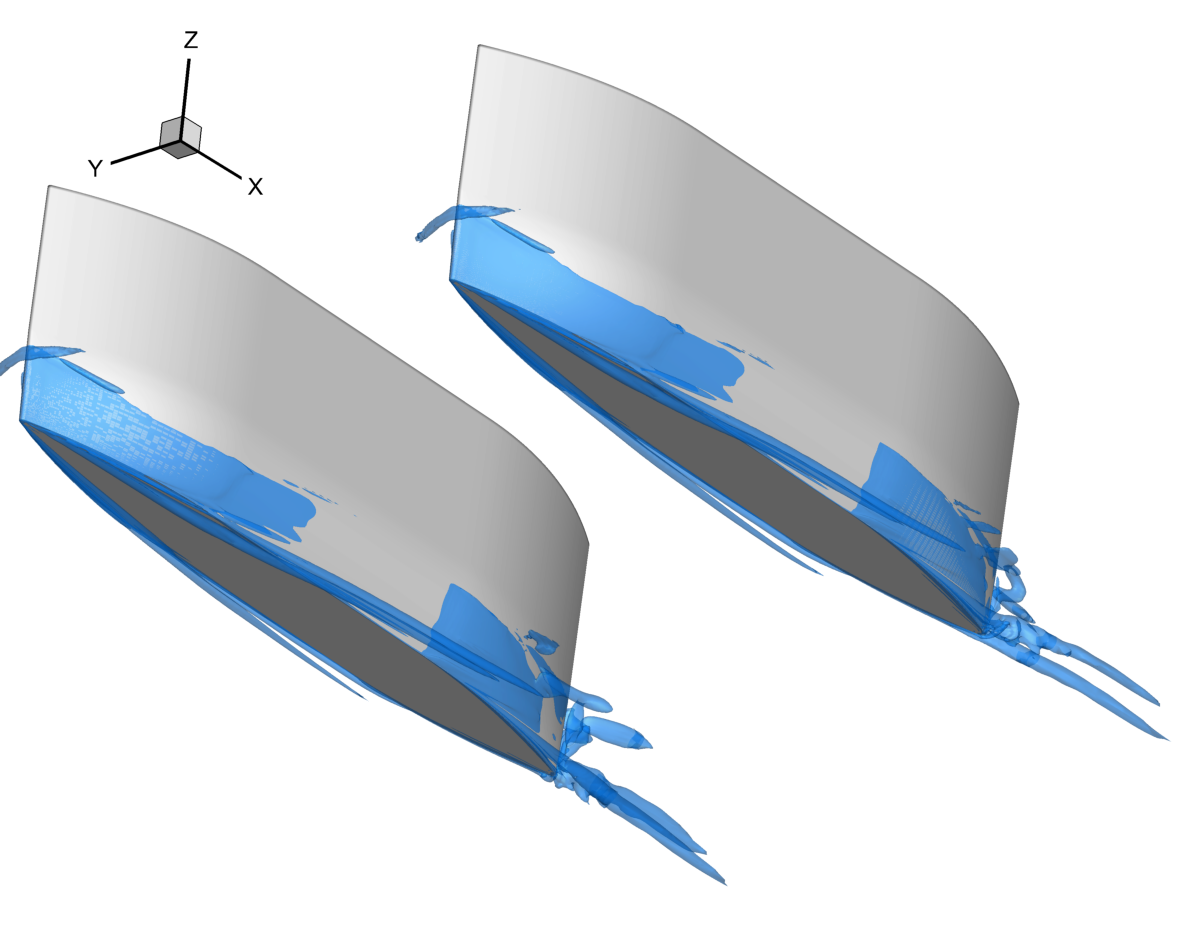}} \hfill   
\subfloat[$\nabla = 58$kg,   $\delta x_G = 3\%$L]{\includegraphics[width=0.33\textwidth]{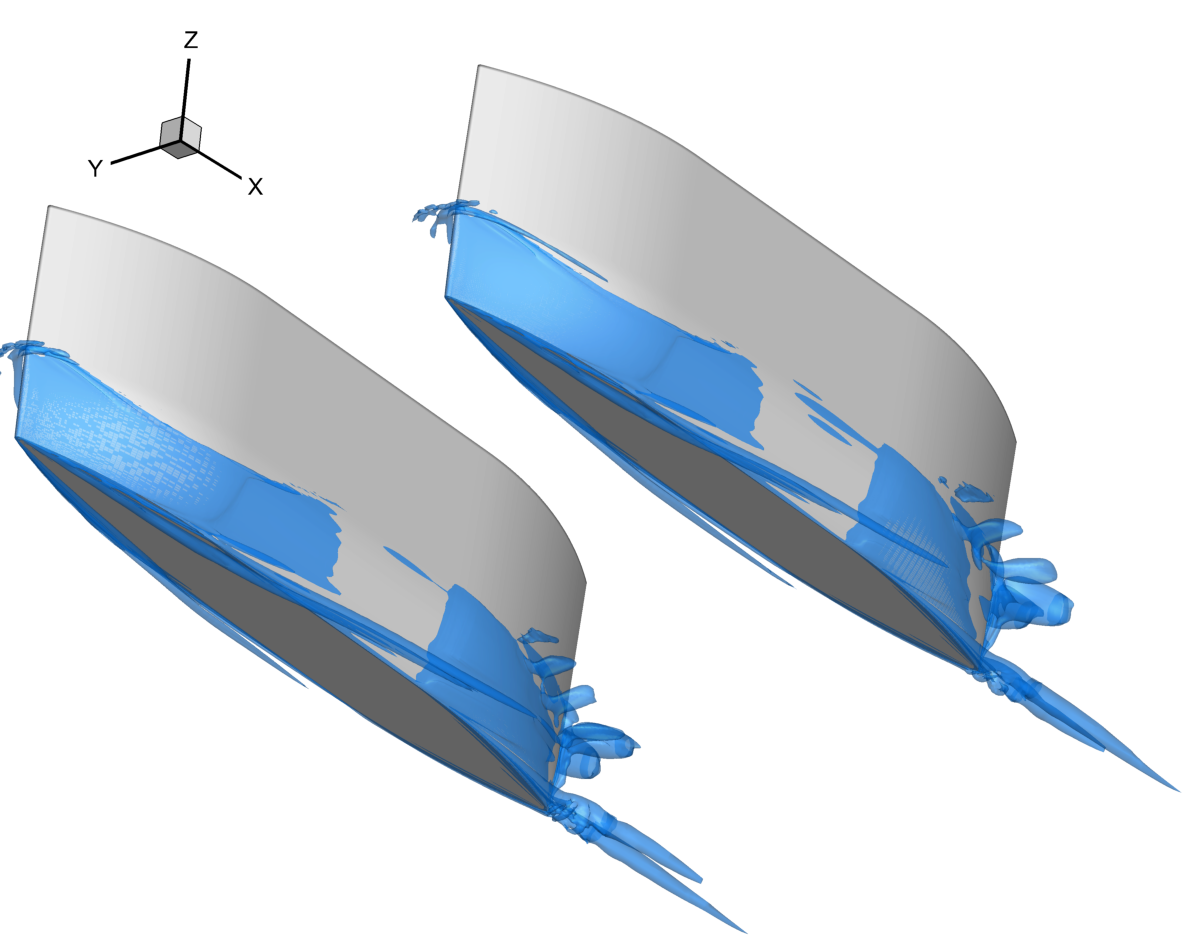}}   \hfill \mbox{}  \\\vspace{-2mm}
\mbox{} \hfill
\subfloat[$\nabla = 37$kg,   $\delta x_G = 7.5\%$L]{\includegraphics[width=0.33\textwidth]{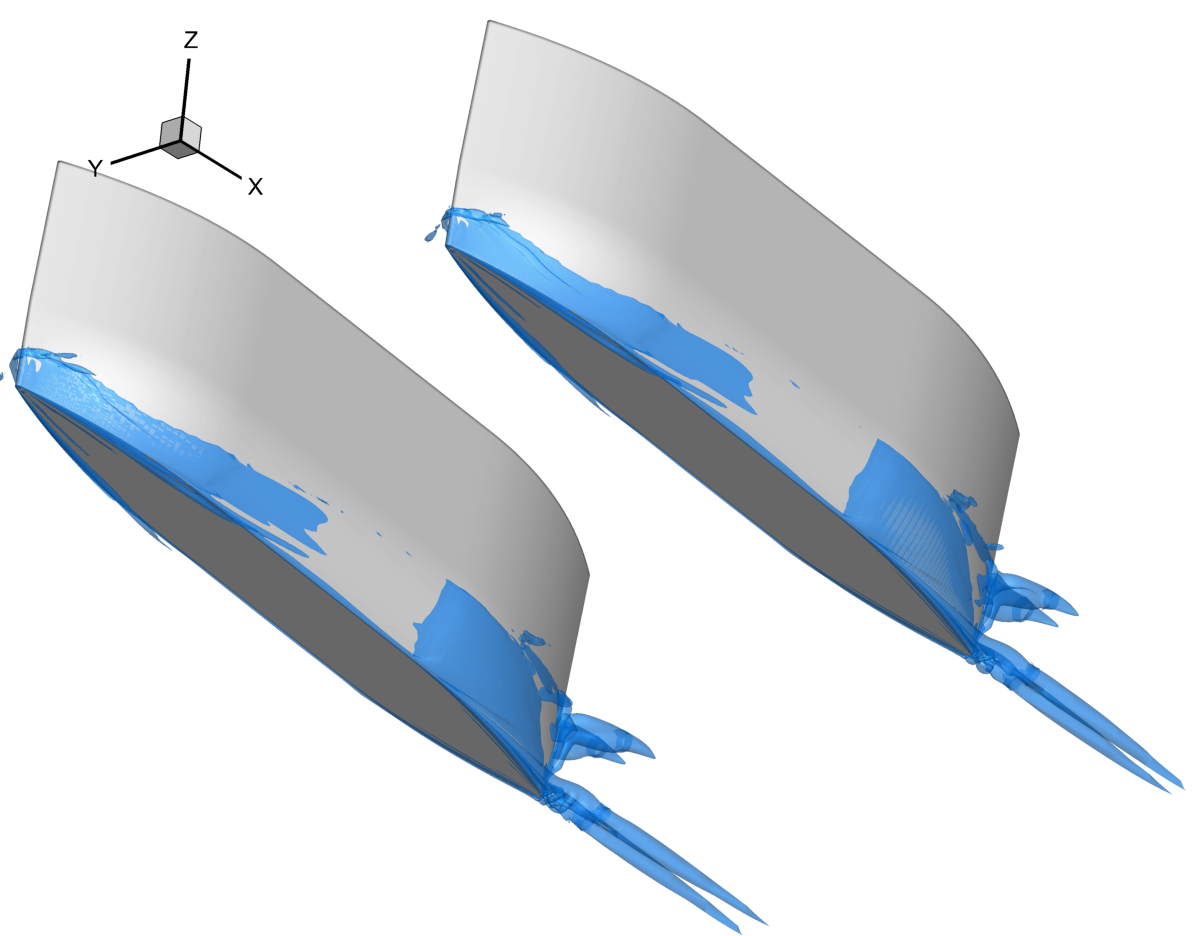}} \hfill  
\subfloat[$\nabla = 47.5$kg, $\delta x_G = 7.5\%$L]{\includegraphics[width=0.33\textwidth]{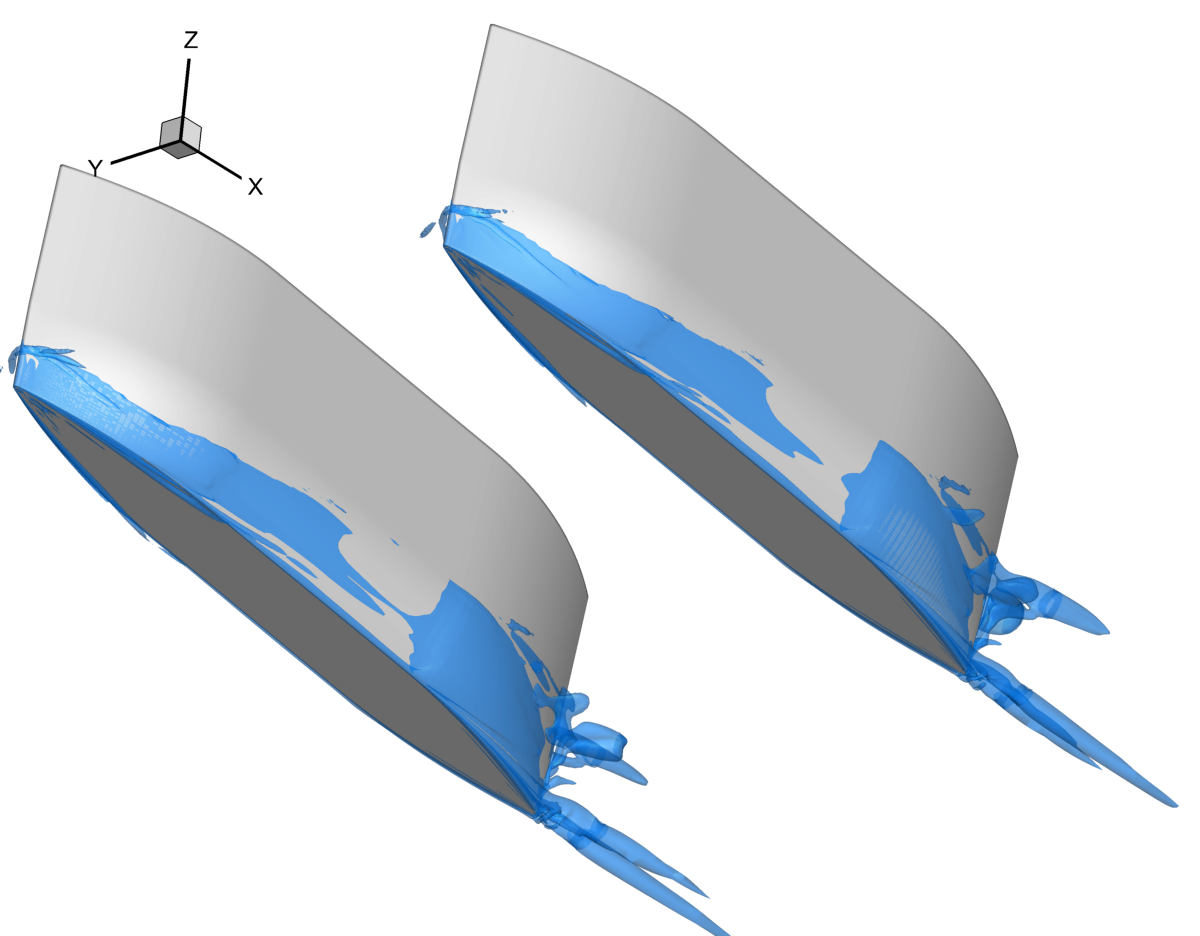}} \hfill   
\subfloat[$\nabla = 58$kg,   $\delta x_G = 7.5\%$L]{\includegraphics[width=0.33\textwidth]{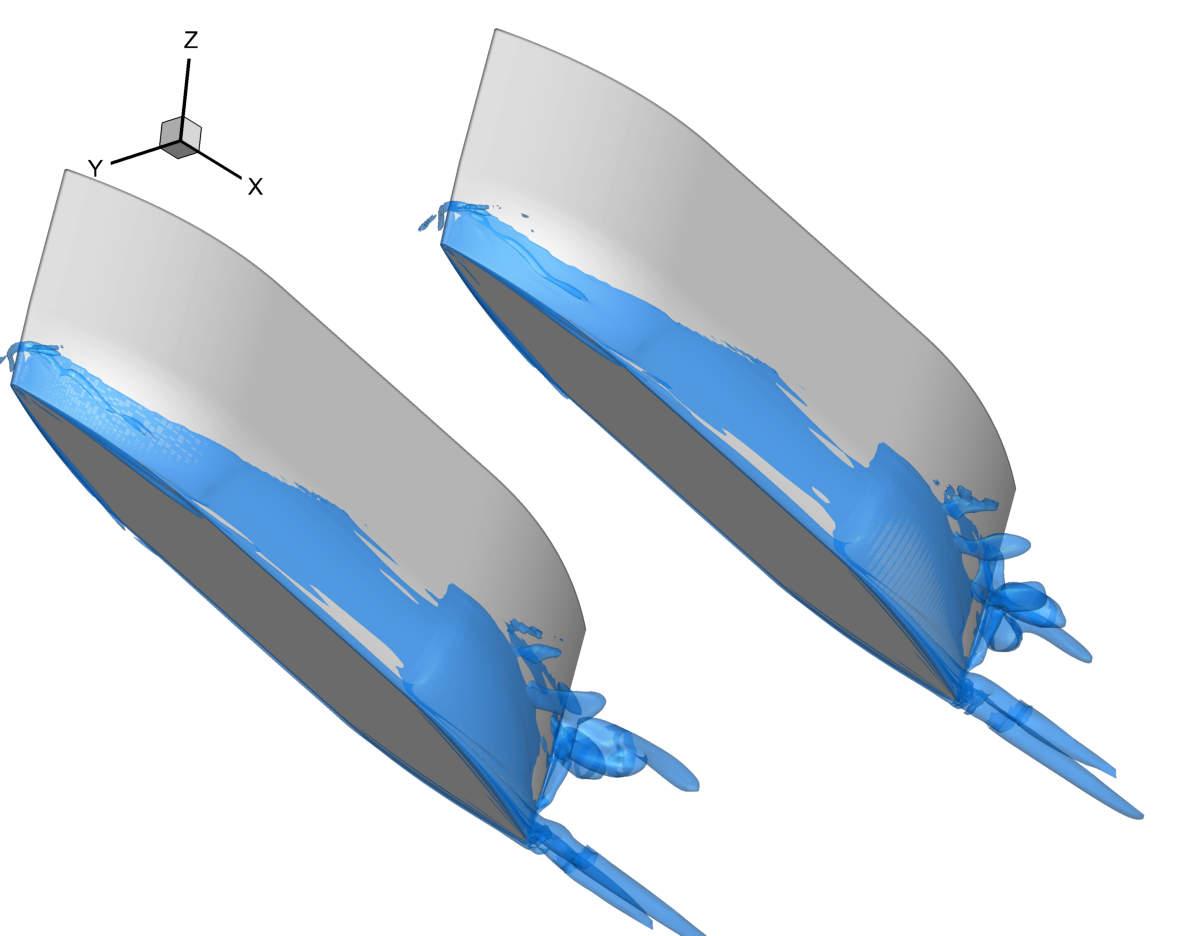}}   \hfill \mbox{}  
\caption{RANSE solution: Q-criterion iso-surface (Q=50) for the twelve configurations of the SWAMP}\label{fig:Qcrit} 
\end{figure}

Finally, Fig. \ref{fig:Qcrit} shows the Q-criterion iso-surface (Q=50) for the twelve configurations of the SWAMP. It is worth noting that for $-3\%\mathrm{L} \leq \delta x_G \leq 3\%\mathrm{L}$ two structures depart from the bow and lie close to the keel. These structures are no longer evident for $\delta x_G = 7.5\%$L, whereas it is evident that two structures are formed in the rear part of the hull and near the keel. These are producing the low pressure stripe that has been discussed in Figs. \ref{fig:PSide0}-\ref{fig:PSide75}. Finally, as the payload and $\delta x_G$ values increase, the vortexes behind the stern also increase.

\begin{figure}[!h] 
\centering 
\mbox{} \hfill
\subfloat[$u$, $x=0$L]{\includegraphics[width=0.31\textwidth]{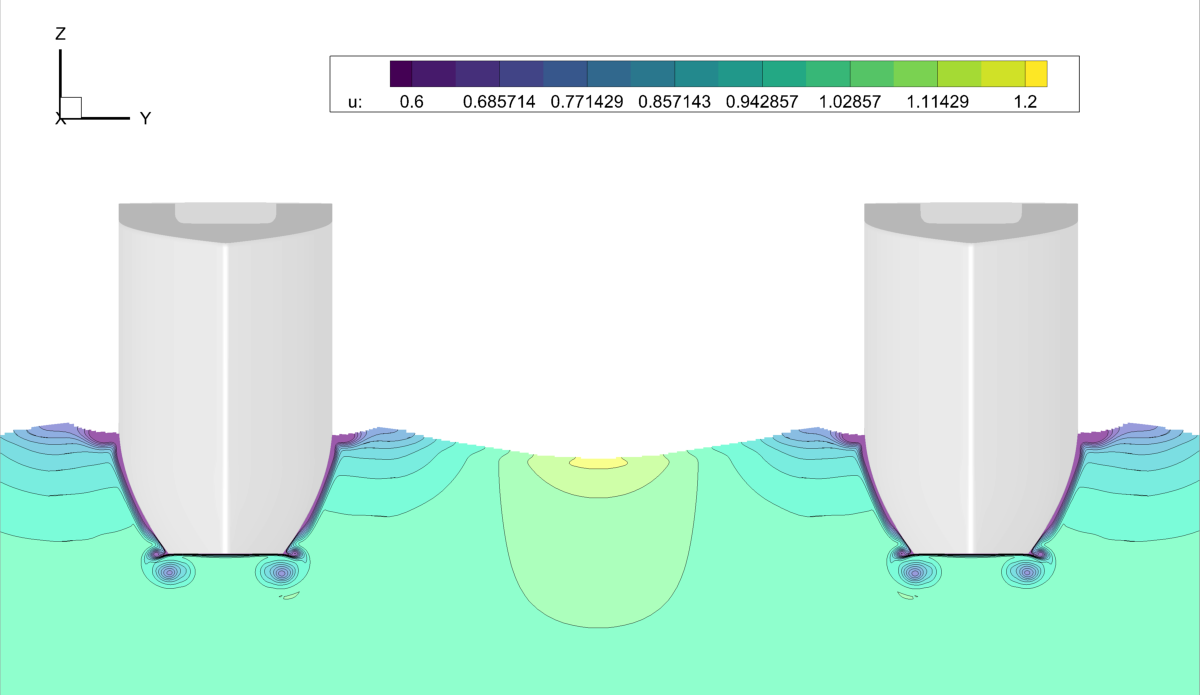}} \hfill 
\subfloat[$u$, $x=0.25$L]{\includegraphics[width=0.31\textwidth]{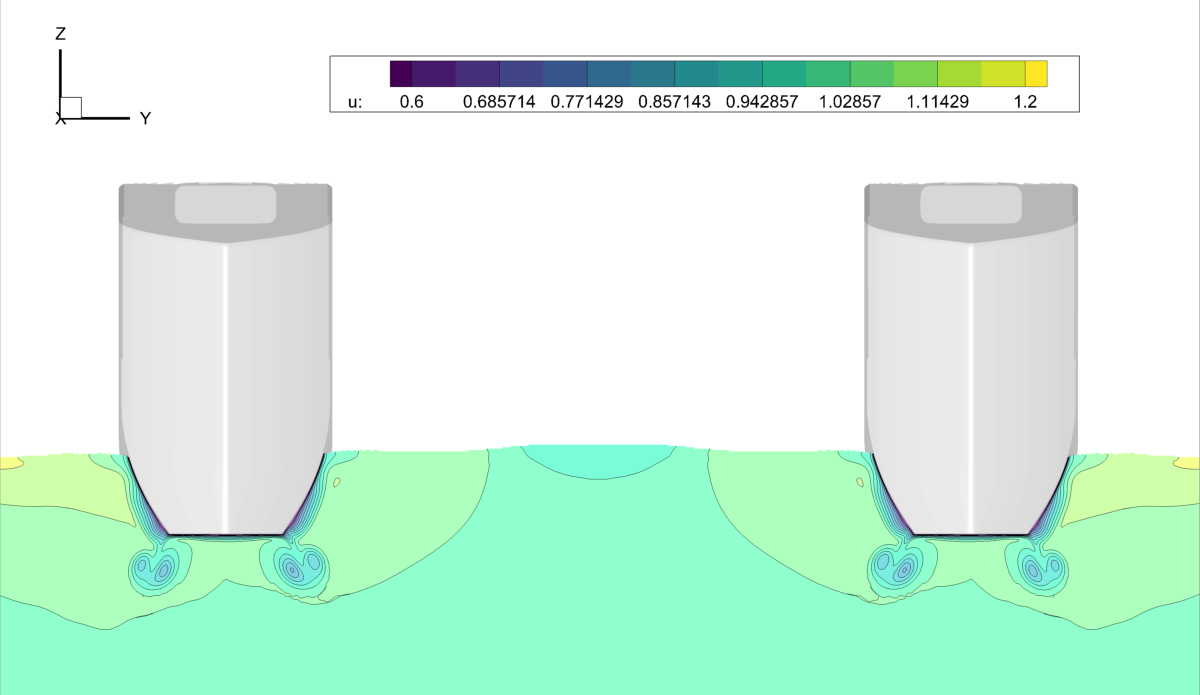}} \hfill  
\subfloat[$u$, $x=0.75$L]{\includegraphics[width=0.31\textwidth]{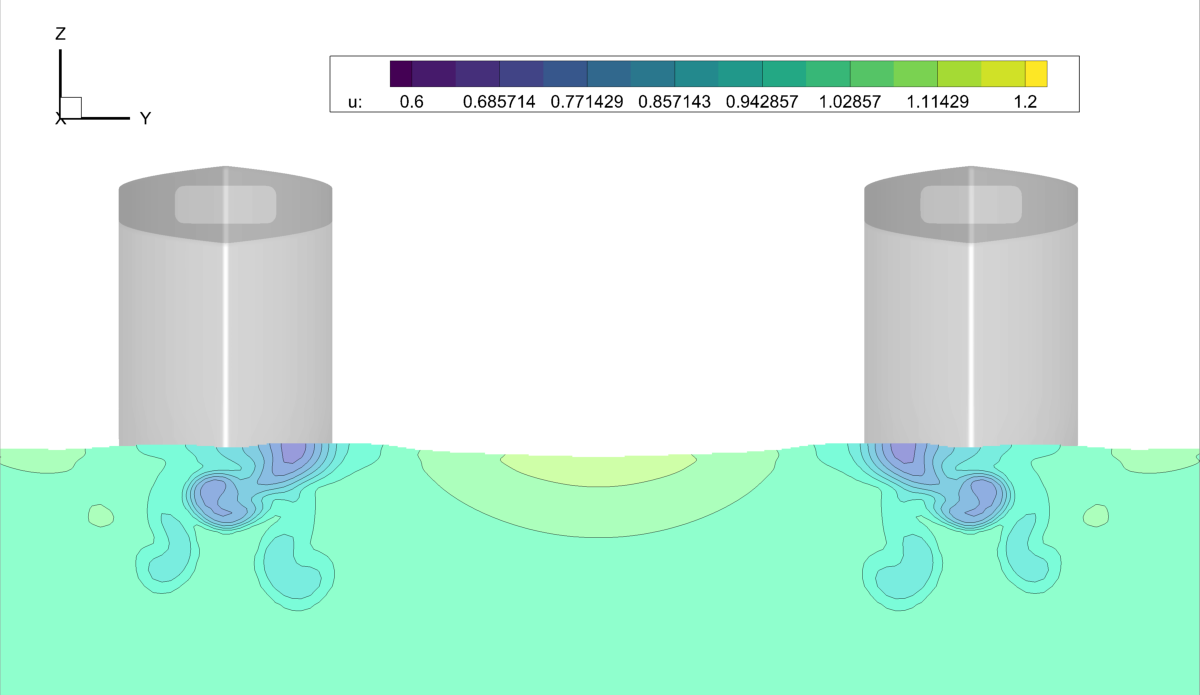}}  \hfill \mbox{}  \\
\mbox{} \hfill
\subfloat[$\Psi$, $x=0$L]{\includegraphics[width=0.31\textwidth]{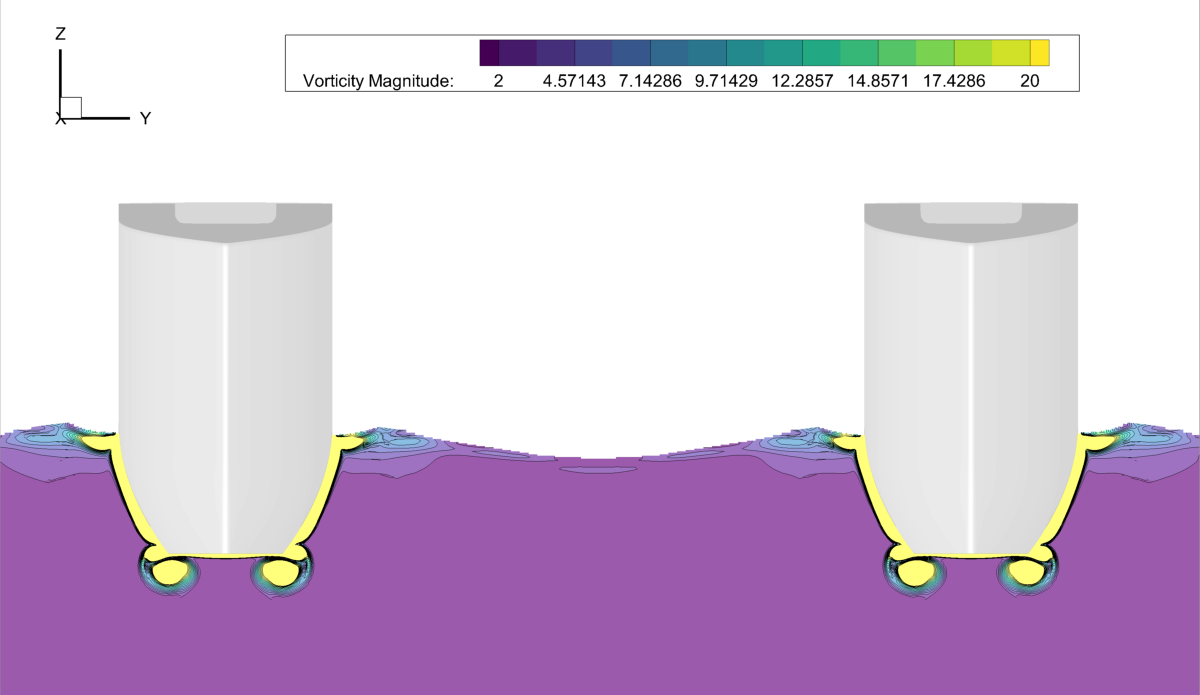}} \hfill  
\subfloat[$\Psi$, $x=0.25$L]{\includegraphics[width=0.31\textwidth]{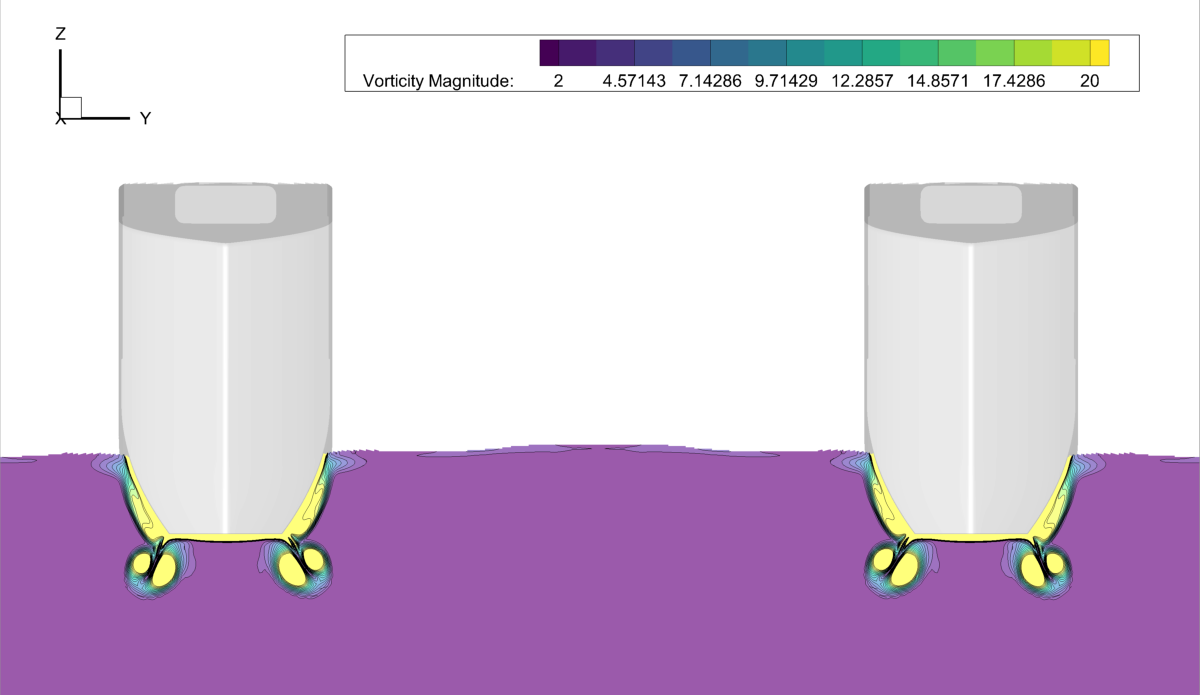}} \hfill 
\subfloat[$\Psi$, $x=0.75$L]{\includegraphics[width=0.31\textwidth]{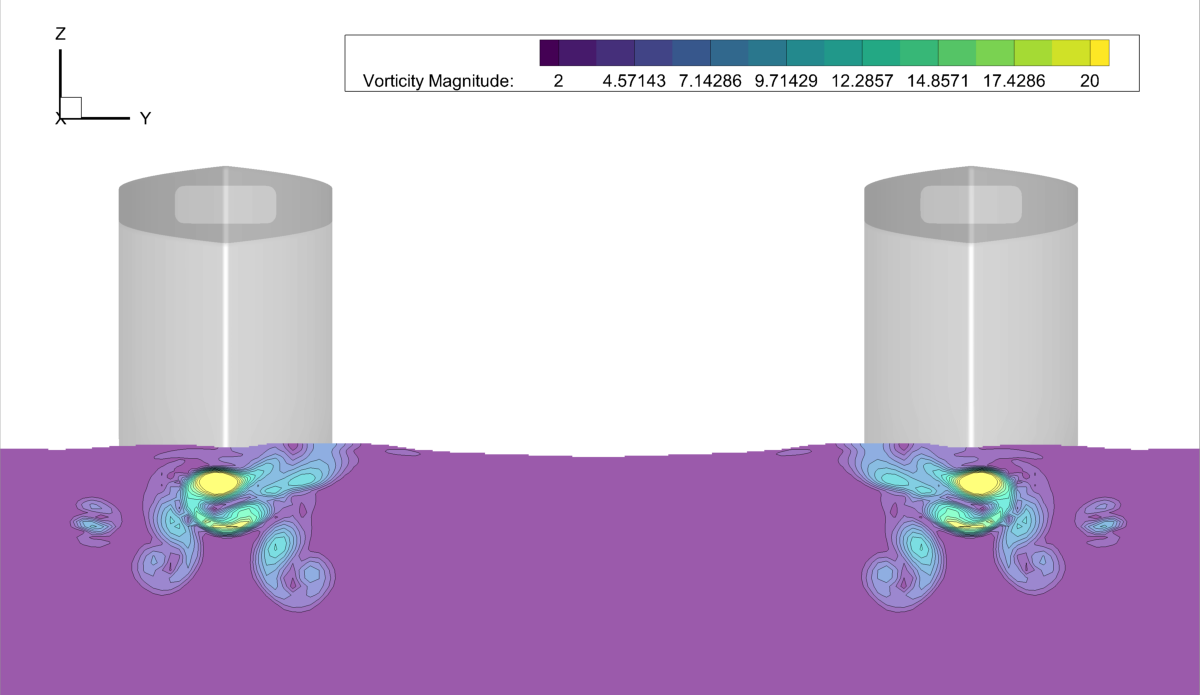}} \hfill \mbox{}   
\caption{RANSE solution, $\nabla = 47.5$kg,   $\delta x_G = -3\%$L condition: Z-Y planes with contour of the $x$-component velocity ($u$) and vorticity magnitude ($\Psi$), at $x=0$L, $x=0.25$L, and $x=0.75$L (the hull extends from $x=-0.5$L to $x=0.5$L)}\label{fig:Slic-3} 
\end{figure}

\begin{figure}[!h] 
\centering 
\mbox{} \hfill
\subfloat[$u$, $x=0$L]{\includegraphics[width=0.31\textwidth]{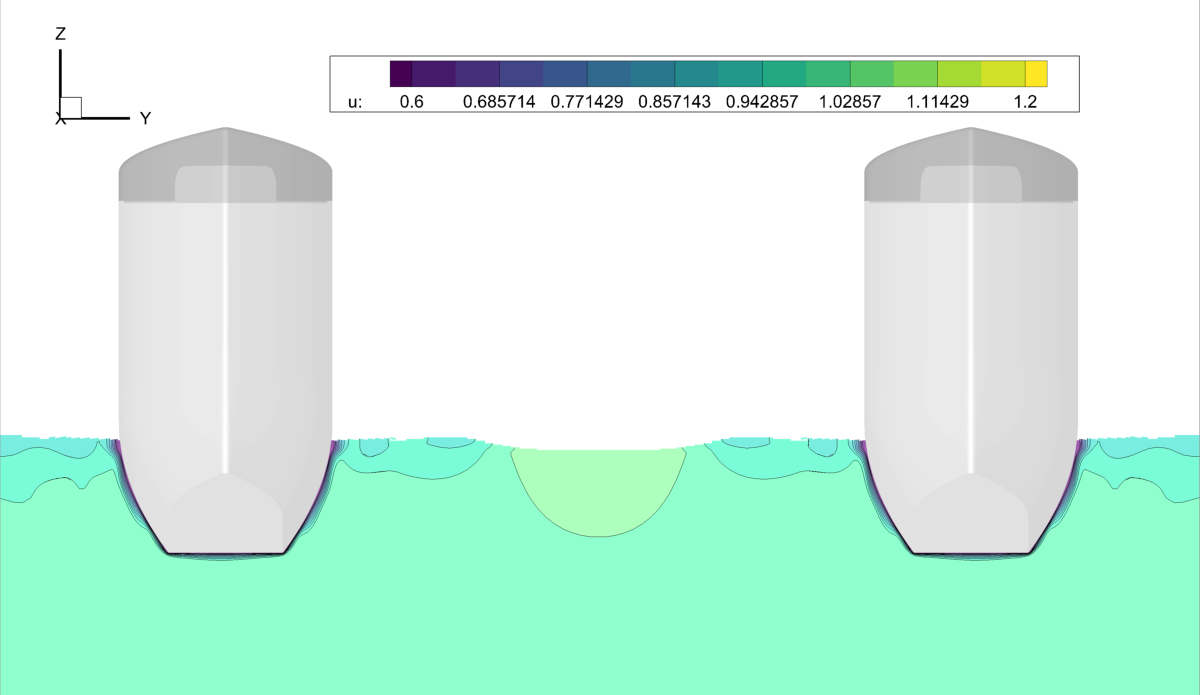}} \hfill 
\subfloat[$u$, $x=0.25$L]{\includegraphics[width=0.31\textwidth]{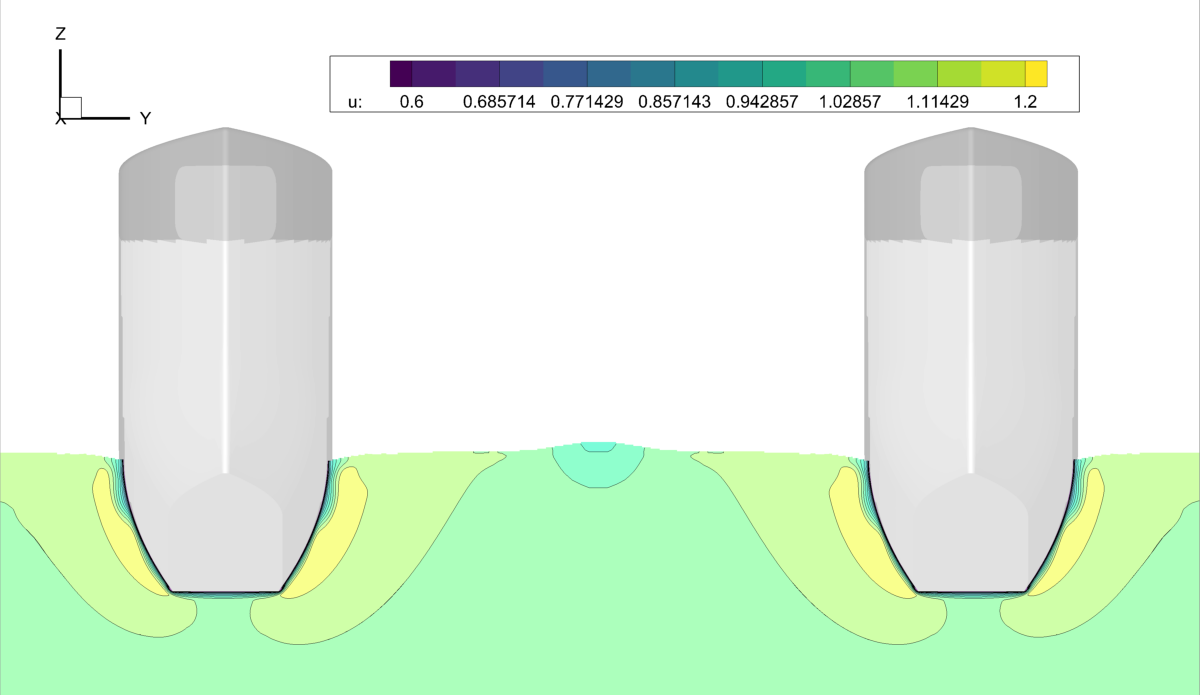}} \hfill  
\subfloat[$u$, $x=0.75$L]{\includegraphics[width=0.31\textwidth]{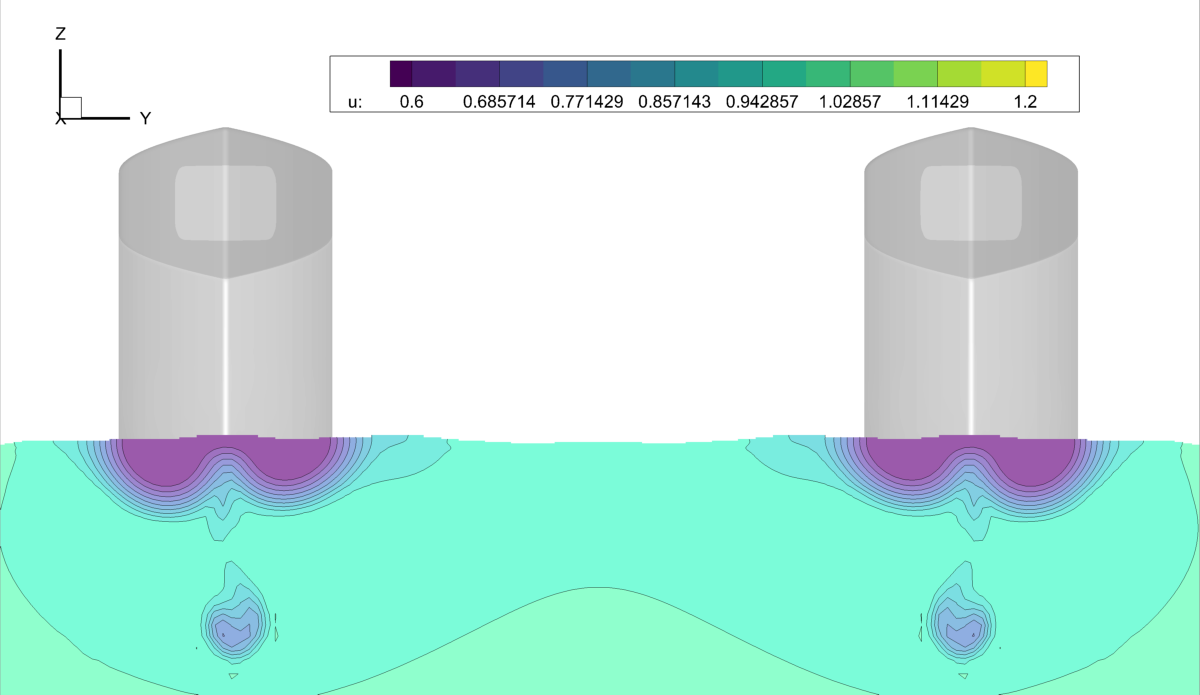}}  \hfill \mbox{}  \\
\mbox{} \hfill
\subfloat[$\Psi$, $x=0$L]{\includegraphics[width=0.31\textwidth]{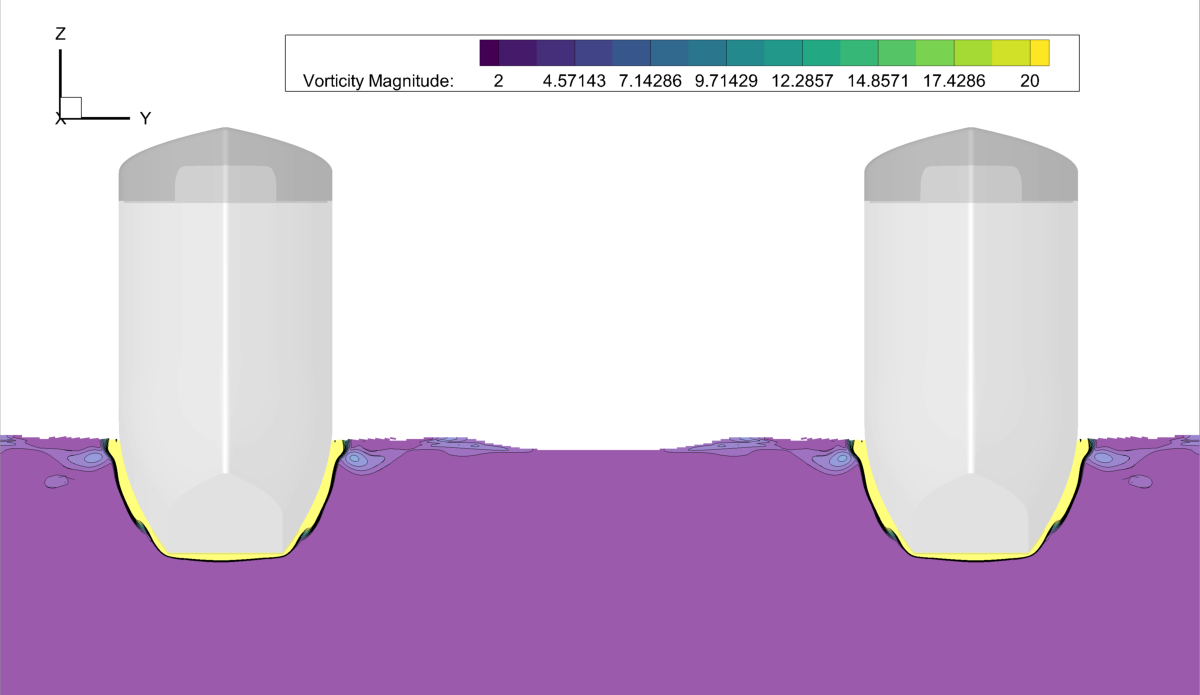}} \hfill  
\subfloat[$\Psi$, $x=0.25$L]{\includegraphics[width=0.31\textwidth]{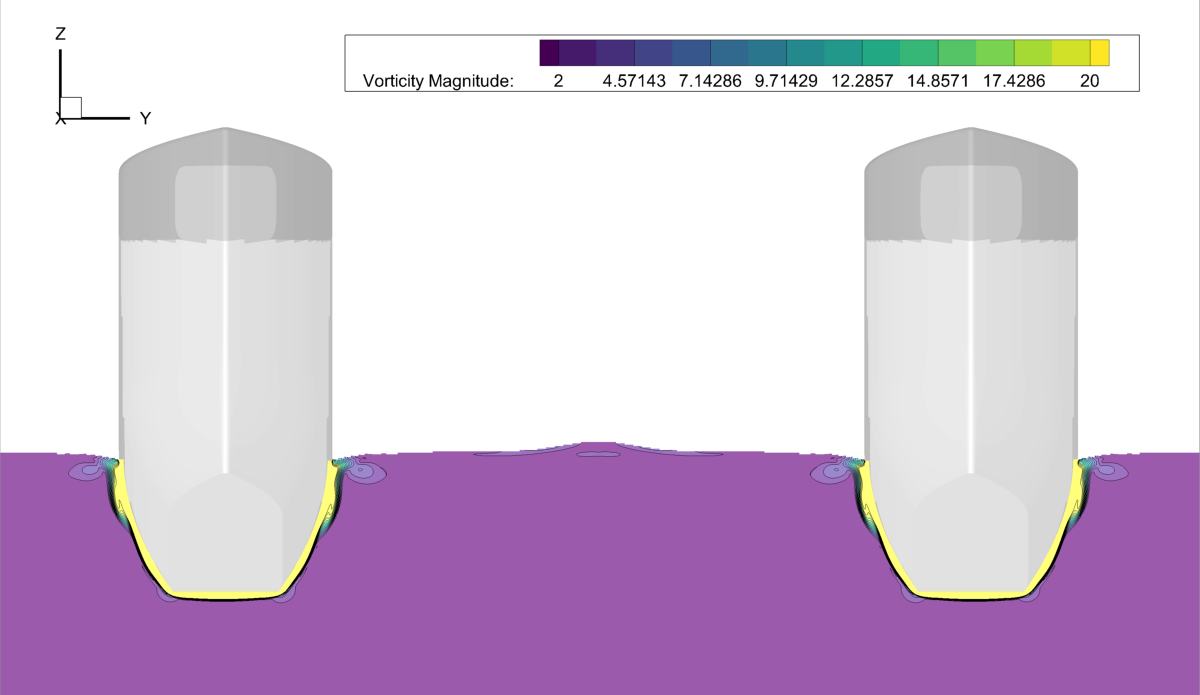}} \hfill   
\subfloat[$\Psi$, $x=0.75$L]{\includegraphics[width=0.31\textwidth]{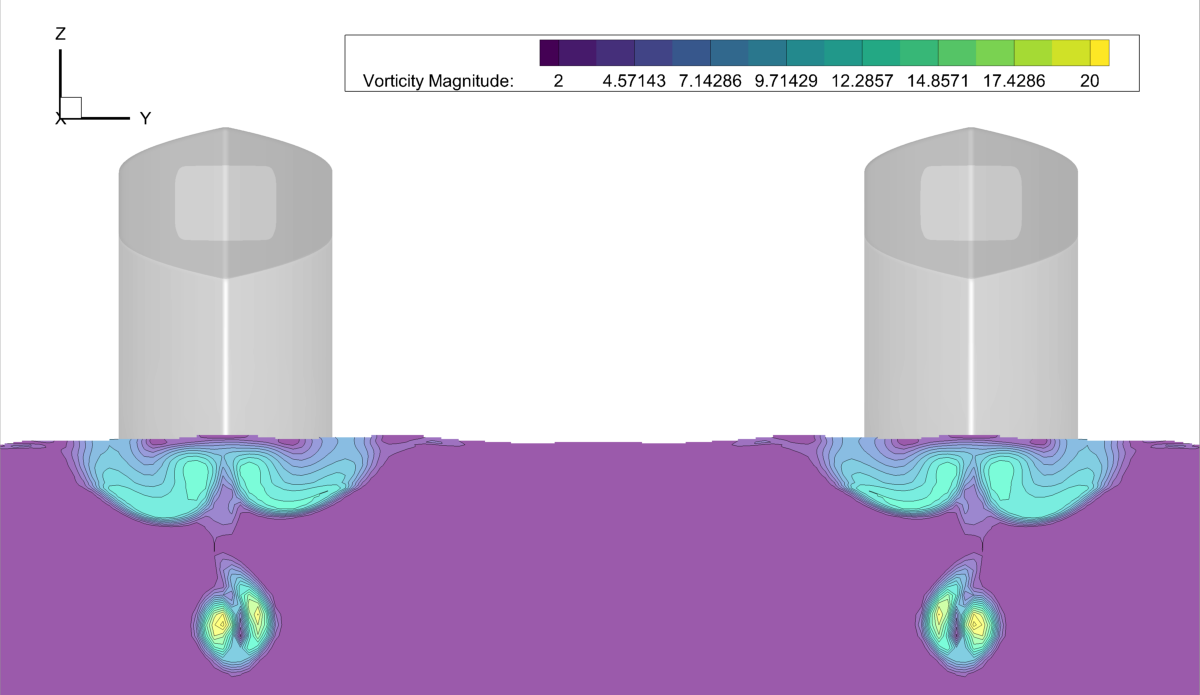}}  \hfill \mbox{}  
\caption{RANSE solution, $\nabla = 47.5$kg,   $\delta x_G = 7.5\%$L condition: Z-Y planes with contour of the $x$-component velocity ($u$) and vorticity magnitude ($\Psi$), at $x=0$L, $x=0.25$L, and $x=0.75$L (the hull extends from $x=-0.5$L to $x=0.5$L)}\label{fig:Slic75} 
\end{figure}

Figures \ref{fig:Slic-3} and \ref{fig:Slic75} show three slices took at $x=0$L, $x=0.25$L, and $x=0.75$L (the hull extends from $x=-0.5$L to $x=0.5$L), for the conditions with $\nabla = 47.5$kg and $\delta x_G = -3\%$L and $\delta x_G = 7.5\%$L, respectively. For both figures, Figs. a-c show the $x$-component of the velocity ($u$) and Figs. d-e the vorticity magnitude ($\Psi$). 
Figures \ref{fig:Slic-3}a-c show that the flow is initially slower near the hull and in the proximity of the free-surface and faster in the middle and then accelerates as it moves towards the stern of the hull. Figure \ref{fig:Slic-3}c shows that the wake is slightly slower than the average flow and tends towards the middle of the SWAMP. 
Figures \ref{fig:Slic-3}d-f, in combination with Fig. \ref{fig:Qcrit}b, show that the two spots near the lower edges of the hulls with a lower axial velocity are due to the vortical structures. These two structures are also present in the wake of the SWAMP. 
Figures \ref{fig:Slic75}a-c show that the flow is significantly more uniform than in Fig. \ref{fig:Slic-3}a-c. Overall, Figs. \ref{fig:Slic75}a-f show that the two vortical structures below the hull are no longer present, whereas a single large vortex is evident in the wake of the SWAMP. Furthermore, the wake looks slightly more symmetrical than in the $\delta x_G = -3\%$L case.

\begin{figure}[!h] 
\centering 
\mbox{} \hfill
\subfloat[RANSE, $\nabla = 37$kg]{\includegraphics[width=0.275\textwidth]{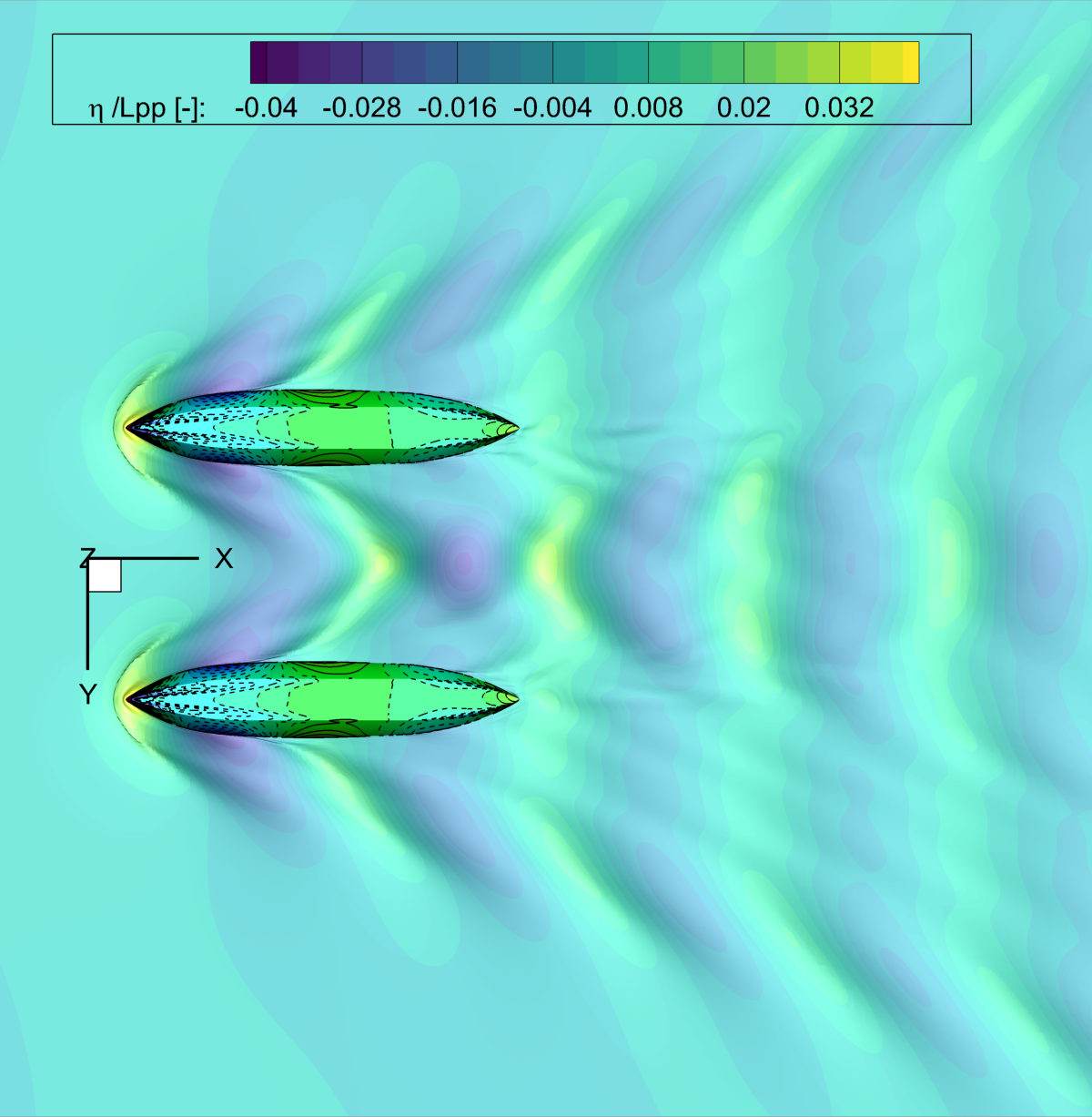}} \hfill
\subfloat[RANSE, $\nabla = 47.5$kg]{\includegraphics[width=0.275\textwidth]{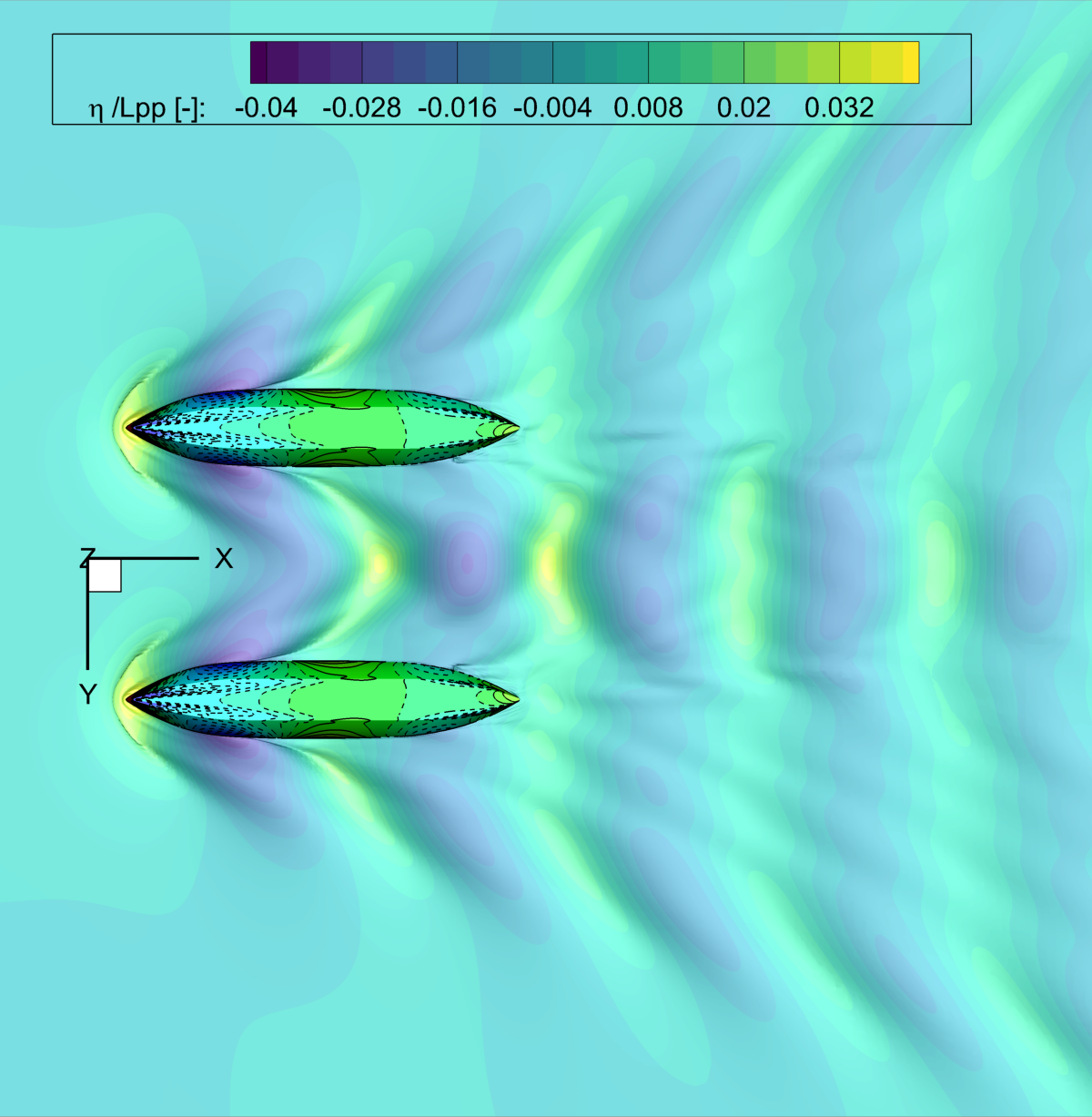}}  \hfill
\subfloat[RANSE, $\nabla = 58$kg]{\includegraphics[width=0.275\textwidth]{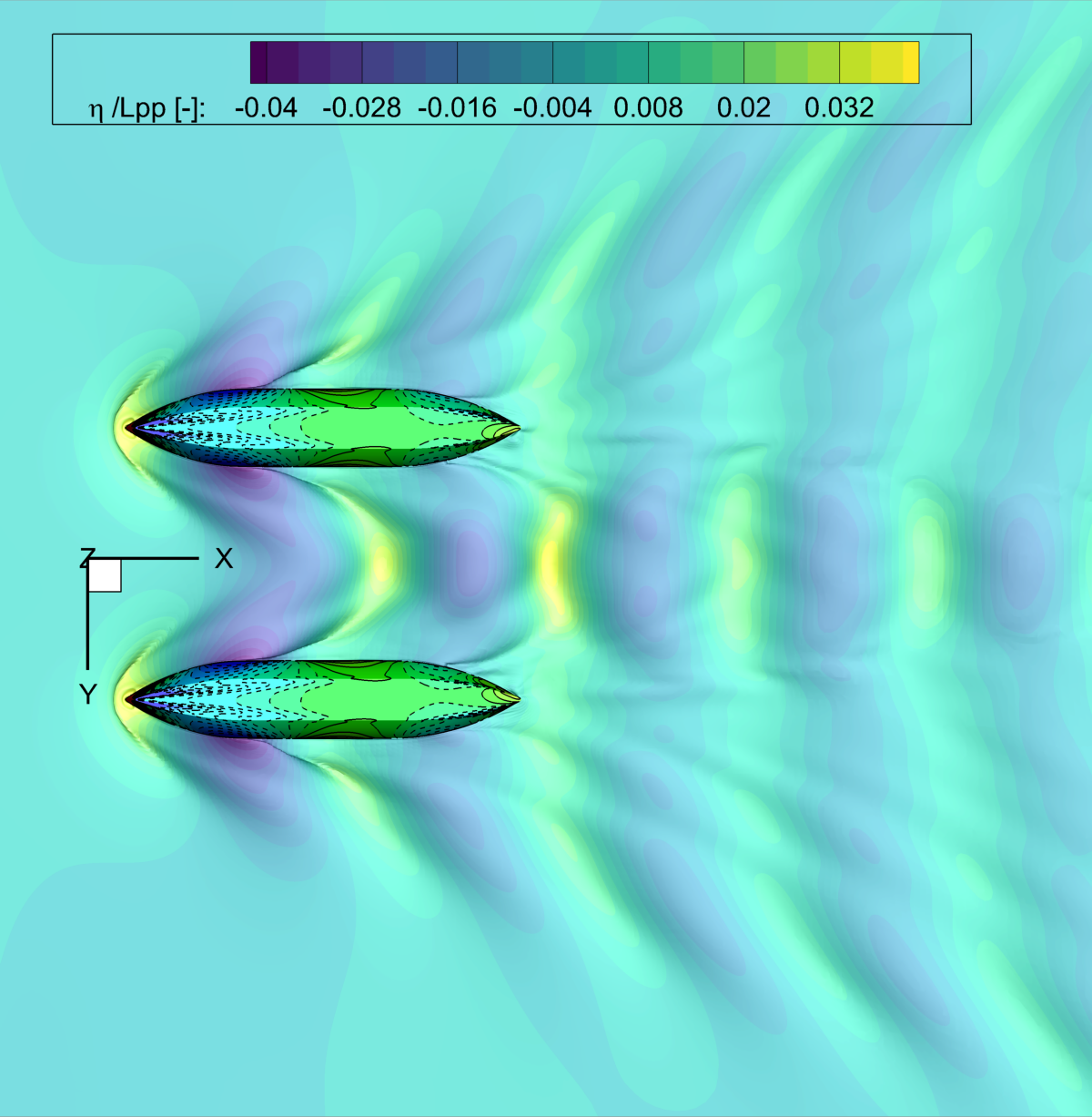}} \hfill \mbox{}  \\ \vspace{-2mm}
\mbox{} \hfill
\subfloat[PF, $\nabla = 37$kg]{\includegraphics[width=0.275\textwidth]{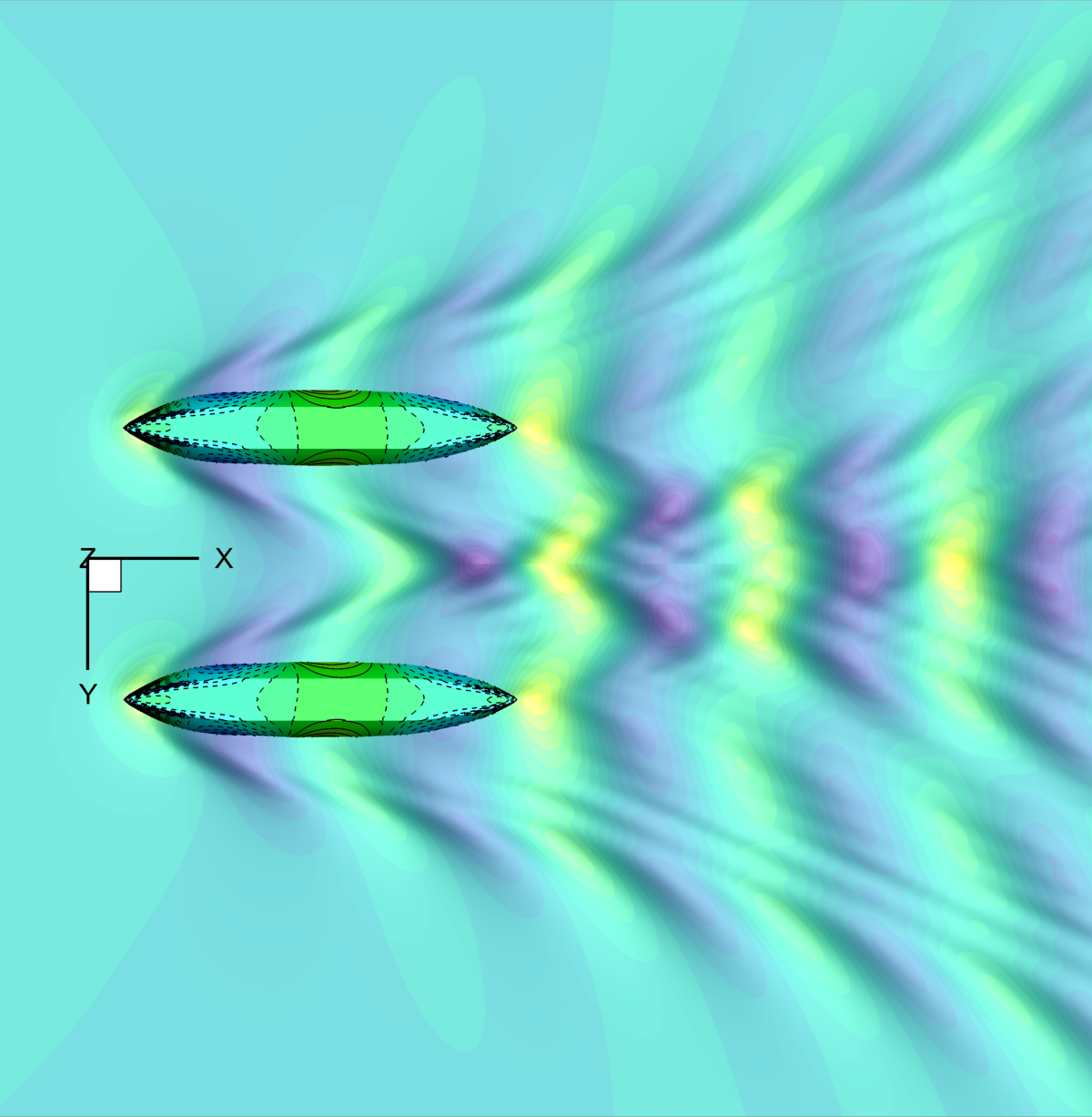}} \hfill
\subfloat[PF, $\nabla = 47.5$kg]{\includegraphics[width=0.275\textwidth]{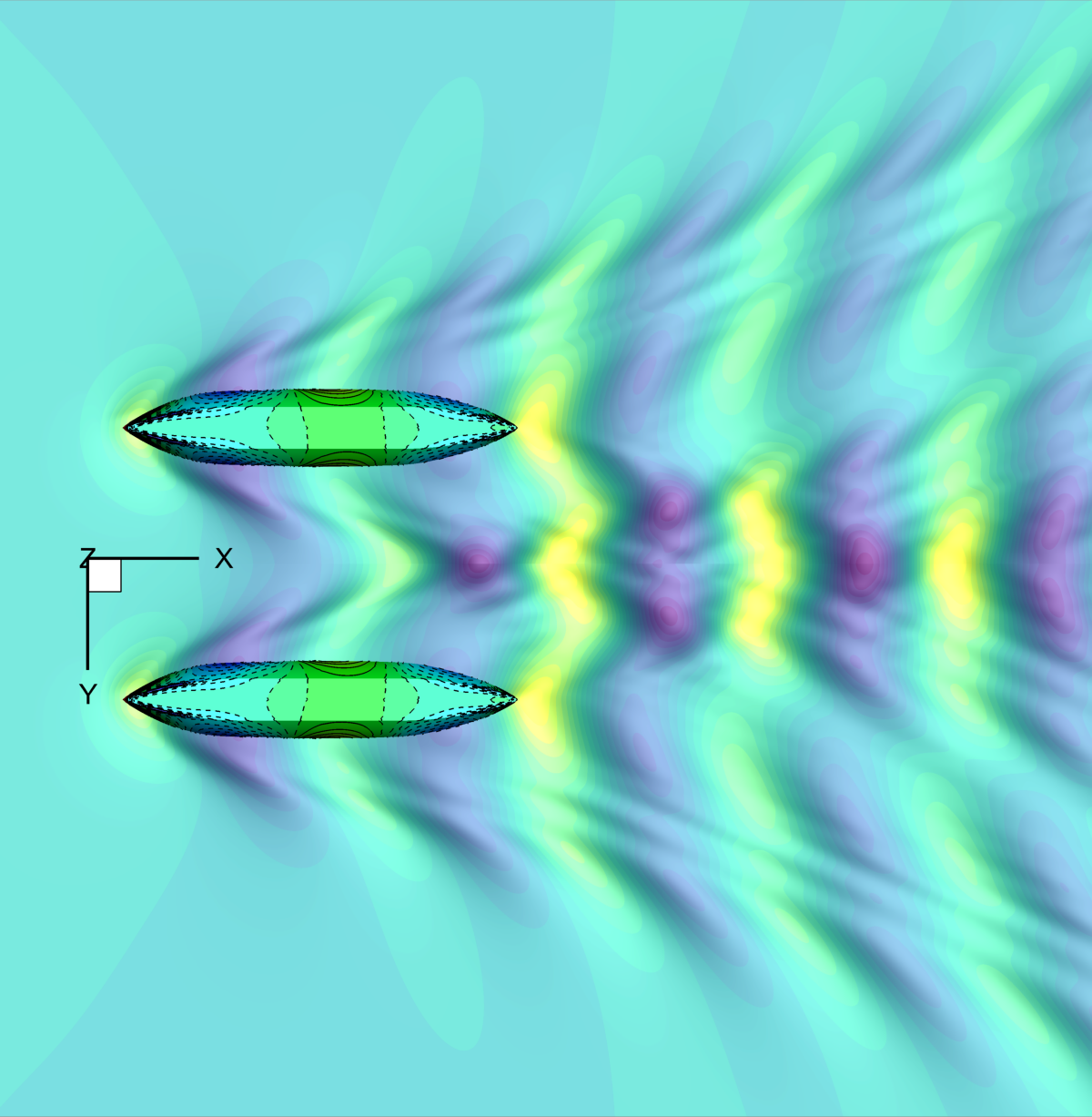}} \hfill
\subfloat[PF, $\nabla = 58$kg]{\includegraphics[width=0.275\textwidth]{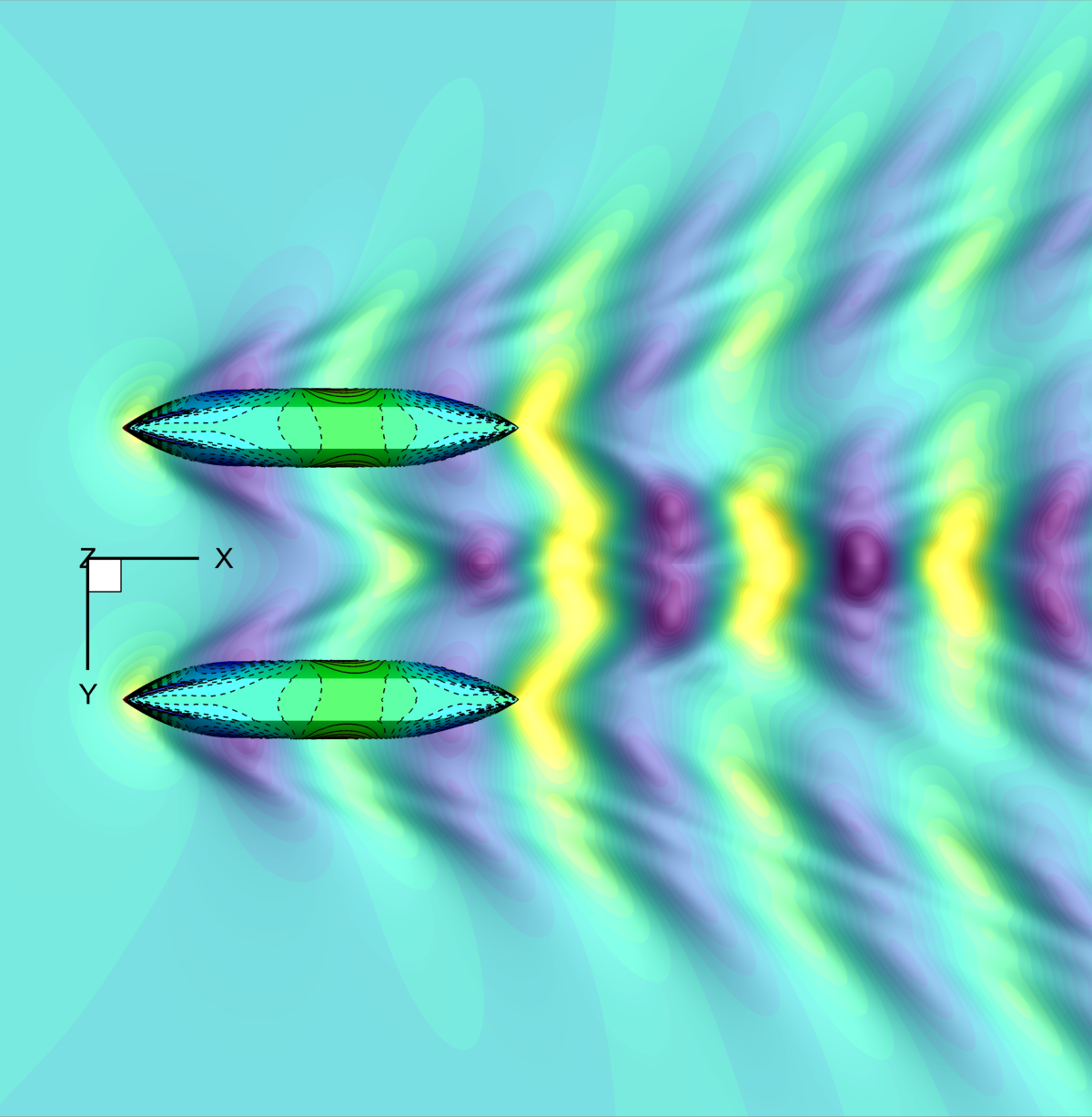}} \hfill \mbox{}  \\ \vspace{-2mm}
\mbox{} \hfill
\subfloat[RANSE, $\nabla = 37$kg]{\includegraphics[width=0.275\textwidth]{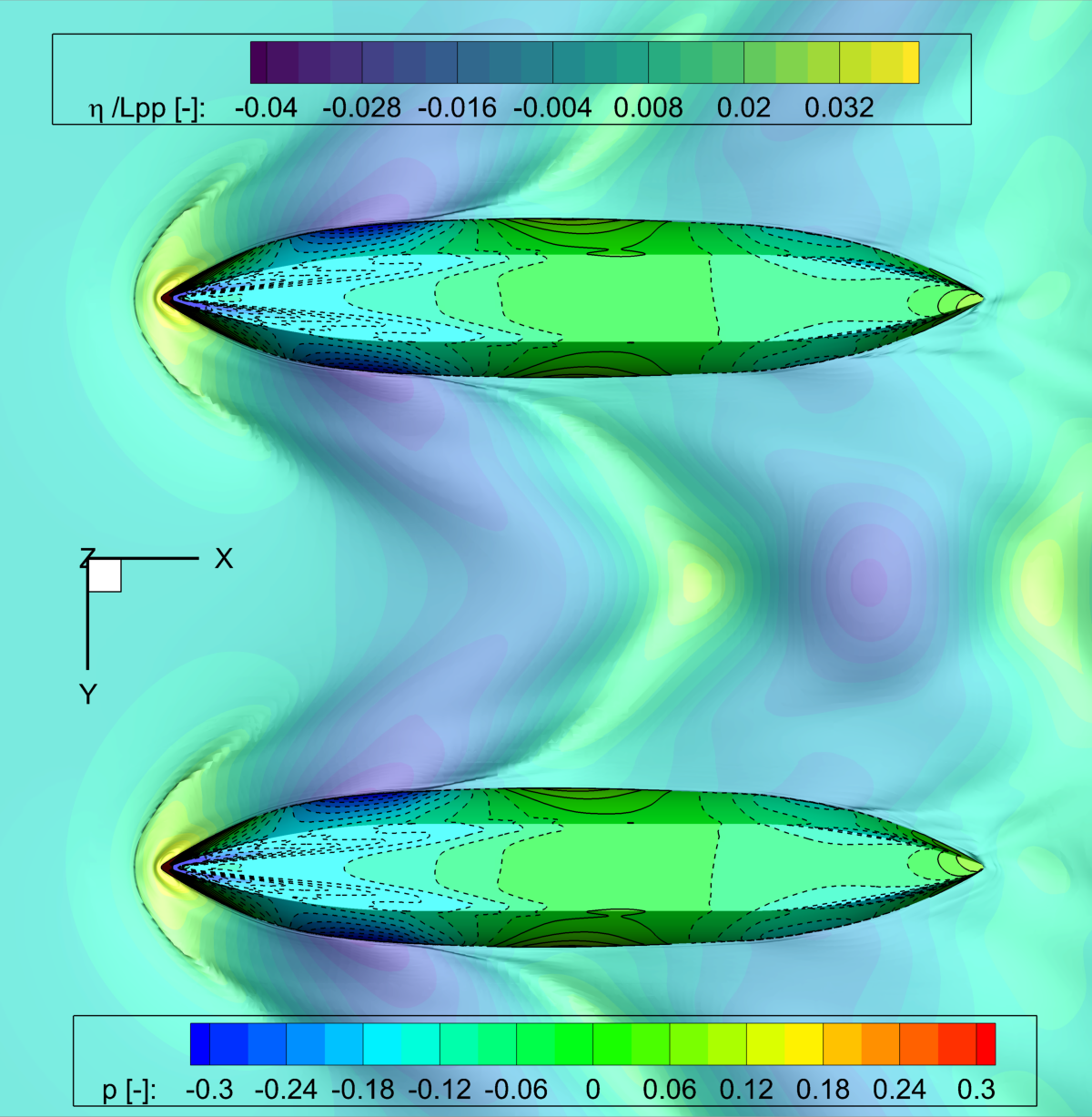}} \hfill
\subfloat[RANSE, $\nabla = 47.5$kg]{\includegraphics[width=0.275\textwidth]{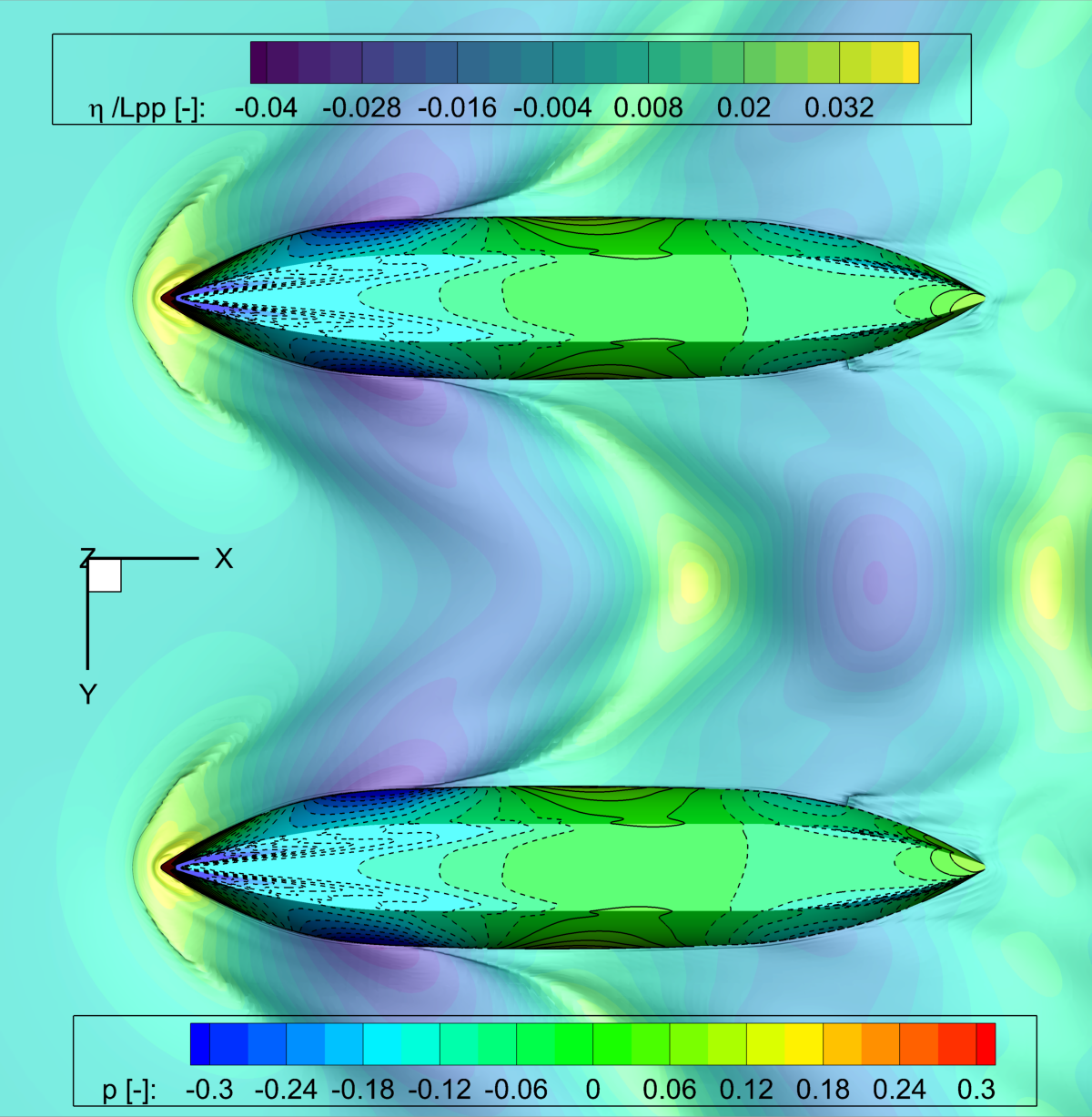}} \hfill
\subfloat[RANSE, $\nabla = 58$kg]{\includegraphics[width=0.275\textwidth]{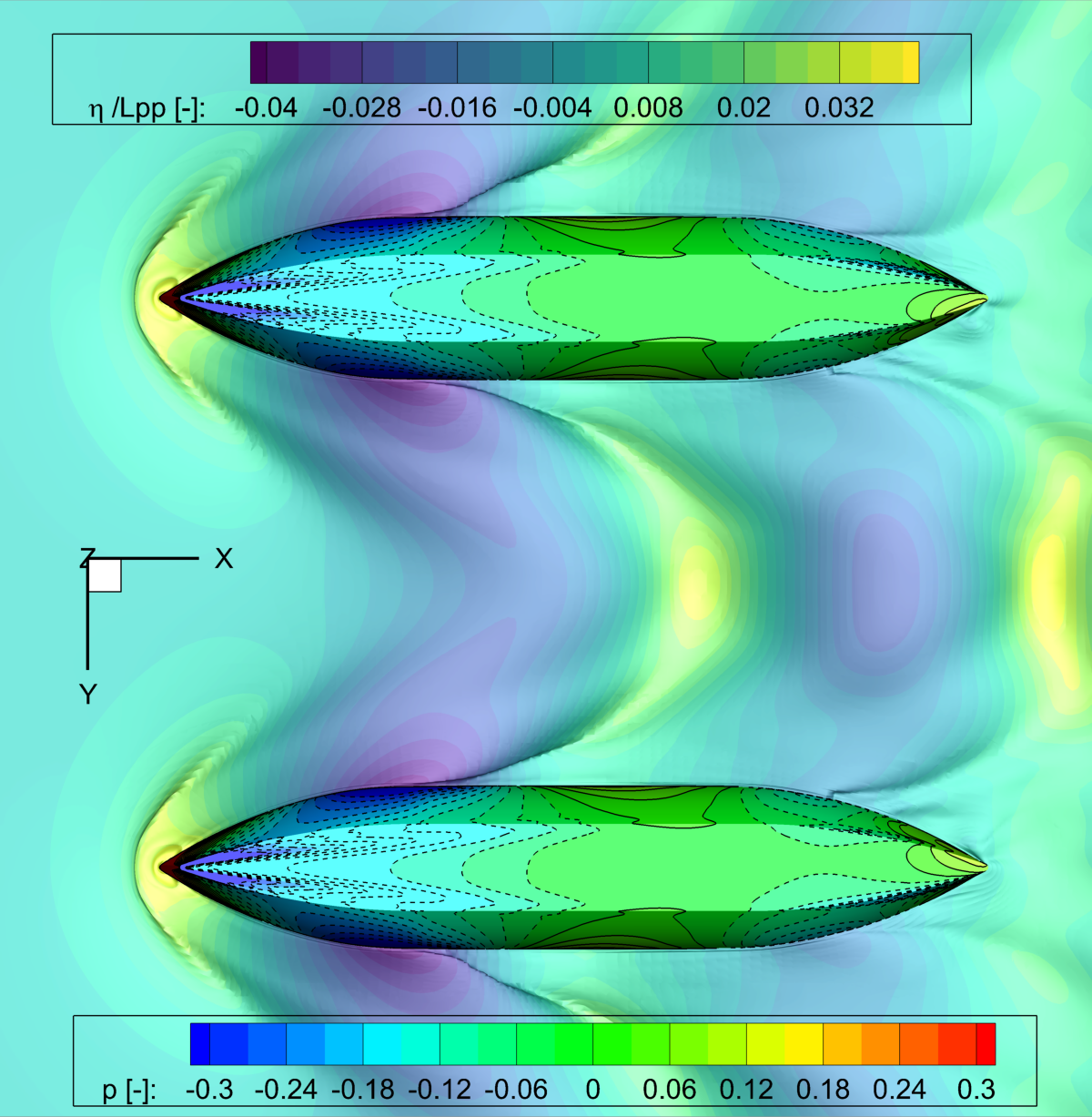}} \hfill \mbox{}  \\ \vspace{-2mm}
\mbox{} \hfill
\subfloat[PF, $\nabla = 37$kg]{\includegraphics[width=0.275\textwidth]{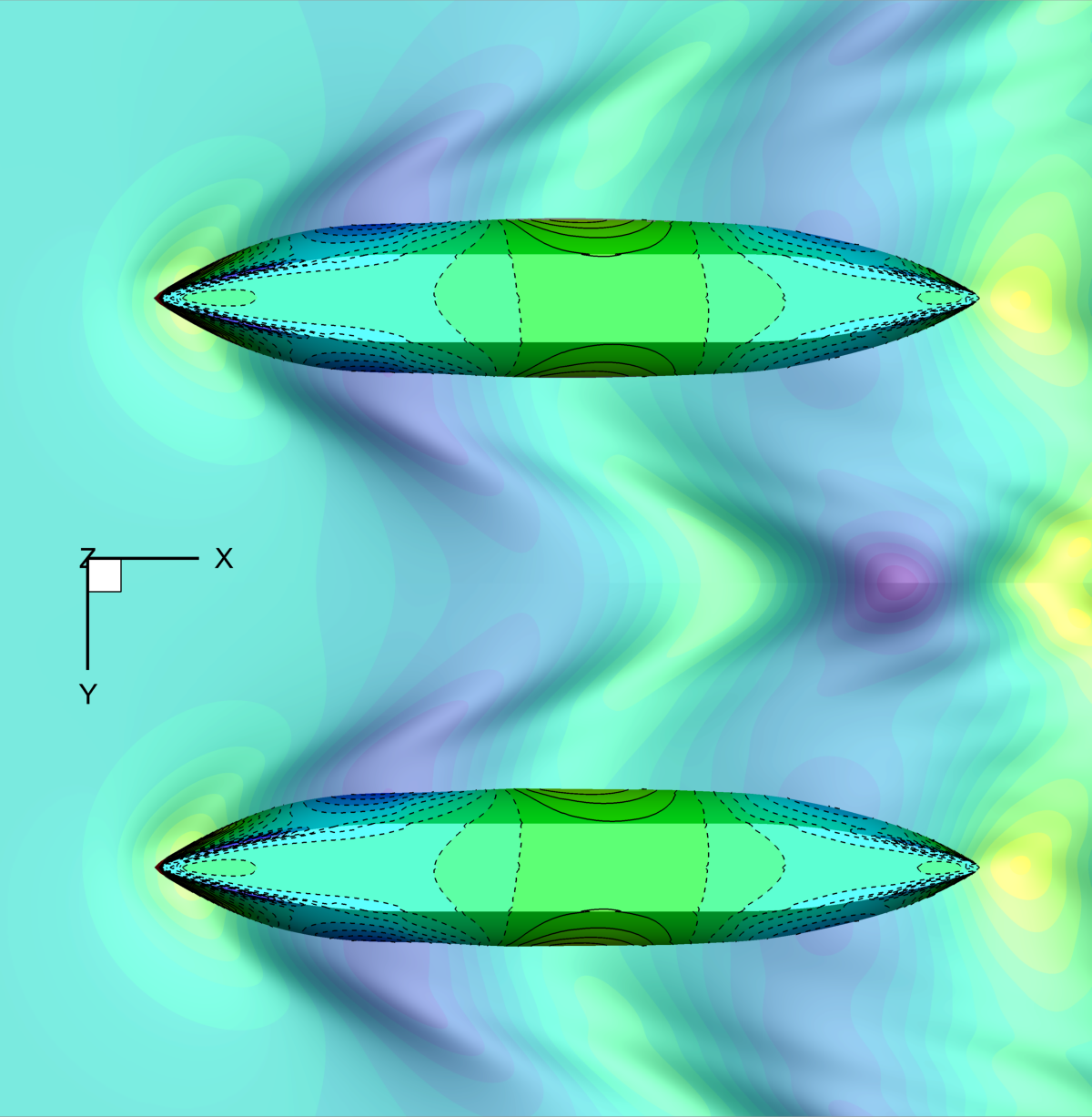}} \hfill
\subfloat[PF, $\nabla = 47.5$kg]{\includegraphics[width=0.275\textwidth]{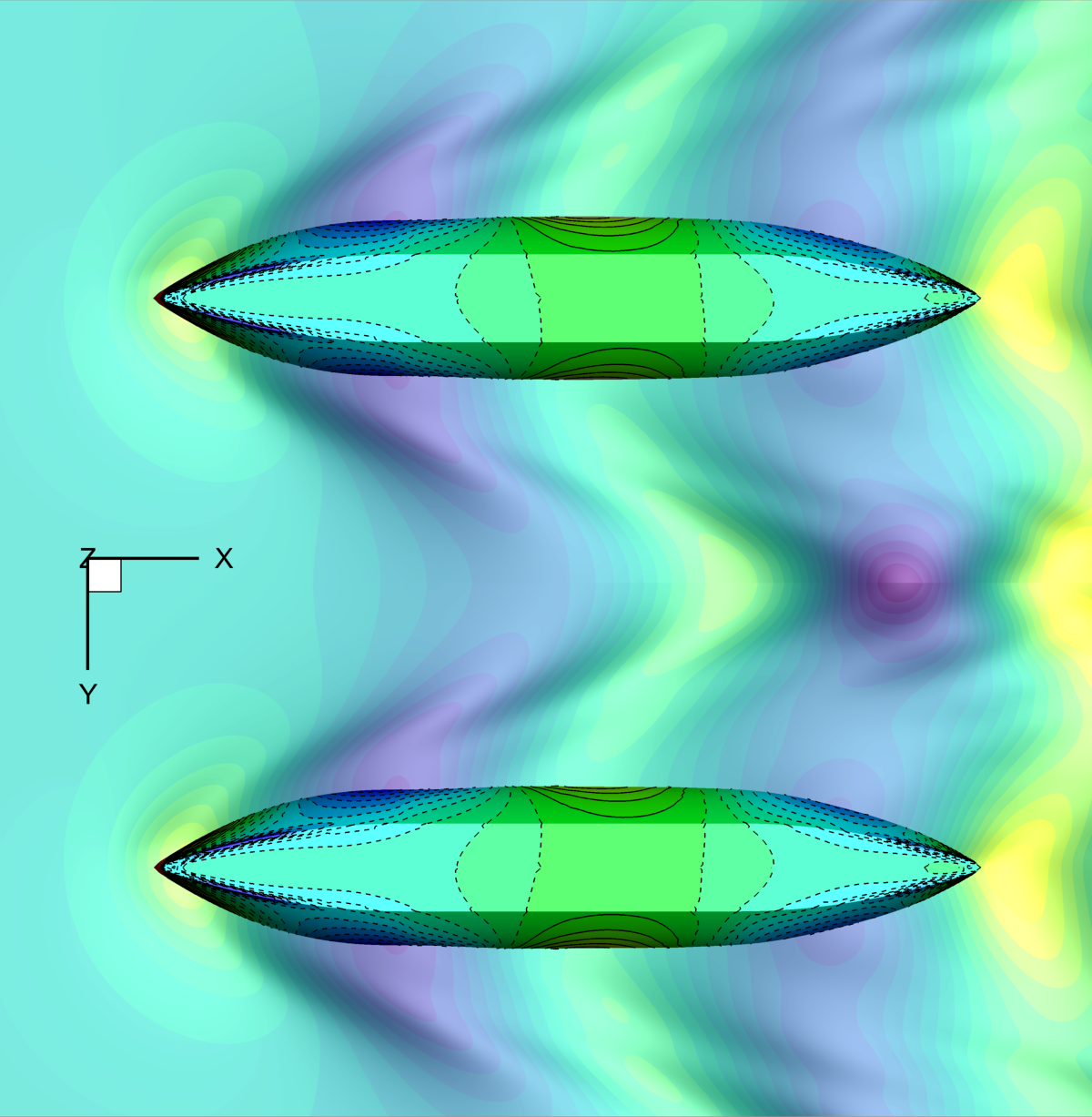}} \hfill
\subfloat[PF, $\nabla = 58$kg]{\includegraphics[width=0.275\textwidth]{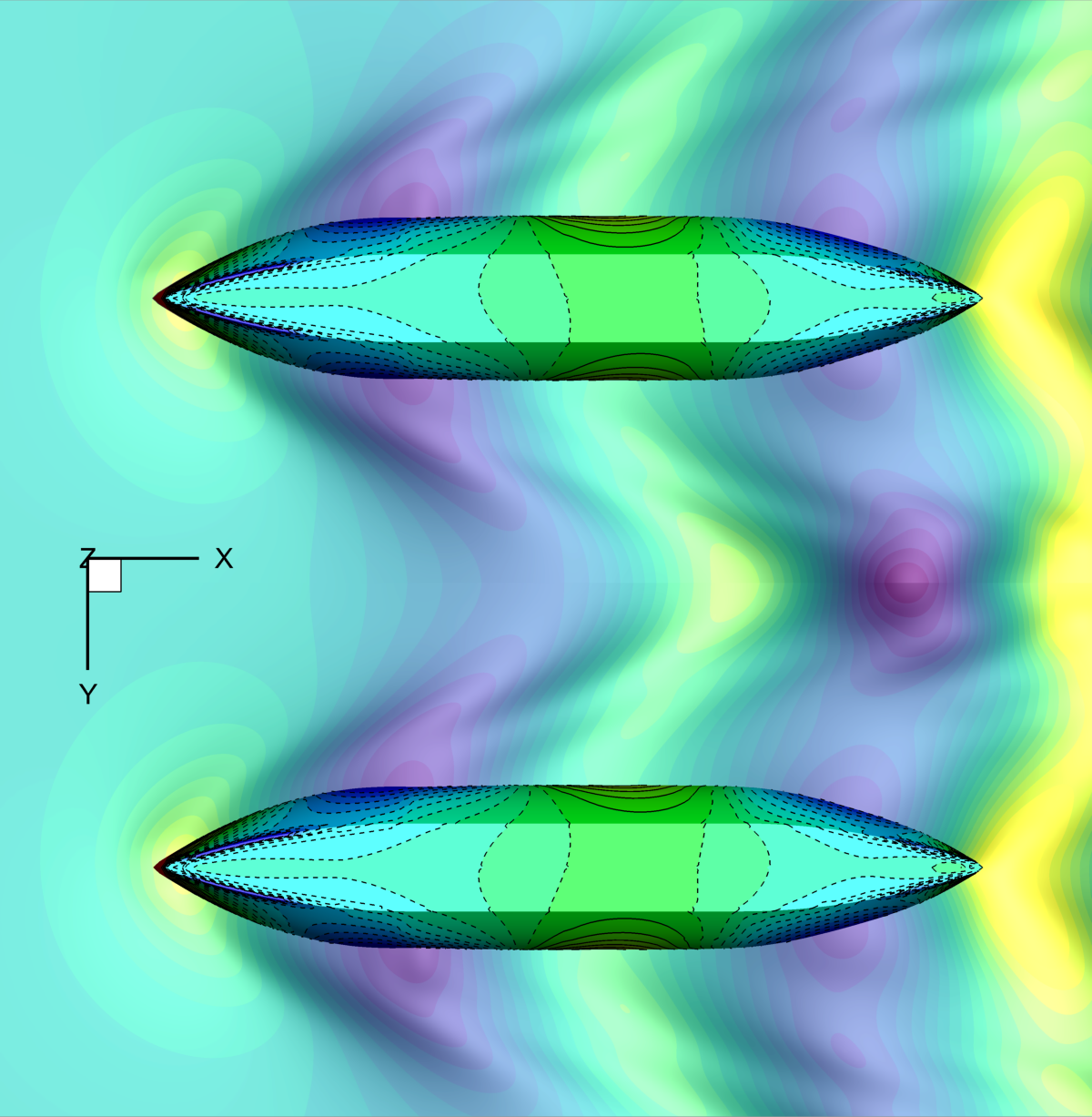}} \hfill \mbox{}  \\ 
\caption{Pressure contour and wave elevation varying the SWAMP payload with $\delta x_G = -3\%$L, bottom view}\label{fig:FS-3} 
\end{figure}

\begin{figure}[!h] 
    \centering 
    \mbox{} \hfill
    \subfloat[RANSE, $\nabla = 37$kg]{\includegraphics[width=0.275\textwidth]{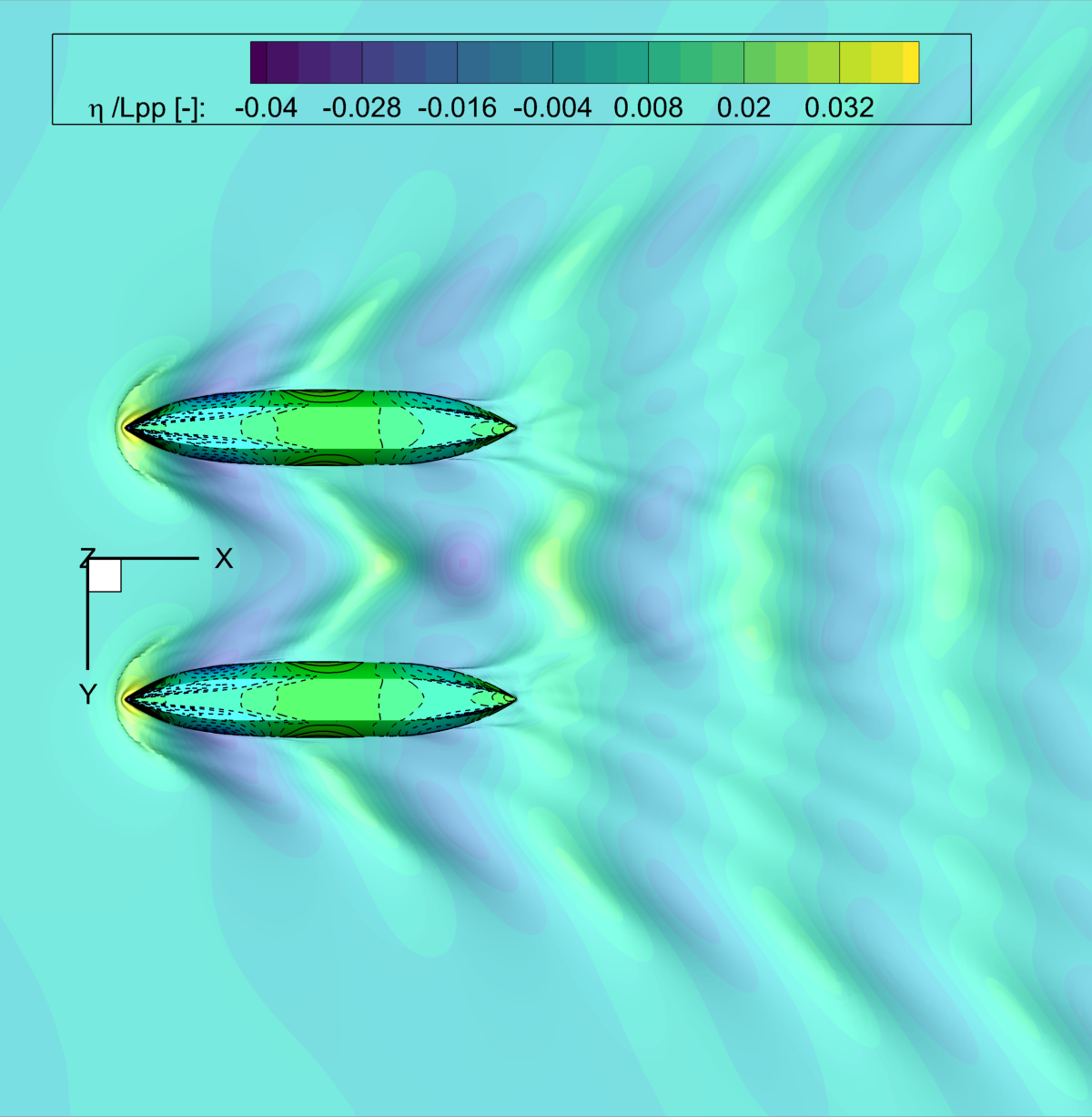}} \hfill 
    \subfloat[RANSE, $\nabla = 47.5$kg]{\includegraphics[width=0.275\textwidth]{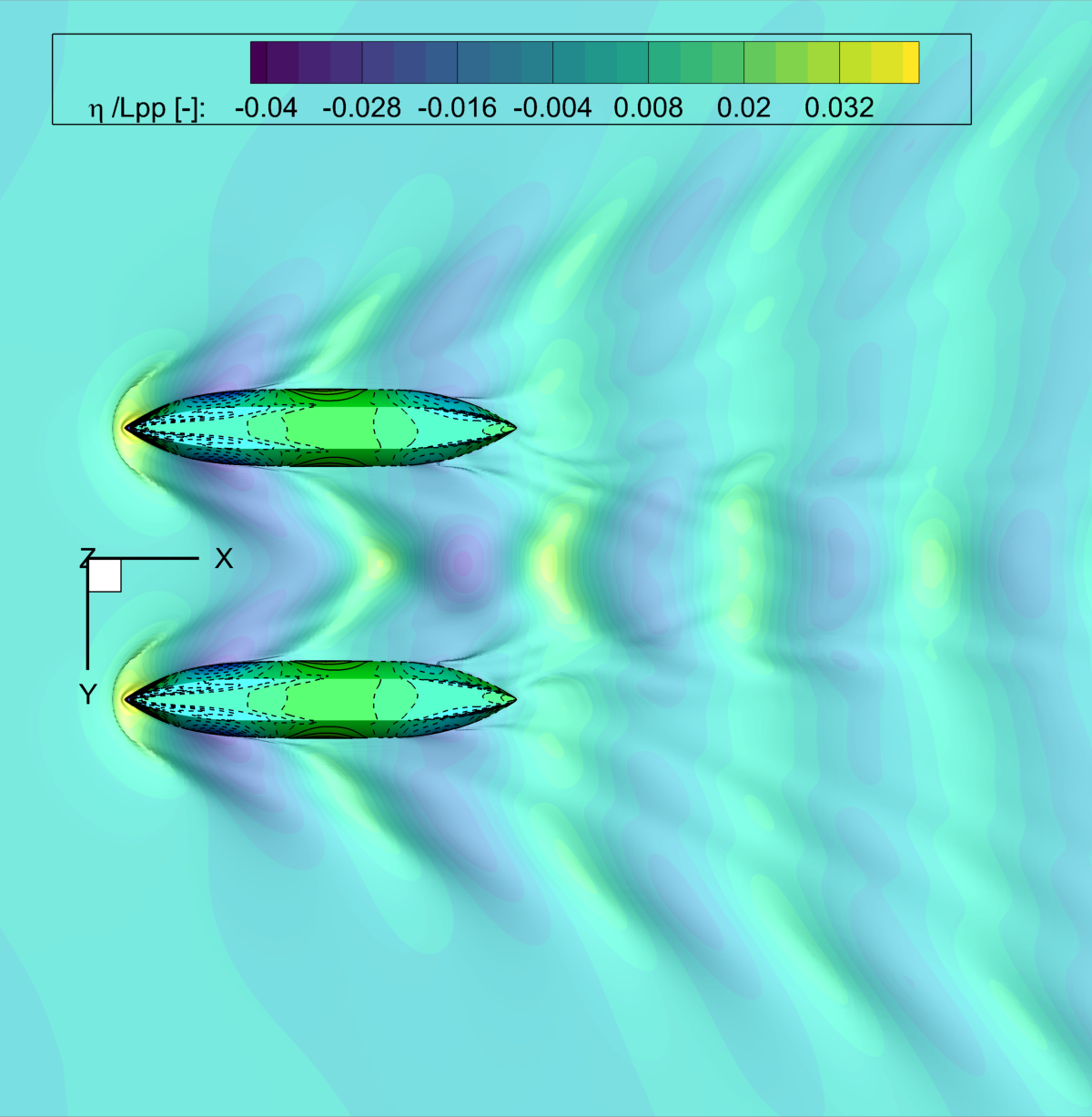}}   \hfill 
    \subfloat[RANSE, $\nabla = 58$kg]{\includegraphics[width=0.275\textwidth]{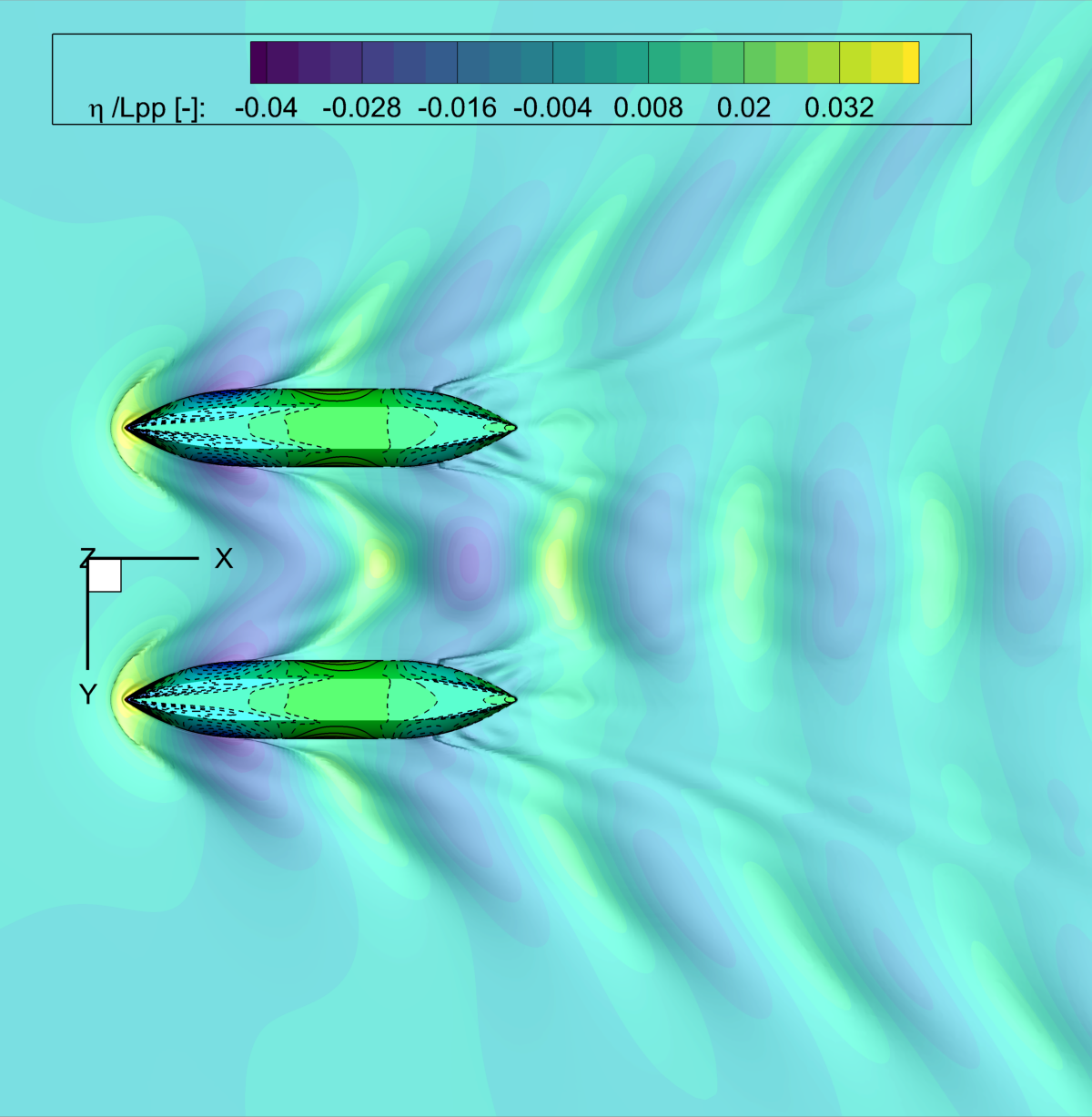}} \hfill \mbox{}  \\\vspace{-2mm}
    \mbox{} \hfill
    \subfloat[PF, $\nabla = 37$kg]{\includegraphics[width=0.275\textwidth]{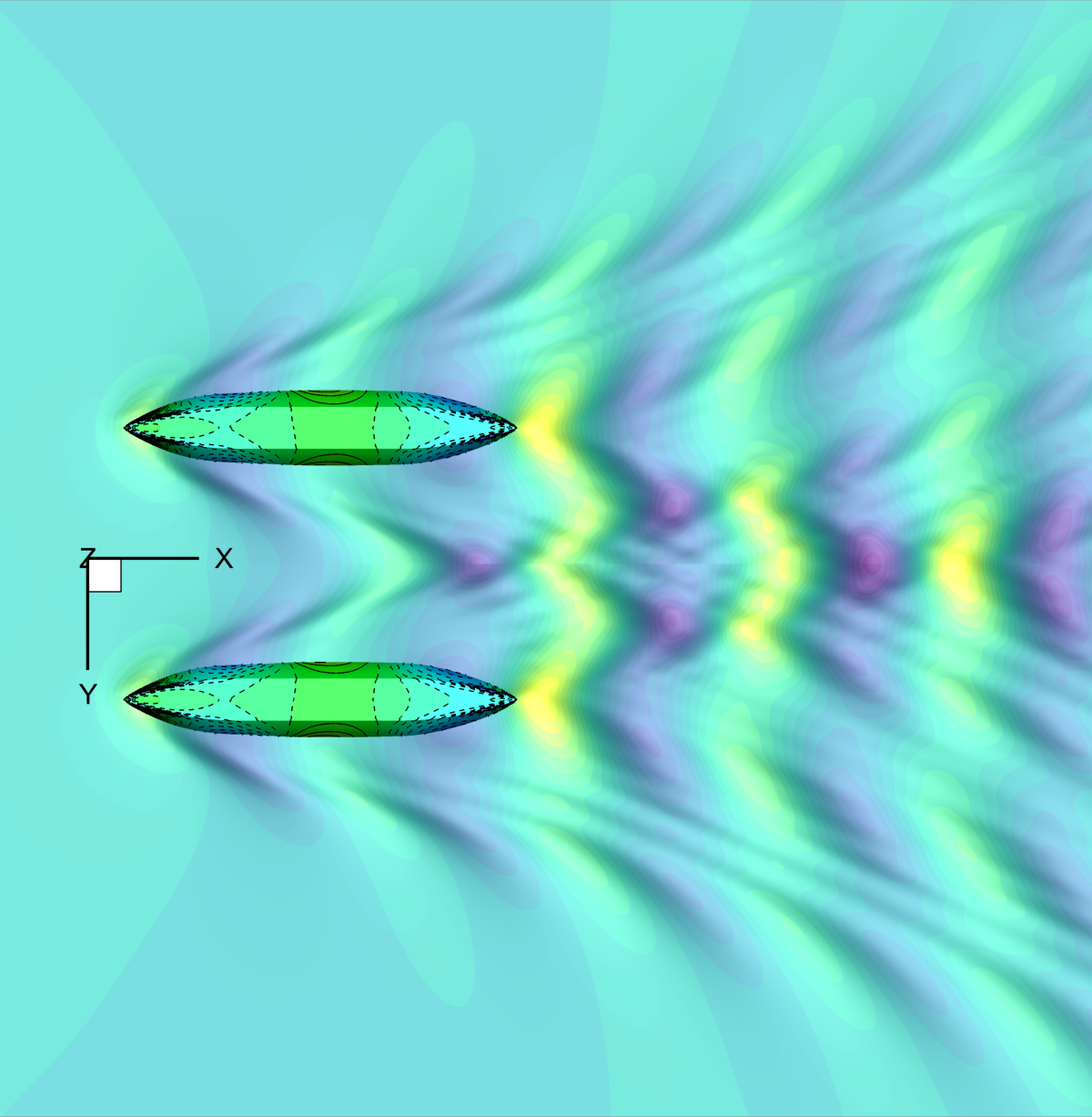}} \hfill 
    \subfloat[PF, $\nabla = 47.5$kg]{\includegraphics[width=0.275\textwidth]{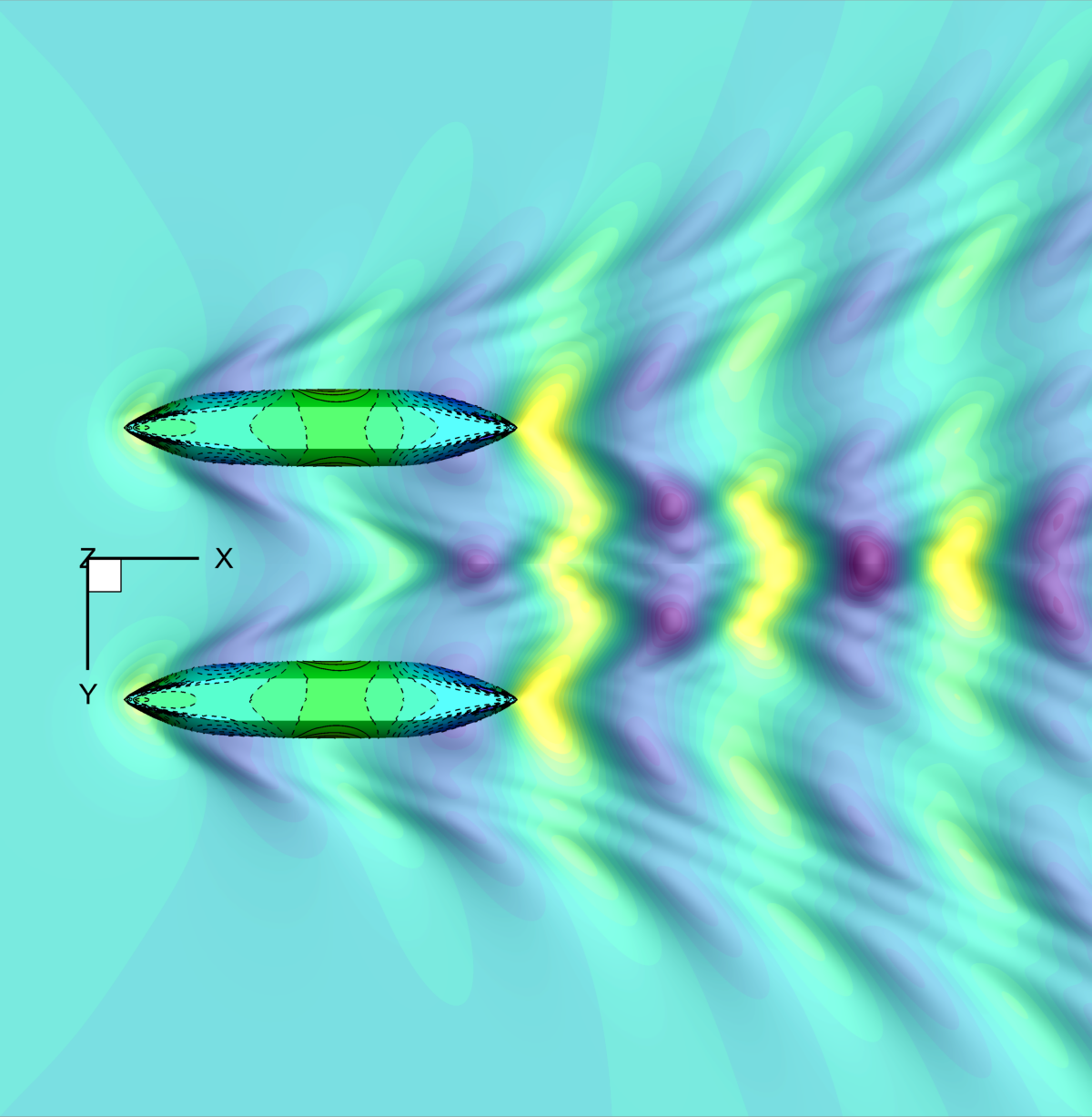}}   \hfill 
    \subfloat[PF, $\nabla = 58$kg]{\includegraphics[width=0.275\textwidth]{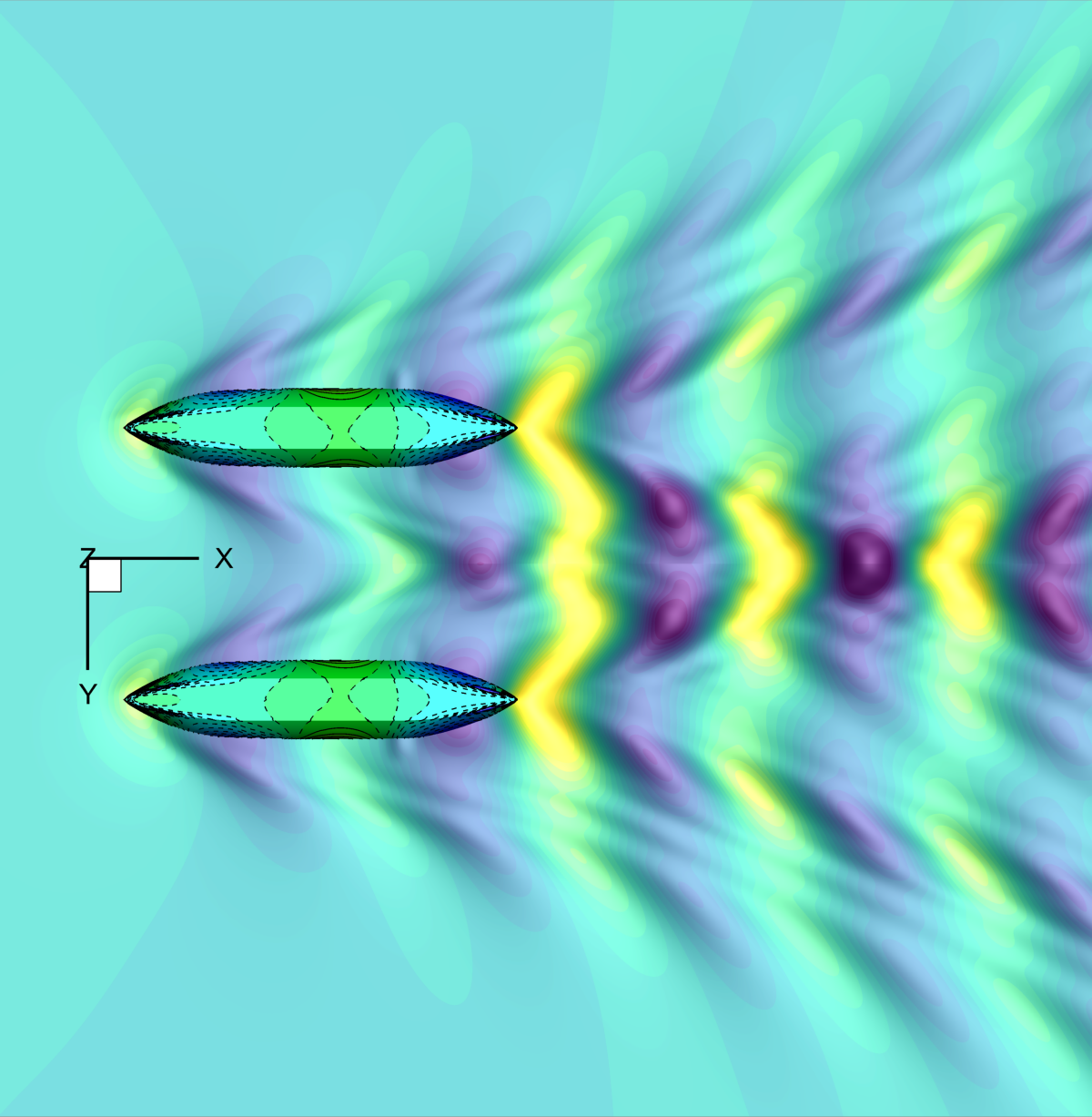}} \hfill \mbox{}  \\\vspace{-2mm}
    \mbox{} \hfill
    \subfloat[RANSE, $\nabla = 37$kg]{\includegraphics[width=0.275\textwidth]{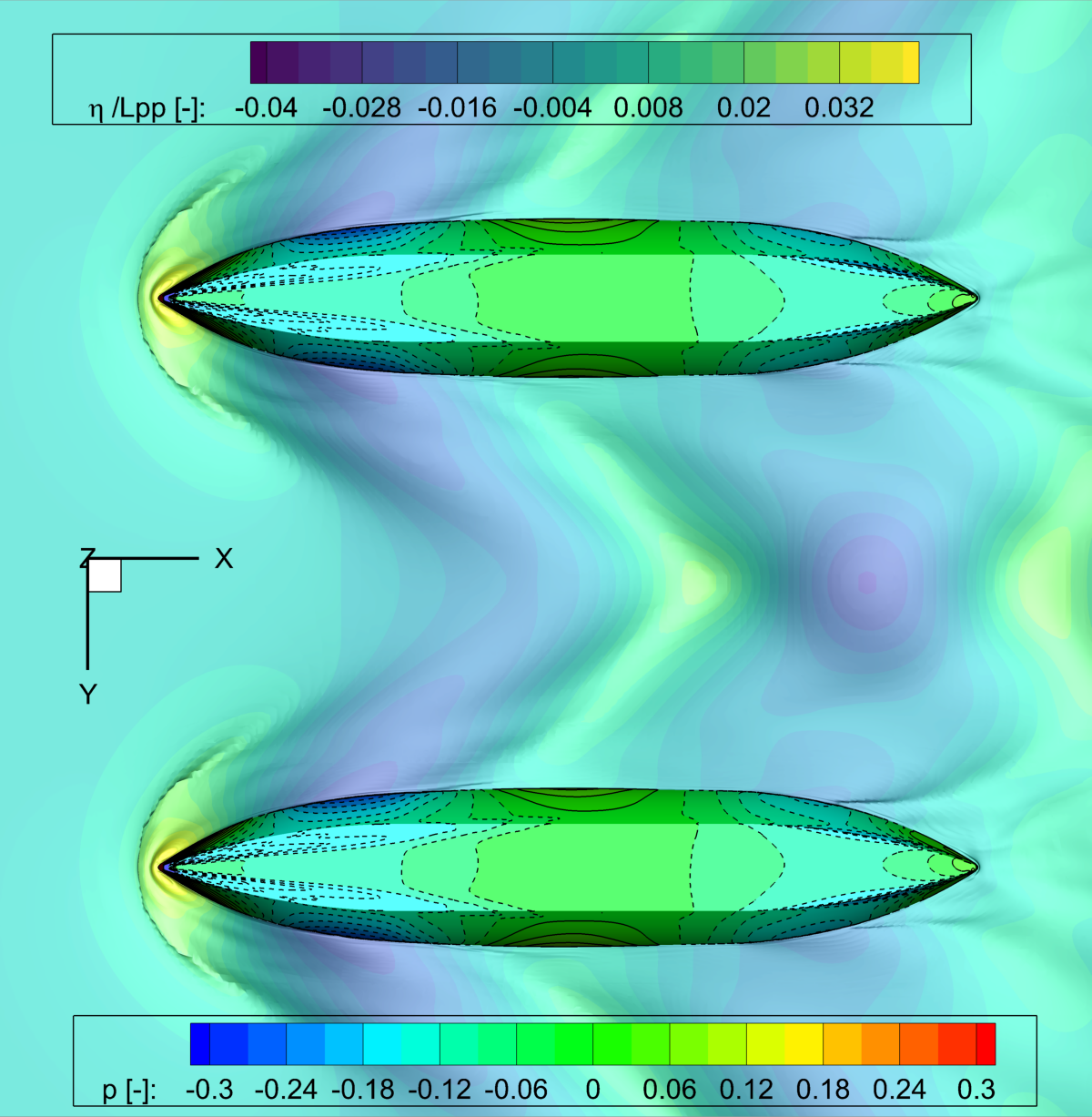}} \hfill 
    \subfloat[RANSE, $\nabla = 47.5$kg]{\includegraphics[width=0.275\textwidth]{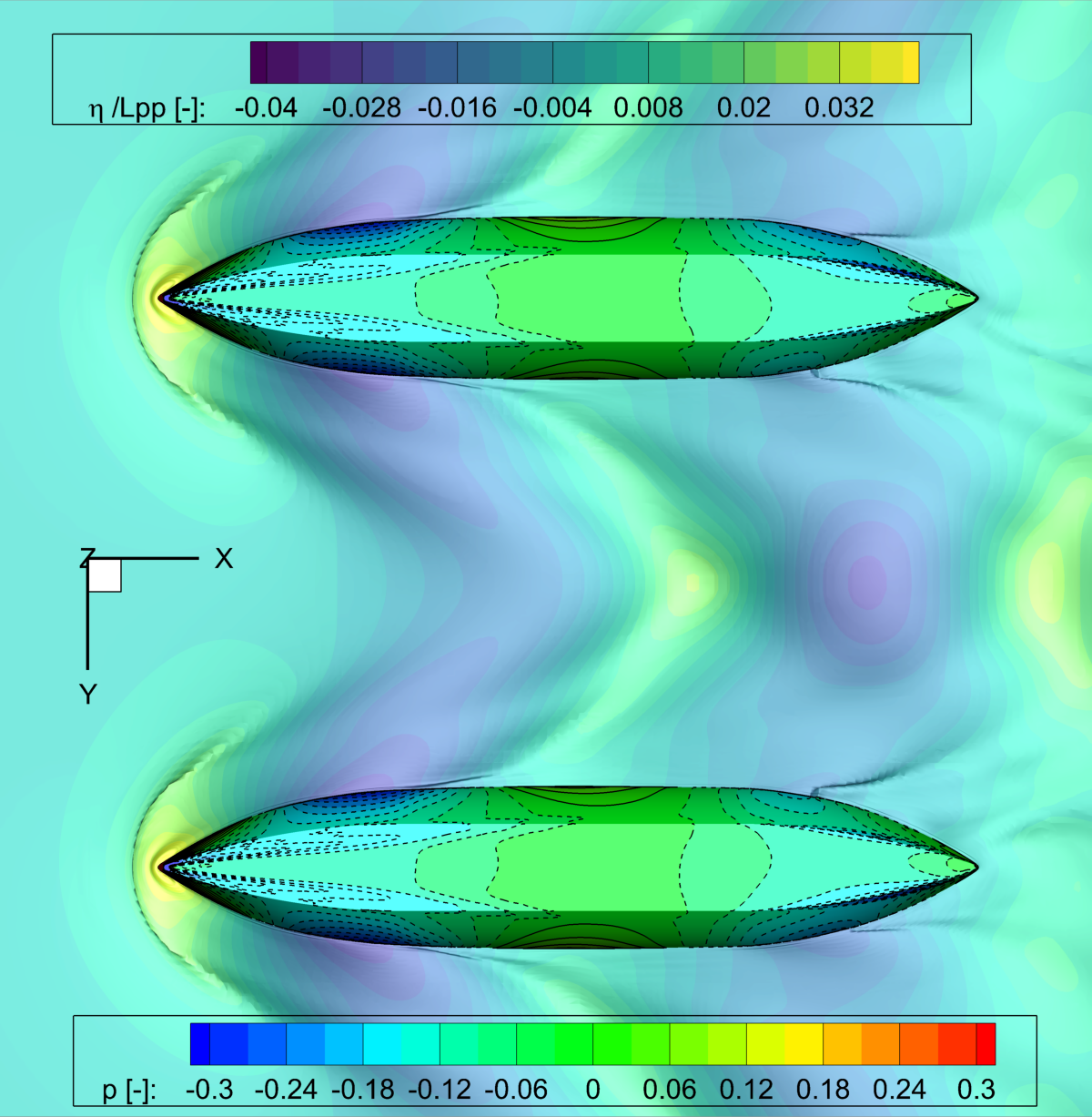}}   \hfill 
    \subfloat[RANSE, $\nabla = 58$kg]{\includegraphics[width=0.275\textwidth]{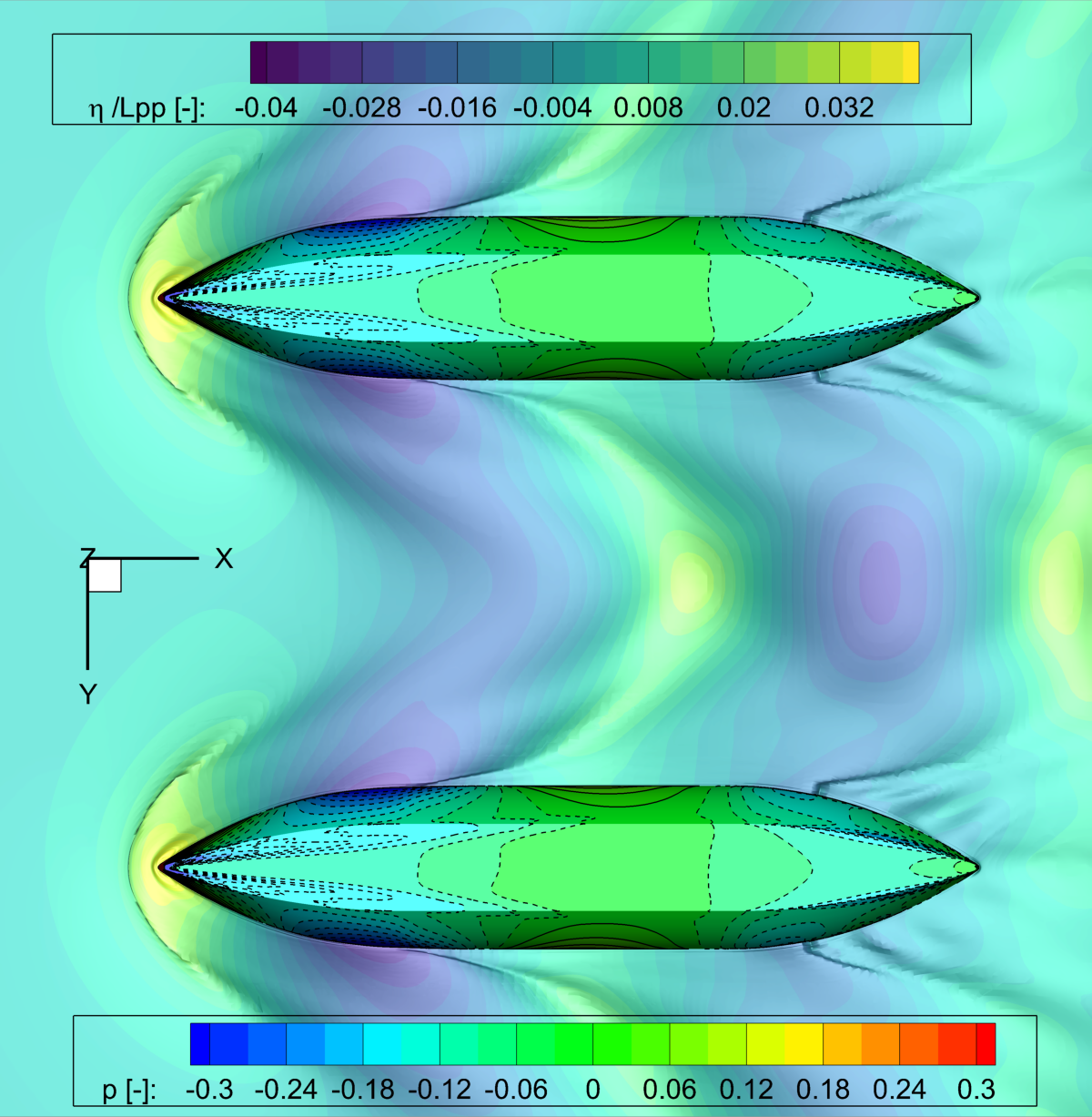}} \hfill \mbox{}  \\\vspace{-2mm}
    \mbox{} \hfill
    \subfloat[PF, $\nabla = 37$kg]{\includegraphics[width=0.275\textwidth]{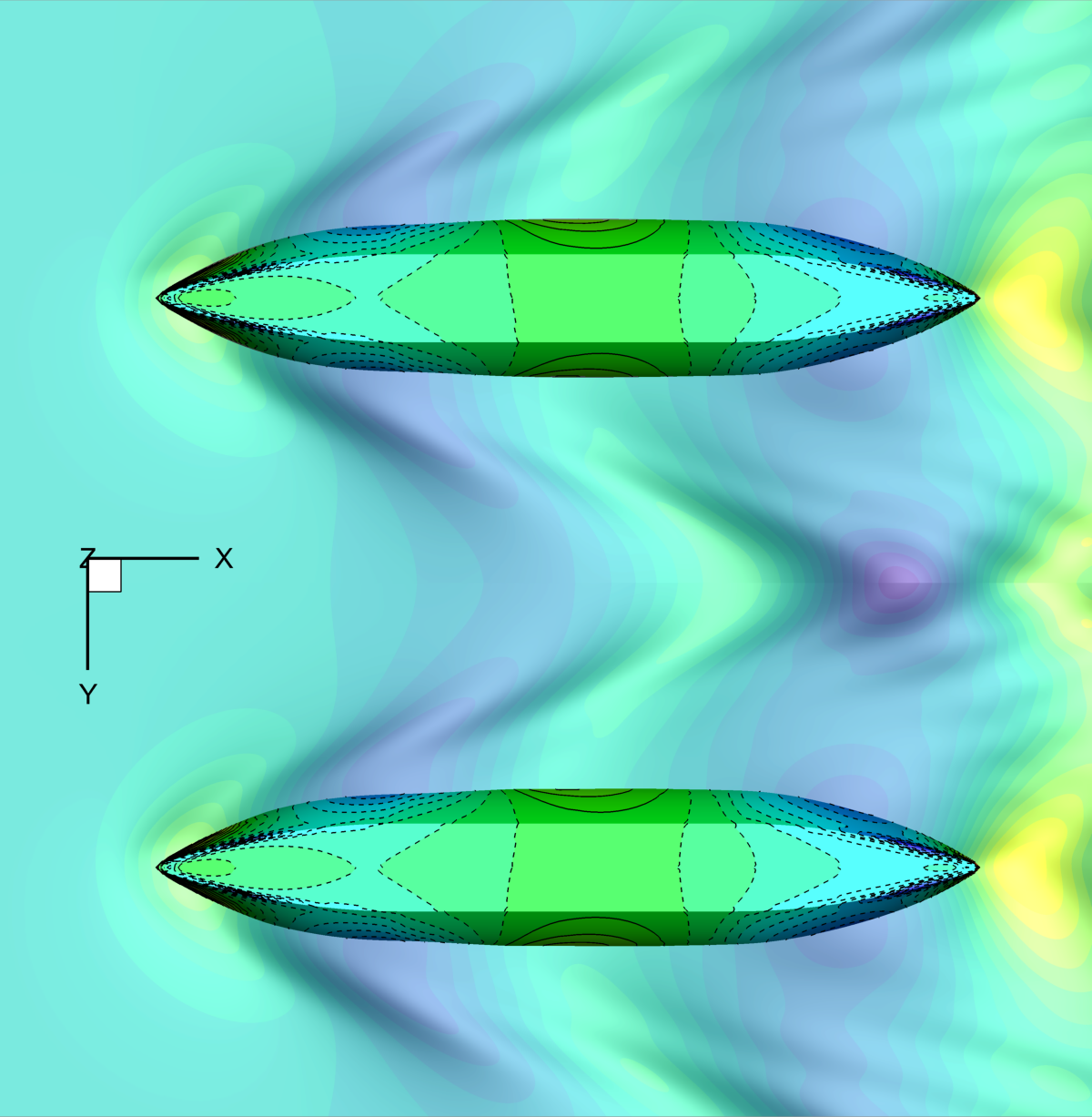}} \hfill 
    \subfloat[PF, $\nabla = 47.5$kg]{\includegraphics[width=0.275\textwidth]{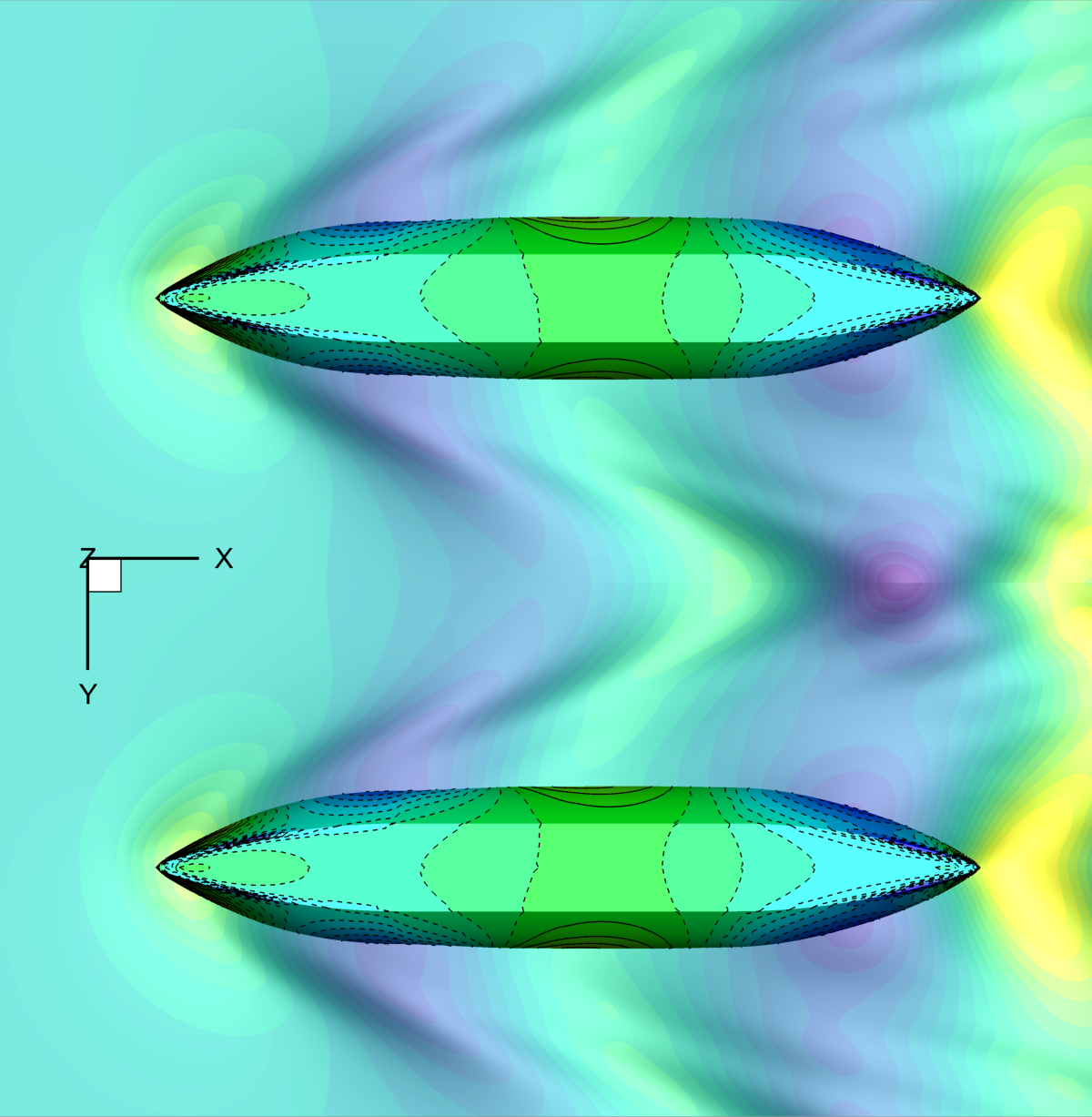}}   \hfill 
    \subfloat[PF, $\nabla = 58$kg]{\includegraphics[width=0.275\textwidth]{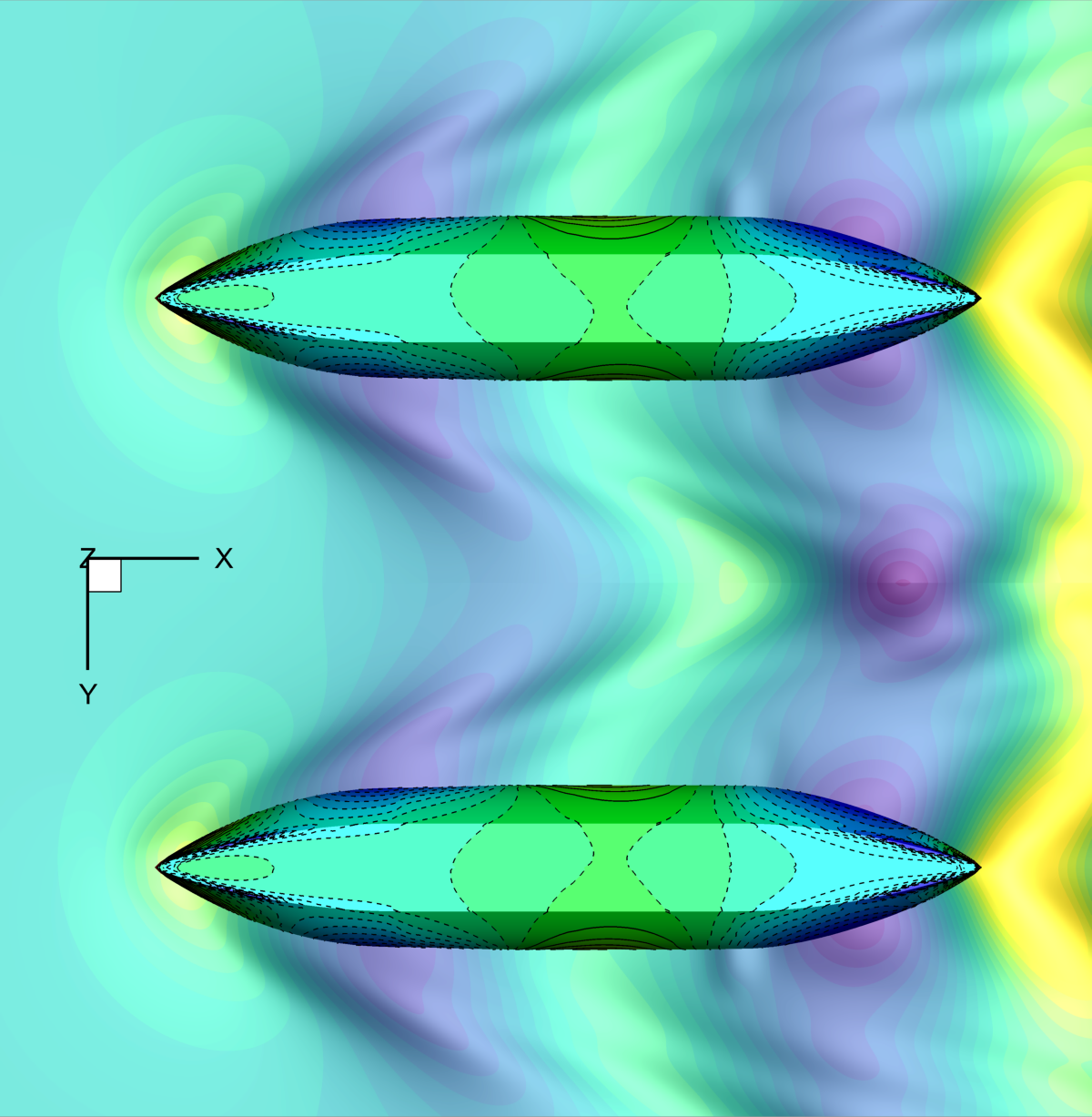}} \hfill \mbox{}  \\
    \caption{Pressure contour and wave elevation varying the SWAMP payload with $\delta x_G = 0\%$L, bottom view}\label{fig:FS0} 
\end{figure}

\begin{figure}[!h] 
    \centering 
    \mbox{} \hfill
    \subfloat[RANSE, $\nabla = 37$kg]{\includegraphics[width=0.275\textwidth]{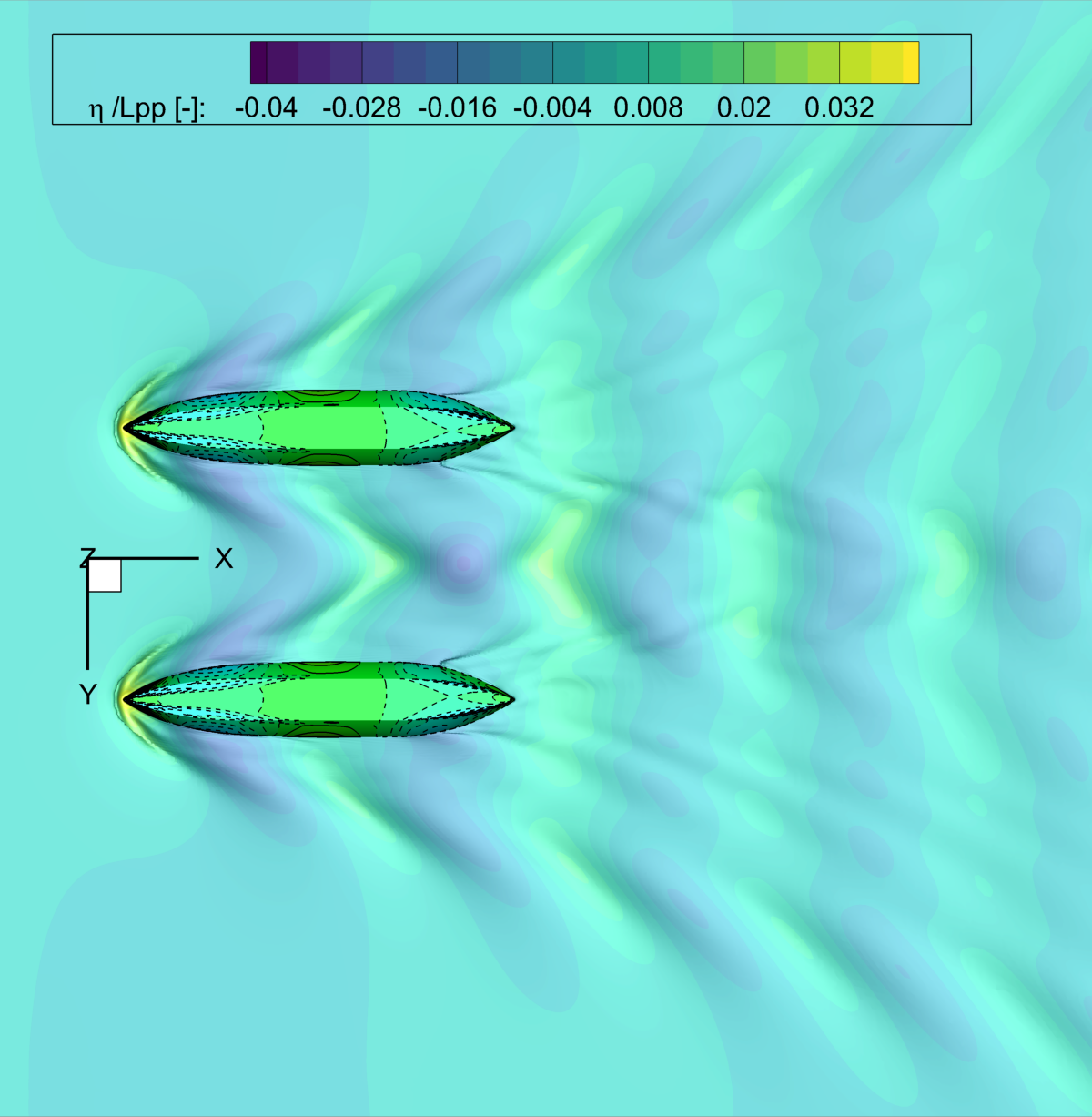}} \hfill 
    \subfloat[RANSE, $\nabla = 47.5$kg]{\includegraphics[width=0.275\textwidth]{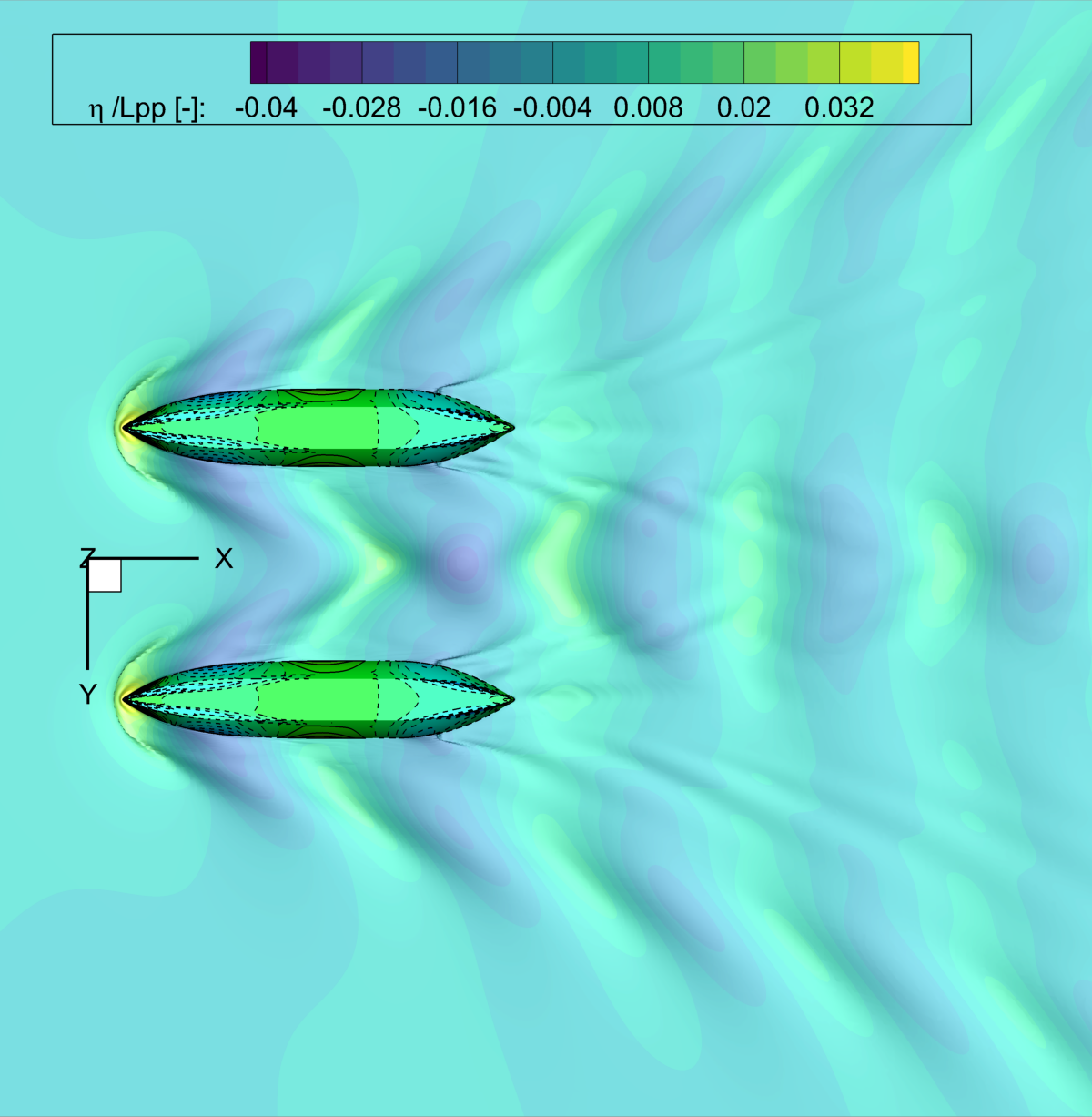}} \hfill   
    \subfloat[RANSE, $\nabla = 58$kg]{\includegraphics[width=0.275\textwidth]{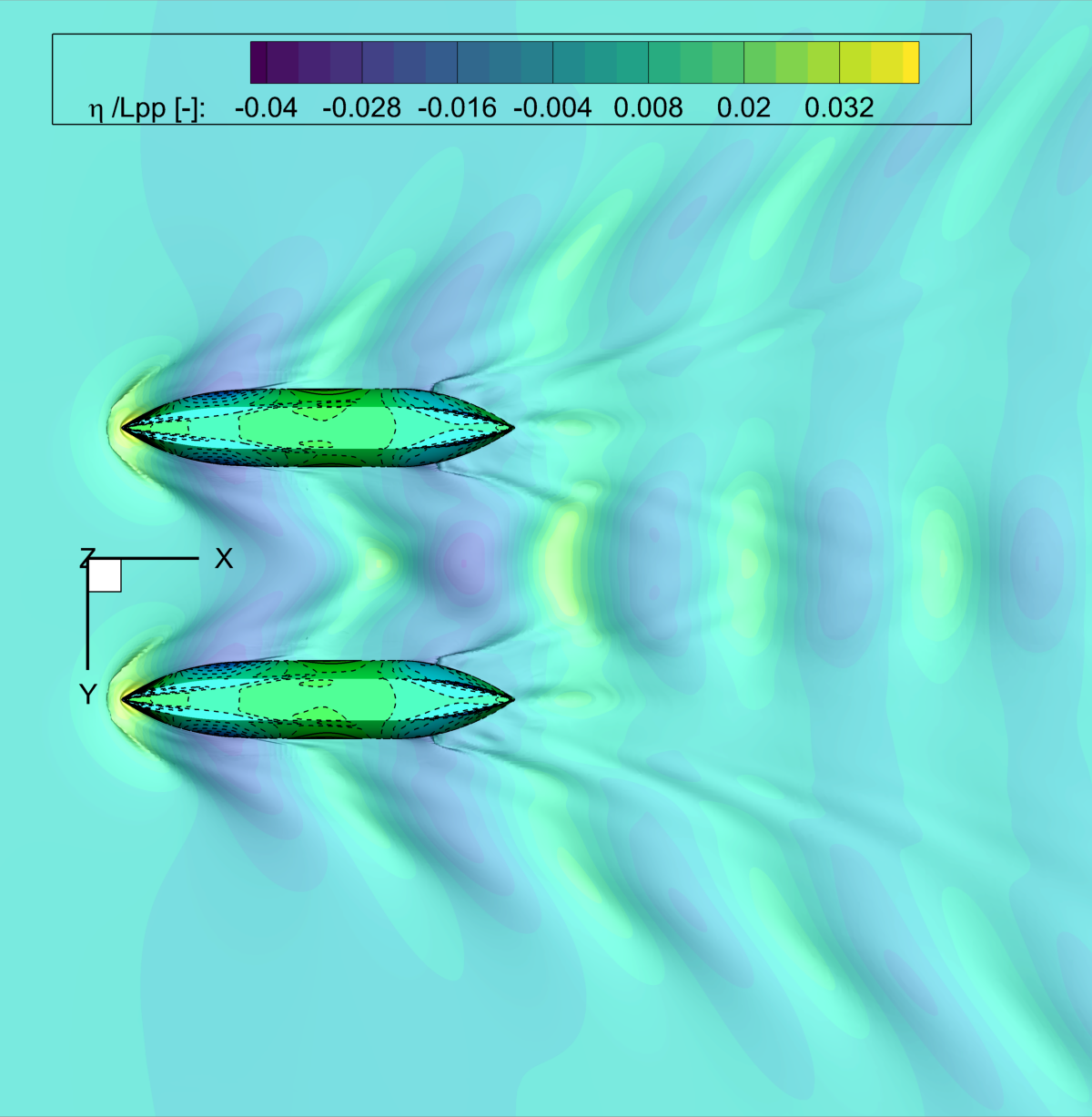}} \hfill \mbox{}  \\\vspace{-2mm}
    \mbox{} \hfill
    \subfloat[PF, $\nabla = 37$kg]{\includegraphics[width=0.275\textwidth]{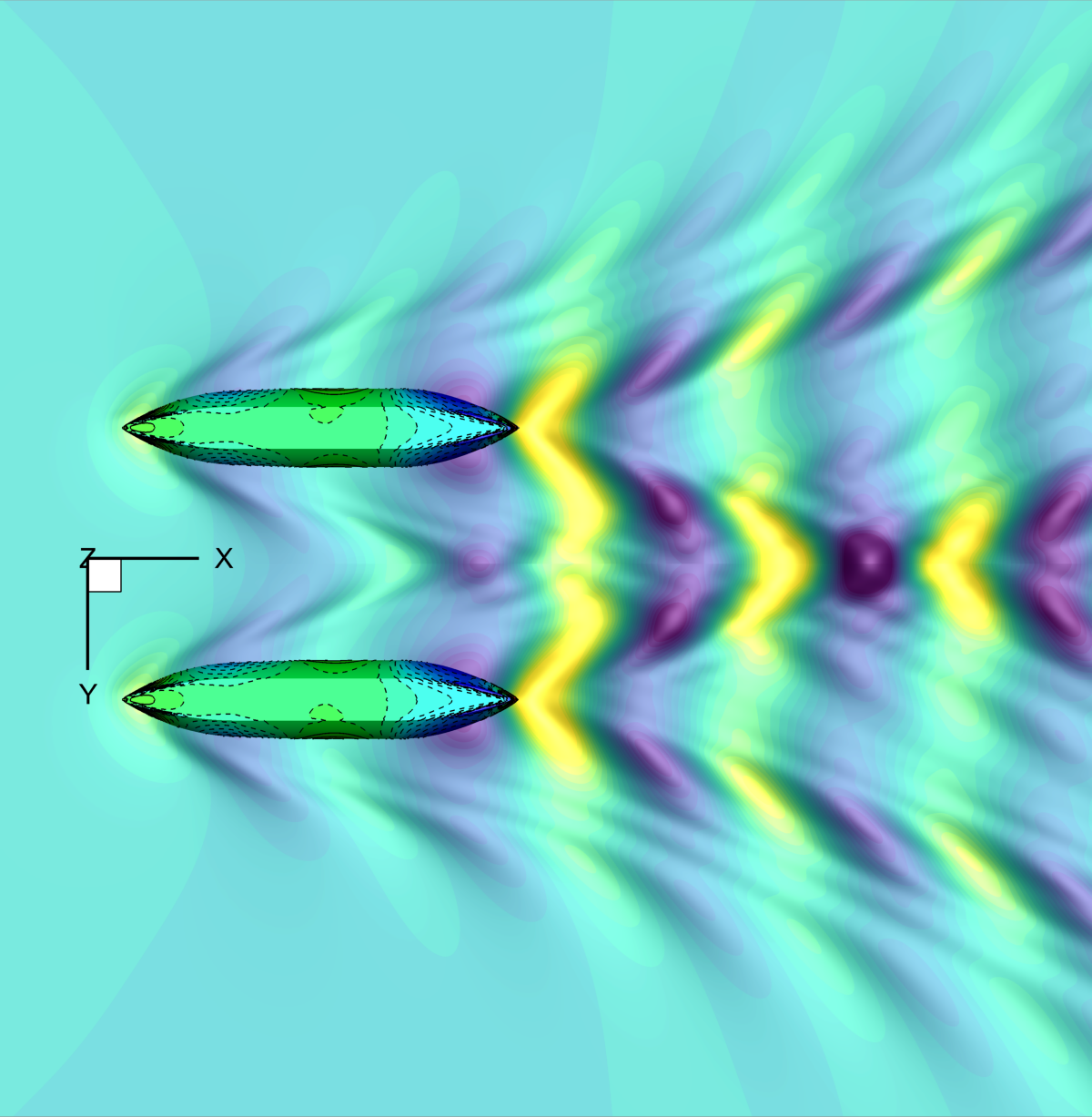}} \hfill 
    \subfloat[PF, $\nabla = 47.5$kg]{\includegraphics[width=0.275\textwidth]{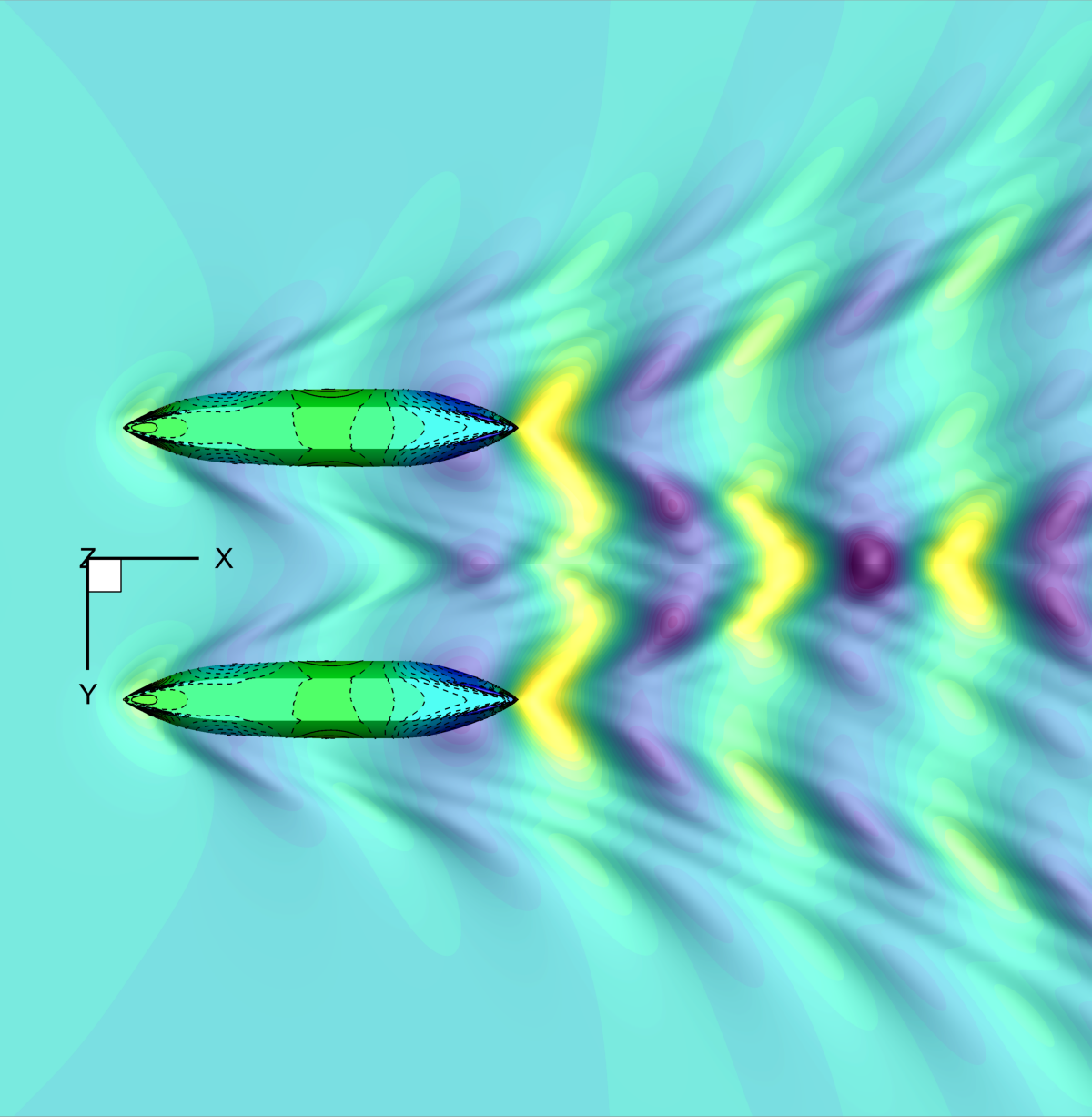}} \hfill   
    \subfloat[PF, $\nabla = 58$kg]{\includegraphics[width=0.275\textwidth]{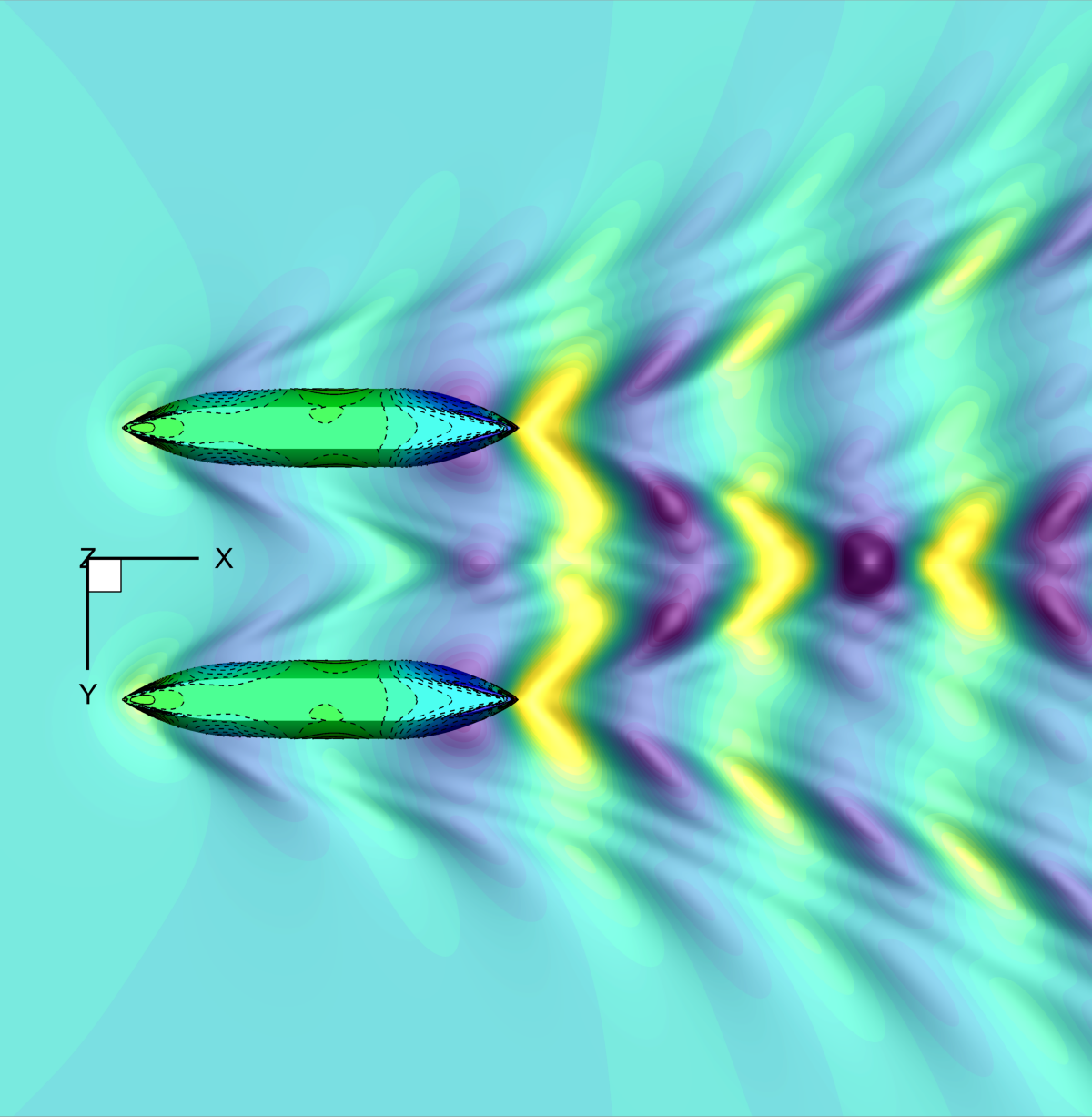}} \hfill \mbox{}  \\\vspace{-2mm}
    \mbox{} \hfill
    \subfloat[RANSE, $\nabla = 37$kg]{\includegraphics[width=0.275\textwidth]{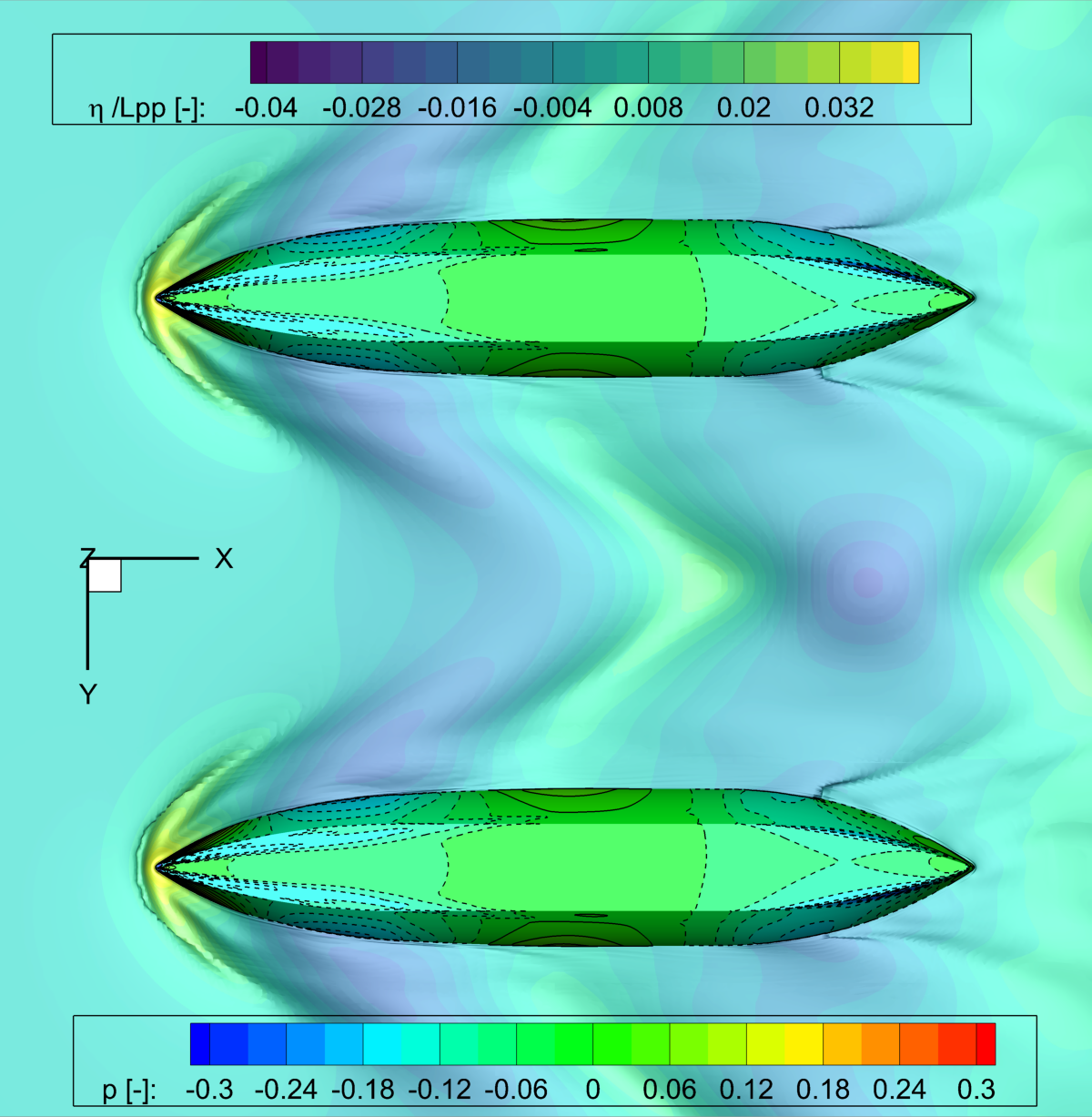}} \hfill 
    \subfloat[RANSE, $\nabla = 47.5$kg]{\includegraphics[width=0.275\textwidth]{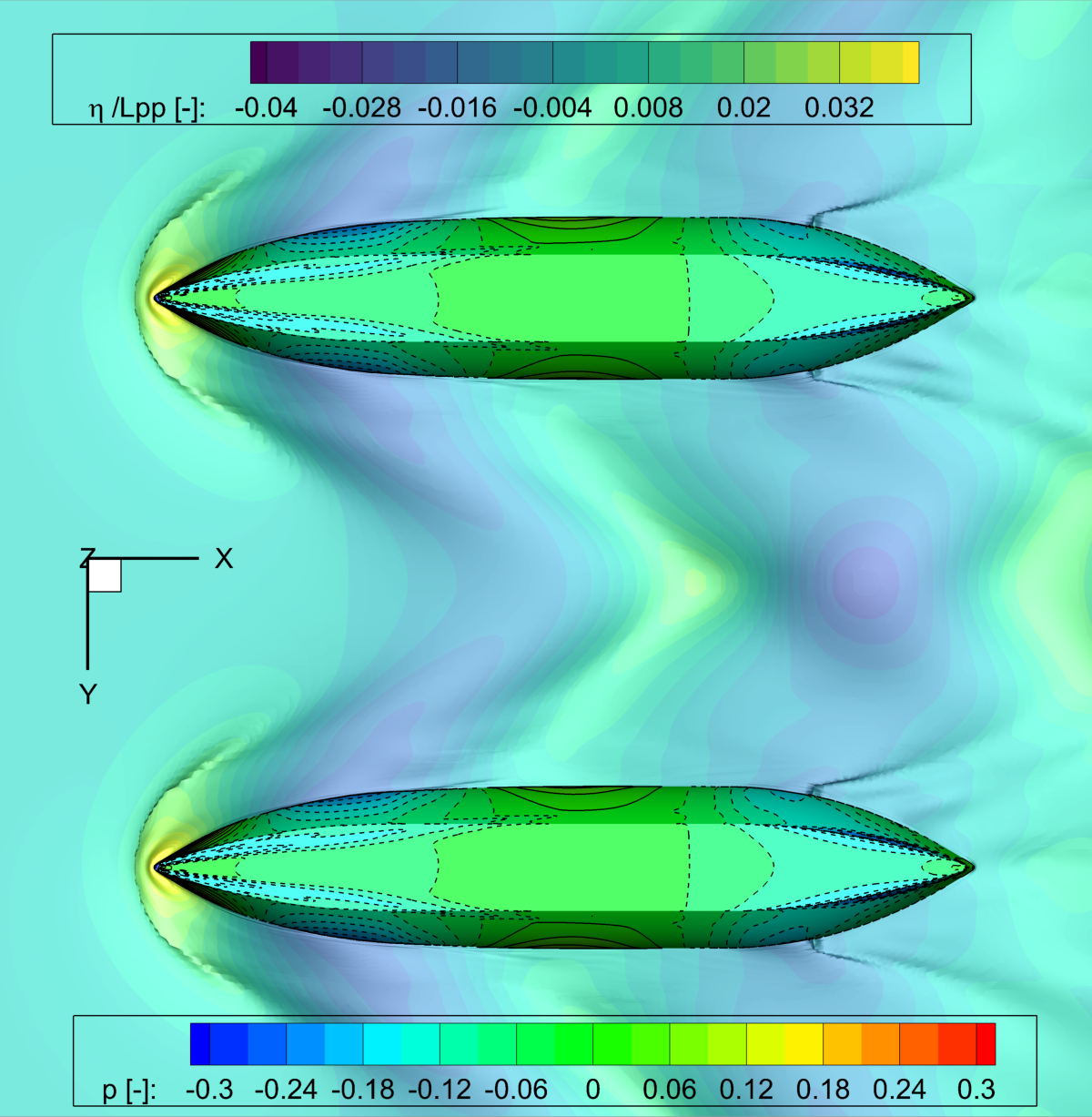}}  \hfill  
    \subfloat[RANSE, $\nabla = 58$kg]{\includegraphics[width=0.275\textwidth]{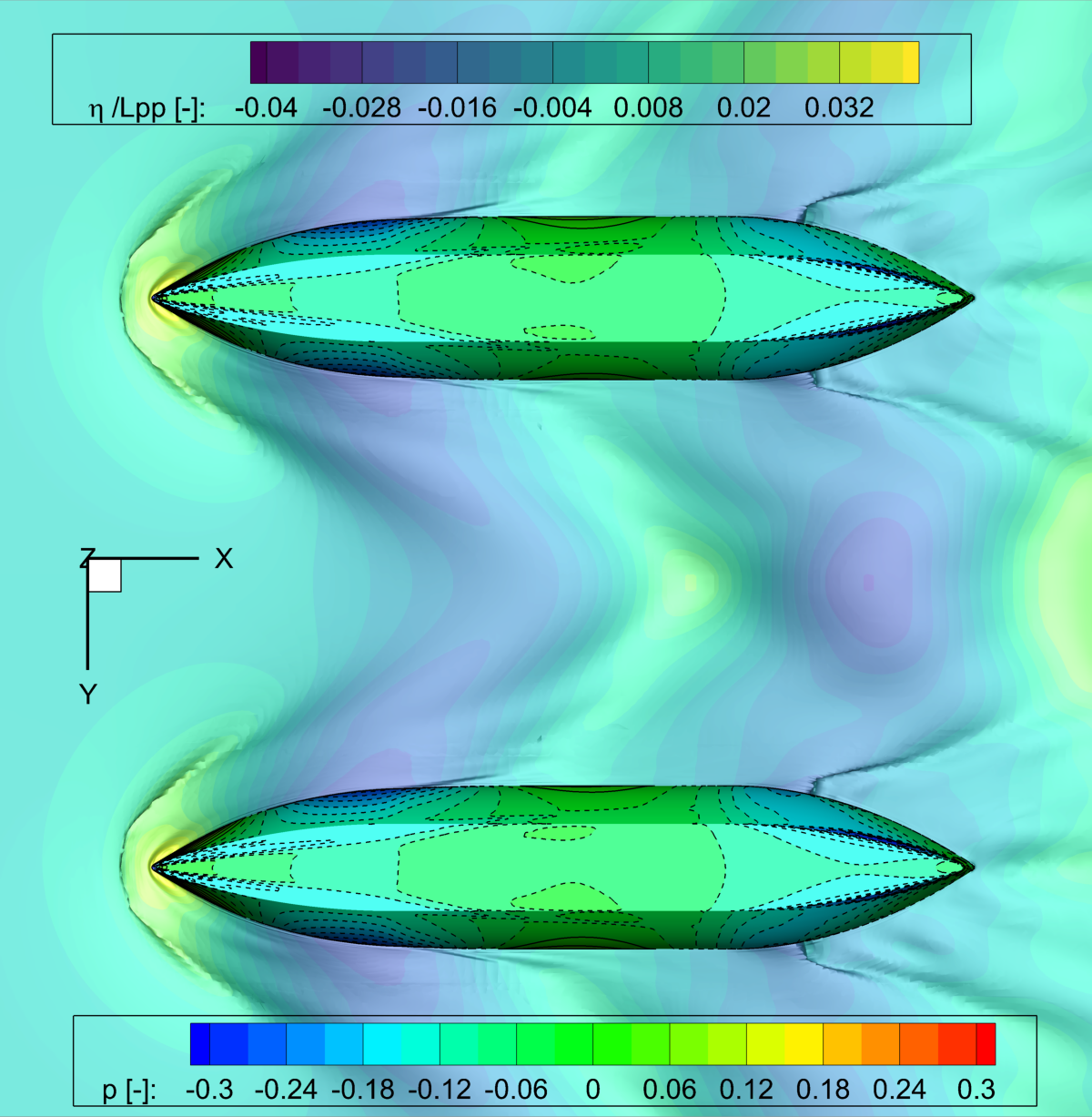}} \hfill \mbox{}  \\\vspace{-2mm}
    \mbox{} \hfill
    \subfloat[PF, $\nabla = 37$kg]{\includegraphics[width=0.275\textwidth]{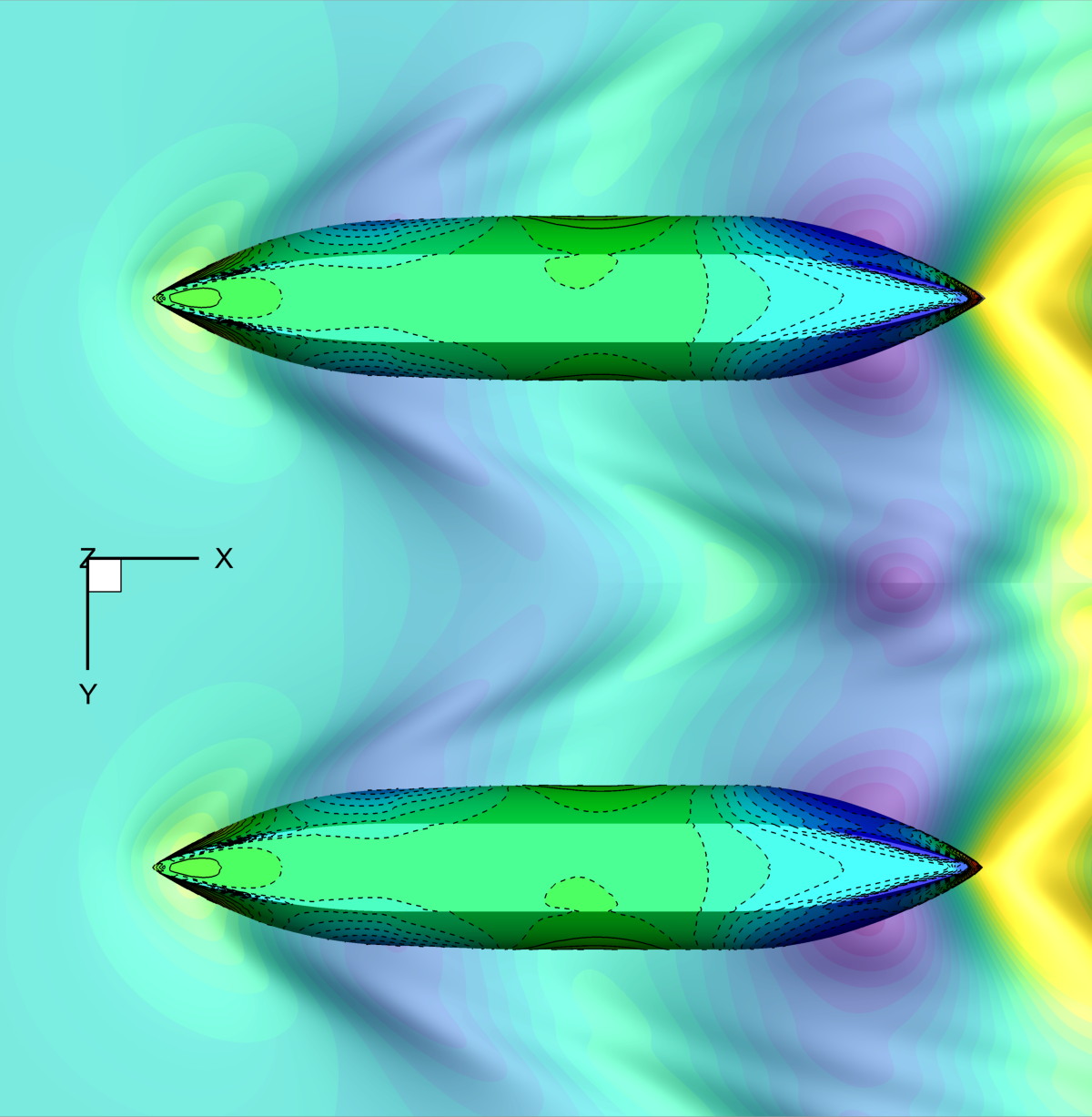}} \hfill 
    \subfloat[PF, $\nabla = 47.5$kg]{\includegraphics[width=0.275\textwidth]{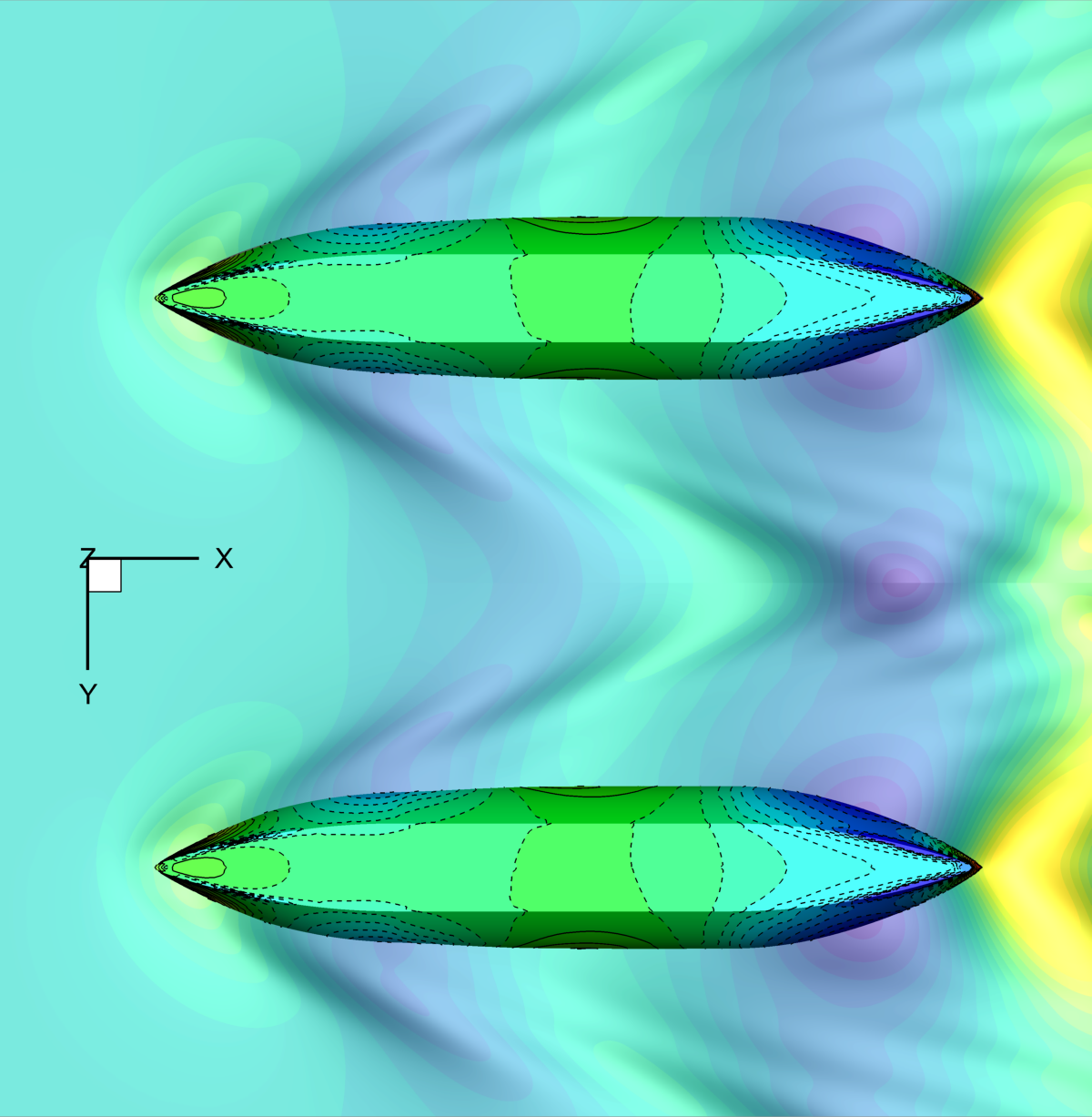}} \hfill   
    \subfloat[PF, $\nabla = 58$kg]{\includegraphics[width=0.275\textwidth]{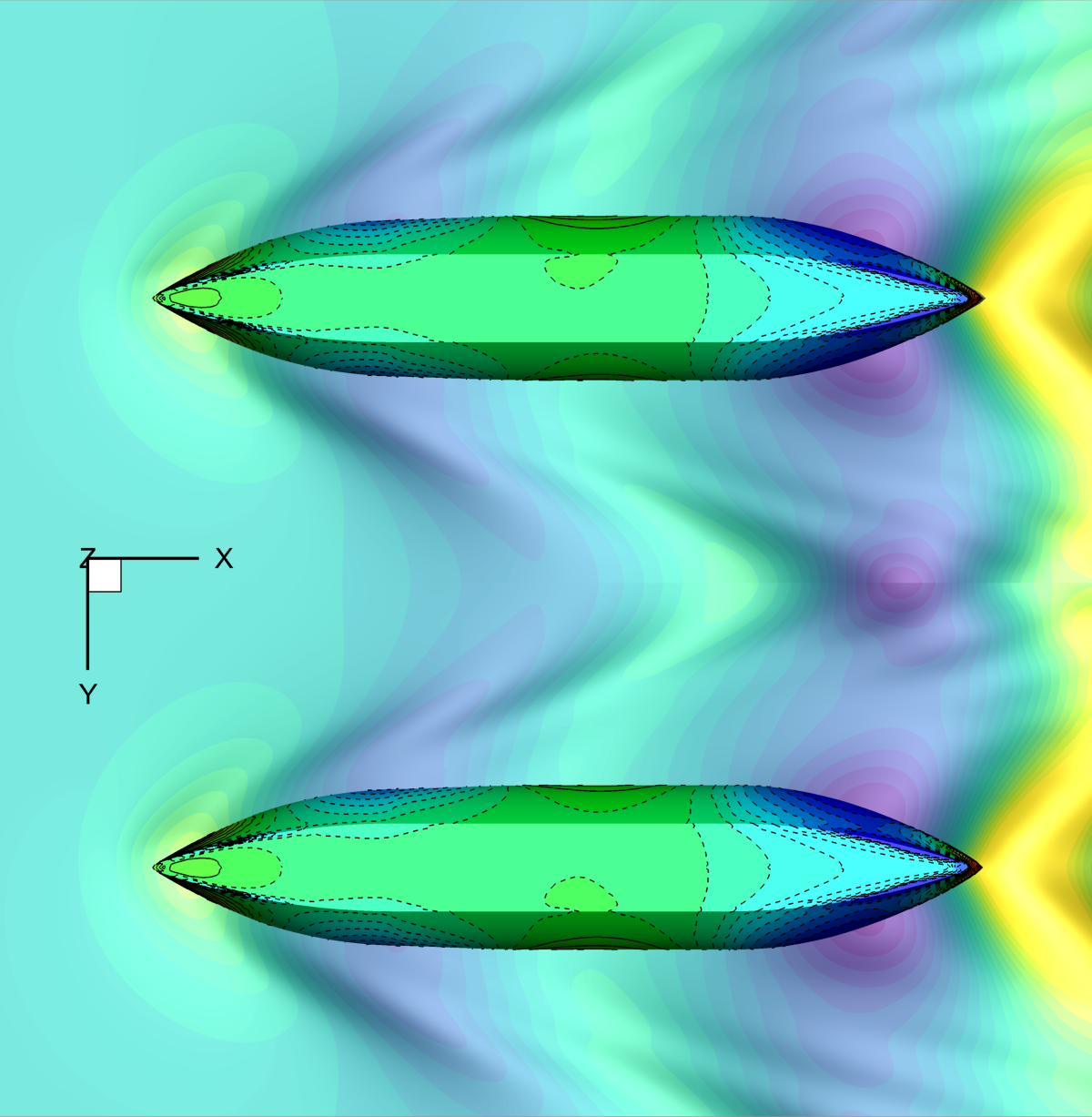}} \hfill \mbox{}  \\
    \caption{Pressure contour and wave elevation varying the SWAMP payload with $\delta x_G = 3\%$L, bottom view}\label{fig:FS3} 
\end{figure}

\begin{figure}[!h] 
    \centering 
    \mbox{} \hfill
    \subfloat[RANSE, $\nabla = 37$kg]{\includegraphics[width=0.275\textwidth]{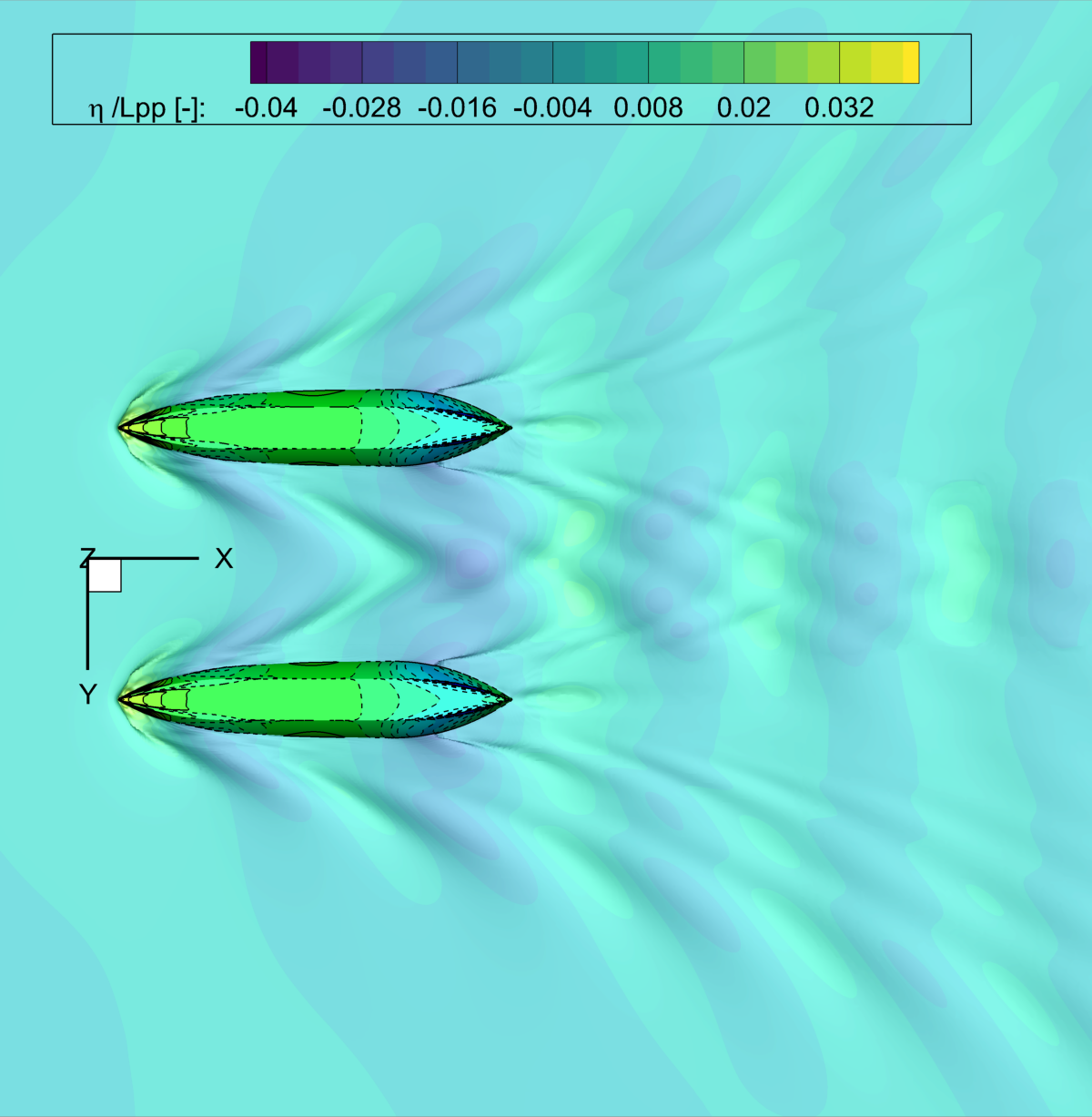}} \hfill 
    \subfloat[RANSE, $\nabla = 47.5$kg]{\includegraphics[width=0.275\textwidth]{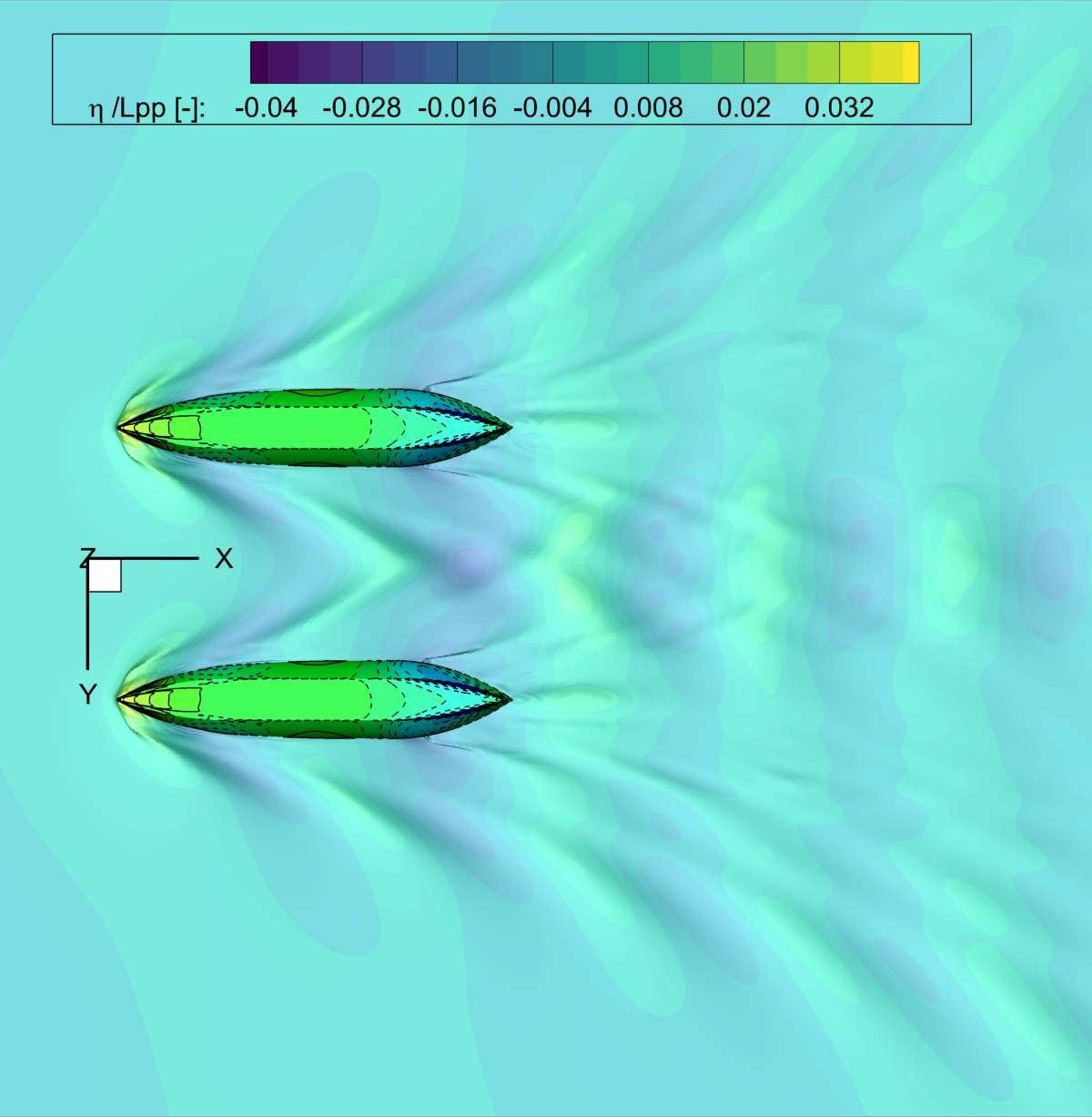}}   \hfill 
    \subfloat[RANSE, $\nabla = 58$kg]{\includegraphics[width=0.275\textwidth]{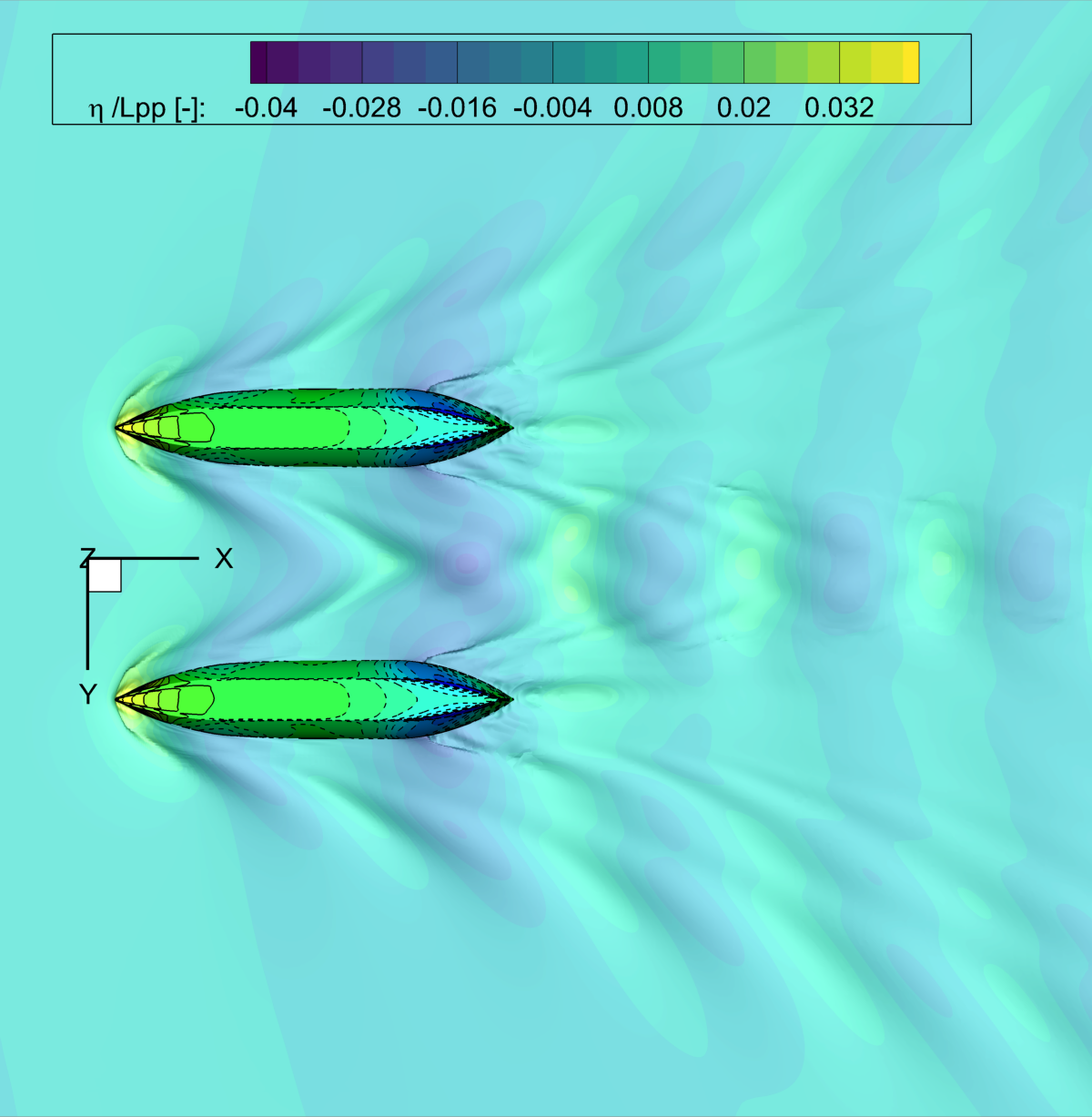}} \hfill \mbox{}  \\\vspace{-2mm}
    \mbox{} \hfill
    \subfloat[PF, $\nabla = 37$kg]{\includegraphics[width=0.275\textwidth]{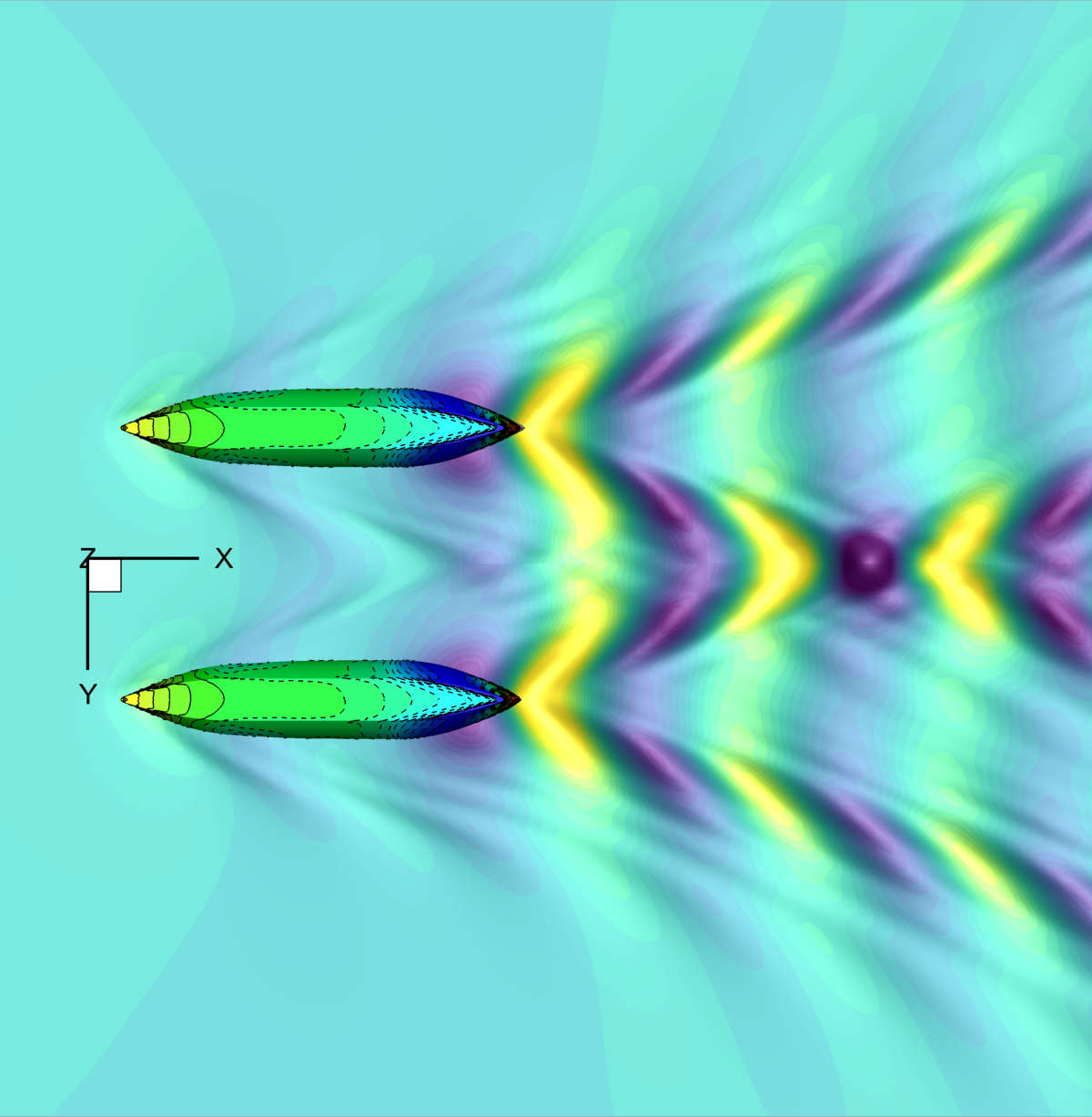}} \hfill 
    \subfloat[PF, $\nabla = 47.5$kg]{\includegraphics[width=0.275\textwidth]{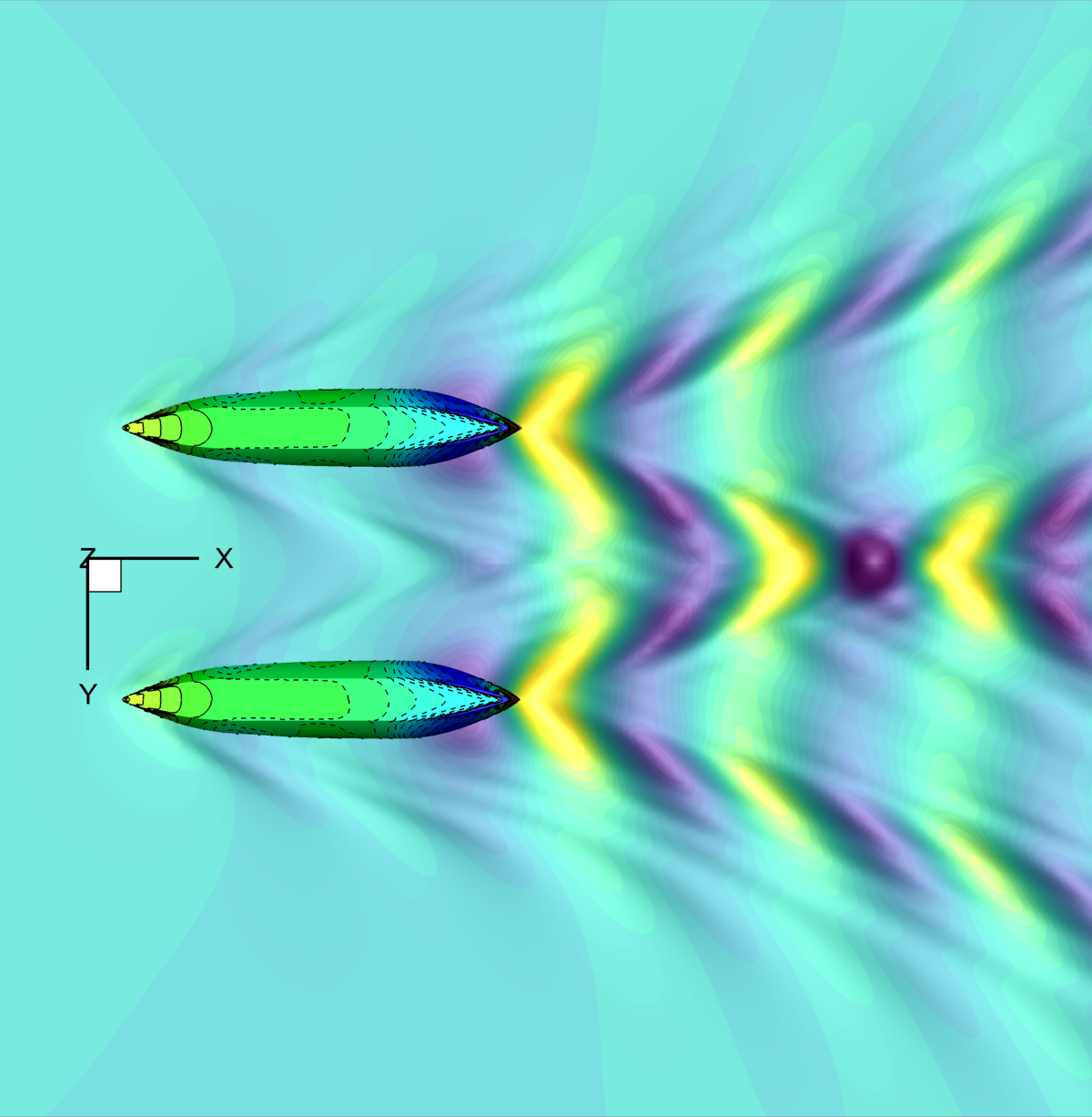}} \hfill   
    \subfloat[PF, $\nabla = 58$kg]{\includegraphics[width=0.275\textwidth]{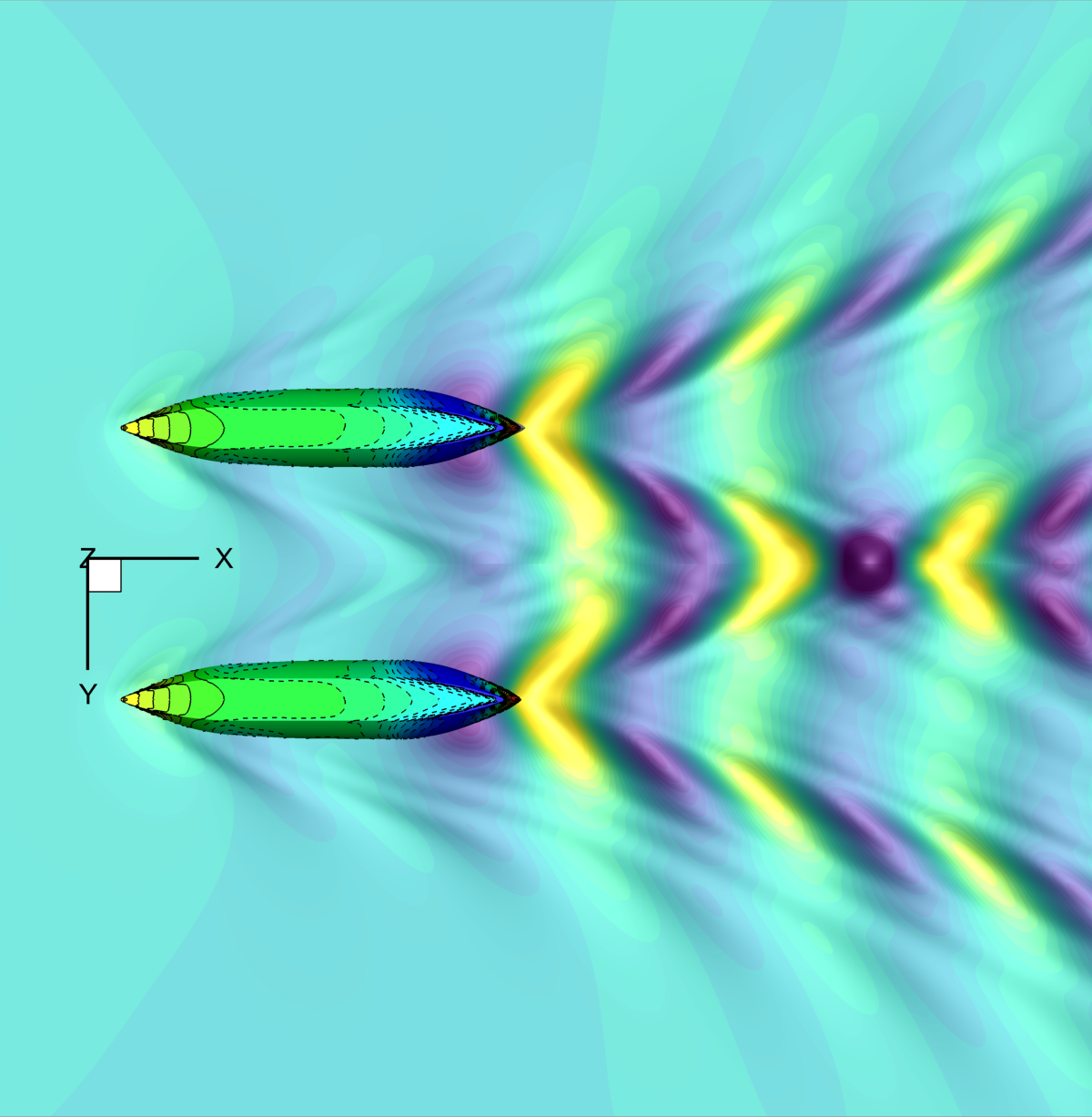}} \hfill \mbox{}  \\\vspace{-2mm}
    \mbox{} \hfill
    \subfloat[RANSE, $\nabla = 37$kg]{\includegraphics[width=0.275\textwidth]{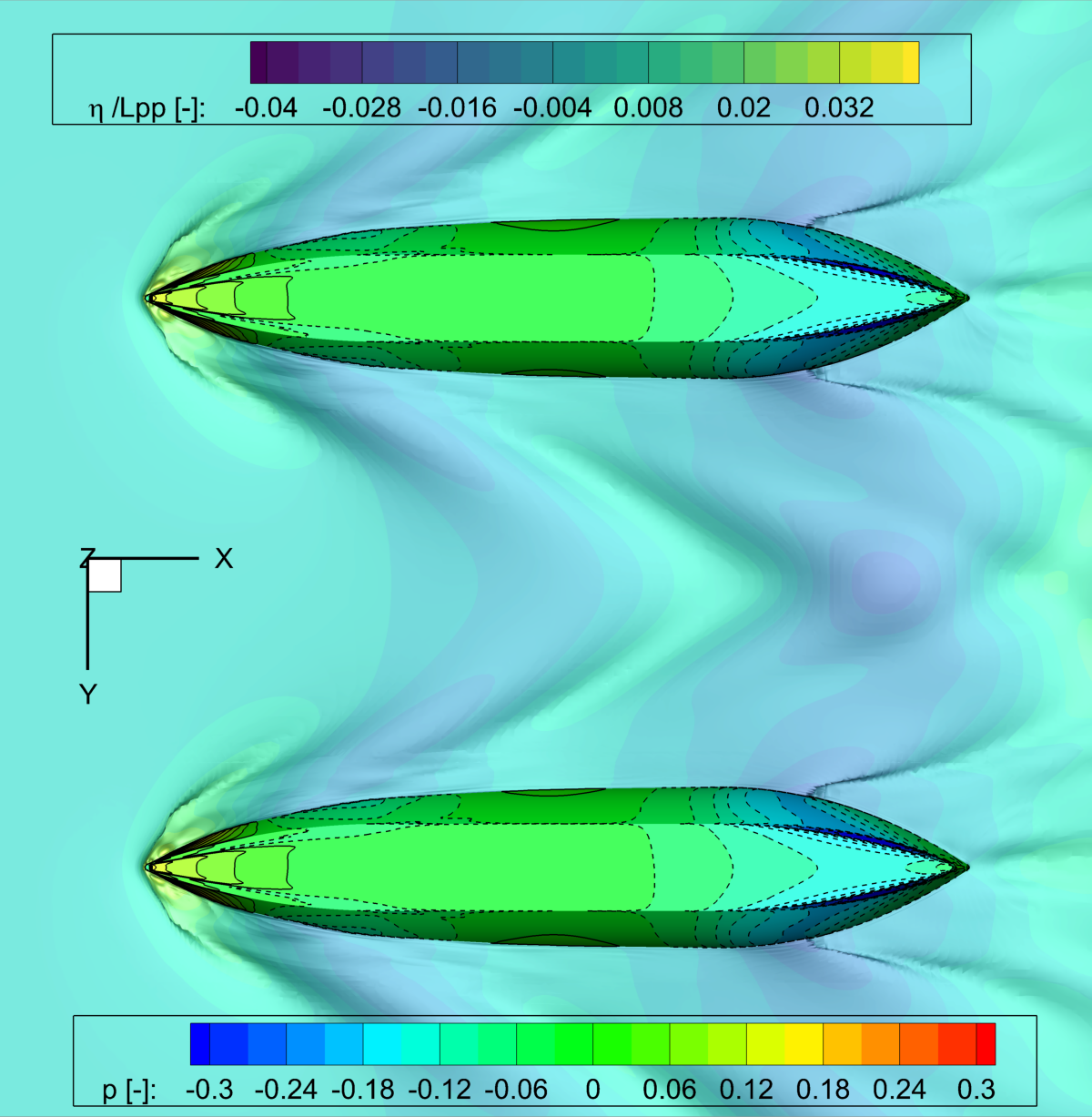}} \hfill 
    \subfloat[RANSE, $\nabla = 47.5$kg]{\includegraphics[width=0.275\textwidth]{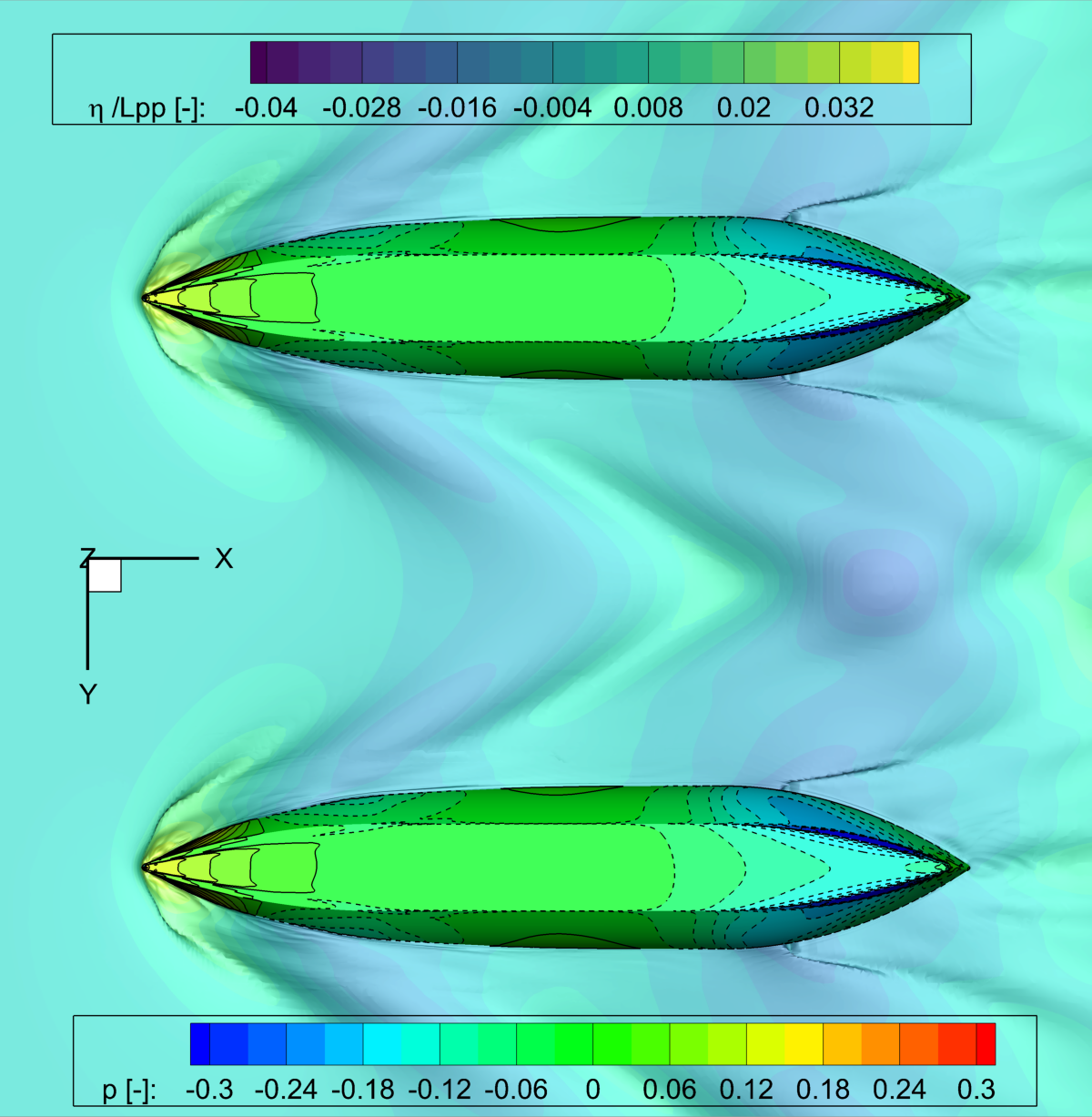}}  \hfill  
    \subfloat[RANSE, $\nabla = 58$kg]{\includegraphics[width=0.275\textwidth]{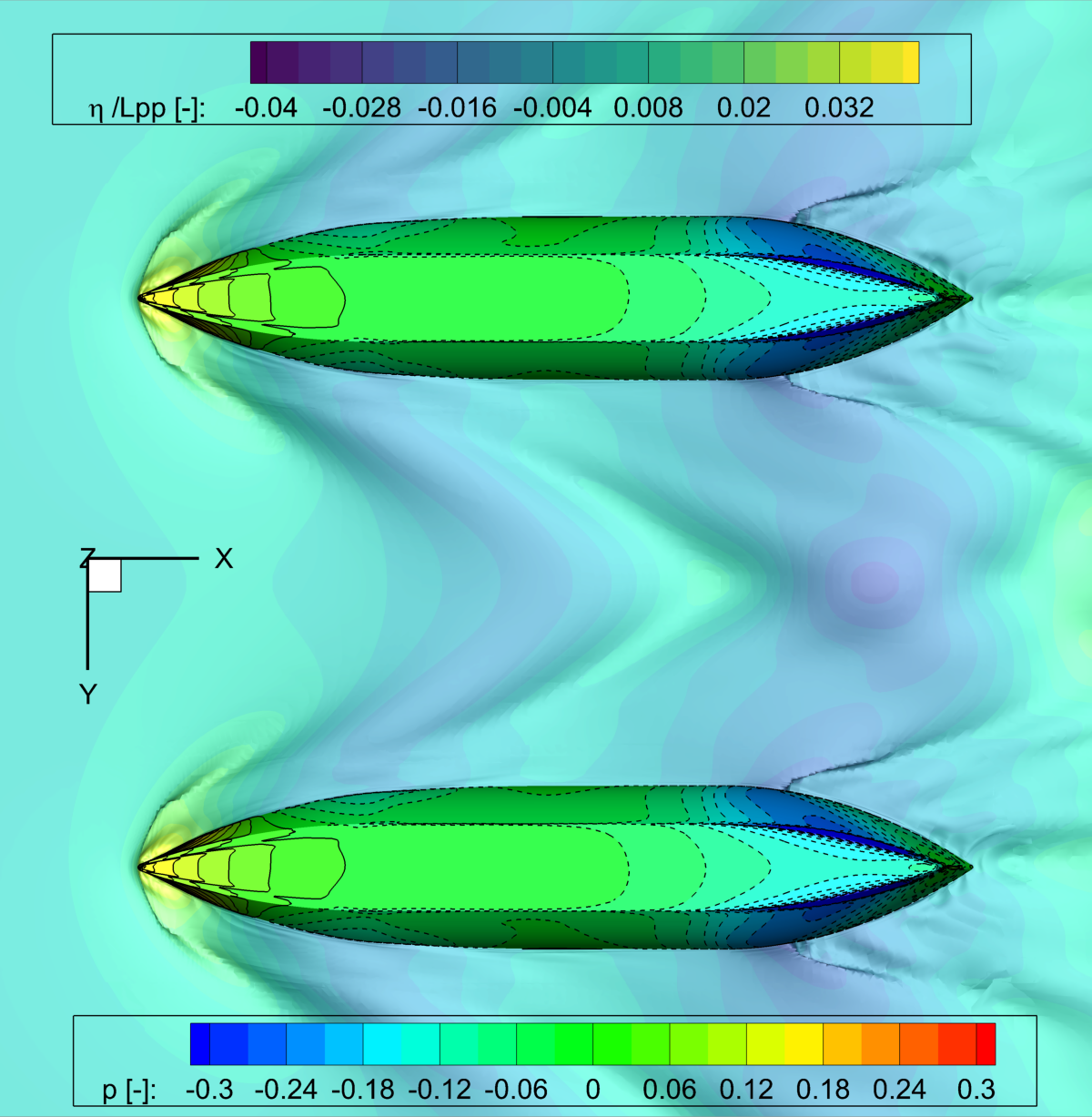}} \hfill \mbox{}  \\\vspace{-2mm}
    \mbox{} \hfill
    \subfloat[PF, $\nabla = 37$kg]{\includegraphics[width=0.275\textwidth]{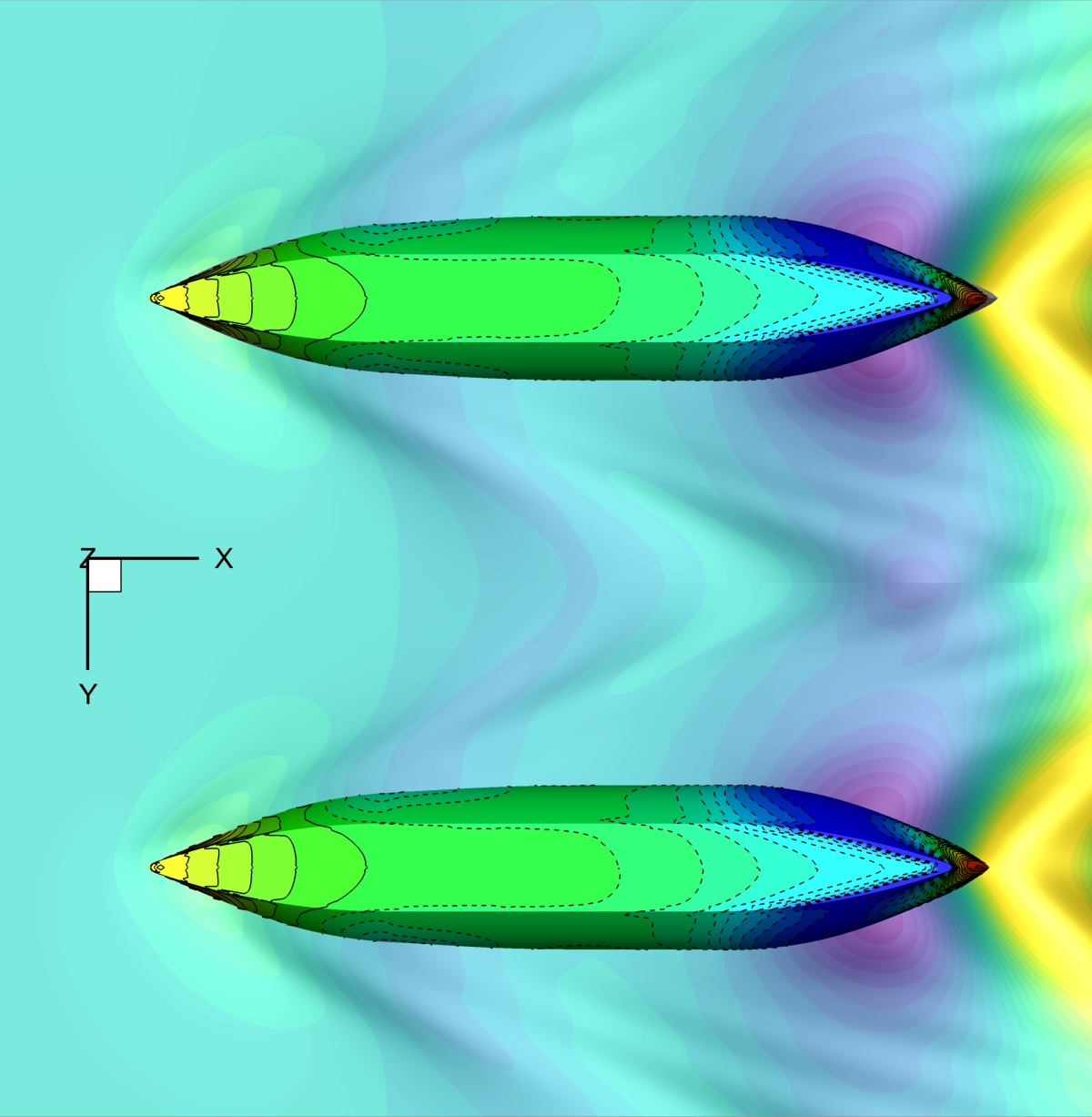}} \hfill 
    \subfloat[PF, $\nabla = 47.5$kg]{\includegraphics[width=0.275\textwidth]{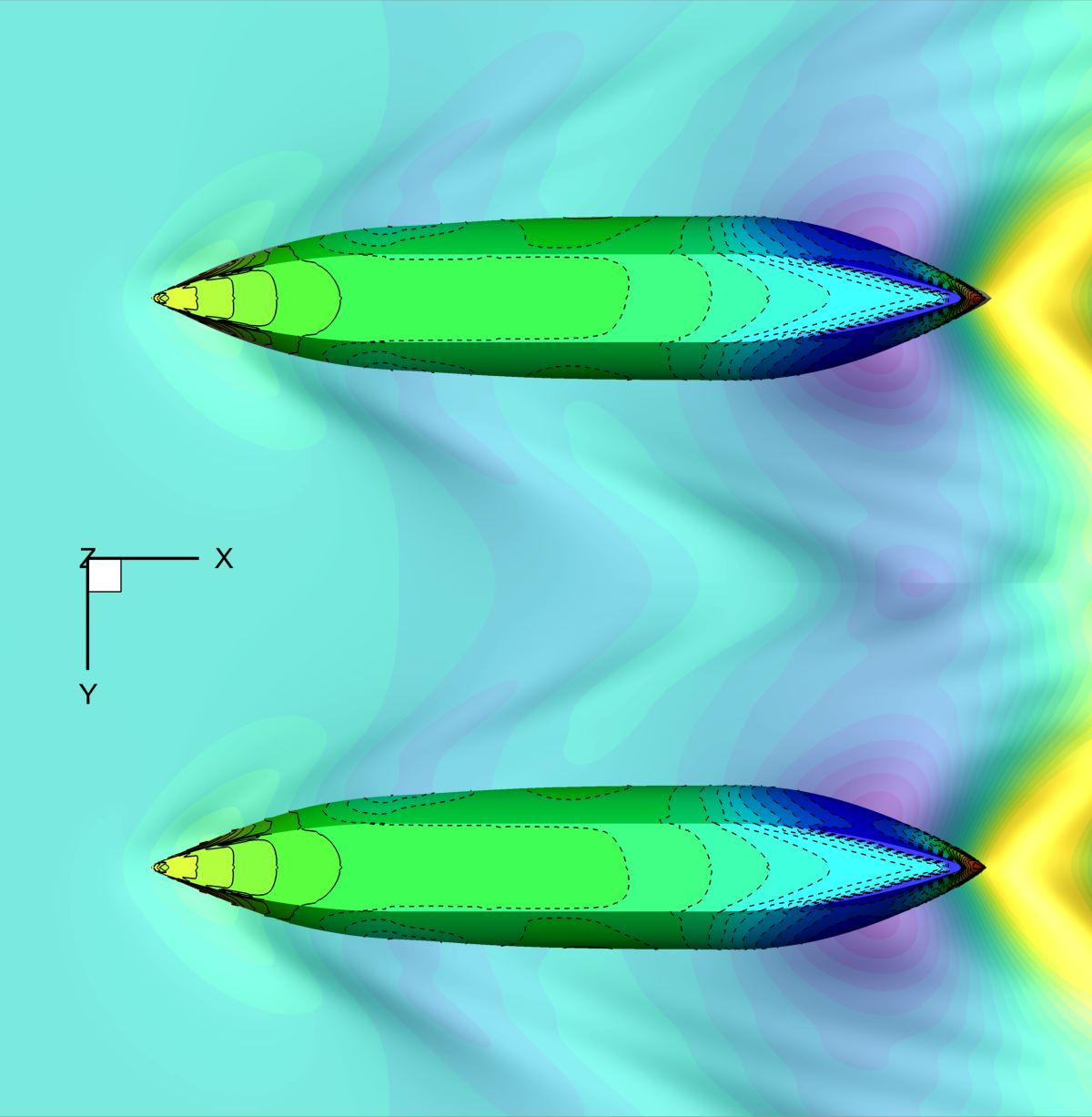}}  \hfill  
    \subfloat[PF, $\nabla = 58$kg]{\includegraphics[width=0.275\textwidth]{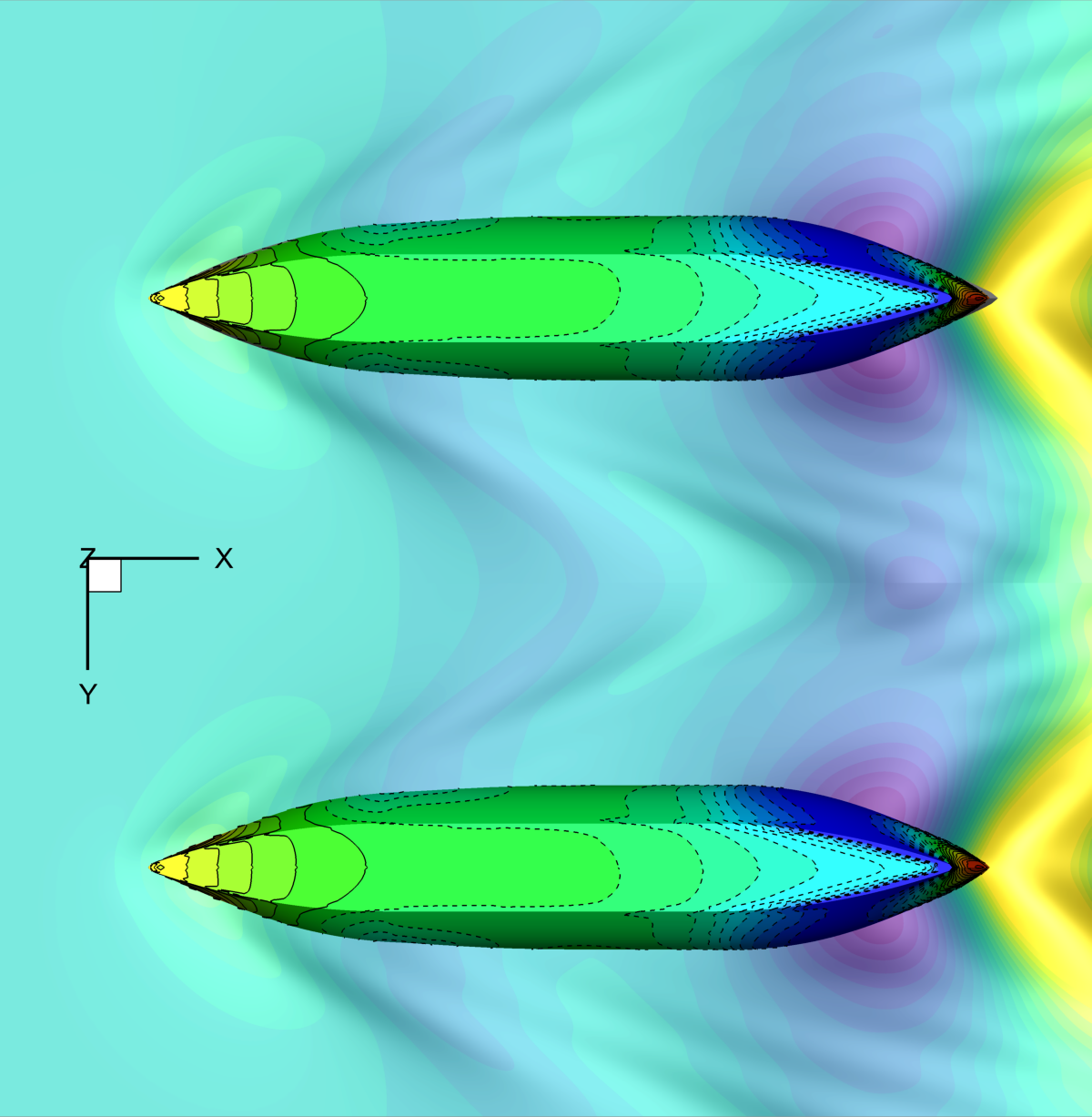}} \hfill \mbox{}  \\
    \caption{Pressure contour and wave elevation varying the SWAMP payload with $\delta x_G = 7.5\%$L, bottom view}\label{fig:FS75} 
\end{figure}

Figures \ref{fig:FS-3}-\ref{fig:FS75} show the wave elevation around the hull and a detail of the wave between the hulls along with the pressure contour on the hull. The results are shown varying the SWAMP payload and keeping $\delta x_G$ constant and equal to $-3\%$L, $0\%$L, $3\%$L, and $7.5\%$L, respectively. 

Figures \ref{fig:FS-3}a-c show that as the payload increases the RANSE solver predicts that the height of the transverse waves between the hulls increases, whereas the height of the diverging waves is almost unchanged, except for the first wave trough at the bow. 
The PF solver predicts a similar wave system (see, Figs. \ref{fig:FS-3}d-f), although the wave height is larger in comparison with the RANSE prediction, especially for the transverse waves. Also, the PF solver predicts an increase of the wave height as the payload increases, but similarly for the transverse and diverging waves.
Figures \ref{fig:FS-3}g-i show that the RANSE solver predicts a significant wave height at the bow of the SWAMP, which increases as the payload increases. The pressure contour shows that the pressure along the hull is slightly affected by the payload variations and by the proximity of the other hull, except in the rear part of the hull where the contour lines highlight differences. At the bow a "V" shaped low pressure structure is evident, likely due to the vortical structures discussed in Fig. \ref{fig:Qcrit}a-i.
The PF solver predicts a significantly smaller wave height at the bow and a smaller wave trough between the hulls (see, Figs. \ref{fig:FS-3}j-l) in comparison with the RANSE prediction. Furthermore, the pressure contour shows that the PF solver cannot predict the pressure distribution as the RANSE solver in the bow region and predicts significantly smaller proximity effects for the pressure in the rear part of the hull.

Figures \ref{fig:FS0}a-c show that as the payload increases the RANSE solver predicts that the height of the transverse waves between the hulls slightly increases, whereas the height of the diverging waves is almost unchanged. In comparison with the $\delta x_G = -3\%$L case, the height of the transverse waves is reduced for each payload.
The PF solver predicts a similar wave system (see, Figs. \ref{fig:FS0}d-f), although the wave height is larger in comparison with RANSE prediction, especially for the transverse waves. Also, the PF solver predicts an increase of the wave height as the payload increases, but similarly for the transverse and diverging waves. In comparison with the $\delta x_G = -3\%$L case, the height of the transverse waves is slightly increased for each payload (differently from the RANSE solver).
Figures \ref{fig:FS0}g-i show that the RANSE solver predicts a significant wave height at the bow of the SWAMP, which increases as the payload increases. Furthermore, the wake of the SWAMP is more evident than for the $\delta x_G = -3\%$L case at the stern of the hull, especially for the highest payload. The pressure contour shows that the pressure along the hull is slightly affected by the payload variations. The proximity of the other hull has a smaller effect than for the $\delta x_G = -3\%$L case, especially in the rear part of the hull. The "V" shaped low-pressure structure at the bow is less evident than for the $\delta x_G = -3\%$L case.
The PF solver predicts a significantly smaller wave height at the bow and a smaller wave trough between the hulls (see, Figs. \ref{fig:FS0}j-l) in comparison with RANSE prediction. The wake of the SWAMP is not modeled by the PF solver. Again, the pressure contour shows that the PF solver cannot predict the pressure distribution as the RANSE solver in the bow region and predicts negligible proximity effects for the pressure in the rear part of the hull.

Figures \ref{fig:FS3}a-c show that as the payload increases the RANSE solver predicts that the height of the transverse waves between the hulls slightly increases, whereas the height of the diverging waves is almost unchanged. The wake of the SWAMP becomes more evident as the payload increases. In comparison with the $\delta x_G = -3\%$L and $\delta x_G = 0\%$L cases, the height of the transverse waves is further reduced for each payload.
The PF solver predicts a different wave system (see, Figs. \ref{fig:FS3}d-f), where the wave height is significantly larger in comparison with the RANSE prediction for the transverse and stern diverging waves. In comparison with the $\delta x_G = -3\%$L and $\delta x_G = 0\%$L cases, the height of the transverse waves is further increased for each payload (differently from the RANSE solver). 
Figures \ref{fig:FS3}g-i show that the RANSE solver predicts a significant wave height at the bow of the SWAMP, which increases as the payload increases. Furthermore, the wake of the SWAMP is more evident than for the $\delta x_G = -3\%$L and $\delta x_G = 0\%$L cases at the stern of the hull, especially for the highest payload. The pressure contour shows that the pressure along the hull is slightly affected by the payload variations. The proximity of the other hull has a smaller effect than for the previous cases, especially in the rear part of the hull. The "V" shaped low-pressure structure is further reduced in intensity than for the previous cases.
The PF solver predicts a significantly smaller wave height at the bow and a smaller wave trough between the hulls (see, Figs. \ref{fig:FS3}j-l) in comparison with the RANSE solver. The wake of the SWAMP is not modeled but a wave trough is forming instead. Again, the pressure contour shows that the PF solver cannot predict the pressure distribution as the RANSE solver in the bow region and predicts small proximity effects for the pressure in the central part of the hull.

\begin{figure}[!h] 
    \centering
    \mbox{} \hfill
    \subfloat[Total resistance $R_T$, MF prediction with RANSE]{\includegraphics[width=0.29\textwidth]{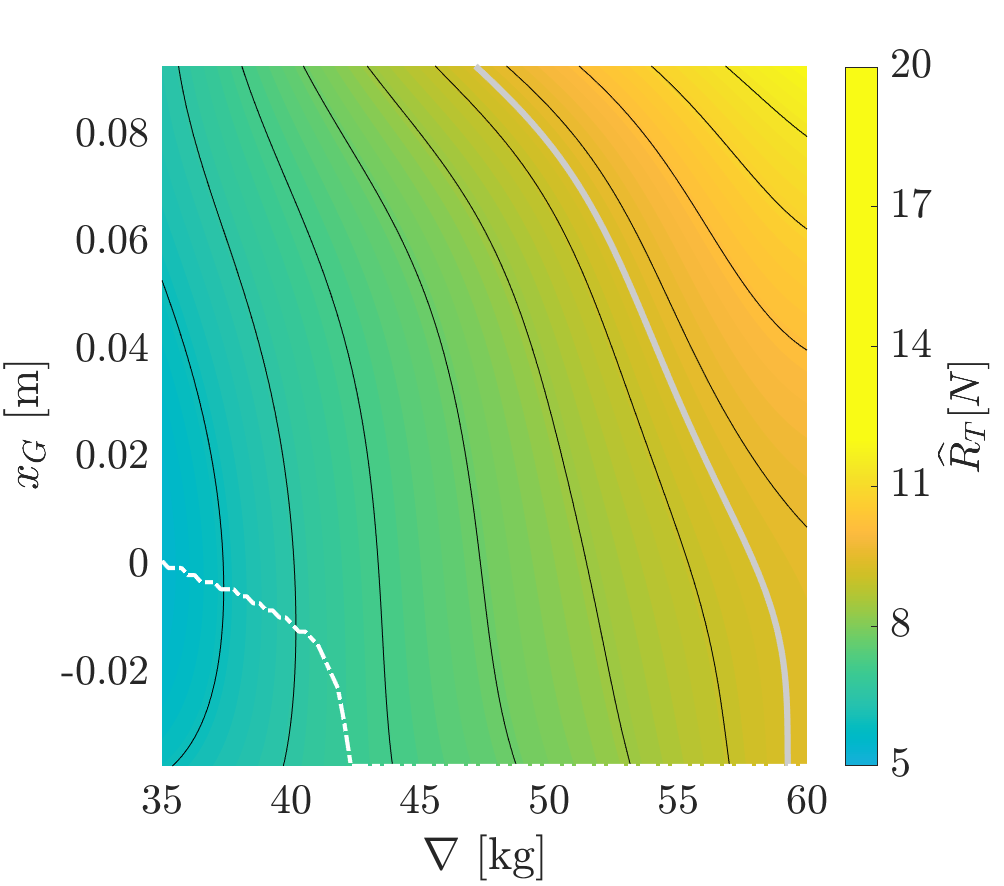}} \hfill 
    \subfloat[Total resistance $R_T$, MF prediction with PF]{\includegraphics[width=0.29\textwidth]{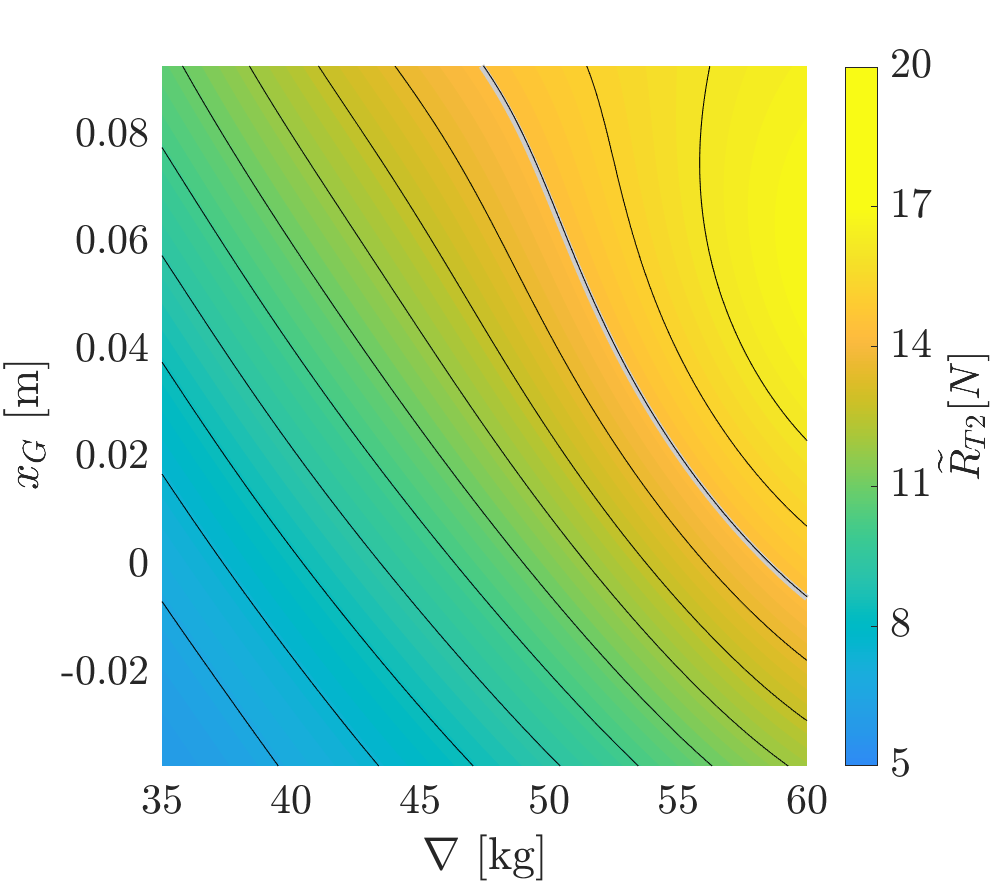}} \hfill 
    \subfloat[Total resistance $R_T$, RANSE-PF error]{\includegraphics[width=0.29\textwidth]{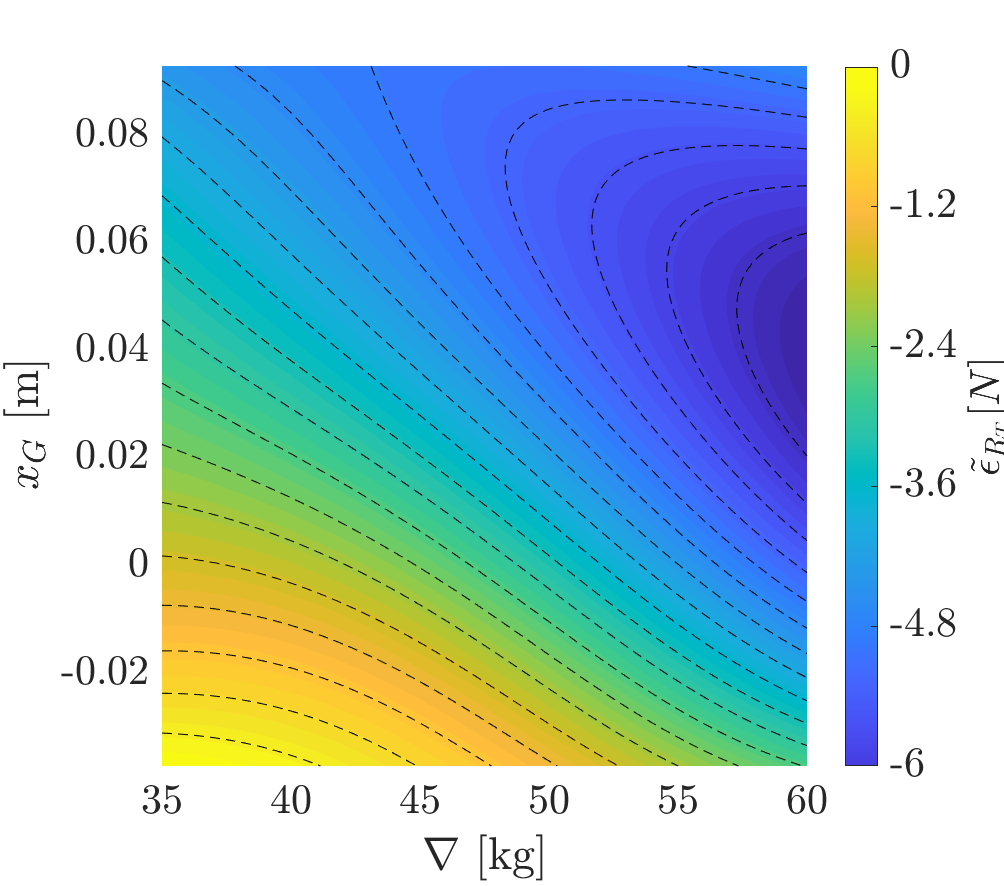}} \hfill \mbox{} \\ \vspace{-3.5mm}
    \mbox{} \hfill
    \subfloat[Sinkage $T$, MF prediction with RANSE]{\includegraphics[width=0.29\textwidth]{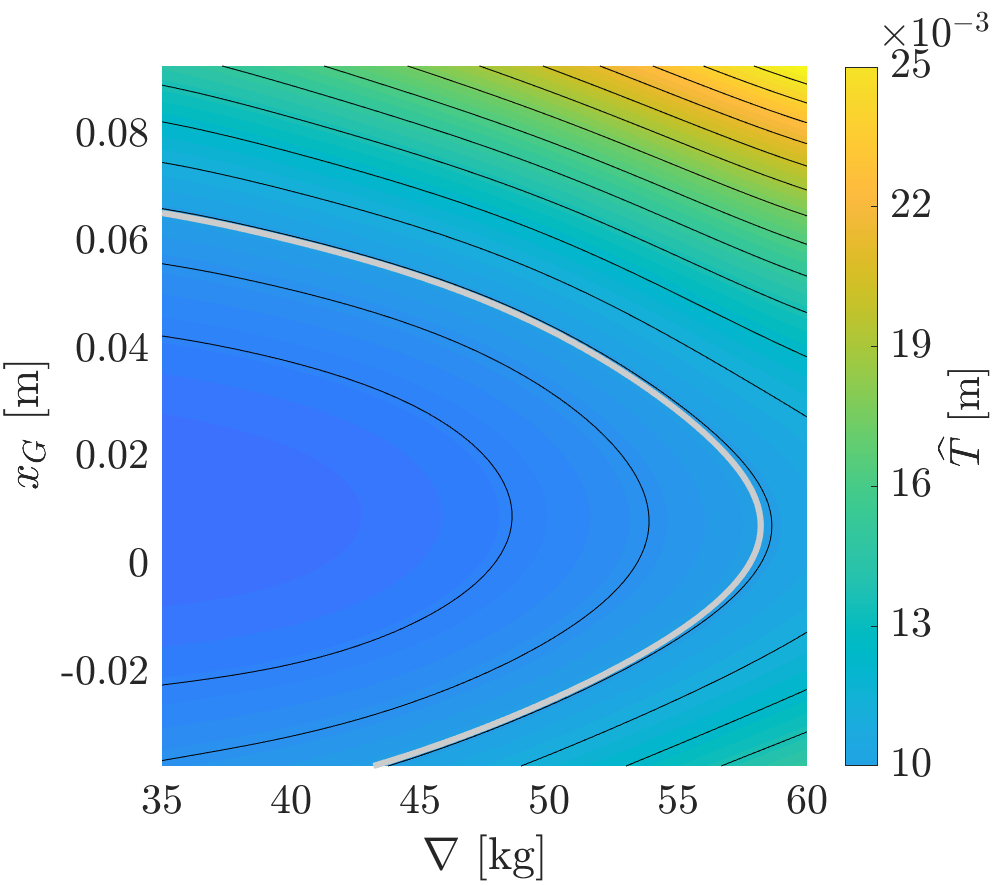}} \hfill 
    \subfloat[Sinkage $T$, MF prediction with PF]{\includegraphics[width=0.29\textwidth]{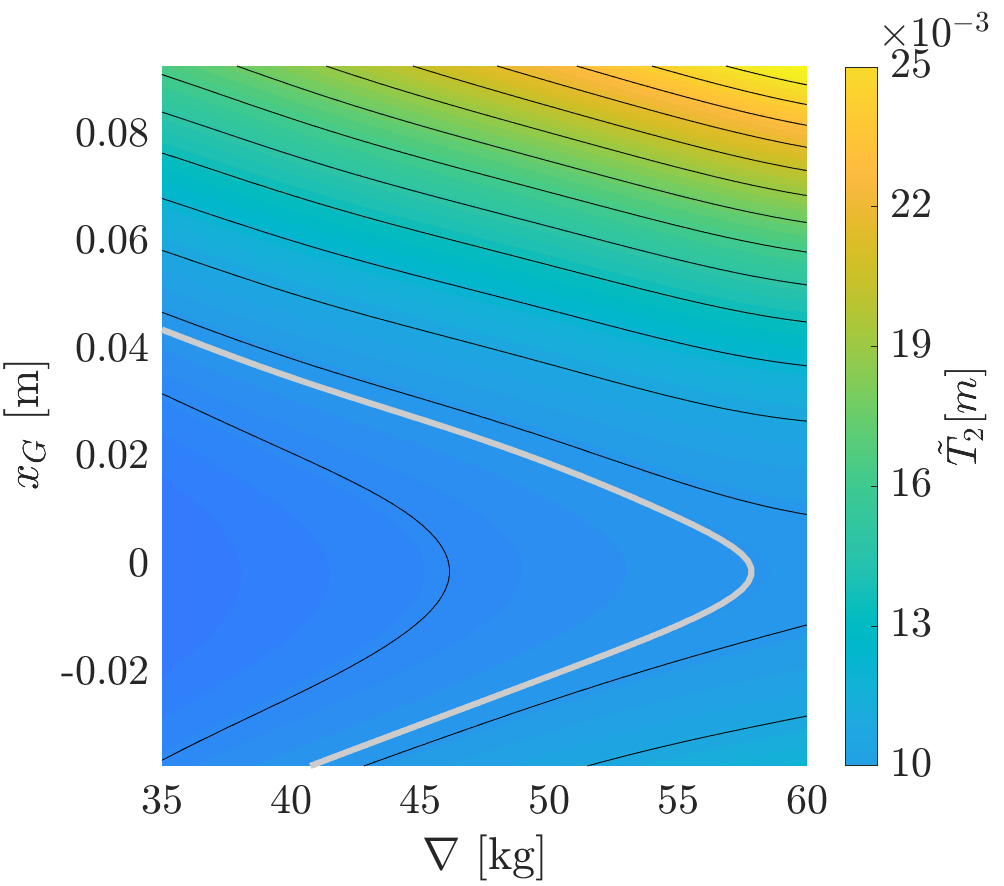}} \hfill 
    \subfloat[Sinkage $T$, RANSE-PF error]{\includegraphics[width=0.29\textwidth]{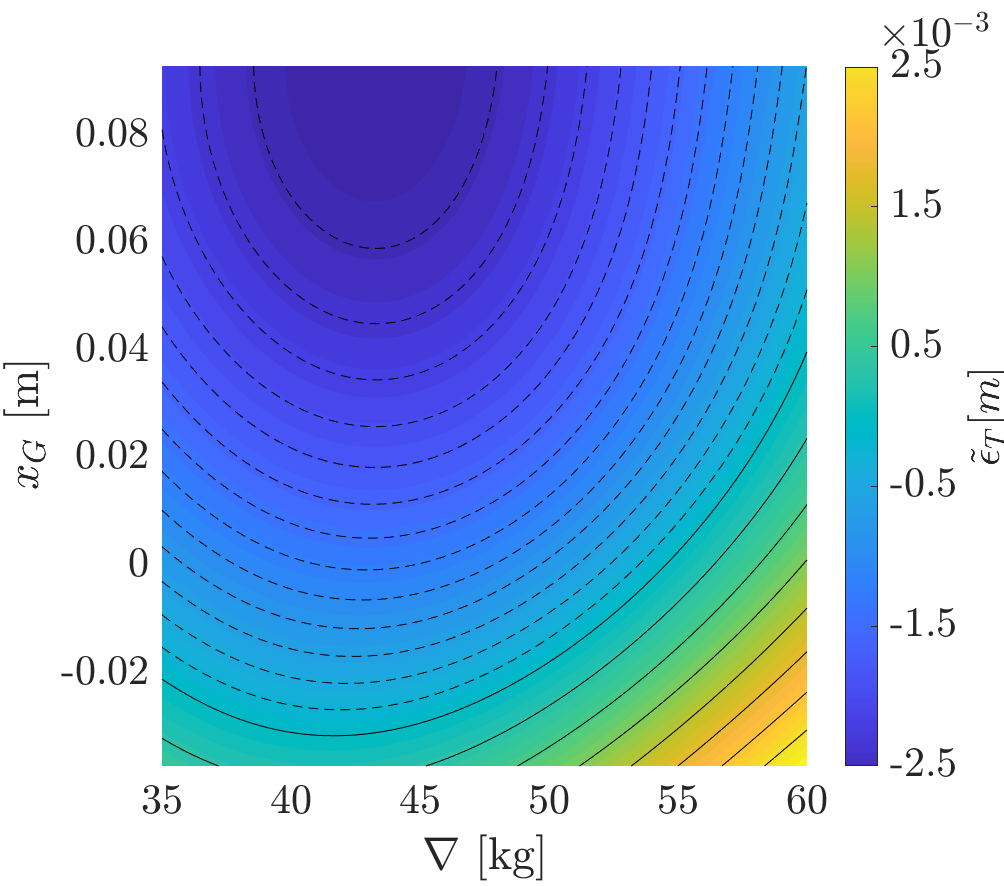}} \hfill \mbox{} \\ \vspace{-3.5mm}
    \mbox{} \hfill
    \subfloat[Trim $\theta$, MF prediction with RANSE]{\includegraphics[width=0.29\textwidth]{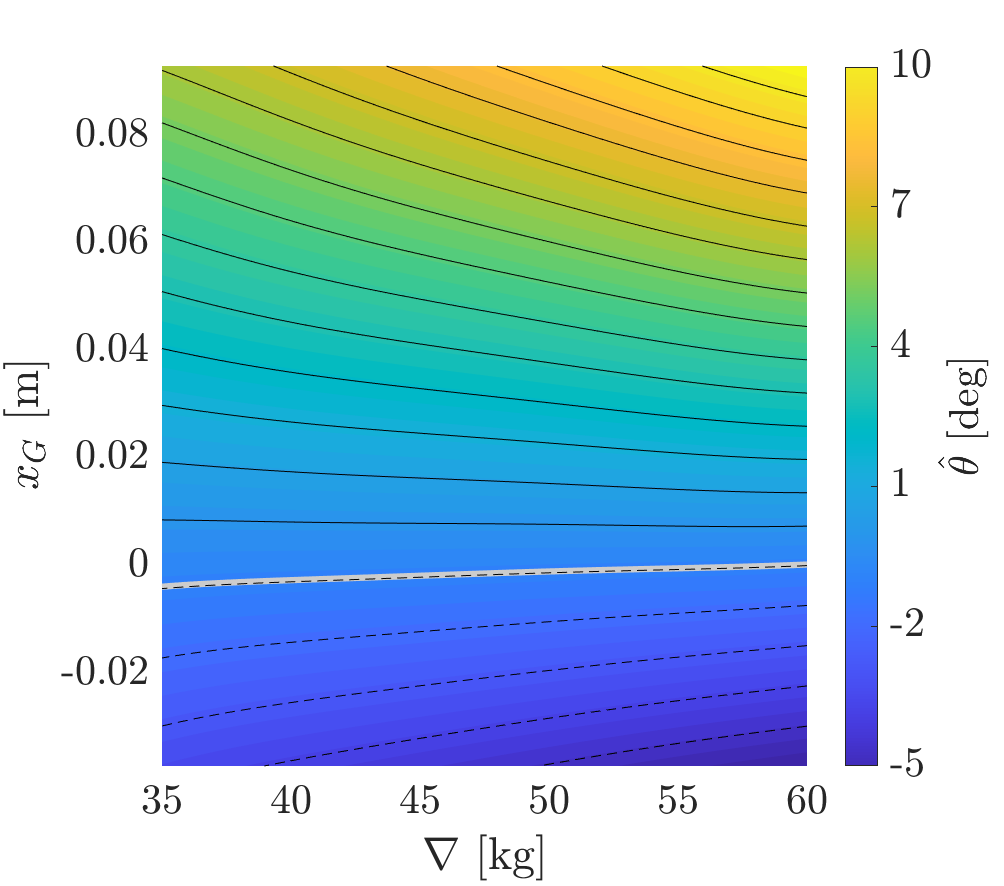}} \hfill 
    \subfloat[Trim $\theta$, MF prediction with PF]{\includegraphics[width=0.29\textwidth]{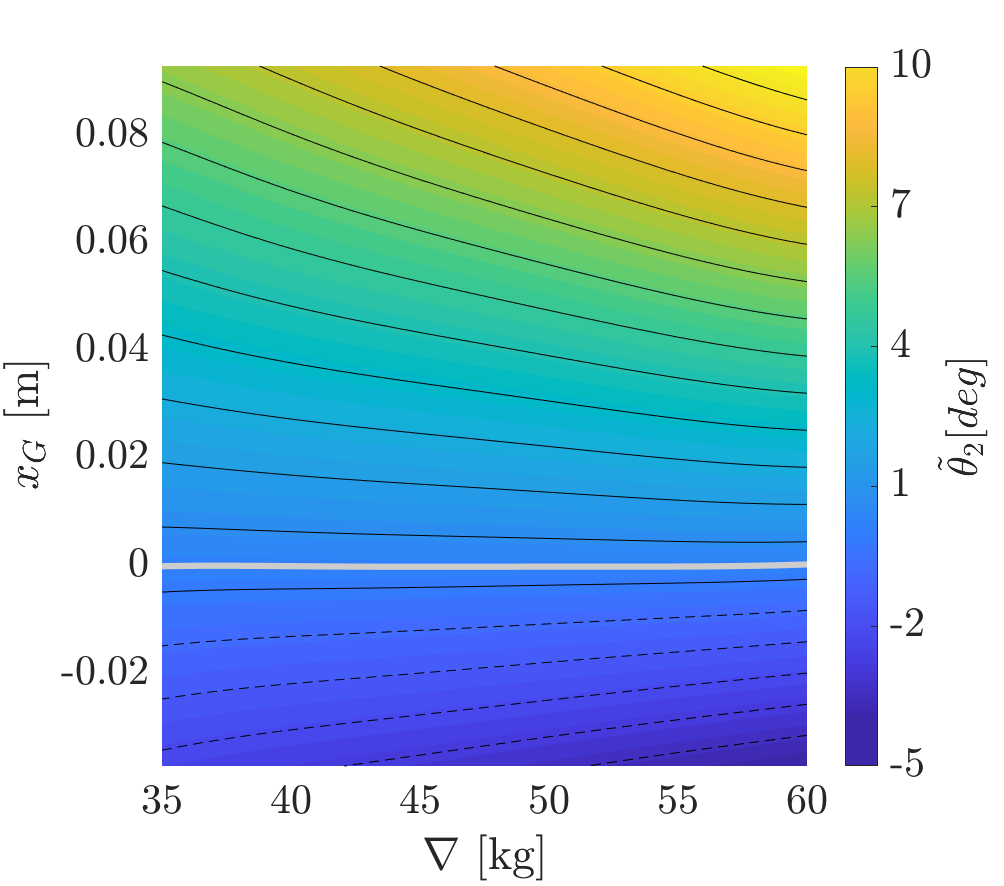}} \hfill 
    \subfloat[Trim $\theta$, RANSE-PF error]{\includegraphics[width=0.29\textwidth]{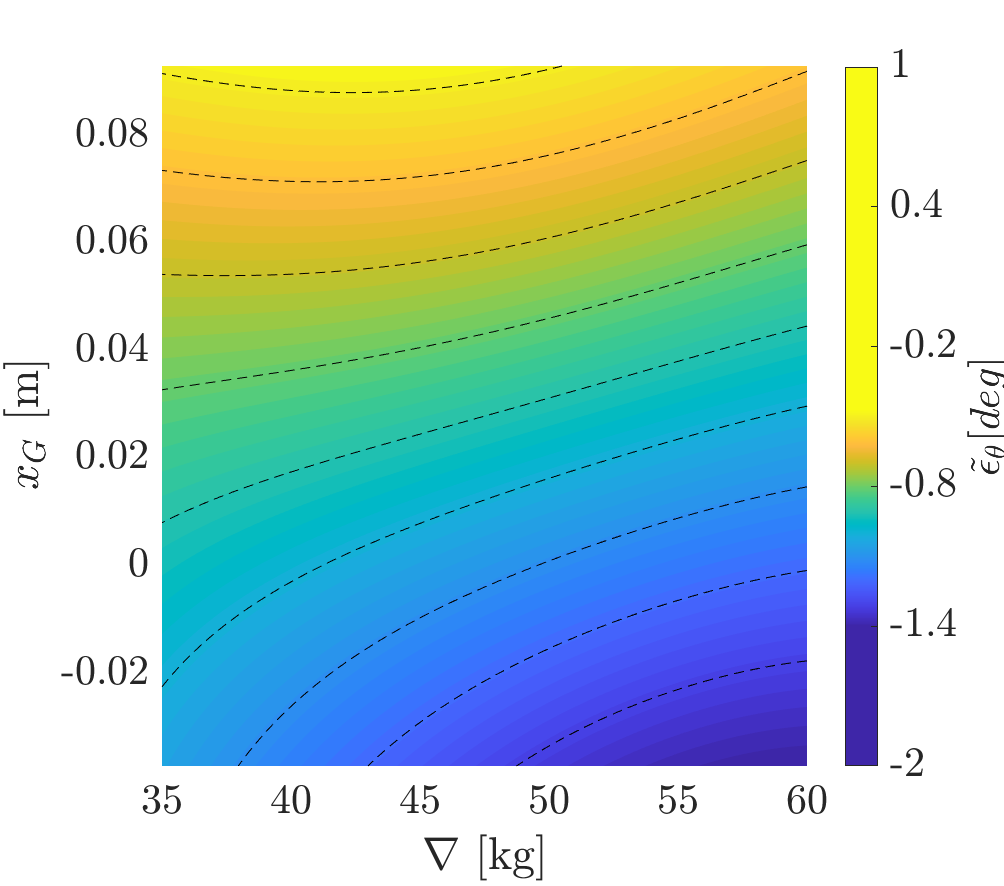}} \hfill \mbox{} \\ \vspace{-3.5mm}
    \mbox{} \hfill
    \subfloat[Free-surface RMS, MF prediction with RANSE]{\includegraphics[width=0.29\textwidth]{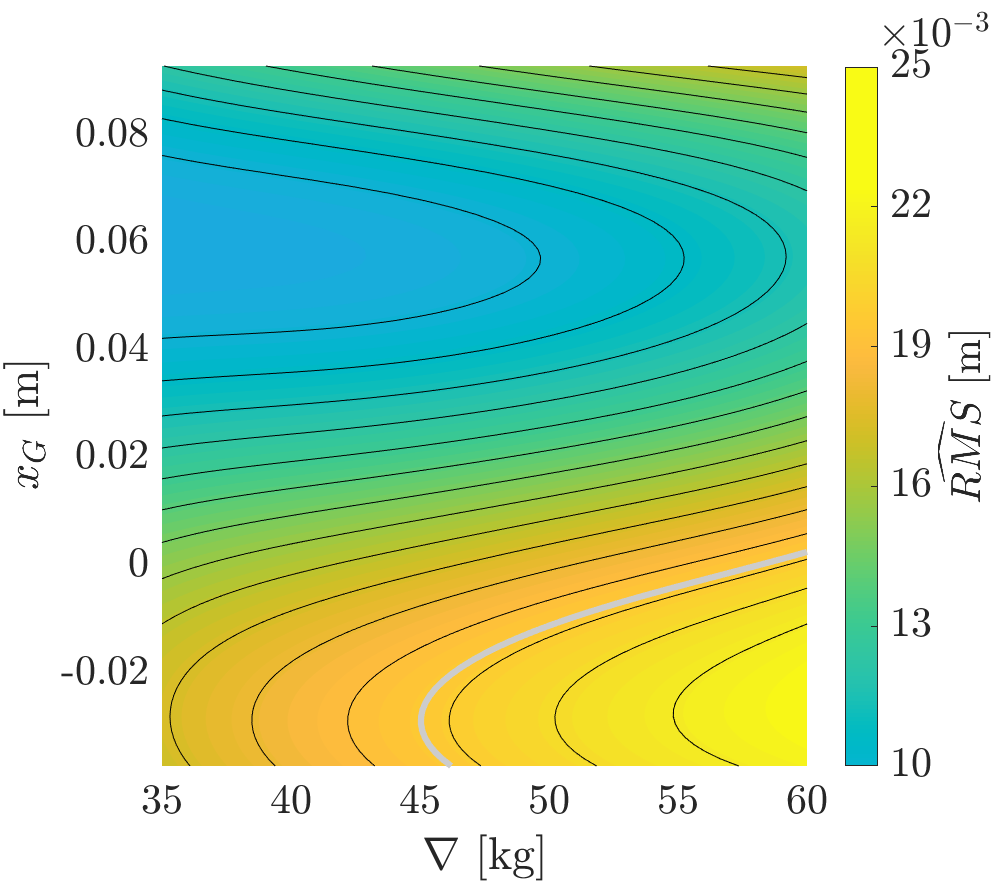}} \hfill 
    \subfloat[Free-surface RMS, MF prediction with PF]{\includegraphics[width=0.29\textwidth]{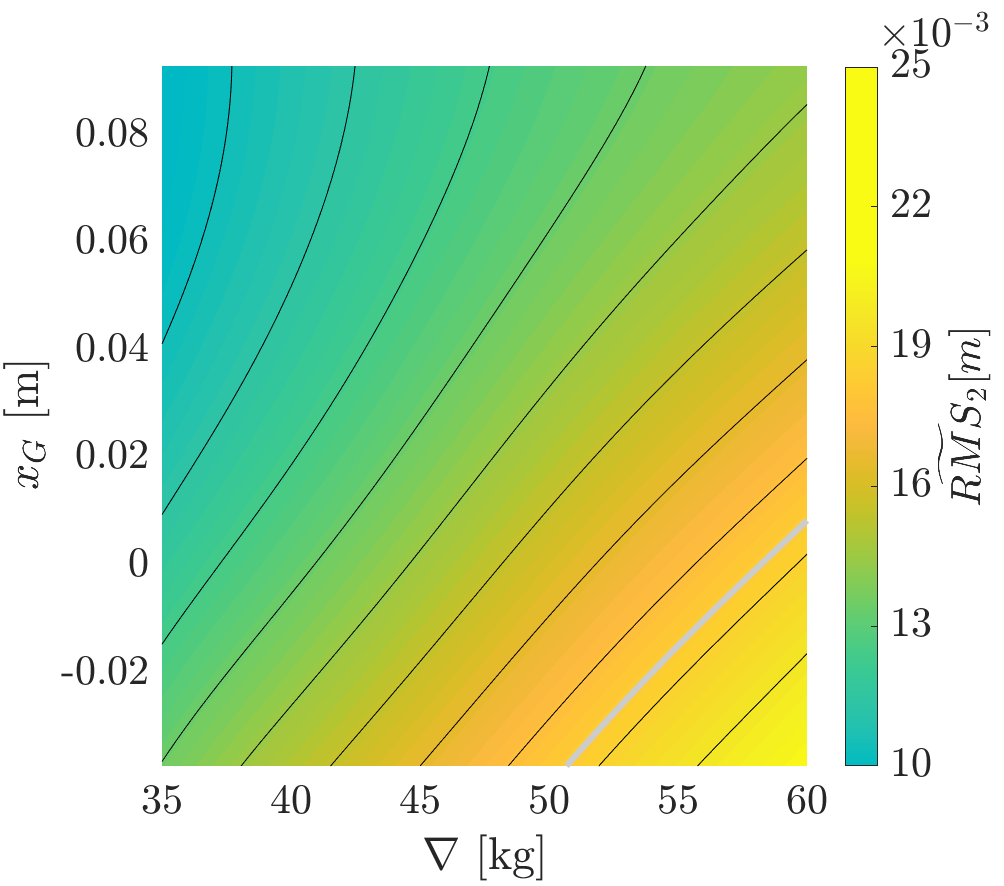}} \hfill 
    \subfloat[Free-surface RMS, RANSE-PF error]{\includegraphics[width=0.29\textwidth]{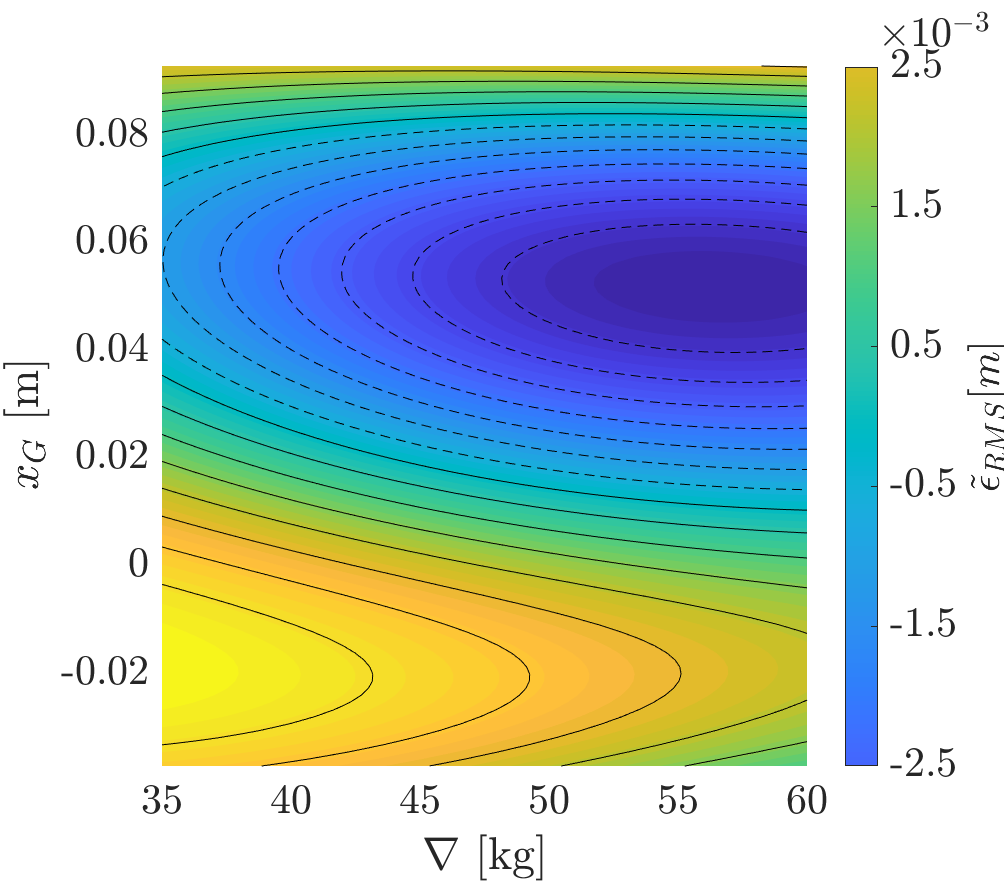}} \hfill \mbox{} \\
    \caption{MF-GP response surfaces, the gray line shows the value of the SWAMP in nominal conditions ($\nabla = 58$kg,   $\delta x_G = 0\%$L); the white line in (a) shows the locus of minimum resistance}\label{fig:RSs} 
\end{figure} 

Figures \ref{fig:FS75}a-c show that as the payload increases the RANSE solver predicts that the height of the transverse waves between the hulls slightly increases, whereas the height of the diverging waves remains unchanged. The wake of the SWAMP becomes more evident as the payload increases. In comparison with the $\delta x_G = -3\%$L, $\delta x_G = 0\%$L, and $\delta x_G = 3\%$L cases the height of the transverse waves is further reduced and the SWAMP wake is now evident for each payload.
The PF solver predicts a different wave system (see, Figs. \ref{fig:FS75}d-f), where the wave height is significantly larger in comparison with RANSE prediction for the transverse and stern diverging waves. This is consistent with the pressure overestimation of the PF solver discussed in Figs. \ref{fig:FS-3}-\ref{fig:FS75}. As for the RANSE solution, the wave height does not show significant changes varying the payload. In comparison with the $\delta x_G = -3\%$L, $\delta x_G = 0\%$L, and $\delta x_G = 3\%$L cases is further increased for each payload (differently from the RANSE solver).
Figures \ref{fig:FS75}g-i show that the RANSE solver predicts a significant wave height at the bow of the SWAMP, which increases as the payload increases. Furthermore, the wake of the SWAMP is evident at the stern of the hull. The pressure contour shows that the pressure along the hull is slightly affected by the payload variations, except at the bow where the pressure increases as the payload increases. The proximity of the other hull has a negligible effect than for the previous cases. The "V" shaped low-pressure structure is not evident anymore.
The PF solver predicts a smaller wave height at the bow and a smaller wave trough between the hulls in comparison with RANSE prediction (see, Figs. \ref{fig:FS75}j-l) in comparison with the RANSE solver. The wake of the SWAMP is not modeled but a wave trough is evident instead. The pressure contours are now similar between the PF and RANSE solutions.

\section{Analysis of the SWAMP performance}

The active learning procedure is initialized with the performance evaluated for the simulations discussed above (plus considering the results from the G3 and G4 grids), as shown in Fig. \ref{fig:FinalTS}. The acquisition function aims at reducing the maximum uncertainty of the resistance MF surrogate model.
The computational budget for the active learning approach is fixed equal to 72h of wall clock time, equivalent to one RANSE simulation. 


Figure \ref{fig:FinalTS} shows the final training set for the MF surrogate models. Most of the PF simulations are focused on the boundaries of the domain. The samples distribution within the domain is reasonably regular, indicating that the MF surrogate model of the resistance is smooth.

Figure \ref{fig:RSs} shows the MF response surface considering RANSE and PF simulations, the MF response surface considering PF solutions only, and their difference (error). For each MF response surface, a gray line is plotted, highlighting the performance of the SWAMP in nominal operating conditions. Furthermore, Fig. \ref{fig:RSs}a shows a white dashed line representing the locus of minimum resistance for each payload value. It is worth noting that for $42 \leq \nabla \leq 58$kg the minimum resistance is achieved for $\delta x_G = -3\%$L, whereas for $\nabla \leq 42$kg the optimal value of $\delta x_G$ goes toward zero. The resistance variation is mostly affected from the payload and from $\delta x_G$ only for larger values.
The sinkage is mostly affected by the payload. Differently, the trim value is mostly affected by $\delta x_G$ and is almost insensitive to the payload.
Finally, the RMS is slightly affected from the payload and mostly from $\delta x_G$. It is worth noting that the region of the minimum RMS lies in the opposite part of the domain than the region of the minimum for the resistance. 
Overall, the MF response surface using only PF solutions approximate reasonably well the sinkage and trim, predicting a change in the pitch angle for a different value of $\delta x_G$ than when also the RANSE solutions are considered (as also evident from Fig. \ref{fig:PSide0}). Greater differences (error) are evident in the resistance variation response surface, predicting the minimum resistance at each payload always for $\delta x_G = 0\%$L. Finally, the greatest error occurs for the RMS response surface, where MF response surface trained only with PF solutions provides a completely different result. This is consistent with the discussion provided above for Figs. \ref{fig:FS-3}-\ref{fig:FS75}, where the PF solutions of the free-surface failed in predicting the SWAMP wake, predicting a smaller wave trough after the bow, and did not correctly predict the wave elevation at the bow.
%


%

\section{Conclusions}

The multi-fidelity hydrodynamic analysis of an autonomous surface vehicle (ASV) designed to perform surveys in remote or difficult to access areas (such as polar regions, rivers, and alpine lakes) and harsh and dangerous environments is presented. 

The SWAMP (Shallow Water Autonomous Multipurpose Platform) is a fully electric catamaran, modular, portable, lightweight, and highly controllable ASV. It can serve for various missions, such as bathymetric analysis, water sampling, and physical and chemical data collection. Therefore, the SWAMP payload (the instrumentation required by the survey) can vary depending on the mission and furthermore the location of the center of mass can vary depending on the payload arrangement.
The considered environmental conditions are calm and deep water. Calm water is considered representative of the real-world operating conditions since the SWAMP is not currently intended to operate with a significant wave height, to guarantee the safety of the equipment and accuracy of the measurements. The deep sea conditions are of interest because, although the SWAMP is developed with specific features to allow survey in shallow water (such as the flat bottom) it is often if not mostly used in deep water (such as during coastal or harbor surveys). 
The analysis was performed evaluating the response surface of the hydrodynamic resistance (which affects the SWAMP autonomy), and the sinkage, trim, and root mean square (RMS) of the free-surface elevation between SWAMP hulls (which affect the accuracy of the measurements), subject to two operational parameters, namely the payload and the location of the center of mass. 
The SWAMP performance was evaluated using a Raynolds-Averaged Navier Stokes Equations (RANSE) and a potential flow (PF) solver, the latter leveraging three grids with different refinements with different accuracy in coupling the hydrodynamic loads with the rigid body equations of motion. 
The solutions provided by the different solvers were combined by a multi-fidelity (MF) surrogate model based on a Gaussian process. After an initial sampling of the operational space the MF surrogate model was refined using an active learning approach aiming to minimize the maximum prediction uncertainty of the MF surrogate model. 

The hydrodynamic analysis focused on identifying the effects of varying the payload and the location of the center of mass to the pressure distribution, wave system, and vortex generation. These quantities affect the SWAMP resistance and its attitude, and the RMS of the free-surface elevation between the catamaran hulls. 
The RANSE solutions were compared with the PF solution, highlighting the limitations of the latter in predicting the correct wave elevation and pressure distribution. Specifically, the PF solver overestimates the wave height, especially when the center of mass is far from the mid ship. The PF solver cannot identify recirculation zones and flow separation, yielding a wrong prediction of the pressure at the stern and in general an overestimate of the pressure values. Furthermore, the PF solver has poorly predicted the pressure at the keel of the SWAMP, where the pump-jets are mounted. This yields that the results provided by the PF solver cannot be used to investigate the effect of the pressure distribution on the hull to the pump-jet performance.
The low-accuracy of the PF solver in predicting the free-surface between the SWAMP hulls yields that such solver should be carefully used when designing the SWAMP operating conditions for a given survey, where the disturbance provided by the free-surface must be considered.
Finally, the PF solver cannot provide information about the vortical structures whereas the RANSE can. This is a significant difference that should be considered when choosing the solver for performing hydrodynamic analysis. 

The MF approach was successful in providing response surfaces with a limited computational cost, effectively exploring the operational space and combining the results provided by the PF solver with different accuracy and the solvers with different physical models (RANSE and PF). 
The response surfaces highlighted that for reduced payloads the location of the center of mass can be calibrated to minimize the resistance. Furthermore, for any payload the location of the center of mass can be tuned to either reduce the resistance or minimize the wave elevation between the hulls, depending on the survey requirements.

Future work will address the extension of the MF hydrodynamic analysis to variable speed, in order to provide detailed information about the best operating conditions for different surveys.

\section*{Acknowledgements}
The authors are particularly grateful to their colleagues Ga. Bruzzone, Gi. Bruzzone, M. Bibuli, M. Giacopelli, and E. Spirandelli, whose contribution has been fundamental for the development of the SWAMP ASV.

CNR-INM is grateful to the Italian Ministry of University and Research for the support under the Italian National Research Program PON "Ricerca e Innovazione" 2014-2020, Action II.2, Specialization area: "Blue Growth", Project n.ARS01\_00682 ARES-Autonomous Robotics for the Extended Ship.

The Multidisciplinary Optimization group at CNR-INM is grateful to Professor Frederick Stern from the University of Iowa for providing the URANSE code CFDShip-Iowa.

\section*{Conflict of interest}

The authors report there are no competing interests to declare.

\bibliographystyle{unsrt}  
\bibliography{biblio}

\end{document}